\documentclass[11pt, a4paper, oneside]{Thesis} % Paper size, default font size
\graphicspath{{Pictures/}} % Specifies the directory where pictures are stored

\usepackage[square,sort,compress,numbers]{natbib}
\usepackage{cite}
\usepackage{mathtools}
\renewcommand{\thefootnote}{\arabic{footnote}}
\usepackage{pifont}
\usepackage{cancel}
\usepackage{perpage}
\usepackage{amsmath,latexsym} 
\usepackage{subfigure}
\usepackage{float}
\usepackage{mathpazo}
\usepackage{amsfonts}
\usepackage{bigints} % used for increasing the size of the integration symbol
\usepackage{mathptmx}
\usepackage{bm}
\usepackage{palatino}
\usepackage{textcomp}
\usepackage{booktabs}
\usepackage{dcolumn}
\usepackage{ragged2e}
\usepackage{hyperref}
\newcommand{\gc}[1]{\widehat{#1}}%metric affine connection
\title{\ttitle} % Defines the thesis title - don't touch this
\DeclareUnicodeCharacter{2212}{-}
\DeclareUnicodeCharacter{2212}{+}
\newcommand\blfootnote[1]{%
     \begingroup
     \renewcommand\thefootnote{}\footnote{#1}%
     \addtocounter{footnote}{-1}%
      \endgroup
    }
\begin{document}

\setstretch{1.3} % Line spacing of 1.3

% Define page headers using FancyHdr package and set up for one-sided printing
\fancyhead{} % Clears all page headers and footers
\rhead{\thepage} % Sets the right side header to show the page number
\lhead{} % Clears the left side page header

% Input all the variables used in the document. Please fill out the
% variables.tex file with all your details.
%-------------------------------------------------------------------------------
%	DOCUMENT VARIABLES
%
%	Fill in the lines below to set the various variables for the document
%-------------------------------------------------------------------------------

%-------------------------------------------------------------------------------
% Your thesis title - this is used in the title and abstract
% Command: \ttitle
\thesistitle{Study of Wormholes in Symmetric Teleparallel Theories of Gravity}
%-------------------------------------------------------------------------------
% The document type: Thesis / report, etc.
% Command: \doctype
\documenttype{\textbf{THESIS}}
%-------------------------------------------------------------------------------
% Your supervisor's name - this is used in the title page
% Command: \supname
\supervisor{\textbf{Prof. Pradyumn Kumar Sahoo}}
%-------------------------------------------------------------------------------
% The supervisor's position - Used on Certificate
% Command: \suppos
\supervisorposition{\textbf{Professor}}
%-------------------------------------------------------------------------------
% Supervisor's institute
% Command: \supinst
\supervisorinstitute{\textbf{BITS-Pilani, Hyderabad Campus}}
%-------------------------------------------------------------------------------
% Your Co-Supervisor's name
% Command: \cosupname
%\cosupervisor{Dr. Pradyumn Kumar Sahoo}
%-------------------------------------------------------------------------------
% Co-Supervisor's Position - Used on Certificate
% Command: \cosuppos
%\cosupervisorposition{Associate Professor}
%-------------------------------------------------------------------------------
% Co-Supervisor's Institute
% Command: \cosupinst
%\cosupervisorinstitute{BITS-Pilani, Hyderabad Campus}
%-------------------------------------------------------------------------------
% Your Examiner's name. Not currently used anywhere.
% Command: \examname
\examiner{}
%-------------------------------------------------------------------------------
% Name of your degree
% Command: \degreename
\degree{Ph.D. Research Scholar}
%-------------------------------------------------------------------------------
% The BITS Course Code for which this report is written
% COmmand: \ccode
\coursecode{\textbf{DOCTOR OF PHILOSOPHY}}
%\coursecode{BITS C799T}
%-------------------------------------------------------------------------------
% The name of the Course
% Command: \cname
\coursename{Thesis}
%-------------------------------------------------------------------------------
% Your name. Extend manually in case of multiple authors
% Command: \authornames
\authors{\textbf{MORESHWAR JAGADEORAO TAYDE}}
%-------------------------------------------------------------------------------
% Your ID Number - used on the Title page and abstract
% Command: \idnum
\IDNumber{2021PHXF0065H}
%-------------------------------------------------------------------------------
% Your address
% Command: \addressnames
\addresses{}
%-------------------------------------------------------------------------------
% Your subject area
% Command: \subjectname
\subject{}
%-------------------------------------------------------------------------------
% Keywords for this report.
% Command: \keywordnames
%-------------------------------------------------------------------------------
% University details
% Command: \univname
\university{\texorpdfstring{\href{http://www.bits-pilani.ac.in/} % URL
                {Birla Institute of Technology and Science, Pilani}} % University name
                {Birla Institute of Technology and Science, Pilani}}
%-------------------------------------------------------------------------------
% University details, in Capitals
% Command: \UNIVNAME
\UNIVERSITY{\texorpdfstring{\href{http://www.bits-pilani.ac.in/} % URL
                {BIRLA INSTITUTE OF TECHNOLOGY AND SCIENCE, PILANI}} % name in capitals
                {BIRLA INSTITUTE OF TECHNOLOGY AND SCIENCE, PILANI}}

%-------------------------------------------------------------------------------
% Campus Name
% Command: \campusname
%\campus{Hyderabad Campus}

%-------------------------------------------------------------------------------
% Campus Name, in capitals
% Command: \CAMPUSNAME
%\CAMPUS{HYDERABAD CAMPUS}

%-------------------------------------------------------------------------------
% Department Details
% Command: \deptname
\department{\texorpdfstring{\href{http://www.bits-pilani.ac.in/pilani/Mathematics/Mathematics} % Your department's URL
                {Mathematics}} % Your department's name
                {Mathematics}}
%-------------------------------------------------------------------------------
% Department details, in Capitals
% Command: \DEPTNAME
\DEPARTMENT{\texorpdfstring{\href{http://www.bits-pilani.ac.in/pilani/Mathematics/Mathematics} % Your department's URL
                {Mathematics}} % Your department's name in capitals
                {Mathematics}}
%-------------------------------------------------------------------------------
% Research Group Details
% Command: \groupname
\group{\texorpdfstring{\href{Research Group Web Site URL Here (include http://)}
                {Research Group Name}} % Your research group's name
                {Research Group Name}}
%-------------------------------------------------------------------------------
% Research Group Details, in Capitals
% Command: \GROUPNAME
\GROUP{\texorpdfstring{\href{Research Group Web Site URL Here (include http://)}
                {RESEARCH GROUP NAME (IN BLOCK CAPITALS)}}
                {RESEARCH GROUP NAME (IN BLOCK CAPITALS)}}
%-------------------------------------------------------------------------------
% Faculty details
% Command: \facname
\faculty{\texorpdfstring{\href{Faculty Web Site URL Here (include http://)}
                {Faculty Name}}
                {Faculty Name}}
%-------------------------------------------------------------------------------
% Faculty details, in Capitals
% Command: \FACNAME
\FACULTY{\texorpdfstring{\href{Faculty Web Site URL Here (include http://)}
                {FACULTY NAME (IN BLOCK CAPITALS)}}
                {FACULTY NAME (IN BLOCK CAPITALS)}}
%-------------------------------------------------------------------------------

%-------------------------------------------------------------------------------
%   NON-CONTENT PAGES
%-------------------------------------------------------------------------------
\maketitle

%-------------------------------------------------------------------------------
%	DEDICATION
%-------------------------------------------------------------------------------
\clearpage
\setstretch{1.3} % Return the line spacing back to 1.3

\pagestyle{empty} % Page style needs to be empty for this page
% Dedication text
\pagenumbering{gobble}
% \Dedicatory{\bf \begin{LARGE}
% Dedicated to
% \end{LARGE} 
% \\
% \vspace{3cm}
%  My Family \\}
% \vspace{1cm}
% \begin{huge}
%\emph{\textit{......}}
% \end{huge}}

\addtocontents{toc}{\vspace{2em}} % Add a gap in the Contents, for aesthetics
%\Declaration
\frontmatter % Use roman numbering style (i, ii...) for the pre-content pages
%\pagenumbering{roman}
\Certificate
%\Quotation{Insert Random Quote here. Publish like a boss.}{Your Name}
\Declaration
\begin{acknowledgements}
I would like to express my heartfelt gratitude to my supervisor, \textbf{Prof. Pradyumn Kumar Sahoo}, Professor in the Department of Mathematics, BITS-Pilani, Hyderabad Campus, for his invaluable guidance, unwavering support, and profound expertise throughout the course of my Ph.D. research. His patience, encouragement, and trust in my abilities motivated me to deepen my understanding of the subject and develop a strong aptitude for scientific inquiry, enabling me to successfully complete this thesis.

I extend my sincere thanks to the members of my Doctoral Advisory Committee (DAC), \textbf{Prof. Bivudutta Mishra} and \textbf{Prof. Sashideep Gutti}, for their insightful feedback, constructive suggestions, and continuous encouragement, which greatly improved the quality of my research work.

I am deeply grateful to the Head of the Department, the DRC convener, faculty members, my colleagues, and the staff of the Department of Mathematics for their constant support and contribution to making this Ph.D. journey memorable and enriching.

I would like to acknowledge \textbf{BITS-Pilani, Hyderabad Campus}, for providing the necessary resources and facilities and the \textbf{University Grants Commission (UGC)}, for awarding me the \textbf{National Fellowship for Scheduled Caste Students (NFSC)} (UGC Ref. No. 201610123801), which enabled me to carry out my research.

I am deeply grateful to my co-authors, especially \textbf{Dr. Zinnat Hassan} and \textbf{Sayantan Ghosh}, whose insightful contributions, collaborative spirit, and constant encouragement were instrumental in shaping this research. I also extend my sincere thanks to my colleagues, particularly \textbf{Dr. Raja Solanki} and \textbf{Dr. Gaurav Gadbail}, for their support and valuable input.

I express my deepest appreciation to my friends \textbf{Akankshya Sahu, Kartik Tathe, Debismita Nayak}, for their unwavering support and encouragement. Also, I would like to give special thanks to \textbf{Prof. Rahul Mapari} for encouraging me to pursue this path.

Finally, I express my sincere gratitude to my family: my parents, my brother \textbf{Akash}, my sisters, and my sweet little niece and nephew, for their unwavering love, support, and encouragement, which have been my constant source of strength throughout this journey.

\vspace{1.4 cm}
Moreshwar Jagadeorao Tayde,\\
ID: 2021PHXF0065H.
\end{acknowledgements}
\begin{abstract}
This thesis provides a detailed investigation of traversable wormhole (WH) solutions within the framework of symmetric teleparallel gravity and its extension, such as $f(Q)$ and $f(Q, T)$ gravity. To set the stage for the investigation, the fundamental concepts and the necessary background are introduced in Chapter \ref{Chapter1}. This chapter delves into the basics of WH geometry, exploring its essential features and properties. Additionally, it provides a comprehensive overview of general relativity. Furthermore, key concepts related to modified theories of gravity are also discussed, offering a broader perspective on theoretical advancements and their relevance to WH studies. \\
Chapter \ref{Chapter2} investigates WHs supported by dark matter models like Pseudo-Isothermal and Navarro-Frenk-White profiles and the role of monopole charge in $f(Q)$ gravity. Linear $f(Q)$ models are studied with various redshift and shape functions, while for nonlinear models, we use the embedding class-I method. The results show that the shape functions satisfy the flare-out conditions, and the monopole parameter $\eta$ plays a key role in violating the Null Energy Condition (NEC). Furthermore, by using the Volume Integral Quantifier method, minimal exotic matter is required for the linear model, but nonlinear models such as $f(Q) = Q + m Q^n$ are incompatible with WH solutions.
Chapter \ref{Chapter3} explores the analysis in the extension of $f(Q)$ gravity, ie $f(Q, T)$ gravity, where the Lagrangian depends on the non-metricity scalar $Q$ and the trace of the energy-momentum tensor $T$. WH solutions are derived under linear barotropic and anisotropic Equation of State (EoS) cases. The linear and nonlinear forms of $f(Q, T)$, including $f(Q, T) = \alpha Q + \beta T$ and $f(Q, T) = Q + \lambda_1 Q^2 + \eta_1 T$, are analyzed. NEC violations are observed in the linear models, while stability analysis using the Tolman-Oppenheimer-Volkoff (TOV) equation highlights the equilibrium of WH configurations under different conditions.\\
Chapter \ref{Chapter4} employs the MIT bag model to further explore WH solutions in $f(Q, T)$ gravity. Specific shape functions are assumed to derive equilibrium and energy conditions, demonstrating that the NEC is violated at the WH throat. The stability of these WH configurations is confirmed through the TOV equation, which offers insights into their physical viability.
Chapter \ref{Chapter5} investigates spherically symmetric static WH solutions in the framework of $f(Q, T)$ gravity under noncommutative geometries inspired by string theory, utilizing Gaussian and Lorentzian distributions. Both analytic and numerical solutions are derived for linear and nonlinear $f(Q, T)$ models, respectively. The flare-out conditions are satisfied for the derived shape functions, and the NEC is consistently violated at the WH throat under noncommutative geometries. Additionally, gravitational lensing properties are explored for a specific WH model. Furthermore, it is followed by concluding remarks in Chapter \ref{Chapter6}.
%This comprehensive study advances our understanding of traversable wormholes in modified gravity frameworks, highlighting the interplay between exotic matter, stability, and the astrophysical implications of $f(Q)$ and $f(Q, T)$ gravity theories. The findings provide a foundation for future theoretical and observational investigations into the nature of wormholes and their role in cosmology.
\end{abstract} 

\Dedicatory{\bf \begin{LARGE}
Dedicated to
\end{LARGE} 
\\
\vspace{0.2cm}
\it My Family\\}

% \Dedicatory{\bf \begin{LARGE}
% Dedicated to
% \end{LARGE} 
% \\
% \vspace{0.2cm}
% \it My Parents\\}

%\chapter{Preface}

%-------------------------------------------------------------------------------
%	LIST OF CONTENTS/FIGURES/TABLES PAGES
%-------------------------------------------------------------------------------

\lhead{\emph{Contents}} % Set the left side page header to "Contents"
\tableofcontents % Write out the Table of Contents
\addtocontents{toc}{\vspace{1em}}
% Set the left side page header to "List of Figures"
\addtocontents{toc}{\vspace{1em}}
 % Set the left side page header to "List of Tables"
\lhead{\emph{List of Tables}}
\listoftables % Write out the List of Tables
\addtocontents{toc}{\vspace{1em}}
\lhead{\emph{List of Figures}}
\listoffigures % Write out the List of Figures
\addtocontents{toc}{\vspace{1em}}
%-------------------------------------------------------------------------------
%	ABBREVIATIONS
%-------------------------------------------------------------------------------

%\clearpage % Start a new page

 % Set the line spacing to 1.5, this makes the following tables easier to read
%\setstretch{1.5}

\lhead{\emph{List of symbols and Abbreviations}}
 % Set the left side page header to "Abbreviations"
\listofsymbols{ll}{
\begin{tabular}{lcl}
$G$&:& Newton's gravitational constant.\\
$c$&:& Speed of light in the vacuum\\
$\kappa$&:&   $8\pi G/c^4$ \\
$g_{\mu\nu}$&:& Metric tensor\\
$g$&:&  Determinant of $g_{\mu\nu}$\\
${R}^{\lambda}_{\;\:\mu\nu\sigma}$&:&  Riemann tensor \\
$R_{\mu\nu}$&:& Ricci tensor \\
$R$&:&   Ricci scalar \\
%$S_{M}:$ \,\,\,\,\, Matter action\\
$T_{\mu\nu}$&:&  Energy-momentum tensor\\
$T$&:& Trace of energy-momentum tensor\\
$\mathcal{L}_m$&:&  Matter Lagrangian density\\
$\tilde{\Gamma}^{\lambda}_{\;\:\mu\nu}$&:&  General affine connection\\
${\Gamma}^{\lambda}_{\;\:\mu\nu}$&:& Levi-Civita connection\\
$K^{\lambda}_{\;\:\mu\nu}$&:&  Contorsion tensor\\
$L^{\lambda}_{\;\:\mu\nu}$&:& Disformation tensor\\
%$\lbrace^{\lambda}_{\mu\nu}\rbrace $\,\,\,\,\,\,\,\, Levi-Civita connection\\
$\nabla_{\gamma}$&:& Covariant derivative\\
%$(ij):$ \,\,\,\, Symmetrization over the indices $i$ and %$j$\\
%$[ij]:$ \,\,\,\,\, Anti-symmetrization over the indices $i$ and $j$\\
$Q_{\lambda\mu\nu}$ &:&  Non-metricity tensor\\
$Q$&:&   Non-metricity scalar \\
$P^{\lambda}_{\;\:\mu\nu}$&:&  Non-metricity conjugate\\
$B$&:& Bag parameter\\
WH&:& Wormhole\\
GR&:& General Relativity\\
TEGR&:& Teleparallel Equivalent of General Relativity\\
STEGR&:& Symmetric Teleparallel Equivalent of General Relativity\\
NEC&:& Null Energy Condition\\
WEC&:& Weak Energy Condition\\
DEC&:& Dominant Energy Condition\\
%DE:\,\,\,\,\, \,  Dark Energy\\
%DM:\,\,\,\,\, \,   Dark Matter\\
\end{tabular}\\
\begin{tabular}{lcl}
SEC&:& Strong Energy Condition\\
$\Lambda$CDM&:& $\Lambda$ Cold Dark Matter\\
%ECs:\,\,\,\,\,  Energy Conditions\\
EoS&:& Equation of State\\
DM&:& Dark Matter\\
NFW&:& Navarro-Frenk-White\\
%BEC:\,\,\,\,\, \,\, Bose-Einstein Condensate\\
PI&:& Pseudo-Isothermal\\
MOND&:& Modified Newtonian Dynamics\\
URC&:& Universal Rotational Curve\\
WMAP&:& Wilkinson Microwave Anisotropy Probe\\
CMB&:& Cosmic Microwave Background\\
COBE&:& Cosmic Background Explorer\\
WIMPs&:& Weakly Interacting Massive Particles\\
TOV&:& Tolman-Oppenheimer-Volkoff\\
VIQ&:& Volume Integral Quantifier\\
\end{tabular}
}

%  Include a list of Abbreviations (a table of two columns)
%{
%\textbf{LAH} : \textbf{L}ist \textbf{A}bbreviations %\textbf{H}ere \\
%\textbf{Acronym}:  \textbf{W}hat (it) \textbf{S}tands %%\textbf{F}or 
%}
\addtocontents{toc}{\vspace{2em}}
%-------------------------------------------------------------------------------
%	PHYSICAL CONSTANTS/OTHER DEFINITIONS
%-------------------------------------------------------------------------------

%\clearpage % Start a new page
%
%% Set the left side page header to "Physical Constants"
%\lhead{\emph{Physical Constants}}
%
% % Include a list of Physical Constants (a four column table)
%\listofconstants{lrcl}
%{
%Speed of Light & $c$ & $=$ & $2.997\ 924\ 58\times10^{8}\ \mbox{ms}^{-\mbox{s}}$ (exact)\\
%% Constant Name & Symbol & = & Constant Value (with units) \\}

%-------------------------------------------------------------------------------
%	SYMBOLS
%-------------------------------------------------------------------------------

\clearpage % Start a new page

%\lhead{\emph{Glossary}} % Set the left side page header to "Symbols"

%\listofnomenclature % List the nomenclature. (We use the glossaries package)
%{
%The standard nomenclatures used in this thesis are listed below. 

%}

%-------------------------------------------------------------------------------
%	THESIS CONTENT - CHAPTERS
%-------------------------------------------------------------------------------

\mainmatter % Begin numeric (1,2,3...) page numbering

\pagestyle{fancy} % Return the page headers back to the "fancy" style

% Include the chapters of the thesis as separate files from the Chapters folder
% Uncomment the lines as you write the chapters
%\part{Preliminaries in a Nutshell}
%\part{Minimal Coupling}
% Chapter 1

\chapter{Introduction} % Main chapter title
\label{Chapter1}
%\ref{Chapter1}  % For referencing the chapter elsewhere, use \ref{Chapter1} 

\lhead{Chapter 1. \emph{Introduction}} % This is for the header on each page - perhaps a shortened title
%\vspace{10 cm}

%\blindtext
%----------------------------------------------------------------------------------------
% \clearpage
% \pagebreak
%----------------------------------------------------------------------------------------
This thesis, titled {\bf Study of Wormholes in Symmetric Teleparallel Theories of Gravity}, is dedicated to exploring wormhole (WH) geometries within the framework of symmetric teleparallel theories of gravity and its extension, particularly $f(Q)$ and $f(Q, T)$ theories. Before addressing the specific problems investigated in this work, the current chapter provides a comprehensive overview of the historical background, mathematical notation, fundamental concepts, and key principles of gravity. This preliminary discussion serves to establish the necessary foundation for understanding the research conducted and its significance in the context of gravitational physics.
%%%%%%%%%%%%%%%%%%%%%%%%%%%%%%%%%%%%%%%%%%%%%%%%%%%%%%%%%%%%%%%%%%%%%%%%%%%%%%%%%%%%%%%%%%%%%%%%%%%%%%%%%%%%
\section{Motivations}
WH is a hypothetical tunnel-like structure that connects two locations in the same Universe or possibly two separate Universes \cite{a1, a2, a3}. WHs are fascinating theoretical entities predicted by Einstein's General Relativity (GR) and represent one of the most intriguing possibilities in modern physics. These structures, characterized by their unique ``throat"- a minimal radius that distinguishes them from black holes, offer a conceptual bridge through space-time \cite{a4}, avoiding the presence of an event horizon. Despite the absence of experimental evidence to confirm their existence, their implications for our understanding of the geometry of the Universe and gravitational interactions require rigorous theoretical scrutiny.\\
The persistent study of WHs serves as both a validation and a challenge to GR. As the most comprehensive and experimentally corroborated theory of gravity, GR provides the foundation for exploring WHs. However, the inability of GR to reconcile certain observations, such as accelerating expansion of the Universe \cite{a14}, galaxy rotation curves \cite{a14} and quantization challenges \cite{a15}, has inspired the development of modified gravity theories. These theories aim to address the limitations of GR and explore gravitational phenomena in detail.\\
A critical aspect of the geometry of the WH involves the violation of the null energy condition in the throat region \cite{a17}, which requires the existence of exotic matter, a hypothetical form of matter not yet observed in astrophysical systems. Modified gravity theories offer a potential resolution to this challenge by altering gravitational dynamics in a manner that eliminates the need for such unobserved entities. These modifications pave the way for reassessing the conditions under which WHs could form and persist.\\
Furthermore, investigating modified symmetric teleparallel theories allows for an innovative exploration of the boundaries of gravity. These theories provide a minimal yet effective deviation from GR, ensuring mathematical simplicity while enabling a systematic analysis of gravitational phenomena. By modeling WHs in the context of these frameworks, this thesis aims to address fundamental questions: \textit{Are traversable WHs feasible within alternative theories of gravity?} \textit{Can modifications to GR provide insights into their existence without resorting to exotic matter?}\\
One of the key focuses within modified gravity frameworks is the theoretical development and mathematical modeling of traversable WHs without relying on the existence of exotic matter. This pursuit revolves around proposing adjustments to the gravitational-field equations, allowing for the stabilization of WHs within these alternative models. By exploring such modifications, researchers aim to uncover the fundamental physics underlying WHs while expanding our comprehension of the nature and behavior of gravity. The study of symmetric teleparallel theories of gravity, along with its extended formulations, represents a significant step in this direction. These theories provide a framework for revisiting gravitational phenomena and space-time structures beyond the scope of GR. This exploration pushes the boundaries of gravity theories and offers new insights into how the Universe works, linking what we already know to the mysteries we have yet to uncover.
%%%%%%%%%%%%%%%%%%%%%%%%%%%%%%%%%%%%%%%%%%%%%%%%%%%%%%%%%%%%%%%%%%%%%%%%%%%%%%%%%%%%%%%%%%%%%%%%%%%%%%%%%%%%
\section{Historical overview of wormholes}
The primary focus of this thesis is the study of WHs. At their core, WHs can be described as shortcuts through both space and time. The term WH is often interchangeable with ``spacewarp", which refers to the warping, bending, or folding of space and potentially time. Interestingly, while physicists generally use the term WH, the science fiction community tends to favor spacewarp. Although there is currently no direct experimental evidence to confirm the existence of such phenomena, many believe that WHs could potentially form in regions with extremely strong gravitational fields. In these scenarios, the highly curved structure of space-time could permit the existence of nontrivial topologies. Given the location of the two regions, WHs are generally classified into two types:
\begin{itemize}
    \item \textbf{Inter-Universe WHs:} These connect our Universe to an entirely different Universe.
    \item \textbf{Intra-Universe WHs:} These link two separate regions within the same Universe.
\end{itemize}
WHs can be broadly classified on the basis of their important characteristic, i. e. traversability, which determines whether they allow safe passage for matter and information. They are as follows
\subsection{Non-traversable wormholes}
\begin{figure}[H]
    \centering
    \includegraphics[scale=0.7]{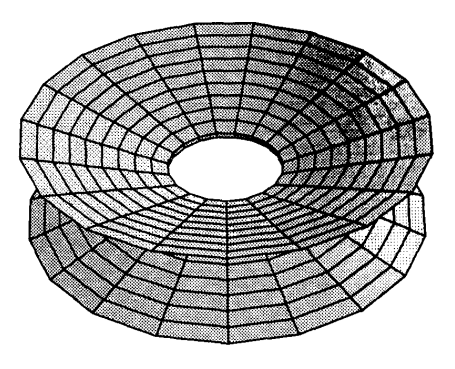}
    \caption{Schematic representation of an Einstein-Rosen bridge. This is a black hole in disguise. The region outside the event horizon is covered twice, while the region inside the event horizon is discarded. This figure is adapted from Ref. \cite{a2}.}
    \label{fig1ch1}
\end{figure}

\begin{figure}
    \centering
    \includegraphics[scale=0.8]{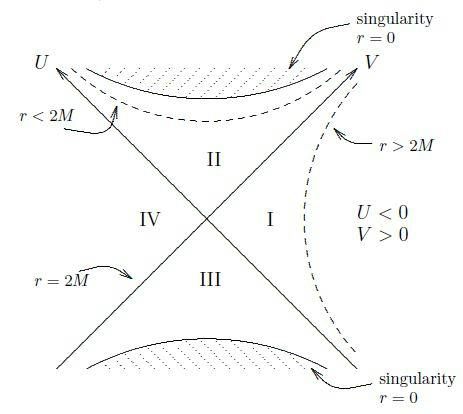}
    \caption{A Kruskal diagram representing space-time is divided into four distinct regions. This figure is adapted from Ref. \cite{aa5}.}
    \label{fig2ch1}
\end{figure}

\begin{figure}[H]
    \centering
    \includegraphics[scale=0.8]{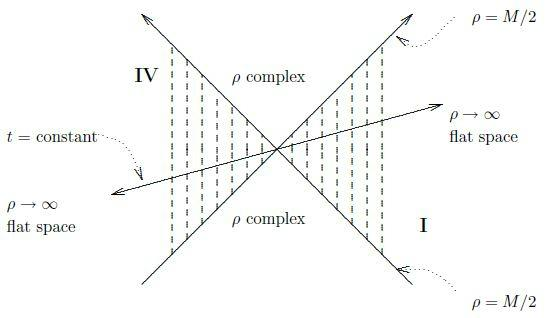}
    \caption{Isotropic coordinates depicted on a Kruskal diagram. Figure taken from Ref. \cite{aa5}.}
    \label{fig3ch1}
\end{figure}

The concept of a WH that cannot be traversed originates from the Schwarzschild solution to Einstein's field equations in the framework of GR. In 1916, Ludwig Flamm \cite{aa1} discovered an alternative interpretation of the Schwarzschild black hole solution, which is now recognized as a white hole. This alternative solution describes a distinct region of space-time \cite{aa1}. In 1935, Albert Einstein and Nathan Rosen \cite{aa2} extended this idea by examining it further in their effort to develop a unified theory of matter and electricity. Their work sought to eliminate field singularities using only the tensors of GR ($g_{\mu\nu}$) and Maxwell theory ($\phi_\mu$). This investigation ultimately gave rise to what we now call the Einstein-Rosen bridge \cite{aa2}, a theoretical construct linking two separate points in space-time (refer Fig. \ref{fig1ch1}). It mathematically describes a non-traversable bridge formed when the gravitational collapse of a massive object creates a singularity, connecting two asymptotically flat regions of space-time. To analyze this structure in mathematical and topological terms, it is essential to delve into the metrics derived from the solutions to the Einstein field equations and the associated coordinate systems. Such an examination provides a deeper understanding of the geometry and physics underlying these intriguing connections in the fabric of space-time.\\
The simplest nontrivial solution to Einstein's field equations is the Schwarzschild geometry. This solution represents the only spherically symmetric vacuum configuration. In the standard Schwarzschild coordinates ($t,\,r,\,\theta,\,\varphi$), this solution can be expressed as (in natural units $c=1,\, G=1$)
\begin{equation}
ds^2 = -\left(1 - \frac{2M}{r}\right) dt^2 + \frac{1}{\left(1 - \frac{2M}{r}\right)} dr^2 + r^2(d\theta^2 + \sin^2\theta d\varphi^2)\,.
\end{equation}
Here, the radial coordinates $r>2M$, $0<\theta<\pi$ and $-\pi<\varphi< \pi$. At first glance, there may appear to be a singularity at $r=2M$; however, this is simply an illusion of the selected coordinate system, which creates the illusion of a singularity where none actually exists. By transitioning to an alternative coordinate system, such as the Eddington–Finkelstein (EF) coordinates, this issue can be resolved. The EF coordinates distinguish between black holes, which are represented by ingoing EF coordinates, and white holes, described by outgoing EF coordinates \cite{aa4}. For black holes, by substituting $t$ with $t = v - r^k$, where $r^k$ is defined through the relationship $dr^k/dr = \left(1 - \frac{2M}{r}\right)^{-1}$, the metric is then modified as follows
\begin{equation}
ds^2 = -\left(1 - \frac{2M}{r}\right) dv^2 + 2dvdr + r^2(d\theta^2 + \sin^2\theta d\varphi^2).
\end{equation}
For white holes, by altering the coordinates to $t = u + r^k$, the metric is consequently modified as
\begin{equation}
ds^2 = -\left(1 - \frac{2M}{r}\right) du^2 - 2dudr + r^2(d\theta^2 + \sin^2\theta d\varphi^2).
\end{equation}
The mathematical framework that connects black holes and white holes is based on a shared coordinate system known as Kruskal-Szekeres (KS) coordinates. In these coordinates, denoted as $(U, V, \theta, \varphi)$, the definitions are given by $U = -e^{-u/4M}$ and $V = e^{v/4M}$. Initially, the metric is defined for $U < 0$ and $V > 0$, however, through an analytical continuation, it is possible to derive the maximally extended Schwarzschild solution,
\begin{equation}
ds^2 = -\frac{32 M^3}{r} e^{-r/2M} dU dV + r^2 (d\theta^2 + sin^2\theta d\varphi^2)\,.
\end{equation}
With $-\infty < U,\,V < \infty$, the radial coordinate $r$ is related to $U$ and $V$ through the following expression
\begin{equation}
UV = -\frac{(r - 2M)}{2M} e^{r/2M}.
\end{equation}
This results in the creation of the Kruskal diagram (see Fig. \ref{fig2ch1}), which divides the entire space-time into four distinct quadrants based on the signs of $U$ and $V$. Region I corresponds to the portion of the space where $r > 2M$ in Schwarzschild coordinates. Both Region I and Region II are bounded by ingoing EF coordinates, crucial for describing black hole geometry. On the other hand, Region I and Region III are covered by outgoing EF coordinates, which are used for white holes. Region IV, a new area, is isometric to Region I under transformation $(U, V) \rightarrow (-U, -V)$. The singularity at $r = 0$ is described by the relation $UV = 1$, while the boundary at $r = 2M$ corresponds to $UV = 0$, indicating that either $U = 0$ or $V = 0$. In this diagram, each point represents a 2-sphere with radius $r$. Another interpretation of the diagram is that it displays the causal structure of the radial motion for fixed polar angles $\theta$ and $\varphi$.\\
In region I, the expression $\frac{U}{V} = e^{-t/2M}$ is satisfied, indicating that constant-time slices in the Schwarzschild coordinates are represented as straight lines through the origin of the Kruskal diagram. These hypersurfaces extend partially across both Region I and Region IV. To gain a clearer understanding of this geometry, it is beneficial to transition to isotropic coordinates $(t, \rho, \theta, \varphi)$ where $\rho$ denotes the modified radial coordinate
\begin{equation}
r = \rho\left(1 +\frac{M}{2\rho}\right)^2\,.
\end{equation}
The metric takes the following form in isotropic coordinates
\begin{equation}
ds^2 =-\left(\frac{1 -\frac{M}{2\rho}}{1 +\frac{M}{2\rho}}\right)^2 dt^2+\left(1 +\frac{M}{2\rho}\right)^4 \left(d\rho^2+\rho^2 \left(d\theta^2 + sin^2\theta d\varphi^2\right)\right)\,.
\end{equation}
For a given $r$, there exist two possible solutions for $\rho$, which are connected through isometry $\rho \to M^2 / (4\rho)$. This transformation has a fixed point on a 2-sphere with a radius of $2M$ and relates regions I and IV, similar to the transformation $(U, V) \to (-U, -V)$. In region I, $\rho > M/2$, while in region IV, $0 < \rho < M/2$. Consequently, the isotropic coordinates are only applicable to regions I and IV, as $\rho$ becomes complex when $r < 2M$ (see Fig. \ref{fig3ch1}).\\
In the spatial geometry of a constant $t$, as $\rho$ approaches $M/2$ from either side, the radius of the 2-sphere decreases, reaching a minimum value of $r = 2M$ at $\rho = M/2$, known as the minimal 2-sphere. There are two asymptotically flat regions: one where $\rho \to \infty$ and the other where $\rho \to 0$. These regions are connected through a throat with a minimum radius equivalent to that of the minimal 2-sphere $2M$. This configuration is referred to as the \textit{Einstein-Rosen bridge}.
While it is tempting to consider this structure as a traversable WH bridging Regions I and IV, its constant depiction proves it is not feasible to cross through, categorizing it as a non-traversable WH. Interest in this area remained minimal for two decades after its discovery. The physicists Misner and Wheeler \cite{wormhole} reignited attention to this phenomenon in 1957, introducing the term ``\textit{wormholes}" in their research. They explored them as topological figures known as geons (solutions to the Einstein-Maxwell equations that, despite being unstable, persist for a considerable duration) and presented the initial depiction of a WH as a tunnel connecting disparate space-time areas. 
%%%%%%%%%%%%%%%%%%%%%%%%%%%%%%%%%%%%%%%%%%%%%%%%%%%%%%%%%%%%%%%%%%%%%%%%%%%%%%%%%%%%%%%%%%%%%%%%%%%%%%%%%%%%
\subsection{Traversable wormholes}
The WH is considered traversable if a particle can enter one side of a WH and exit through the other. The year 1973 marked a pivotal advancement in the study of traversable WHs, with groundbreaking research independently conducted by Homer G. Ellis \cite{aa6} and Kirill A. Bronnikov \cite{aa7}. Their work demonstrated that traversable WHs could theoretically exist within the GR framework. Homer G. Ellis \cite{aa6} introduced a model widely regarded as the first comprehensive representation of such a WH. This model describes a static, spherically symmetric solution to the Einstein's vacuum field equations, incorporating a scalar field $\Phi$ that interacts minimally with the space-time geometry but features a reversed coupling (negative instead of positive) \cite{aa6}
\begin{equation}
R_{\mu\nu}-\frac{1}{2}R g_{\mu\nu}=8\pi\left(\nabla_{\mu}\Phi\nabla_{\nu}\Phi-\frac{1}{2}g_{\mu\nu}\nabla^{\sigma}\Phi\nabla_{\sigma}\Phi\right).
\end{equation}
In this case, the metric is expressed as
\begin{eqnarray}
ds^2 &=& -dt^2 + (d\rho - f(\rho)dt)^2 + r^2(\rho)(d\theta^2 + \sin^2\theta d\varphi^2),\nonumber\\
&=&  -[1 - f^2(\rho)]dT^2 + \frac{1}{1 - f^2(\rho)} d\rho^2 + r^2(\rho)(d\theta^2 + \sin^2\theta d\varphi^2),
\end{eqnarray}
where 
\begin{equation*}
  dT = dt + \frac{f(\rho)}{1 - f^2(\rho)} d\rho.
\end{equation*}
%\( T = t + \int \frac{Zf(\rho)}{1 - f^2(\rho)} d\rho\),
The functions \(r(\rho)\) and \(f(\rho)\) must be derived from the corresponding field equations. The specified range for the coordinates is as follows
\begin{equation}\nonumber
-\infty<t<\infty; \quad -\infty<\rho<\infty; \quad 0<\theta<\pi; \quad -\pi<\varphi<\pi.
\end{equation}
The behavior of solutions depends on two parameters, $m$ and $n$, which must satisfy the condition $0 \leq m < n$. Apart from this restriction, the parameters are otherwise unconstrained. The functions are expressed as follows
\begin{equation}
f(\rho)=-\sqrt{1 - e^{-\left(\frac{2m}{n}\right)\tau(\rho)}}
\end{equation}
and
\begin{equation}
r(\rho)=\sqrt{(\rho - m)^2 + a^2}e^{\left(\frac{m}{n}\right)\tau(\rho)}\,.
\end{equation}
Here, \(\tau(\rho) = \frac{n}{a}\left[\frac{\pi}{2}-\tan^{-1}\left(\frac{\rho-m}{a}\right)\right]\) and \(a = \sqrt{n^2 - m^2}\) \cite{aa6}.\\
The motion of particles through this WH resembles the flow of liquid along chaotic draining curves. Depending on the initial velocity direction of each particle, various trajectories can be observed \cite{aa6}. For further information about Ellis's \textit{drainhole} theory, one can refer to Ref. \cite{aa6}.\\
Later Kip Thorne, Michael Morris and Uri Yertsever \cite{a17, MorrisThrone} made remarkable development in 1988 by presenting the concept of the Morris-Thorne WH in two pivotal papers. They proposed nine defining characteristics for a stable and traversable WH. Some of these features are indispensable for its formation, while others were adopted to simplify the computational framework:
\begin{itemize}
    \item[1.] The metric should be static and spherically symmetric.
   \item[2.] The solution must satisfy Einstein's field equations everywhere, as described by
   \begin{equation}\label{Einstein field equations}
   G_{\mu\nu}\equiv R_{\mu\nu} - \frac{1}{2}g_{\mu\nu}R =\frac{8\pi G}{c^4} T_{\mu\nu}\,,
   \end{equation}
  where\\
   \begin{minipage}[t]{0.5\textwidth}
        \hspace{1cm} \begin{tabular}{lcl}
             $\bullet$ $G_{\mu\nu}$ &$\longrightarrow$& Einstein tenor\\
             $\bullet$ $R_{\mu\nu}$ &$\longrightarrow$& Ricci tensor\\
             $\bullet$ $R$ &$\longrightarrow$& Ricci scalar\\
             $\bullet$ $g_{\mu\nu}$ &$\longrightarrow$& Metric tensor\\
             $\bullet$ $G$ &$\longrightarrow$& Newton's gravitational constant\\
             $\bullet$ $c$ &$\longrightarrow$& Speed of light in the vacuum\\
             $\bullet$ $T_{\mu\nu}$ &$\longrightarrow$& Energy-momentum tensor.
         \end{tabular}    
    \end{minipage}\\
   \item[3.] The WH must possess a throat that links two separate regions of space-time, each of which is asymptotically flat.
   \item[4.] For two-way travel to be possible, the WH must be free of horizons.
   \item[5.] Travelers must experience tidal gravitational forces that are small enough to be bearable.
   \item[6.] The WH must allow a traveler to traverse it in a finite and relatively short proper time.
   \item[7.] The matter and energy fields responsible for the space-time curvature must have a stress-energy tensor that aligns with physical expectations.
   \item[8.] The solution must remain stable when subjected to small perturbations.
   \item[9.] The WH should be physically possible to construct, meaning that the required mass should be much smaller than the total mass of the Universe, and the time needed for its formation should be much less than the age of the Universe.
\end{itemize}
The authors classified the first four conditions as basic WH criteria, while the fifth and sixth conditions, which pertain to making the WH suitable for human travelers, were referred to as usability criteria.\\
Based on all the above conditions, the Morris-Thorne WH metric is given as
\begin{equation}\label{1ch1}
    ds^2 = e^{2\phi(r)}dt^2 - \frac{dr^2}{\left(1 - \frac{b(r)}{r}\right)} - r^2d\theta^2- r^2\sin^2\theta d\varphi^2\,.
\end{equation}
In this context, $b(r)$ defines the shape of the WH and is referred to as the shape function, while $\phi(r)$ governs the gravitational redshift, known as the redshift function.\\
From a mathematical perspective, ensuring that a WH is traversable requires the shape function $b(r)$ and the redshift function $\phi(r)$ to meet the following conditions:
\begin{itemize}
  \item[(1)] For $r>r_0$, that is, out of the throat, $1-\frac{b(r)}{r}>0$, and in the throat of the WH, that is, $r=r_0$, $b(r)$ must satisfy the condition $b(r_0)=r_0$, where $r_0$ denotes the radius of the throat.
  \item[(2)] The shape function $b(r)$ has to meet the flare-out requirement at the throat, i.e., $b'(r_0)<1$.
  \item[(3)] For the asymptotic flatness condition, the limit $\frac{b(r)}{r}\rightarrow 0$ is required as $r\rightarrow \infty$.
  \item[(4)] The redshift function $\phi(r)$ should be finite everywhere.
\end{itemize}
Compliance with these conditions guarantees the potential existence of exotic matter at the throat of the WH within the framework of Einstein's GR.
% To simplify the calculation of the Einstein tensor components, we switch to a different coordinate system that makes the process more straightforward. In this new system, we adopt a set of orthonormal vectors that represent the ``proper reference frame of observers who remain stationary at fixed coordinates \((r, \theta, \varphi)\)"
% \begin{equation}
% e_t = e^{-\phi}e_{t0}, \quad e_r = \left(1 - \frac{b}{r}\right)^{-\frac{1}{2}} e_{r_0}, \quad e_\theta = \frac{1}{r} e_{\theta_0}, \quad e_\varphi = \frac{1}{r \sin \theta} e_{\varphi_0},
% \end{equation}
% % \[
% % e_t = e^{-\Phi}e_{t0}, \quad e_r = \left(1 - \frac{b}{r}\right)^{-\frac{1}{2}} e_{r0}, \quad e_\theta = \frac{1}{r} e_{\theta0}, \quad e_\phi = \frac{1}{r \sin \theta} e_{\phi0}
% % \]
% where $e_{t_0}, e_{r_0}, e_{\theta_0}, e_{\varphi_0}$ are the initial orthonormal basis vectors given by
% \begin{equation}
% e_{t_0} = \frac{\partial}{\partial t}, \quad e_{r_0} = \frac{\partial}{\partial r}, \quad e_{\theta_0} = \frac{\partial}{\partial \theta}, \quad e_{\varphi_0} = \frac{\partial}{\partial \varphi}.
% \end{equation}
% In this newly adopted basis, the metric coefficients reduce to their standard form, as seen in special relativity
% \[
% g_{\mu \nu} = e_\mu \cdot e_\nu = \eta_{\mu \nu} =
% \begin{pmatrix}
% -1 & 0 & 0 & 0 \\
% 0 & 1 & 0 & 0 \\
% 0 & 0 & 1 & 0 \\
% 0 & 0 & 0 & 1
% \end{pmatrix}.
% \]
% This representation simplifies the calculations by aligning the metric coefficients with those of flat space-time.
Now, for the Einstein tensor components, we compute the Riemann tensor \(R^{\lambda}_{\;\:\mu\nu\sigma}\), the Ricci tensor \(R_{\mu\nu}\) and the Ricci scalar \(R\), and by solving the Einstein's field equations, we obtain the following results
\begin{equation}
G_{tt} = \frac{b'(r)}{r^2},
\end{equation}
\begin{equation}
G_{rr} = -\frac{b(r)}{r^3} + 2\left(1 - \frac{b(r)}{r}\right) \frac{\phi'(r)}{r}\,,
\end{equation}
and
\begin{equation}
G_{\theta\theta} = \left(1 - \frac{b(r)}{r}\right)\left[\phi''(r) - \frac{rb'(r)-b(r)}{2r(r - b(r))} \phi'(r) + \phi'(r)^2 + \frac{\phi'(r)}{r} - \frac{rb'(r)-b(r)}{2r^2(r - b(r))}\right] = G_{\varphi\varphi}.
\end{equation}
Further, the components of the stress-energy tensor are expressed as follows \cite{a17}
\begin{equation}
T_{tt} = \rho(r), \quad T_{rr} = -\tau(r), \quad T_{\theta\theta} = T_{\varphi\varphi} = p(r).
\end{equation}
Here, $\rho$ is the energy density, $\tau(r)$ is the radial tension per unit area, and $p(r)$ is the lateral pressure. Now, by using the equation \(G_{\mu\nu} = 8\pi T_{\mu\nu}\) (in natural units $c=1,\, G=1$), we find
\begin{equation}\label{p1}
\rho(r) = \frac{1}{8\pi}\left[\frac{b'(r)}{ r^2}\right],
\end{equation}
\begin{equation}\label{p2}
\tau(r)=\frac{1}{8\pi} \left[\frac{b(r)}{r^3} - 2\left(1 - \frac{b(r)}{r}\right) \frac{\phi'(r)}{r}\right]\,,
\end{equation}
and
\begin{equation}\label{p3}
p(r) = \frac{1}{8\pi}\left(1 - \frac{b(r)}{r}\right)\left[\phi''(r) - \frac{rb'(r)-b(r)}{2r(r - b(r))} \phi'(r) + \phi'(r)^2 + \frac{\phi'(r)}{r} - \frac{rb'(r)-b(r)}{2r^2(r - b(r))}\right].
\end{equation}
Furthermore, one can derive $\rho(r)$ and $\tau(r)$ by choosing suitable forms for $b(r)$ and $\phi(r)$. In the throat of the WH, the most significant condition requires the radial pressure to exceed the energy density, expressed as $ \rho_0-\tau_0 <0$. Materials that exhibit such characteristics are referred to as exotic. This scenario poses unique challenges, especially when analyzed from the perspective of an observer traveling through the WH at nearly the speed of light, who would perceive a negative energy density. Similarly, a stationary observer could also detect a negative energy density $\rho_0 < 0$, which naturally follows from these conditions. 
%The next subsection will examine specific cases in which this phenomenon occurs.
%%%%%%%%%%%%%%%%%%%%%%%%%%%%%%%%%%%%%%%%%%%%%%%%%%%%%%%%%%%%%%%%%%%%%%%%%%%%%%%%%%%%%%%%%%%%%%%%%%%%%%%%%%%%
\subsubsection{Rotating wormhole}
In 1998, Teo constructed the stationary and axially
symmetric generalization of the Morris-Thorne WH \eqref{1ch1} and physically described rotating WHs \cite{ec4}. Considering this case was significant for several reasons. Firstly, it represents one of the most general extensions of the Morris-Thorne WH that remains analytically tractable, barring scenarios with no space-time symmetries- which would be highly challenging to study with current methods. Furthermore, such an analysis could facilitate the derivation or modeling of explicit WH solutions, both in the classical regime \cite{F. Schein} and within semiclassical approximations \cite{D. Hochberg}, which are relevant in various theoretical contexts. Moreover, if an arbitrarily advanced civilization \cite{MorrisThrone} were to construct a traversable WH, it is highly probable that the structure would be either spherical or rotating. Such configurations could allow interstellar travelers to circumvent the direct interaction with exotic matter necessary to sustain the WH.\\
The metric for a stationary, axisymmetric traversable WH can be written as
\begin{equation}
    ds^2 =  N^2 dt^2 - \frac{dr^2}{\left(1 - \frac{b}{r}\right)} - r^2 K^2 \left[d\theta^2 + \sin^2\theta (d\varphi-\omega dt)^2\right],
\end{equation}
where $N$, $b$, $K$ and $\omega$ are functions of $r$ and $\theta$, such that it is regular on the symmetry axis $\theta=\{0,\pi\}$. This metric describes a rotating traversable WH, where the off-diagonal term $g_{t\varphi}$ encodes the rotational effects, similar to the Kerr metric for rotating black holes. It describes two identical, asymptotically flat regions joined together at the throat. $N$ serves as the counterpart to the redshift function found in Eq. \eqref{1ch1} and must remain finite and nonzero to guarantee the absence of event horizons or curvature singularities. Further, $K$ defines the proper radial distance, while $\omega$ regulates the angular velocity of the WH. Moreover, in 2018, Shaikh \cite{R. Shaikh} investigated the shadows produced by a specific class of rotating WHs and highlighted the essential role that the rotating WH throat plays in shaping the structure of the shadow. Additionally, Chew \cite{Chew} explored rotating WH solutions sustained by a complex phantom scalar field with a quartic self-interaction, where the phantom field drives the space-time rotation. These solutions were both regular and asymptotically flat. Additionally, they identified a subset of solutions that correspond to static yet non-spherically symmetric WHs.
%%%%%%%%%%%%%%%%%%%%%%%%%%%%%%%%%%%%%%%%%%%%%%%%%%%%%%%%%%%%%%%%%%%%%%%%%%%%%%%%%%%%%%%%%%%%%%%%%%%%%%%%%%%%
\section{Energy conditions}\label{Energy conditions}
To better comprehend the properties of matter within the WH, Morris and Thorne \cite{a17} proposed a dimensionless function, defined as $\xi = \frac{(\tau - \rho)}{|\rho|}$. Using Eqs. \eqref{p1} and \eqref{p2}, the corresponding expression can be derived
\begin{equation}\label{p4}
\xi=\frac{(\tau-\rho)}{|\rho|}=\frac{b/r-b'-r(1-b/r)\phi'}{|b'|}.
\end{equation}
By combining Eq. \eqref{p4} with the flare-out condition $\frac{d^2r}{dz^2} = \frac{b - r b'}{b^2} > 0$, the exotic function $\xi(r)$ is obtained as
\begin{equation}
\xi= \frac{2b^2}{r|b'|} \frac{d^2r}{dz^2}-\frac{r(1-b/r)\phi'}{|b'|}.
\end{equation}
Since $\rho$ and $b'$ are finite, and noting that $(1 - b/r)\phi' \to 0$ at the throat, the following relation is derived
\begin{equation}
\xi(r_0)=\frac{\tau_0-\rho_0}{|\rho_0|} > 0.
\end{equation}
The condition $\tau_0 > \rho_0$ poses a considerable challenge, implying that the radial tension in the throat must exceed the energy density. To describe matter satisfying this criterion, Morris and Thorne referred to it as exotic matter. Further examination reveals that such matter not only violates the null energy condition but also fails to satisfy all other energy conditions.\\
To investigate the energy conditions, we focus on the case where the energy-momentum tensor is diagonal, expressed as
\begin{equation}
T_{\mu}^{\,\,\,\nu} = \text{diag}(\rho, -p_1, -p_2, -p_3).
\end{equation}
Here, \(p_i\) denotes the pressure components along $x_i$ coordinates. For an isotropic matter distribution, we can take $p_1=p_2=p_3$.
The energy conditions in GR are formulated as follows:
\begin{itemize}
  \item[\ding{114}] \textbf{Null Energy Condition (NEC):} The NEC requires that for any null vector \(k^\mu\)
  \begin{equation}
   T_{\mu\nu}k^\mu k^\nu \geq 0.   
  \end{equation}
  In the case of a diagonal energy-momentum tensor, the NEC simplifies to 
  \begin{equation}
     \rho + p_i \geq 0,\quad \forall \,\,i.
  \end{equation} 
This essentially means that the energy density measured by any null observer with a four-velocity \(k^{\mu}\) must be non-negative.
\end{itemize}
\begin{itemize}
    \item[\ding{114}] \textbf{Weak Energy Condition (WEC):} The WEC states that for any time-like vector \(U^\mu\), the following inequality \begin{equation}
    T_{\mu\nu}U^\mu U^\nu \geq 0\,,
    \end{equation}
must hold. This implies that the energy density measured by any time-like observer must be positive, which can be expressed as
\begin{equation}
  \rho \geq 0 \quad \text{and} \quad \rho + p_i \geq 0, \quad \forall \,\,i.
\end{equation}
Importantly, the WEC inherently encompasses the NEC as a special case.
\end{itemize}
\begin{itemize}
    \item[\ding{114}] \textbf{Strong Energy Condition (SEC):} The SEC requires that for any time-like vector \(U^\mu\), the relation
    \begin{equation}
    \left(T_{\mu\nu} - \frac{T}{2}g_{\mu\nu}\right)U^\mu U^\nu \geq 0
    \end{equation}
    hold, where $T$ represents the trace of the energy-momentum tensor. For a diagonal energy-momentum tensor, this condition can be written as
    \begin{equation}
    \rho + p_i \geq 0 \quad \text{and}\quad \rho + \sum_i p_i \geq 0.
    \end{equation}
The SEC is fundamentally related to preserving the attractive nature of gravity. It implies the NEC but does not necessarily encompass the WEC.
\end{itemize}
\begin{itemize}
    \item[\ding{114}] \textbf{Dominant Energy Condition (DEC):} The DEC requires that for any time-like vector \(U^\mu\), the relation
    \begin{equation}
   T_{\mu\nu}U^\mu U^\nu \geq 0 
    \end{equation}
    and $T_{\mu\nu}U^\nu$ being not a space-like to be satisfied. These conditions ensure that the locally observed energy density is positive, and the energy flux is either time-like or null. For a diagonal tensor, this results in
    \begin{equation}
      \rho \geq 0 \quad \text{and}\quad p_i \in [-\rho, +\rho]\,.
    \end{equation}
This ensures that the speed of sound remains below the speed of light, a condition often linked to the stability of the system.
\end{itemize}
To conclude, we analyze four fundamental energy conditions, i.e., NEC, WEC, SEC and DEC, within the context of diagonal energy-momentum tensor. These conditions are represented as follows.
\begin{itemize}
    \item \textbf{NEC:}\,\,\, \(\rho + p_r \geq 0\), \quad \(\rho + p_t \geq 0\).
    \item \textbf{WEC:}\,\,\, \(\rho + p_r \geq 0\), \quad\(\rho + p_t \geq 0\), \quad\(\rho \geq 0\).
    \item \textbf{SEC:}\,\,\, \(\rho + p_r \geq 0\), \quad\(\rho + p_t \geq 0\),\quad \(\rho + p_r + 2 p_t \geq 0\).
    \item \textbf{DEC:}\,\,\, \(\rho-|p_r| \geq 0\),\quad \(\rho-|p_t| \geq 0\), \quad\(\rho \geq 0\),
\end{itemize}
where $p_r$ is the radial pressure and $p_t$ is the tangential pressure.\\
The Einstein field equations, when applied at the WH's throat along with the assumption of a finite redshift function, lead to the flare-out condition, which implies $(\rho + p_r)|_{r_0} < 0$. This condition directly violates the NEC and, consequently, all the energy conditions. Although classical matter is generally expected to comply with these energy conditions, certain quantum fields, such as those observed in the Casimir effect, have been shown to violate them. Therefore, the flare-out condition necessitates a breach of the NEC at the throat, indicating that the WH's geometry must be supported by exotic matter. Although negative energy densities are not strictly required, negative pressures in the throat, such as $p_r(r_0) = -1/(8\pi r^2_0)$, are crucial to maintaining the structure of the throat. Although violations of the NEC may seem unlikely from a classical point of view, quantum effects that have been experimentally observed suggest that minimal violations are possible. However, whether such violations are substantial enough to sustain a traversable WH remains an open question. To address this issue, various strategies have been proposed, including evolving WH geometries, rotating configurations, thin-shell constructions using the cut-and-paste method, and modifications to existing gravitational theories \cite{ec1,ec2,ec3,ec4,ec5,ec6,ec7,a19,a20,ec11,ec12}.
\section{Dark matter}
Our current understanding of the Universe covers only about 5\% of its total mass-energy. According to the $\Lambda$ Cold Dark Matter ($\Lambda$CDM) model \cite{dm1}, this 5\% corresponds to baryonic matter- stars, gas and dust. The remaining 95\% consists of dark energy (68\%) and dark matter (DM) (27\%) \cite{dm2,dm3}. DM is thought to be nonbaryonic, cold (non-relativistic in radiation-matter equality) and collisionless (interacting only through gravity), which makes it difficult to study \cite{dm4,dm5,dm6}.\\
Weakly Interacting Massive Particles (WIMPs) were among the first proposed DM candidates \cite{dm8,dm9}. With masses near 100 GeV and interactions similar to the weak nuclear force, WIMPs were likely created thermally in the early Universe. Their abundance aligns with theoretical predictions, a concept known as the WIMPs miracle \cite{dm11}. Despite extensive searches, no direct evidence of WIMPs has been found, leading to reduced confidence in their role as DM \cite{dm12,dm13}.\\
Axions, initially proposed to solve the strong charge-parity problem in quantum chromodynamics, have emerged as another strong candidate. They are predicted to weakly interact with Standard Model particles, remain cold, and have masses between $10^{-5}$ and $10^{-2}$ eV, satisfying DM requirements \cite{dm14,dm15,dm16}. Numerous other candidates have also been suggested; for details, see Refs. \cite{dm17,dm18}.\\
Many experiments aim to detect DM particles directly or indirectly, customized to specific hypotheses \cite{dm19,dm20,dm21,dm22}. However, no conclusive detection has been made \cite{dm12}. In the absence of direct evidence, researchers analyze the effects of DM on visible baryonic matter to constrain its properties and refine the search \cite{dm16}.
%%%%%%%%%%%%%%%%%%%%%%%%%%%%%%%%%%%%%%%%%%%%%%%%%%%%%%%%%%%%%%%%%%%%%%%%%%%%%%%%%%%%%%%%%%%%%%%%%%%%%%%%%%%%
\subsection{Evidence of dark matter}
The term DM is often attributed to Zwicky’s work in the 1930s \cite{dm26, dm27, dm28}. During the study of the Coma cluster, Zwicky found that its observed velocity dispersions ($\sim$ 1000 km/s) were much higher than predicted $\sim$ 80 km/s, suggesting that the cluster was about 400 times more massive than its visible matter indicated. He proposed the existence of nonluminous DM, a hypothesis later supported by similar findings in the Virgo cluster \cite{dm29}.\\
Despite Zwicky’s early work, DM gained significant attention only in the 1970s when Rubin and Ford’s studies on galaxy rotation curves \cite{dm30} provided compelling evidence. Observations of the Andromeda galaxy and others using advanced spectroscopy showed that rotation curves flattened at large radii instead of declining as predicted by Keplerian models. This indicated that galaxies are embedded in extended DM halos, revealing the pervasive role of DM in the Universe \cite{dm31, dm32, dm33}.\\
Further evidence for DM comes from gravitational lensing, where the bending of light by massive foreground objects reveals the presence of unseen mass \cite{dm37, dm38}. On cosmological scales, data from the Cosmic Microwave Background (CMB) have been instrumental. Temperature fluctuations in the CMB, mapped by missions such as the Wilkinson Microwave Anisotropy Probe (WMAP), Cosmic Background Explorer (COBE) and Planck \cite{dm2, dm3}, confirm density variations in the early universe, aligned with the $\Lambda$CDM model. Combining CMB data with studies of galaxy clustering and abundances further constrains DM properties and underscores its role in cosmic structure formation \cite{dm52c, dm53c, dm54c}.
%%%%%%%%%%%%%%%%%%%%%%%%%%%%%%%%%%%%%%%%%%%%%%%%%%%%%%%%%%%%%%%%%%%%%%%%%%%%%%%%%%%%%%%%%%%%%%%%%%%%%%%%%%%%
\subsection{Dark matter halos}
DM halos are dense regions of DM that form through virialization \cite{dm55c}. These regions typically have densities hundreds of times greater than the average matter density in the Universe, with their boundaries often defined by the virial radius \cite{dm56c}, which evolves with cosmic time and depends on cosmology.\\
In the hierarchical structure formation model, most of the Universe's mass is concentrated in these halos, making them essential for studying large-scale matter distribution. Halos trace the underlying density field and their clustering provides insight into cosmic structure formation. However, halos exhibit biased clustering relative to DM, which requires careful interpretation of clustering statistics.\\
The size of a halo is often described by its virial radius, $r_{200}$, the radius where the density is 200 times the critical density of the Universe at $z = 0$. The virial mass, $M_{200}$, enclosed within this radius is given by
\begin{equation}
M_{200} = \frac{4}{3}\, \pi \,\,200\, \rho_{\text{crit}}\,\, r_{200}^3,
\end{equation}
where $\rho_{\text{crit}}$ is the critical density. The virial mass is linked to the concentration of the halo via the mass-concentration relation \cite{dm57c,dm58c,dm59c}.
%%%%%%%%%%%%%%%%%%%%%%%%%%%%%%%%%%%%%%%%%%%%%%%%%%%%%%%%%%%%%%%%%%%%%%%%%%%%%%%%%%%%%%%%%%%%%%%%%%%%%%%%%%%%
\subsection{Some dark matter profiles}
Various DM density profiles have been introduced in the scientific literature to describe the distribution of DM within halos. These profiles aim to capture the variation in density from dense central regions to diffuse outer regions. In this section, we will provide a brief overview of some of the most widely studied and commonly used DM density profiles.
\subsubsection{Navarro-Frenk-White profile}
The density profile, $\rho(r)$, describes the internal structure of the DM halos. Dubinski and Carlberg (1991) \cite{dm50} showed that halos in N-body simulations could be modeled using the Hernquist profile \cite{dm51}
\begin{equation}
\rho(r) = \rho_s \left(\frac{r}{r_s}\right)^{-\gamma} \left(1 + \frac{r}{r_s}\right)^{\gamma - \beta},
\end{equation}
where $\rho_s$ is the central density, $r_s$ the scale radius, $\gamma$ the inner slope ($\gamma = 1$) and $\beta$ the outer slope ($\beta = 4$). Later, Navarro, Frenk and White \cite{dm52,dm53} introduced the widely-used Navarro-Frenk-White (NFW) profile, adjusting $\beta$ to 3
\begin{equation}\label{4aaa1}
\rho_{\text{NFW}}(r) = \rho_s \left(\frac{r}{r_s}\right)^{-1} \left(1 + \frac{r}{r_s}\right)^{-2}.
\end{equation}
Although the NFW profile is simple and effective, particularly for small halos such as dwarfs \cite{dm58c}, higher resolution simulations have shown variations in $\gamma$ with radius, challenging its universal applicability \cite{dm54,dm55,dm56}. The Einasto profile \cite{dm57}, with three parameters, has demonstrated better accuracy in the fitting of the simulation results \cite{dm54}, but is less frequently used due to its complexity.
%%%%%%%%%%%%%%%%%%%%%%%%%%%%%%%%%%%%%%%%%%%%%%%%%%%%%%%%%%%%%%%%%%%%%%%%%%%%%%%%%%%%%%%%%%%%%%%%%%%%%%%%%%%%
\subsubsection{Pseudo-Isothermal profile}
An important class of DM models is connected to modified gravity, such as Modified Newtonian Dynamics (MOND) \cite{Begeman}, often utilize the Pseudo-Isothermal (PI) density profile
\begin{equation}\label{4aa1}
\rho_{\text{PI}}(r)=\frac{\rho_s}{1+\left(\frac{r}{r_s}\right)^2}\,.
\end{equation}
Unlike the NFW profile, which predicts a central cusp, the Pseudo-Isothermal (PI) profile has a flat core, aligning better with observed rotation curves in low surface brightness and dwarf galaxies \cite{Begeman}. Blok et al. \cite{Frenk4} showed that the PI profile resolves the core-cusp problem and provides an improved fit to observed data. Gentile et al. \cite{pi1} also demonstrated its consistency across spiral galaxy rotation curves, while Oh et al. \cite{pi2} effectively applied it to dwarf galaxies without additional modifications. Furthermore, in WH geometry, the PI profile in Ref. \cite{pi3} shows that DM density around axisymmetric traversable WHs mimics a black hole spike but depends on spin direction. This highlights the utility of the PI profile in both observational and theoretical studies.
%%%%%%%%%%%%%%%%%%%%%%%%%%%%%%%%%%%%%%%%%%%%%%%%%%%%%%%%%%%%%%%%%%%%%%%%%%%%%%%%%%%%%%%%%%%%%%%%%%%%%%%%%%%%
\subsubsection{Universal Rotation Curve profile}
The Universal Rotation Curve (URC) model suggests that the density of DM in a galaxy decreases with distance from its center unlike visible matter, which is concentrated near the core. Observations show that galaxy rotation curves remain flat at large radii, contradicting expectations if only visible matter were present. This indicates the presence of an invisible DM that contributes to the high rotation speeds of stars and gas in the outer regions.\\
The URC DM density profile, valid across galactic halos, is expressed as \cite{dm70}
\begin{equation}\label{urc1}
\rho_{\text{URC}}(r) = \frac{\rho_s\, r_s^3}{(r + r_s)(r^2 + r_s^2)}\,.
\end{equation}
For the Milky Way, $r_s = 9.11$ kpc and $\rho_s = 5\times 10^{-24}\,g/cm^3$, respectively.\\
In inner regions ($r \ll r_s$), the URC profile approximates the isothermal profile
\begin{equation}
\rho_{\text{PI}}(r)=\frac{\rho_s}{1+r^2/r_c^2},
\end{equation}
where $r_s=r_c$, the core radius. Unlike the isothermal profile, where the mass diverges with $r$, the URC profile results in a logarithmic mass divergence, consistent with predictions from cosmological CDM models, such as Navarro et al. \cite{dm71}.\\
The URC framework also explains galaxy rotation curves, a key indicator of DM. Persic et al. \cite{dm72} analyzed rotation curves using $H_\alpha$ and radio data, demonstrating their applicability across various types of galaxy and luminosities. This led to the concept of the URC, underscoring its wide applicability.
The URC model provides valuable insights into the DM distribution and its influence on galaxy dynamics, although rotation curve variations across galaxies require tailored models for accurate interpretation.
%%%%%%%%%%%%%%%%%%%%%%%%%%%%%%%%%%%%%%%%%%%%%%%%%%%%%%%%%%%%%%%%%%%%%%%%%%%%%%%%%%%%%%%%%%%%%%%%%%%%%%%%%%%%
\section{Relevant physical background}
In this section, we dive into several important concepts that play a significant role in modern physics, including the MIT bag model and noncommutative geometry. We provide a comprehensive overview of their theoretical foundations and applications.
%%%%%%%%%%%%%%%%%%%%%%%%%%%%%%%%%%%%%%%%%%%%%%%%%%%%%%%%%%%%%%%%%%%%%%%%%%%%%%%%%%%%%%%%%%%%%%%%%%%%%%%%%%%%%
\subsection{Equation of state for the MIT bag model}\label{MIT}
A theoretical model was developed at the Massachusetts Institute of Technology (MIT) to explain the properties of elementary particles \cite{A. Chodos1, A. Chodos2}. Several works have been connecting phenomenological astrophysical effects to the presence of strange quark matter, as was beautifully described in the work of Witten \cite{Witten}. It is emphasized that if a star composed of quark matter is considered and the contributions from their masses are neglected, the pressure-density relationship takes the form $p=\frac{1}{3}(\rho - 4B)$, which is called the MIT bag model. In this model, the vacuum pressure $B$ in the bag wall plays a crucial role in stabilizing the confinement of the quarks. The deconfinement phase transition depends on the temperature and the baryon number density of the system at high densities. This equation of state (EoS) has been widely applied in the literature, as we can see in Refs. \cite{Prasad, Aguirre, Burgio, Liu, Peng, Chakrabarty 1, Chakrabarty 2, Chakrabarty 3, N. Itoh}. In order to justify the form of EoS for the MIT bag model, let us carefully revise the discussions presented in Ref. \cite{Deb/2022}. 

As it is known, the energy density and pressure of a system of particles can be determined from a general theory, and the physical features of such a system of particles may be derived from the following partition function
\begin{equation} \label{ch4eq1}
\mathcal {Z}=\sum _{N_i, \epsilon } e^{ - \zeta (E_{N_i, \epsilon } - \sum _i N_i \mu _i) } \,,
\end{equation}
where $\zeta = \frac{1}{ k_B T}$, $N_i$ stands for the particle number and $\mu_i$ is the chemical potential. We can realize that in general, the microscopic energy $E_{N_i, \epsilon }$ is a function of the particle number $N_i$, the masses of the particles, the volume of the system $V$, and other quantum numbers $\epsilon$. Therefore, we may write $E_{N_i, \epsilon } = f(N_i, m_i, V, \epsilon )$. The pressure of the system is derived from the following relation
\begin{eqnarray}\label{ch4eq2}
p_0 = -\Xi = \frac{1}{\zeta V} \ln \mathcal {Z}\,,
\end{eqnarray}
where $\Xi$ represents the thermodynamic potential which depend on chemical potential $\mu_i$, the mass of the particle $m_i$, and the temperature $T$. Furthermore, the statistical average of the energy density is such that
\begin{eqnarray}\label{ch4eq3}
\bar{E} = - \frac{\partial }{\partial \zeta } \ln \mathcal {Z} + \sum _i \bar{N}_i \mu _i\,,
\end{eqnarray}
where the particle number $N_i$ has the form
\begin{eqnarray}\label{ch4eq4}
\bar{N}_i = \frac{1}{\zeta } \left( \frac{\partial }{\partial \mu _i} \ln \mathcal {Z} \right) _{T, V, m_j} =-V \left( \frac{\partial \Xi }{\partial \mu _i} \right) _{T, m_j}\,.
\end{eqnarray}
We can also observe that the energy of the system is evaluated as
\begin{eqnarray}\label{ch4eq5}
E_0 = \Xi + \sum _i n_i \mu _i\,,
\end{eqnarray}
and also that number density of particles is such that
\begin{equation}\label{ch4eq6}
n_i = \frac{\bar{N}_i }{V} =- \left( \frac{\partial \Xi }{\partial \mu _i}\right) _{T, m_j} \,.
\end{equation}
In the case of the MIT bag model, the internal energy of the strange matter system is described as follows
\begin{equation}\label{ch4eq7}
E_{N_i, \epsilon }^{Bag}= E_{N_i, \epsilon } + B V\,,
\end{equation}
with the correspondent partition function
\begin{eqnarray}\label{ch4eq8}
\mathcal {Z}^{Bag} =\mathcal {Z} e^{- \zeta B V}\,.
\end{eqnarray}
In this case, the number density of particles has the form
\begin{equation}\label{ch4eq9}
n_i^{Bag} = \frac{\bar{N}_i}{V} =  - \left( \frac{\partial }{\partial \mu _i} (\Xi +B) \right) _{T, m_j, E_{N_i, \epsilon }, B }\,,
\end{equation}
where the bag pressure and energy are
\begin{equation}\label{ch4eq10}
p^{Bag} = - (\Xi +B)\,; \qquad E^{Bag} = (\Xi +B) + \sum _i n_i \mu _i\,.
\end{equation}
So, by combining these previous equations with \eqref{ch4eq2} and \eqref{ch4eq5}, we find that
\begin{equation}
    p^{Bag}= p_0 - B\,; \qquad E^{Bag} = (E_0 +B)\,.
\end{equation}
As is known for a relativistic fluid $p_0=E_0/3$. Consequently, by defining that $E^{Bag}=\rho$, we may write $p^{Bag}$ as
\begin{equation}
    p^{Bag} = \frac{1}{3}\,\left(\rho-4\,B\right)\,,
\end{equation}
proving the EoS for the MIT bag model.
%%%%%%%%%%%%%%%%%%%%%%%%%%%%%%%%%%%%%%%%%%%%%%%%%%%%%%%%%%%%%%%%%%%%%%%%%%%%%%%%%%%%%%%%%%%%%%%%%%%%%%%%%%%%%
\subsection{Noncommutative geometry}
The noncommutative geometry arises from insights provided by string theory, which suggests the existence of a fundamental lower limit to distance measurements. At extremely small length scales, the classical notion of coordinates breaks down and they begin to behave as noncommutative operators on a D-brane, effectively discretizing space-time. This discretization is characterized by a commutator of the form
$$[x^\mu,x^\nu]= i \Theta^{\mu\nu},\,$$
where $\Theta^{\mu\nu}$ is an anti-symmetric tensor. An intriguing physical implication of this framework is that, at these small scales, particles are no longer treated as point-like entities but instead as smeared objects. These are modeled using Gaussian or Lorentzian distributions.\\
The geometric contribution of noncommutative geometry can be understood as a self-gravitating source modeled by an anisotropic fluid, with its energy density profile uniquely determined by the tensor $\Theta^{\mu\nu}$ and the effective mass of the system. Here, $\Theta^{\mu\nu}$ is represented as a second-order antisymmetric matrix with dimensions of $\textit{length}^2$, analogous to how the Planck constant governs the democratization of phase space in quantum mechanics. In particular, the standard concept of mass density, typically expressed using the Dirac delta function, becomes invalid in noncommutative space. Instead, the distribution of mass is described by Gaussian or Lorentzian profiles, with their characteristic length scale governed by $\sqrt{\Theta}$, where $\Theta$ is the non-commutativity parameter.\\
For a static, spherically symmetric gravitational source, the geometry associated with a Gaussian distribution reflects the smeared, particle-like nature of the source in a noncommutative setting. The maximum mass $M$ remains a key feature, while the density profile accommodates the diffused nature of the object. For Lorentzian distributions, the total mass $M$ can also be interpreted as a form of unified diffused object. Instead, the mass is treated as an entity diffused over the noncommutative length scale $\sqrt{\Theta}$. These distributions help address the physical implications of short-scale divergences arising from noncommutative coordinates, offering valuable insights into the behavior of black holes, WHs and other astrophysical objects. The noncommutative parameter $\Theta$ plays a crucial role in shaping these effects, providing a deeper understanding of space-time geometry at the smallest scales. \\
This effective approach can be viewed as a significant enhancement to semiclassical gravity, providing a deeper understanding of noncommutative effects. Building on this perspective, various models for noncommutative geometry-inspired black holes have been developed. In particular, Nicolini \cite{Nicolini/2006}, Smailagic \cite{Smailagic1/2003} and Spallucci \cite{Ansoldi/2007} introduced the concept of Schwarzschild-Tangherlini black holes. These models were further extended to include Reissner-Nordström black holes in four dimensions and generalized to higher-dimensional space-times by Rizzo \cite{Rizzo/2006}. This framework was also adapted to study charged black holes in higher dimensions and BTZ black holes in three dimensions. In addition, Ghosh \cite{Ghosh/2018} has investigated noncommutative black holes within the framework of Gauss-Bonnet gravity, offering further insights into the interplay between noncommutative effects and higher-order curvature corrections. The thermodynamic properties of noncommutative black holes have also been thoroughly explored. For example, Banerjee et al. \cite{Banerjee/2008} demonstrated that the classical Bekenstein-Hawking entropy formula could be recovered in noncommutative space-time. These results highlight the reliability of the noncommutative geometry approach in replicating classical results while incorporating the effects of quantum gravity at small scales.\\
The exploration of noncommutative geometry in the astrophysical context was first initiated by Nicolini \cite{Nicolini} and was further developed in subsequent work \cite{Nicolini2}. In the comprehensive review article by Nicolini \cite{Nicolini}, the motivation for incorporating non-commutativity in the study of black holes is discussed in detail, providing a foundational understanding of its relevance and implications. Subsequent research extended this approach to the study of WHs, as explored by Rahaman \cite{Islam, Banerjee} and Hassan \cite{Hassan1}, where noncommutative geometry was used to investigate WH solutions and assess their physical viability. Further advances were made in the context of gravastars, as discussed in Ref.  \cite{Sayantan/Gravastar}, where the effects of noncommutativity were analyzed in these alternative compact objects.\\
The energy density for a spherically symmetric, static, smeared, particle-like gravitational source has been provided in Refs. \cite{Nicolini/2006, Smailagic1/2003}
 \begin{equation}\label{21}
 \rho_{G} =\frac{M e^{-\frac{r^2}{4 \Theta }}}{8 \pi ^{3/2} \Theta ^{3/2}}\,.
 \end{equation}
 The particle mass $M$, instead of being concentrated at a single point, is spread over a region with a characteristic scale of $\sqrt{\Theta}$.\\
 Moreover, the energy density for the Lorentzian case is given in Refs. \cite{Nicolini/2006, Mehdipour11}
 \begin{equation}\label{22}
 \rho_{L} =\frac{\sqrt{\Theta } M}{\pi ^2 \left(\Theta +r^2\right)^2}\,.
 \end{equation}
%%%%%%%%%%%%%%%%%%%%%%%%%%%%%%%%%%%%%%%%%%%%%%%%%%%%%%%%%%%%%%%%%%%%%%%%%%%%%%%%%%%%%%%%%%%%%%%%%%%%%%%%%%%%%
\section{Geometric insights into space-time}
In GR, geometry plays a pivotal role in describing the nature of space-time, with differential geometry serving as the mathematical framework. A key concept in differential geometry is the manifold, which can be understood as a mathematical structure that locally resembles the familiar Euclidean space $\mathbb{R}^n$. Within GR, space-time is modeled as a four-dimensional manifold, equipped with a non-degenerate symmetric metric tensor $g_{\mu\nu}$, where the determinant $g$ is nonzero.\\
A fundamental concept associated with the manifold is the connection, denoted by $\Gamma^{\lambda}_{\;\:\mu\nu}$, which governs how vectors are transported along curves in the manifold while maintaining consistency with the geometric structure of space-time (for further details, see Ref. \cite{a4}). This connection is integral in defining the covariant derivative, $\nabla$, which is adapted for curved space-time. For a tensor $V^{\mu}_{\;\:\nu}$, the covariant derivative is expressed as
\begin{equation}
\nabla_{\gamma} V^{\mu}_{\;\:\nu} = \partial_{\gamma} V^{\mu}_{\;\:\nu} +  V^{\delta}_{\;\:\nu} \Gamma^{\mu}_{\;\:\delta\gamma}- V^{\mu}_{\;\:\delta} \Gamma^{\delta}_{\;\:\nu\gamma} .
\end{equation}
In the context of GR, the connection $\Gamma^\gamma_{\;\:\alpha\beta}$ is specifically related to the metric $g_{\alpha\beta}$ and is referred to as the Levi-Civita connection. This connection is uniquely defined by two key properties: It is torsion-free and preserves the metric under parallel transport. These conditions ensure that\\
$\bullet$ The connection is symmetric in its lower indices
\begin{equation} 
\Gamma^{\lambda}_{\;\:\mu\nu} = \Gamma^{\lambda}_{\;\:\nu\mu}\,. 
\end{equation}
$\bullet$ The metric remains invariant under covariant differentiation
\begin{equation}
    \nabla_\gamma g_{\mu\nu} = 0\,.
\end{equation}
Under these constraints, the Levi-Civita connection can be explicitly determined and is given by
\begin{equation}\label{cr1}
\Gamma^{\lambda}_{\;\:\mu\nu} = \frac{1}{2} g^{\lambda\delta} \left( \partial_\mu g_{\delta\nu} + \partial_{\nu} g_{\mu\delta} - \partial_\delta g_{\mu\nu} \right)\,.
\end{equation}
The curvature of space-time is described by the Riemann tensor, which quantifies the deviation of manifold from flatness. This tensor is computed using the connection as follows
\begin{eqnarray}
R^{\lambda}_{\;\:\mu\nu\sigma} = \partial_\nu \Gamma^{\lambda}_{\;\:\mu\sigma} - \partial_\sigma \Gamma^{\lambda}_{\;\:\mu\nu} + \Gamma^\tau_{\;\:\mu\sigma} \Gamma^{\lambda}_{\;\:\tau\nu} - \Gamma^\tau_{\;\:\mu\nu} \Gamma^{\lambda}_{\;\:\tau\sigma}\,.
\end{eqnarray}
From the Riemann tensor, the Ricci tensor is obtained by contracting the first and third indices
\begin{equation}
R_{\mu\nu} = R^\lambda_{\;\:\mu\lambda\nu} = -R^\lambda_{\;\:\mu\nu\lambda}\,.
\end{equation}
Finally, the Ricci scalar represents the overall curvature can be derived by contracting the Ricci tensor with the metric
\begin{equation}
R = g^{\mu\nu} R_{\mu\nu}. 
\end{equation}
These mathematical constructs- the metric, connection, Riemann tensor, Ricci tensor and Ricci scalar- form the foundation of the geometric formulation of GR, providing a framework to describe the curvature of space-time and its interaction with matter and energy.
%%%%%%%%%%%%%%%%%%%%%%%%%%%%%%%%%%%%%%%%%%%%%%%%%%%%%%%%%%%%%%%%%%%%%%%%%%%%%%%%%%%%%%%%%%%%%%%%%%%%%%%%%%%%
\section{General Relativity}
The dynamics of the gravitational field in GR can be compared to the Poisson equation, $\nabla^2 \Psi = 4\pi\rho$, which governs the gravitational potential in Newtonian gravity. In an empty space, GR simplifies to $R_{\mu\nu} = 0$, analogous to Laplace's equation $\nabla^2 \Psi = 0$. Extending this analogy to scenarios involving matter requires considering the metric as the fundamental field in GR, with all other fields treated as matter fields influencing gravity. The energy-momentum tensor $T_{\mu\nu}$ serves as the gravitational field source, similar to mass density in Newtonian gravity. For a detailed discussion, see Ref. \cite{sp2}, and for challenges in defining this quantity, refer to Refs. \cite{sp3,gr8}.\\
% GR relies on several key principles
% \begin{itemize}
%     \item The connection is symmetric: $\Gamma^\mu_{\nu\gamma} = \Gamma^\mu_{\gamma\nu}$.
%     \item The metric is covariantly conserved: $\nabla_\gamma g_{\mu\nu} = 0$.
%     \item Gravity depends only on the metric, a second-rank tensor field.
%     \item The field equations are second-order covariant partial differential equations.
% \end{itemize}
Further, Einstein's field Eq. \eqref{Einstein field equations} can be derived from the Einstein-Hilbert action
\begin{equation}\label{gr10}
S=\frac{1}{16\pi}\int d^4x\sqrt{-g} R+\int d^4x\sqrt{-g} \mathcal{L}_m\,,
\end{equation}
with $\mathcal{L}_m$ denotes the matter Lagrangian density. On the right-hand side of the equation mentioned above, the first term represents gravity, while the second term corresponds to matter.
%%%%%%%%%%%%%%%%%%%%%%%%%%%%%%%%%%%%%%%%%%%%%%%%%%%%%%%%%%%%%%%%%%%%%%%%%%%%%%%%%%%%%%%%%%%%%%%%%%%%%%%%%%%%%%%%%%%%%%%%%%%%%%%%%%%%
\section{Why modify General Relativity?}
Modifying GR in the context of wormholes is often motivated by the goal of allowing traversable wormholes, which could theoretically enable shortcuts through space-time. In standard GR, traversable wormholes are not naturally supported because they require exotic matter. However, classical GR does not provide a natural source for such exotic matter, making traversable wormholes seemingly unphysical. This limitation has led researchers to explore modifications of GR, which could allow stable wormhole solutions without the need for exotic matter, opening new possibilities in theoretical physics and potential applications in astrophysics and cosmology. Apart from these, GR also faces several cosmological challenges, and they are as follows.\\
The Universe is considered homogeneous and isotropic on large scales, described by the flat Friedmann Lema\^{i}tre Robertson Walker (FLRW) metric
\begin{equation}
ds^2 = dt^2 - a^2(t)d\bar{x}^2 = a^2(\eta)\left[d\eta^2 - d\bar{x}^2\right],
\end{equation}
where $\bar{x}$ represents the 3-space, and $t$ and $\eta$ are cosmic and conformal time, respectively.\\
The $\Lambda$CDM model, the standard cosmological framework, assumes that GR governs cosmic-scale gravity. It attributes the energy composition of the Universe to dark energy (via the cosmological constant $\Lambda$) and DM. $\Lambda$CDM aligns well with most observations, while it raises unresolved issues.\\
$\bullet$ \textbf{Fine-tuning Problem:} The observed energy density of $\Lambda$ $(\sim 10^{-47} \,\, \text{GeV})$ is far smaller than the prediction of quantum field theory $(\sim 10^{74} \,\, \text{GeV})$.\\
$\bullet$ \textbf{Coincidence Problem:} The current dark energy density parameter $\Omega_\Lambda^{(0)} \sim 0.7$ and matter density parameter $\Omega_m^{(0)} \sim 0.3$ are unexpectedly comparable, despite evolving differently over time.\\
$\bullet$ \textbf{$H_0$ Tension:} Discrepancies in the estimates of the Hubble constant $(H_0)$ between early and late universe data ($5\sigma$ tension) and differences in large-scale structure parameters $(3\sigma)$ challenge the validity of the model.\\
To address these challenges, various modified gravity theories, such as those involving geometric alterations and scalar-tensor frameworks, have been proposed. Although some describe late-time cosmic evolution successfully, none have outperformed the $\Lambda$CDM model in matching observational data, highlighting the need for more refined approaches.
%%%%%%%%%%%%%%%%%%%%%%%%%%%%%%%%%%%%%%%%%%%%%%%%%%%%%%%%%%%%%%%%%%%%%%%%%%%%%%%%%%%%%%%%%%%%%%%%%%%%%%%%%%%%%%%%%%%%%%%%%%%%%%%%%%%%
\section{The geometrical trinity of gravity}
A fundamental feature of metric-affine geometry is that when a metric is introduced, the components of a general affine connection, denoted by $\tilde{\Gamma}^{\lambda}_{\;\:\mu\nu}$, can be uniquely decomposed into three distinct terms. This decomposition is expressed as
\begin{equation}\label{gt1} \tilde{\Gamma}^{\lambda}_{\;\:\mu\nu} = \Gamma^{\lambda}_{\;\:\mu\nu} + K^{\lambda}_{\;\:\mu\nu} + L^{\lambda}_{\;\:\mu\nu}\,, 
\end{equation}
where the individual terms represent different geometrical contributions.
\begin{itemize}
    \item \textbf{Levi-Civita connection ($\Gamma^{\lambda}_{\;\:\mu\nu}$):} The term $\Gamma^{\lambda}_{\;\:\mu\nu}$ corresponds to the Levi-Civita connection, which is defined as the torsion-free and metric compatible connection. It is given by the well-known Eq. \eqref{cr1}.
% \begin{equation} 
% \Gamma^{\lambda}_{\;\:\mu\nu} = \frac{1}{2} g^{\lambda\delta} \left( \partial_\mu g_{\delta\nu} + \partial_{\nu} g_{\mu\delta} - \partial_\delta g_{\mu\nu} \right)\,.
% \end{equation}
\item \textbf{Contortion tensor ($K^{\lambda}_{\;\:\mu\nu}$):} The second term, $K^{\lambda}_{\;\:\mu\nu}$, is the contortion tensor, which encodes the effects of torsion in the geometry. It is expressed as
\begin{equation} 
K^{\lambda}_{\;\:\mu\nu} = \frac{1}{2} g^{\lambda\delta} \left(\mathcal{T}_{\nu\delta\mu}+\mathcal{T}_{\mu\delta\nu}-\mathcal{T}_{\delta\mu\nu}\right)\,,
\end{equation}
where the torsion tensor $\mathcal{T}^{\lambda}_{\;\:\:\:\mu\nu}$ is defined as $\mathcal{T}^{\lambda}_{\;\:\:\:\mu\nu}= \gc{\Gamma}^{\lambda}_{\;\:\mu\nu}-\gc{\Gamma}^{\lambda}_{\;\:\nu\mu}$\,,
with $\gc{\Gamma}$ represents teleparallel connection.
\item \textbf{Disformation tensor ($L^{\lambda}_{\;\:\mu\nu}$):} The final term, $L^{\lambda}_{\;\:\mu\nu}$, is associated with non-metricity and is given by
\begin{equation}
L^{\lambda}_{\;\:\mu\nu} = \frac{1}{2} g^{\lambda\delta} \left( Q_{\delta\mu\nu}-Q_{\mu\nu\delta} - Q_{\nu\delta\mu} \right),
\end{equation}
where $Q_{\lambda\mu\nu}$ represents the non-metricity tensor, which quantifies how the connection fails to preserve the metric under parallel transport.
\end{itemize}
The decomposition of the affine connection in Eq. \eqref{gt1} highlights the interaction between these geometric aspects in a metric-affine space.\\
Moreover, this decomposition has significant implications for the curvature tensor. Specifically, the Riemann curvature tensor $\gc{R}^{\lambda}_{\;\:\mu\nu\sigma}$ associated with the general connection $\tilde{\Gamma}^{\lambda}_{\;\:\mu\nu}$ can also be expressed as a sum of contributions. This takes the form
\begin{equation}\label{eq:gennaffcurvdec} \gc{R}^\lambda_{\;\:\mu\nu\sigma} = R^\lambda_{\;\:\mu\nu\sigma} + \nabla_\nu \gc{\mathcal{D}}^\lambda_{\;\:\mu\sigma} - \nabla_\sigma \gc{\mathcal{D}}^\lambda_{\;\:\mu\nu} + \gc{\mathcal{D}}^\lambda_{\;\:\tau\nu} \gc{\mathcal{D}}^\tau_{\;\:\mu\sigma} - \gc{\mathcal{D}}^\lambda_{\;\:\tau\sigma} \gc{\mathcal{D}}^\tau_{\;\:\mu\nu}, 
\end{equation}
where $\nabla_\gamma$ denotes the covariant derivative w.r.t. the Levi-Civita connection, and $\gc{\mathcal{D}}^{\lambda}_{\;\:\mu\nu}$ is the distortion tensor.\\
The distortion tensor $\gc{\mathcal{D}}^{\lambda}_{\;\:\mu\nu}$ is defined as the deviation of the general affine connection $\tilde{\Gamma}^{\lambda}_{\;\:\mu\nu}$ from the Levi-Civita connection $\Gamma^{\lambda}_{\;\:\mu\nu}$
\begin{equation} 
\gc{\mathcal{D}}^{\lambda}_{\;\:\mu\nu} = \tilde{\Gamma}^{\lambda}_{\;\:\mu\nu} - \Gamma^{\lambda}_{\;\:\mu\nu} = K^{\lambda}_{\;\:\mu\nu} + L^{\lambda}_{\;\:\mu\nu}. 
\end{equation}
This framework enables a unified treatment of curvature, torsion and non-metricity, offering a versatile approach for exploring the dynamics of metric-affine geometry and its applications in gravitational theories.\\
Two significant scenarios arise in metric-affine geometry when the curvature tensor $\gc{R}^{\lambda}_{\;\:\mu\nu\sigma}$ vanishes:
\begin{itemize}
    \item \textbf{Metric Teleparallel Connection:} In this connection, the Riemann tensor can then be expressed solely in terms of the contortion tensor $K^{\lambda}_{\;\:\mu\nu}$
    \begin{equation} 
    R^\lambda_{\;\:\mu\nu\sigma} = K^\lambda_{\;\:\tau\sigma} K^\tau_{\;\:\mu\nu} - K^\lambda_{\;\:\tau\nu} K^\tau_{\;\:\mu\sigma} + \nabla_\sigma K^\lambda_{\;\:\mu\nu} - \nabla_\nu K^\lambda_{\;\:\mu\sigma}\,. 
    \end{equation}
    \item \textbf{Symmetric Teleparallel Connection:} In this scenario, the affine connection is symmetric and characterized by vanishing torsion. The Riemann tensor, in this case, can be formulated in terms of disformation tensor $L^{\lambda}_{\;\:\mu\nu}$, which accounts for non-metricity
\begin{equation} 
R^\lambda_{\;\:\mu\nu\sigma} = L^\lambda_{\;\:\tau\sigma} L^\tau_{\;\:\mu\nu} - L^\lambda_{\;\:\tau\nu} L^\tau_{\;\:\mu\sigma} + \nabla_\sigma L^\lambda_{\;\:\mu\nu} - \nabla_\nu L^\lambda_{\;\:\mu\sigma}. 
\end{equation}
\end{itemize}
These two cases illustrate the versatility of metric-affine geometry in accommodating different gravitational frameworks. The metric teleparallel approach emphasizes torsion, while the symmetric teleparallel framework focuses on non-metricity, providing distinct paths for constructing modified theories of gravity. Both cases highlight the underlying geometric richness and the potential for new physical insights.\\
From the aforementioned decompositions, the Ricci scalar in the metric teleparallel framework can be expressed as
\begin{equation}\label{gt2}
{R} = K^{\lambda}{}_{\nu\mu}K^{\nu\mu}{}_{\lambda} - K^{\lambda}{}_{\nu\lambda}K^{\nu\mu}{}_{\mu} - 2{\nabla}_{\lambda}K^{\lambda\mu}{}_{\mu} = \mathcal{T} + 2{\nabla}_{\lambda}\mathcal{T}_{\mu}{}^{\mu\lambda}\,,
\end{equation}
where $\mathcal{T}$ represents the torsion scalar. Similarly, in the symmetric teleparallel framework, the Ricci scalar becomes
\begin{equation}\label{gt3} 
{R} = {L}^{\lambda\mu\nu}{L}_{\nu\lambda\mu} - {L}^{\lambda}{}_{\lambda\nu}{L}^{\nu\mu}{}_{\mu} + {\nabla}_{\mu}{L}_{\lambda}{}^{\lambda\mu} - {\nabla}_{\lambda}{L}^{\lambda\mu}{}_{\mu} = {Q} + {\nabla}_{\mu}{Q}_{\lambda}{}^{\lambda\mu} - {\nabla}_{\lambda}{Q}^{\lambda\mu}{}_{\mu}\,, 
\end{equation}
where $Q$ is the non-metricity scalar.\\
The Ricci scalar serves as a foundational element in the geometric part of the Einstein-Hilbert action of GR, which is defined as
\begin{equation}\label{gt4} 
S_{EH} = \frac{1}{16\pi} \int d^4x \sqrt{-g} \, R\,. 
\end{equation}
By adopting either the metric teleparallel or symmetric teleparallel geometry, the Einstein-Hilbert action can be reformulated into alternative equivalent forms. Notably, both Ricci scalar representations are in Eqs. \eqref{gt2} and \eqref{gt3} contain total divergence terms. These terms contribute boundary components when inserted into the Einstein-Hilbert action \eqref{gt4}. However, neglecting the boundary terms leads to two alternative actions:
\begin{itemize}
    \item \textbf{Teleparallel Equivalent of General Relativity (TEGR):}
\begin{equation}
S_{TEGR} = \frac{1}{16\pi} \int d^4x \sqrt{-g} \, \mathcal{T}\,.
\end{equation}
    \item \textbf{Symmetric Teleparallel Equivalent of General Relativity (STEGR):}
\begin{equation} \label{STEGR}
S_{STEGR} = \frac{1}{16\pi} \int d^4x \sqrt{-g} \, Q\,. 
\end{equation}
\end{itemize}
These formulations, collectively referred to as the \textit{geometric trinity of gravity} \cite{sp6}, establish the equivalence of GR, TEGR and STEGR in terms of their dynamics, despite their distinct geometric interpretations. In all three frameworks, the metric remains the sole fundamental variable, underscoring their shared foundation. This equivalence highlights the flexibility of gravitational theories in providing alternative but consistent descriptions of the same physical phenomena.
%%%%%%%%%%%%%%%%%%%%%%%%%%%%%%%%%%%%%%%%%%%%%%%%%%%%%%%%%%%%%%%%%%%%%%%%%%%%%%%%%%%%%%%%%%%%%%%%%%%%%%%%%%%%%%%%%%%%%%%%%%%%%%%%%%%%
\section{Some modified gravity theories}
Modified gravity alters the geometric side of Einstein's equations rather than the stress-energy tensor, modifying the structure of gravity itself. Key approaches include:\\
$\bullet$ \textbf{Higher-order equations:} Some theories, like $f(R)$ gravity \cite{mg2}, introduce higher-order derivatives in the field equations, avoiding instabilities such as Ostrogradski's instability \cite{mg1}.\\
$\bullet$ \textbf{Higher dimensions:} Theories like Kaluza-Klein \cite{mg3} and Dvali-Gabadadze-Porrati \cite{mg4} extend GR's (3+1) framework by adding extra spatial or temporal dimensions.\\
$\bullet$ \textbf{Torsion and non-metricity:} Models incorporate torsion ($\mathcal{T}$) \cite{mg5} or non-metricity ($Q$) \cite{sp6}, with modified versions such as $f(\mathcal{T})$ or $f(Q)$ gravity altering gravitational dynamics.\\
$\bullet$ \textbf{Additional fields:} Introducing scalar, vector, or tensor fields leads to models like the Einstein-Aether theory \cite{mg7}, quintessence \cite{mg8} and Born-Infeld \cite{mg9}, with some mimicking dark energy effects.\\
These approaches are central to many modified gravity theories, which will be discussed in further detail in the following subsections.
%%%%%%%%%%%%%%%%%%%%%%%%%%%%%%%%%%%%%%%%%%%%%%%%%%%%%%%%%%%%%%%%%%%%%%%%%%%%%%%%%%%%%%%%%%%%%%%%%%%%%%%%%%%%%%%%%%%%%%%%%%%%%%%%%%%%
\subsection{Symmetric teleparallel gravity as an equivalent description of General Relativity and beyond}
In this subsection, we delve into the theory of symmetric teleparallel gravity, which serves as an alternative geometric formulation of gravity. This theory offers an equivalent description of GR and provides a foundation for exploring alternative theories of gravity. The key ingredient in symmetric teleparallelism is the concept of non-metricity, which replaces curvature as the central geometric quantity driving gravitational interactions.\\
The geometry in this framework is defined by a Lorentzian metric $ds^2$ and a general linear connection $\tilde{\Gamma}$. The metric $ds^2$ is represented in terms of co-frame as
\begin{equation}
ds^2 = g_{\mu\nu}dx^\mu \otimes dx^\nu = \eta_{ab}\vartheta^a \otimes \vartheta^b,
\end{equation}
where $\eta_{ab}$ is the Minkowski metric, representing the local flat space-time structure, and $\vartheta^a$ is the dual co-frame corresponding to the general frame $e_a$. The co-frame and frame satisfy the orthogonality condition $\vartheta^b(e_a) = \delta^b_a$, ensuring consistency between the metric and the local frame field.\\
The general linear connection $\tilde{\Gamma}^a_b$, which governs the covariant derivative, can be decomposed into three distinct components as follows \cite{fq0}
\begin{equation}\label{fdd}
\tilde{\Gamma}^a_{\;\:b} = \omega^a_{\;\:b} + K^a_{\;\:b} + L^a_{\;\:b},
\end{equation}
where \\
$\bullet$ $\omega^a_{\;\:b}$ denotes the Levi-Civita 1-form connection, corresponding to the torsion-free and metric-compatible part of the connection.\\
$\bullet$ $K^a_{\;\:b}$ represents the contorsion 1-form, which accounts for torsion contributions.\\
$\bullet$ $L^a_{\;\:b}$ is the disformation 1-form, which encapsulates the non-metricity of the connection.

This decomposition highlights the versatility of the general connection $\tilde{\Gamma}$, as it separates the contributions of curvature, torsion and non-metricity to space-time geometry. In the case of symmetric teleparallel gravity, the contorsion and curvature components are set to zero, leaving only non-metricity as the driver of gravitational interactions.\\
% The non-metricity 1-form, which quantifies the failure of the metric to remain covariantly constant under the connection, is defined as
% \begin{equation}
% Q_{ab} = \frac{1}{2}D(\tilde{\Gamma})\eta_{ab} = \tilde{\Gamma}_{(ab)} = -A^c_{\;\:b}\eta_{ac} - A^c_{\;\:a}\eta_{cb},
% \end{equation}
% where $A^a_{\;\:b} = K^a_{\;\:b} + L^a_{\;\:b}$. Additionally, the condition $D(\omega)\eta_{ab} = 0$ ensures that the Levi-Civita connection maintains metric compatibility.\\
For the specific scenario where the contorsion vanishes ($K^a_{\;\:b}=0$), the components of the non-metricity tensor can be expressed as
\begin{equation}\label{ga2}
Q_{\lambda\mu\nu} = \nabla_\lambda g_{\mu\nu} = -L^\alpha_{\;\:\lambda\mu}g_{\alpha\nu} - L^\alpha_{\;\:\lambda\nu}g_{\alpha\mu}.
\end{equation}
Here, the non-metricity scalar $Q$ is defined as
\begin{equation}
Q = -g^{\mu\nu}\left(L^\alpha_{\;\:\beta\nu}L^\beta_{\;\:\mu\alpha} - L^\beta_{\;\:\alpha\beta}L^\alpha_{\;\:\mu\nu}\right).
\end{equation}
This formulation demonstrates that the dynamical behavior of gravity in STEGR is governed entirely by the scalar $Q$, which encapsulates the effects of non-metricity. As a result, STEGR serves as an equivalent description of GR, but with the geometric properties of space-time characterized by non-metricity rather than curvature or torsion.\\
To facilitate further analysis, we introduce the non-metricity conjugate $P^{\lambda}_{\;\:\mu\nu}$, which is defined as
\begin{equation}
P^{\lambda}_{\;\:\mu\nu}=-\frac12L^\lambda_{\;\:\mu\nu}+\frac{1}{4}\left( Q^\lambda-\tilde{Q}^\lambda \right)g_{\mu\nu}-\frac14\delta^\lambda_{\;\:(\mu}Q_{\nu)}\,.
\end{equation}
Here, $Q_{\lambda}=Q_{\lambda}\,^{\mu}\,_{\mu}$ and $\tilde{Q}_\lambda=Q^\mu_{\;\lambda\mu}$ represent the two independent traces of the non-metricity tensor. Using this definition, the non-metricity scalar $Q$ can be rewritten in terms of $P^\lambda_{\;\mu\nu}$ and the non-metricity tensor $Q_{\lambda\mu\nu}$ as
\begin{equation}\label{ga3}
Q = -g^{\mu\nu}\left(L^\alpha_{\;\:\beta\nu}L^\beta_{\;\:\mu\alpha} - L^\beta_{\;\:\alpha\beta}L^\alpha_{\;\:\mu\nu}\right) = -P^{\lambda\mu\nu}Q_{\lambda\mu\nu}.
\end{equation}
% To derive the field equations, we take the variation of the Lagrangian in Eq. \eqref{fq1234} with respect to the metric $g$. The resulting equation in differential form is
% \begin{equation}
% \left[ D(\omega)\left( L^{ab}-Q^{ab} \right)+L^a_{\;c}\wedge L^{cb} \right]\wedge h_{kab}=2\kappa\tau_k,
% \end{equation}
% where $\tau_k$ denotes the energy-momentum 3-form associated with the matter content. In component form, this equation becomes
% \begin{equation}
% 2\bar{\nabla}_\alpha P^\alpha_{\;\mu\nu} - \left(L^\alpha_{\beta\nu}L^\beta_{\mu\alpha} - L^\beta_{\alpha\beta}L^\alpha_{\mu\nu}\right) - \frac{1}{2}g_{\mu\nu}Q = \kappa T_{\mu\nu}\,.
% \end{equation}
% Here, $\bar{\nabla}_\alpha$ represents the covariant derivative with respect to the Levi-Civita connection $\omega$, and $T_{\mu\nu}$ is the energy-momentum tensor of the matter fields.\\
In $f(Q)$ gravity, the replacement of the non-metricity scalar $Q$ with an arbitrary function $f(Q)$ in the gravitational action (Eq. \eqref{STEGR}) allows for a generalization of the standard STEGR framework. Adding the matter Lagrangian $\mathcal{L}_m$, the total action becomes \cite{fq1}
\begin{equation}
S= \frac{1}{16\pi}\int\sqrt{-g}f(Q)d^4x+\int \mathcal{L}_m\,\sqrt{-g}\,d^4x\,.
\end{equation}
The variation of this action with respect to the metric leads to the modified field equation for $f(Q)$ gravity as
% \begin{equation}
% 2\bar{\nabla}_\alpha f_Q P^\alpha_{\;\mu\nu} - f_Q\left(L^\alpha_{\beta\nu}L^\beta_{\mu\alpha} - L^\beta_{\alpha\beta}L^\alpha_{\mu\nu}\right) - \frac{1}{2}g_{\mu\nu}f = \kappa T_{\mu\nu},
% \end{equation}
% where $f_Q = \frac{df(Q)}{dQ}$. Rewriting the term $\bar{\nabla}_\alpha$ in a more unified form by utilizing $\nabla_\alpha$, the covariant derivative corresponding to the general connection $\tilde{\Gamma}$, the field equations take the form
\begin{equation}\label{ga44}
    -\frac{2}{\sqrt{-g}}\nabla_\alpha \left( \sqrt{-g}f_QP^\alpha_{\;\:\mu\nu} \right)-f_Q\left(P_{\mu\alpha\beta}\,Q_\nu^{\;\:\;\:\alpha\beta}-2\,Q^{\alpha\beta}_{\;\:\;\:\;\:\mu}\, P_{\alpha\beta\nu}\right)-\frac{1}{2}g_{\mu\nu}f= 8\pi T_{\mu\nu}\,,
\end{equation}
where $f_Q = \frac{df}{dQ}$.
Moreover, varying the action with respect to the connection, we obtain an additional constraint 
\begin{equation}\label{FQ6}
\nabla_{\mu}\nabla_{\gamma}\left( \sqrt{-g}f_Q P^{\gamma}_{\;\:\mu\nu}\right)=0\,.
\end{equation}
%%%%%%%%%%%%%%%%%%%%%%%%%%%%%%%%%%%%%%%%%%%%%%%%%%%%%%%%%%%%%%%%%%%%%%%%%%%%%%%%%%%%%%%%%%%%%%%%%%%%%%%%%%%%
\subsubsection{The covariant formulation of $f(Q)$ gravity}\label{ch3sec2}
In this framework, the connection defining the geometry is characterized by the absence of both curvature and torsion. This constraint arises from the conditions $R^a_{\;\:b}(\tilde{\Gamma})=0$ and $T^a(\tilde{\Gamma}) = 0$. However, under these constraints, the complete connection $\tilde{\Gamma}$ cannot typically be uniquely specified. This ambiguity necessitates the development of a systematic method to define $\tilde{\Gamma}$ or its individual components within any chosen reference frame before delving into the specific case of spherically symmetric configurations, which is a key focus of this work. It has been demonstrated, as highlighted in Ref. \cite{g1}, that one can always adopt a gauge in which $\tilde{\Gamma}^a_{\;\:b} = 0$. This choice ensures that, alongside the vanishing of $K^a_{\;\:b}$, the conditions $R^a_{\;\:b}(\tilde{\Gamma}) = 0$ and $T^a(\tilde{\Gamma}) = 0$ are simultaneously satisfied. Such a gauge choice is commonly employed in studies of STEGR, as noted in Refs. \cite{sp5, g3}, where it is widely referred to as the coincident gauge \cite{fq1}. This approach has proven particularly useful in simplifying the mathematical framework while maintaining consistency with the fundamental principles of STEGR.\\
It is crucial to emphasize that although $K^a_{\;\:b}$ is a tensor and its components are zero in all reference frames, the connection $\tilde{\Gamma}^a_{\;\:b}$ does not share this property because it is not a tensor. Consequently, the condition $\tilde{\Gamma}^a_{\;\:b} = 0$ does not imply that the components $\tilde{\Gamma}^{\lambda}_{\;\mu\:\nu}$ disappear universally in all coordinate systems. Instead, the relation $\tilde{\Gamma}^{\lambda}_{\;\mu\:\nu}=0$, and, by extension, $L^\lambda_{\;\:\mu\nu} = -{\Gamma}^{\lambda}_{\;\:\mu\nu}$ holds true only within a specific class of frames.\\
Under a general diffeomorphism or frame transformation $\chi(x)$, the components of the connection transform according to the expression provided in Ref. \cite{fq1}
\begin{equation}\label{cv1}
\tilde{\Gamma}^\lambda_{\;\:\mu\nu} =\frac{\partial x^\lambda}{\partial \chi^\alpha}\partial_\mu\partial_\nu\chi^\alpha.
\end{equation}
This equation demonstrates the pivotal role of $\tilde{\Gamma}^{\lambda}_{\;\mu\:\nu}$ in the covariant formulation of $f(Q)$ gravity. It ensures that the framework remains Lorentz covariant while operating within the non-metricity paradigm. However, this also raises the fundamental question of identifying the specific frame in which $\tilde{\Gamma}^{\lambda}_{\;\mu\:\nu}=0$ is valid, a challenge that requires careful consideration. To address this issue, one may draw valuable insights from analogous scenarios in the $f(T)$ gravity counterpart, as discussed in Ref. \cite{g4}.\\
For local Lorentz transformations of the form $\chi^\mu = \Lambda^\mu_{\;\:\nu} x^\nu$, Eq. \eqref{cv1} simplifies to
\begin{equation}\label{cv2}
\tilde{\Gamma}^\lambda_{\;\:\mu\nu} = \Lambda^\lambda_{\;\:\alpha} \partial_\mu(\Lambda^{-1})^\alpha_\nu,
\end{equation}
as expressed in Eq. \eqref{cv2}. This result demonstrates that the connection $\tilde{\Gamma}$ is purely inertial and arises solely from the transformation properties of the reference frame. In the non-metricity framework, the gravitational effects are entirely encoded within the non-metricity tensor $Q_{\lambda\mu\nu}$, which encapsulates all and only the information related to gravity. Consequently, when gravity is canceled, represented by $Q_{\lambda\mu\nu}|_{G=0} = 0$, the gravitational contribution vanishes, and we obtain 
\begin{equation}
  \left. L^\lambda_{\;\:\mu\nu}\right|_{G=0}=\left.\left( \frac12Q^\lambda_{\;\:\mu\nu}-Q_{\;\:(\mu\nu)}^{\lambda} \right)\right|_{G=0}=0,
\end{equation}
where $L^\lambda_{\mu\nu}$ reflects the gravitational disformation contributions.\\
Substituting this result into the expression for the full connection, we find that
\begin{equation}
\left. \tilde{\Gamma}^\lambda_{\;\:\mu\nu}\right|_{G=0}=\left.{\Gamma}^{\lambda}_{\;\:\mu\nu}\right|_{G=0}+\left. L^\lambda_{\mu\nu}\right|_{G=0}=\left.{\Gamma}^{\lambda}_{\;\:\mu\nu}\right|_{G=0}.
\end{equation}
This equation signifies that, when gravity is absent, the full connection $\tilde{\Gamma}^\lambda_{\;\:\mu\nu}$ reduces to the Levi-Civita connection $\Gamma^\lambda_{\;\:\mu\nu}$, as the non-metricity effects vanish.\\
Moreover, since $R^a_{\;\:b}(\tilde{\Gamma})=0$ and $T^a(\tilde{\Gamma}) = 0$, and all gravitational phenomena are attributed to $Q_{\lambda\mu\nu}$, the inertial connection $\tilde{\Gamma}^\lambda_{\;\:\mu\nu}$ remains unaffected by the presence or absence of gravity. This can be expressed as
\begin{equation}
 \tilde{\Gamma}^\lambda_{\;\:\mu\nu}=\left. \tilde{\Gamma}^\lambda_{\;\:\mu\nu}\right|_{G=0}=\left.{\Gamma}^{\lambda}_{\;\:\mu\nu}\right|_{G=0}.
\end{equation}
This relationship provides a practical approach to determining the components of the connection
$\tilde{\Gamma}^\lambda_{\;\:\mu\nu}$. For example, in a Cartesian coordinate frame, the inertial connection components satisfy $\tilde{\Gamma}^\lambda_{\;\:\mu\nu}=\left.{\Gamma}^{\lambda}_{\;\:\mu\nu}\right|_{G=0}=0$. This property is often encountered in spatially flat cosmological models, where Cartesian coordinates simplify the connection to its trivial form.\\
In other coordinate frames, the components of $\tilde{\Gamma}^\lambda_{\;\:\mu\nu}$ can be computed from Eq. \eqref{cv1} or Eq. \eqref{cv2} relative to the Cartesian frame or derived directly using $\left.{\Gamma}^{\lambda}_{\;\:\mu\nu}\right|_{G=0}$. This flexibility highlights the adaptability of the connection in describing different geometries while maintaining consistency within the non-metricity framework.
%%%%%%%%%%%%%%%%%%%%%%%%%%%%%%%%%%%%%%%%%%%%%%%%%%%%%%%%%%%%%%%%%%%%%%%%%%%%%%%%%%%%%%%%%%%%%%%%%%%%%%%%%%%%%
\subsection{$f(Q, T)$ gravity}
The extension of $f(Q)$ gravity has been proposed, such as $f(Q, T)$ gravity \cite{fqt}, where the function $f$ depends on both the non-metricity scalar $Q$ and the trace of the energy-momentum tensor $T$. This introduces an explicit coupling between geometry (via $Q$) and matter (via $T$), leading to additional modifications to the field equations.
This framework is based on the following action
\begin{equation}\label{fqt1}
S=\frac{1}{16\pi}\int f(Q, T)\sqrt{-g}\,d^4x+\int \mathcal{L}_m\,\sqrt{-g}\,d^4x\,.
\end{equation}
To derive the gravitational field equations, one varies the action with respect to the metric. This variation leads to the following equations of motion
\begin{equation}\label{fqt2}
\frac{-2}{\sqrt{-g}}\bigtriangledown_\alpha\left(\sqrt{-g}\,f_Q\,P^\alpha_{\;\:\mu\nu}\right)-\frac{1}{2}g_{\mu\nu}f
+f_T \left(T_{\mu\nu} +\Xi_{\mu\nu}\right)
-f_Q\left(P_{\mu\alpha\beta}\,Q_\nu^{\;\:\alpha\beta}-2\,Q^
{\alpha\beta}_{\;\:\;\:\;\:\mu}\,P_{\alpha\beta\nu}\right)=8\pi T_{\mu\nu},
\end{equation}
where $f_Q=\frac{\partial f}{\partial Q}$ and $f_T=\frac{\partial f}{\partial T}$.\\
%The tensor $P^{\alpha}_{\,\,\,\mu\nu}$ in the above equation is related to the non-metricity $Q$ and encapsulates the geometric structure of the theory.\\
The energy-momentum tensor for matter fields in the space-time is defined as
\begin{equation}\label{fqt3}
T_{\mu\nu}=-\frac{2}{\sqrt{-g}}\frac{\delta\left(\sqrt{-g}\,\mathcal{L}_m\right)}{\delta g^{\mu\nu}},
\end{equation}
while $\Xi_{\mu\nu}$, which arises due to the variation of $T_{\mu\nu}$ with respect to the metric, is given by
\begin{equation}\label{fqt4}
\Xi_{\mu\nu}=g^{\alpha\beta}\frac{\delta T_{\alpha\beta}}{\delta g^{\mu\nu}}.
\end{equation}
In certain cases, $\Xi_{\mu\nu}$ can be expressed in a simplified form. Following the analysis presented in Refs. \cite{fqt, Harko/2011}, the simplified form of $\Xi_{\mu}^{\,\,\,\nu}$ is
\begin{equation}\label{ch3eq1}
\Xi_{\mu}^{\,\,\,\nu}=\delta_{\mu}^{\,\,\,\nu}\,\mathcal{L}_m-2\,T_{\mu}^{\,\,\,\nu}\,.
\end{equation}
This formulation illustrates the coupling between geometry and matter through the function $f(Q, T)$, providing a platform to explore modified gravity theories. The dependence on both non-metricity $Q$ and the trace $T$ allows for richer phenomenology compared to traditional gravitational models. Such a theoretical framework has implications for understanding cosmic evolution, including late-time acceleration, and it invites further exploration into the dynamics of matter and geometry.\\
Moreover, various extensions of $f(Q)$ gravity have been proposed, including $f(Q, C)$ (where $C$ represents the boundary term) \cite{A. De}, $f(Q, \mathcal{L}_m)$ \cite{Ayush}, etc.
%%%%%%%%%%%%%%%%%%%%%%%%%%%%%%%%%%%%%%%%%%%%%%%%%%%%%%%%%%%%%%%%%%%%%%%%%%%%%%%%%%%%%%%%%%%%%%%%%%%%%%%%%%%%
\section{Conclusions}
In this chapter, we provide a comprehensive introduction to the key concepts underlying our study. We began with a detailed examination of the historical evolution of WH geometry, tracing its conceptual origins, and highlighting significant developments that have shaped our current understanding. In addition, we explored the concept of DM, focusing on widely accepted DM density profiles and their relevance to astrophysical phenomena.\\
Furthermore, we discussed the physical foundations of the MIT bag model and its applications, along with the framework of noncommutative geometry and its implications for modern theoretical physics. To establish a strong foundation, we revisit the fundamental principles of GR and critically examine its limitations, which have motivated the development of alternative approaches. This naturally led to an exploration of various modified theories of gravity, each of which offers potential solutions to the challenges posed by GR.\\
The chapter concluded with a summary of these modifications and their significance in advancing our understanding of WH geometry and related phenomena. In the following chapters, we will expand on this foundation, addressing specific challenges in WH geometry, and demonstrating how the theoretical insights introduced here can be applied to resolve them.
%%%%%%%%%%%%%%%%%%%%%%%%%%%%%%%%%%%%%%%%%%%%%%%%%%%%%%%%%%%%%%%%%%%%%%%%%%%%%%%%%%%%%%%%%%%%%%%%%%%%%%%%%%

% Chapter 2

\chapter{Wormhole formations in the galactic halos supported by dark matter models and monopole charge within $f(Q)$ gravity} % Main chapter title
\label{Chapter2} 
% For referencing the chapter elsewhere, use \ref{Chapter1} 

\lhead{Chapter 2. \emph{Wormhole formations in the galactic halos supported by dark matter models and monopole charge within $f(Q)$ gravity}} % This is for the header on each page - perhaps a shortened title
\blfootnote{*The work in this chapter is covered by the following publication:\\
\textit{Wormhole formations in the galactic halos supported by dark matter models and monopole charge within $f(Q)$ gravity}, European Physical Journal Plus, \textbf{84}, 643 (2024).}
%----------------------------------------------------------------------------------------
This chapter investigates the possibility of traversable WHs in the galactic region, supported by DM models and the monopole charge within the framework of $f(Q)$ gravity. The detailed study of the work is described as follows:
\begin{itemize}
    \item WH solutions are analyzed using various redshift functions under different $f(Q)$ models, with shape functions derived for PI and NFW DM profiles under linear $f(Q)$ gravity.
    \item For nonlinear $f(Q)$ models, an embedding class- I approach is employed, demonstrating that shape functions satisfy the flare-out conditions under asymptotic backgrounds for each DM profile.
    \item The energy conditions at the throat of the WH are examined, revealing the influence of the monopole parameter $\eta$ on the violation of the energy conditions, particularly the NEC, in the case of a linear model. 
    \item Certain nonlinear $ f(Q)$ models, such as $f(Q)=Q+mQ^n$ and $f(Q)=Q+\frac{\beta}{Q}$ are found to be incompatible with WH solutions. In addition, the Volume Integral Quantifier (VIQ) technique is followed to calculate the amount of exotic matter needed, confirming that only a minimal amount of exotic matter is required to sustain traversable WHs.

\end{itemize}

%\blindtext
%----------------------------------------------------------------------------------------
%\clearpage
%----------------------------------------------------------------------------------------
\section{Introduction}
\indent WHs serve as valuable frameworks for exploring and testing various gravitational theories \cite{S. Capozziello 1, S. Capozziello 2}, offering insights into space-time structure and exotic matter. Among these, a notable contender is $f(Q)$ gravity, introduced by Jimenez et al. \cite{fq1}, wherein the gravitational interaction is governed by the non-metricity $Q$. Over the past years, $f(Q)$ gravity has undergone extensive observational scrutiny, with Lazkoz et al. \cite{R. Lazkoz} proposed intriguing constraints on this theory. Furthermore, Mandal et al. \cite{S. Mandal} successfully demonstrated the viability of $f(Q)$ gravity models with respect to energy conditions. Moreover, Hassan et al. \cite{Zinnat 2} have systematically explored how the Generalized Uncertainty Principle (GUP) influences the structure and properties of the Casimir WH space-time within the framework of $f(Q)$ gravity. Significantly, numerous captivating astrophysical investigations have been undertaken within the $f(Q)$ gravity and its extended gravity theory. Interested readers can dive deeper into this rich literature, with references \cite{L. Heisenberg, F. Parsaei, O. Sokoliuk 1, O. Sokoliuk 3, S. Pradhan, Debasmita} providing valuable sources for additional exploration.\\
\indent It is well known that, unlike the electric charge, a magnetic monopole can not exist for Maxwell's equations in $\mathbb{R}^4$ (for simply connected manifolds). However, Dirac \cite{P. A. M. Dirac} first pointed out that if one introduces a quantization condition with a magnetic monopole (changing the topology of the manifold by taking a Dirac string out), then one can show that the electric charges are quantized. Although it was a great and natural way to explain the discreteness of the electric charge, it required a non-singular transformation (like removing the Dirac string); it was later discovered by Hooft \cite{G.'t Hooft} and Polyakov \cite{A. M. Polyakov} that for non-abelian gauge theories, there can be monopole. Arising from spontaneous symmetry breaking and ground state having non-trivial topology. Also, unlike the Dirac monopole, there is a singular transformation required to get these monopoles, so it was conjectured in their paper that such a monopole could naturally arise during the early Universe when the GUT phase transitions were taking place. Later, Barriola and Valenkin \cite{M. Barriola} said that a scalar field having $SO(3)$ symmetry near the Schwarchuild metric could go through a spontaneous symmetry-breaking mechanism and give a monopole. In this chapter, we study the geometry of the WH under the influence of the monopole and galactic halo in $f(Q)$ gravity. Although the study of WHs in the context of monopole \cite{Tayde 6, Rahaman11, S. Sarkar, P. Das} and galactic halo \cite{pi3, Rahaman11, G. Mustafa 2, Tayde 5, B7} has been studied before. Here, we give a more general picture with $f(Q)$ gravity.\\
\indent The structure of this chapter unfolds as follows. Sec. \ref{ch2sec2} outlines the conditions necessary for a traversable WH in the presence of a monopole and presents the formalism associated with $f(Q)$ gravity. Sec. \ref{ch2sec3} focuses on the derivation of the field equations within the context of a linear model. Following this, in Sec. \ref{ch2sec4}, we introduce two distinctive DM profiles and meticulously examine the requisite conditions for the existence of a traversable WH while also scrutinizing the energy conditions under different redshift functions. Sec. \ref{ch2sec5} expands our analysis to incorporate the nonlinear formulation of $f(Q)$ regarding DM halo profiles. Sec. \ref{ch2sec6} is dedicated to using a VIQ method to evaluate the quantity of exotic matter necessary. Finally, in the concluding Sec. \ref{ch2sec7}, our findings are analyzed and accompanied by an in-depth discussion of the implications of the results of this study.

\section{Wormhole field equations with monopole charge in $f(Q)$ gravity}
\label{ch2sec2}
Here, we consider the Morris-Thorne WH metric in the Schwarzschild coordinates $(t,r,\theta,\varphi)$ and given by Eq. \eqref{1ch1}.
% \begin{equation}\label{1ch1}
%     ds^2 = e^{2\phi(r)}dt^2 - \frac{dr^2}{\left(1 - \frac{b(r)}{r}\right)} - r^2d\theta^2- r^2\sin^2\theta d\varphi^2\,.
% \end{equation} 
% From a mathematical perspective, ensuring a WH is traversable requires the shape function $b(r)$ to meet the following conditions:
% \begin{itemize}
%   \item[(1)] For $r>r_0$, i.e., out of throat, $1-\frac{b(r)}{r}>0$, and at the WH's throat i.e., $r=r_0$, $b(r)$ must satisfy the condition $b(r_0)=r_0$, where $r_0$ denotes the throat radius.
%   \item[(2)] The shape function $b(r)$ has to fulfill the flare-out requirement at the throat i.e., $b'(r_0)<1$.
%   \item[(3)] For asymptotic flatness condition, the limit $\frac{b(r)}{r}\rightarrow 0$ as $r\rightarrow \infty$ is required.
% \end{itemize}
% Compliance with these conditions guarantees the potential existence of exotic matter at the WH's throat within the framework of Einstein's GR.\\
The action for symmetric teleparallel gravity with a monopole charge, based on a four-dimensional action while neglecting the cosmological constant incorporating a minimally coupled triplet scalar field, is formulated as follows
\begin{equation}\label{ch2eq1}
S=\frac{1}{16\pi}\int\,f(Q)\sqrt{-g}\,d^4x+\int (\mathcal{L}_m+\mathcal{L}_{\text{monopole}})\,\sqrt{-g}\,d^4x\,.
\end{equation}
In the provided context, $\mathcal{L}_{\text{monopole}}$ represents the Lagrangian density of the monopole.\\
The Lagrangian density  of the monopole for a scalar triplet, denoted as $\Phi^a$, which includes self-interactions, is expressed as follows \cite{Jusufi1}
\begin{equation}\label{ch2eq2}
\mathcal{L}_{\text{monopole}}= - \frac{\lambda }{4}(\Phi ^2 -\eta ^2)^2 -\frac{1}{2} \sum _a g^{ij} \partial _i \Phi ^a \partial _j \Phi ^a.
\end{equation}
In this formulation, $a = 1, 2, 3$, while $\eta$ and $\lambda$ denote the scales associated with gauge symmetry breaking and self-interaction terms, respectively. The monopole's field configuration is represented by the following expression
\begin{equation}\label{ch2eq3}
\Phi ^a = \frac{\eta }{r} F(r) x^a\,,
\end{equation}
where the variable $x^a$ is defined as $(r sin\theta cos\varphi , r sin\theta sin\varphi , rcos\theta )$, ensuring that $\sum _a x^a x^a = r^2$.\\
The Lagrangian density of the monopole, incorporating the field configuration, can be reformulated in terms of $F(r)$ as follows
\begin{equation}\label{ch2eq4}
\mathcal{L}_{\text{monopole}} = -\left( 1-\frac{b(r)}{r}\right) \frac{\eta ^2 (F')^2}{2} -\frac{\eta ^2 F^2}{r^2}-\frac{\lambda \eta ^4}{4}(F^2-1)^2\,.
\end{equation}
For the field $F(r)$, the corresponding Euler-Lagrange equation is given by
\begin{equation}\label{ch2eq5}
\left( 1-\frac{b(r)}{r}\right) F'' + F' \left[ \left( 1-\frac{b(r)}{r}\right) \frac{2}{r}+\frac{1}{2}\left( \frac{b-b' r}{r^2}\right) \right] \\
-F\left[ \frac{2}{r^2}+\lambda \eta ^2(F^2-1)\right] =0 .
\end{equation}
Also, the energy-momentum tensor is derived from Eq. \eqref{ch2eq2} and is expressed as
\begin{equation}\label{ch2eq6}
\bar{T}_{ij}=\partial _i\Phi ^a\partial _j\Phi ^a -\frac{1}{2} g_{ij}g^{\mu \nu }\partial _\mu \Phi ^a\partial _\nu \Phi ^a -\frac{g_{ij}\lambda }{4}(\Phi ^2 -\eta ^2)^2.
\end{equation}
Consequently, all four components of the energy-momentum tensor can be determined as
\begin{equation}\label{ch2eq7}
\bar{T}^t_ t=-\eta ^2\left[ \frac{F^2}{r^2}+\left( 1-\frac{b(r)}{r}\right) \frac{(F')^2}{2} +\frac{\lambda \eta ^2}{4}(F^2-1)^2\right],
\end{equation}
\begin{equation}\label{ch2eq8}
\bar{T}^r_ r=-\eta ^2\left[ \frac{F^2}{r^2}+\left( 1-\frac{b(r)}{r}\right) \frac{(F')^2}{2} +\frac{\lambda \eta ^2}{4}(F^2-1)^2\right]\,,
\end{equation}
and
\begin{equation}\label{ch2eq9}
\bar{T}^\theta _ \theta=\bar{T}^\varphi _\varphi =-\eta ^2\left[ \left( 1-\frac{b(r)}{r}\right) \frac{(F')^2}{2} +\frac{\lambda \eta ^2}{4}(F^2-1)^2\right].
\end{equation}
It is difficult to get a precise analytical solution, as shown in Eq. \eqref{ch2eq5}. Hence, approximating the area outside the WH is sufficient to simplify and obtain the result. As a result, the components of the reduced energy-momentum tensor are expressed as
\begin{equation}\label{ch2eq10}
\bar{T}^t_t=\bar{T}^r_r=-\frac{\eta ^2}{r^2}\,, \,\,\,\,\,\,\bar{T}^\theta_\theta =\bar{T}^\varphi _\varphi = 0.
\end{equation}
% Furthermore, the non-metricity tensor is defined by the following equation \cite{fq1}.
% \begin{equation}\label{ch2eq11}
% Q_{\lambda\mu\nu}=\bigtriangledown_{\lambda} g_{\mu\nu}\,.
% \end{equation}
% Additionally, the superpotential, also known as the non-metricity conjugate, has the following formal definition
% \begin{equation}\label{ch2eq12}
% \hspace{-0.2cm}P^\alpha\,_{\mu\nu}=\frac{1}{4}\left[-Q^\alpha\;_{\mu\nu}+2Q_{(\mu}\;^\alpha\;_{\nu)}+Q^\alpha g_{\mu\nu}-\tilde{Q}^\alpha g_{\mu\nu}-\delta^\alpha_{(\mu}Q_{\nu)}\right].
% \end{equation}
% Also, traces of the non-metricity tensor can be provided by
% \begin{equation}
% \label{ch2eq14}
% \tilde{Q}_\alpha=Q^\mu\;_{\alpha\mu}\,,\;Q_{\alpha}=Q_{\alpha}\;^{\mu}\;_{\mu}.
% \end{equation}
% The non-metricity scalar is given by \cite{fq1}
% \begin{eqnarray}
% \label{ch2eq15}
% Q &=& -P^{\alpha\mu\nu}\,Q_{\alpha\mu\nu}\\
% &=& g^{\mu\nu}\left(L^\beta_{\,\,\,\alpha\beta}\,L^\alpha_{\,\,\,\mu\nu}-L^\beta_{\,\,\,\alpha\mu}\,L^\alpha_{\,\,\,\nu\beta}\right),
% \end{eqnarray}
% The disformation tensor, denoted as $L^\beta_{\,\,\,\mu\nu}$, is defined as follows
% \begin{equation}\label{ch2eq16}
% L^\beta_{\,\,\,\mu\nu}=\frac{1}{2}Q^\beta_{\,\,\,\mu\nu}-Q_{(\mu\,\,\,\,\,\,\nu)}^{\,\,\,\,\,\,\beta}.
% \end{equation}
The motion equation for $f(Q)$ gravity with the monopole charge is derived by varying the action with respect to the metric tensor $g_{\mu\nu}$. This equation is formulated as Eq. \eqref{ga44}.
% \begin{equation}\label{ch2eq17}
% \frac{-2}{\sqrt{-g}}\bigtriangledown_\alpha\left(\sqrt{-g}\,f_Q\,P^\alpha_{\;\;\;\mu\nu}\right)-\frac{1}{2}g_{\mu\nu}f -f_Q\left(P_{\mu\alpha\beta}\,Q_\nu^{\;\;\;\alpha\beta}-2\,Q^
% {\alpha\beta}_{\;\;\;\;\;\mu}\,P_{\alpha\beta\nu}\right)=8\pi T_{\mu\nu}\,,
% \end{equation}
In this case, $T_{\mu\nu}$ represents the combination of the energy-momentum tensor of the anisotropic fluid and the matter field. Thus, it can be expressed as
\begin{equation}\label{ch2eq18}
T_{\mu \nu } =\mathbb{T}_{\mu \nu } + \bar{T}_{\mu \nu }\,.
\end{equation}
% The representation of the energy-momentum tensor in the context of a fluid description of space-time is given by
% \begin{equation}\label{8}
% T_{\mu\nu}=-\frac{2}{\sqrt{-g}}\frac{\delta\left(\sqrt{-g}\,\mathcal{L}_m\right)}{\delta g^{\mu\nu}},
% \end{equation}
% Moreover, In this paper, we assume the diagonal energy-momentum tensor for an anisotropic fluid, which can be read as
Further, we consider the matter content to be characterized by an anisotropic energy-momentum tensor in order to investigate the WH solutions, which is expressed as \cite{a17, a2}
\begin{equation}\label{ch2eq18a}
\mathbb{T}_{\mu}^{\,\,\,\nu}\,=\left(\rho+p_t\right)u_{\mu}\,u^{\nu}-p_t\,\delta_{\mu}^{\,\,\,\nu}+\left(p_r-p_t\right)v_{\mu}\,v^{\nu},
\end{equation}
where $u_{\mu}$ and $v_{\mu}$ are the four-velocity vector and unitary space-like vectors, respectively. Also, both satisfy the conditions $u_{\mu}u^{\nu}=-v_{\mu}v^{\nu}=1$. The anisotropic energy-momentum tensor was introduced by Letelier \cite{Letelier/1980} as a way to investigate a two-fluid model in plasma physics. Moreover, it has been applied in several different scenarios for modeling magnetized neutron stars \cite{Debabrata/2021}.\\
Thus, by following the procedure in \cite{R.H. Lin, A. Banerjee}, the nonzero components of $Q_{\lambda\mu\nu}$ and $L^{\lambda}_{\,\,\,\mu\nu}$ are given as
\begin{equation}
Q_{rtt}=2\,e^{2\phi}\phi^{'},\,\,\,\,\,Q_{rrr}=-\frac{(rb^{'}-b)}{(r-b)^2},\,\,\,\,\,Q_{\theta r \theta}=Q_{\theta\theta r}=-\frac{rb}{r-b},\,\,\,\,Q_{\varphi r \varphi}=Q_{\varphi\varphi\,r}=-\frac{rb\sin^2\theta}{r-b},
\end{equation}
and
\begin{equation}
L^{t}_{\,\,t\,r}=L^{t}_{\,\,r\,t}=-\phi^{'},\,\,\,\,L^{r}_{\,\,\theta\theta}=-b,\,\,\,L^{r}_{\,\,t\,t}=-\frac{(r-b)}{r}e^{2\phi}\phi{'},\,\,\,L^{r}_{\,\,r\,r}=-\frac{(rb^{'}-b}{2\,r(r-b)},\,\,L^{r}_{\,\,\varphi\varphi}=-b\sin^2\theta,
\end{equation}
respectively. Here, $'$ denotes differentiation with respect to the radial coordinate $r$.\\
% The components of the energy-momentum tensor for the anisotropic fluid are given by
% \begin{equation}\label{ch2eq19}
% \mathcal{T}_{\mu}^{\,\,\,\nu} =\text{diag}(\rho,-p_r,-p_t,-p_t)\,.
% \end{equation}
Corresponding to the metric \eqref{1ch1}, the non-metricity scalar $Q$ is defined as per the Ref. \cite{A. Banerjee}
\begin{equation}\label{ch2eq20}
Q=-\frac{b}{r^2}\left[2\phi^{'}+\frac{rb^{'}-b}{r(r-b)}\right].
\end{equation}\\
Therefore, the field equations governing $f(Q)$ gravity related to WHs, incorporating the monopole charge, are formulated as
\begin{equation}\label{ch2eq21}
8 \pi  \rho =\frac{(r-b)}{2 r^3} \left[f_Q \left(\frac{(2 r-b) \left(r b'-b\right)}{(r-b)^2}+\frac{b \left(2 r \phi '+2\right)}{r-b}\right)
+\frac{f r^3}{r-b}+\frac{2 b r f_{\text{QQ}} Q'}{r-b}\right]-\frac{8 \pi  \eta ^2}{r^2},
\end{equation}
\begin{equation}\label{ch2eq22}
8 \pi  p_r=-\frac{(r-b)}{2 r^3} \left[-f_Q \left(\frac{b }{r-b}\left(\frac{r b'-b}{r-b}+2+2 r \phi '\right)
-4 r \phi '\right)-\frac{f r^3}{r-b}-\frac{2 b r f_{\text{QQ}} Q'}{r-b}\right]+\frac{8 \pi  \eta ^2}{r^2}\,,
\end{equation}
and
\begin{multline}\label{ch2eq23}
8 \pi  p_t=\frac{(r-b)}{4 r^2} \left[f_Q \left(\frac{\left(r b'-b\right) \left(\frac{2 r}{r-b}+2 r \phi '\right)}{r (r-b)}+\frac{4 (2 b-r) \phi '}{r-b}-4 r \left(\phi '\right)^2-4 r \phi ''\right)+\frac{2 f r^2}{r-b}
\right.\\\left.
-4 r f_{\text{QQ}} Q' \phi '\right].
\end{multline}
These specific field equations allow for a comprehensive exploration of various WH solutions within the framework of $f(Q)$ gravity models.
% \subsection{Energy conditions}
% The classical energy conditions, which are derived from the Raychaudhuri equations, provide a foundational framework for discussing physically plausible matter configurations. Among the four commonly recognized energy conditions- namely weak, null, strong, and dominant-  the NEC holds particular importance, especially within the realm of GR when considering WH solutions. This significance arises from its direct correlation to the energy density required to sustain the openness of the WH's throat. Any deviation from the NEC in the vicinity of the WH's throat suggests the presence of exotic matter characterized by negative energy density, a feature not typically associated with conventional matter sources. Energy conditions serve to constrain the stress-energy tensor, which describes the distribution of matter and energy in space-time and can be given as follows\\
% $\bullet$ The weak energy condition (\textbf{WEC}) :
% $\rho\geq0$,\,\, $\rho+p_t\geq0$,\,\, and \,\, $\rho+p_r\geq0$.\\
% $\bullet$ The NEC (\textbf{NEC}) : $\rho+p_t\geq0$\,\, and \,\, $\rho+p_r\geq0$.\\
% $\bullet$ The dominant energy condition (\textbf{\textbf{DEC}}) :  $\rho\geq0$,\,\, $\rho+p_t\geq0$,\,\, $\rho+p_r\geq0$,\,\, $\rho-p_t\geq0$,\,\, and \,\, $\rho-p_r\geq0$.\\
% $\bullet$ The strong energy condition (\textbf{SEC}) :
%  $\rho+p_t\geq0$,\,\, $\rho+p_r\geq0$,\,\, and \,\, $\rho+p_r+2p_t\geq0$.\\
% In conclusion, energy conditions provide crucial constraints on the behavior of matter in the Universe and play a crucial role in our exploration of WHs.
\section{Linear form of $f(Q)$ model}\label{ch2sec3}
Now, in the current work, we assume a linear form of $f(Q)$ gravity, which is expressed as
\begin{equation}\label{ch2eq24}
f(Q)=\alpha Q\,,
\end{equation}
where $\alpha\neq0$ is a free model parameter. This model parameter can be easily adjusted to retrieve the GR for $\alpha=1$. This model has been used in \cite{G. Mustafa} examined traversable WH inspired by noncommutative geometries that exhibit conformal symmetry and found that WH solutions exist under this noncommutative geometry with viable physical properties. Hence, with an arbitrary redshift function $\phi(r)$, the reduced field equations are given by
\begin{equation}\label{ch2eq25}
\rho=\frac{\alpha  b'-8 \pi  \eta ^2}{8 \pi  r^2},
\end{equation}
\begin{equation}\label{ch2eq26}
p_r = \frac{2 \alpha  r (b-r) \phi '+\alpha  b+8 \pi  \eta ^2 r}{8 \pi  r^3}\,,
\end{equation}
and
\begin{equation}\label{ch2eq27}
p_t = \frac{\alpha  \left(r \phi '+1\right) \left(r b'+2 r (b-r) \phi '-b\right)+2 \alpha  r^2 (b-r) \phi ''}{16 \pi  r^3}.
\end{equation}
Using the above field equations, we obtain the shape function $b(r)$ and study the energy conditions with the help of different DM profiles. These DM profiles will be elaborated in the next Sec. \ref{ch2sec4}.
\section{Different dark matter profiles}\label{ch2sec4}
Within this section, we examine two distinct profiles of DM, namely, the PI and the NFW profiles.
\subsection{Pseudo-Isothermal profile}\label{subsec1}
\begin{figure}[t]
    \centering
    \includegraphics[width=14.5cm,height=6cm]{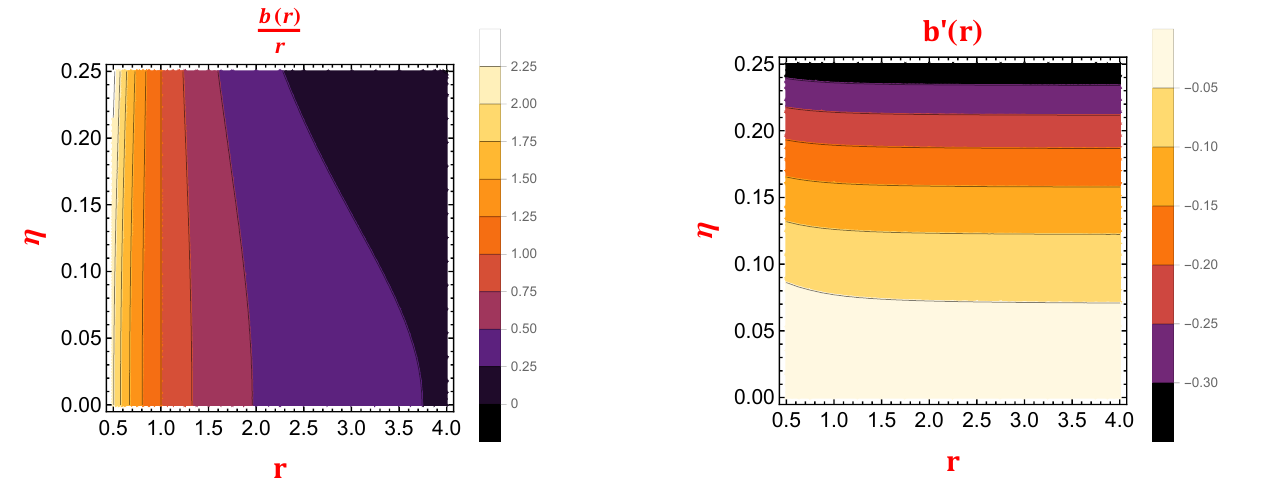}
    \caption{The figure displays PI profile with the variations in the asymptotic flatness condition \textit{(on the left)} and the flare-out condition \textit{(on the right)} as a function of the radial coordinate `$r$'. Furthermore, we keep other parameters fixed at constant values, including $\alpha=-5,\, r_s=0.5,\, \rho_s=0.02,\, \text{and} \, r_0 = 1$.}
    \label{ch2fig.1}
\end{figure}

\begin{figure}[t]
\centering
\includegraphics[width=13.5cm,height=6cm]{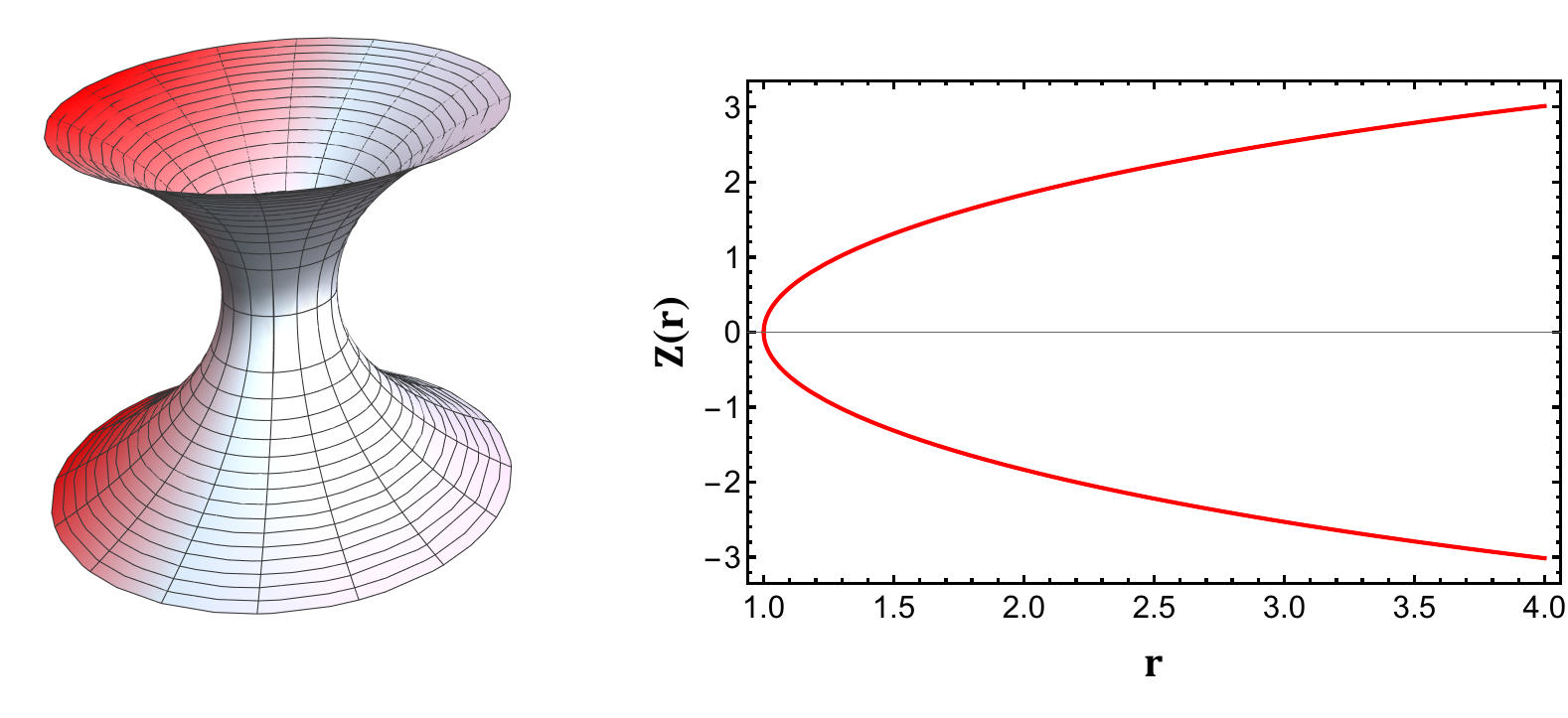}
\caption{The figure displays the embedding diagram for the PI profile. Furthermore, we keep other parameters fixed at constant values, including $\alpha=-5,\, r_s=0.5,\, \rho_s=0.02,\, \eta=0.15,\,\text{and} \, r_0 = 1$.}
\label{ch2fig.9}
\end{figure}

One significant class of DM models is linked to modified gravity, including MOND. Within the MOND framework, the density profile of DM is characterized by the PI profile \cite{Begeman} as given in Eq. \eqref{4aa1}.
% \begin{equation}\label{4aa1}
% \rho=\frac{\rho_s}{\left(\frac{r}{r_s}\right)^2+1}\,.
% \end{equation}
% Here, $\rho_s$ represents the central DM density, and $r_s$ denotes the scale radius. 
In Ref. \cite{pi3}, the author explored the possibility of traversable WH formation in the DM halos and obtained the exact solutions of the traversable WH of spherical symmetry with isotropic pressure condition and found that WEC and NEC are violated at the throat of the WH. Moreover, Paul \cite{B. C. Paul} explored the existence of traversable WHs in Einstein's general theory of relativity with density profile obtained from MOND with or without a scalar field.\\
Now, by comparing the energy densities, i.e. Eqs. \eqref{ch2eq25} and \eqref{4aa1}, the exact form of the shape function using the throat condition $b(r_0)=r_0$ can be obtained as
\begin{equation}\label{ch2eq29}
b(r)=\frac{1}{\alpha }\left(8 \pi  \rho_s {r_s}^3 \left(\tan ^{-1}\left(\frac{r_0}{{r_s}}\right)-\tan ^{-1}\left(\frac{r}{{r_s}}\right)\right)+r_0 \left(\alpha -8 \pi  \left(\eta ^2+\rho_s {r_s}^2\right)\right)+8 \pi  r \left(\eta ^2+\rho_s {r_s}^2\right)\right)\,.
\end{equation}
Next, we investigate the visual representation of the shape function and analyze the essential conditions for the existence of a WH. To do so, we carefully select the relevant parameters. The graphical depiction of the asymptotic behavior of the shape function, along with the verification of the flare-out condition, is presented in Fig. \ref{ch2fig.1} for the parameter $\eta$. The plot on the left of Fig. \ref{ch2fig.1} offers insight into the asymptotic nature of the shape function with respect to the parameter $\eta$. As the radial distance increases, the ratio $\frac{b(r)}{r}$ tends toward $0$, confirming the asymptotic flatness behavior of the shape function. Moreover, the corresponding right graph vividly demonstrates the fulfillment of the flare-out condition, where $b'(r_0) < 1$, at the throat of the WH. Here, we consider the throat of the WH at $r_0=1$.\\
Further, following the formulation proposed by Morris and Thorne \cite{a17}, the WH's embedding surface is described by the function $Z(r)$, which is governed by the following differential equation
\begin{equation}\label{ch2eq30}
\frac{dz}{dr}=\pm \frac{1}{\sqrt{\frac{r}{b(r)}-1}}.
\end{equation}
The equation highlights that $\frac{dz}{dr}$ experiences divergence at the throat of the WH, indicating that the embedding surface assumes a vertical orientation at this crucial point. Moreover, Eq. \eqref{ch2eq30} establishes the following relationship
\begin{equation}\label{ch2eq31}
Z(r)=\pm \bigintss_{r_0}^{r} \frac{dr}{\sqrt{\frac{r}{b(r)}-1}}.
\end{equation}
It is noted that the above integral cannot be solved analytically. Therefore, we resort to numerical techniques to generate the shape of the WH. Furthermore, we present the embedding diagram $Z(r)$ using Eq. \eqref{ch2eq31}, as illustrated in Fig. \ref{ch2fig.9}.\\
In this investigation, we explore three distinct forms of the redshift function to study energy conditions.

\subsubsection{Constant redshift function $\phi(r)=k$}\label{subsubsec1}
The function $\phi(r) = k$ \cite{N. Godani, Zinnat 1}, where $k$ is a constant, is considered a suitable choice as a redshift function due to its consistent behavior concerning the radial coordinate $r$. It is worth noting that when $k = 0$, this function is termed the tidal force. For our analysis, we will proceed with positive values of $k$. Employing this redshift function alongside the shape function provided in Eq. \eqref{ch2eq29} allows for the simultaneous derivation of explicit expressions for radial and tangential pressures, given by
\begin{multline}\label{ch2eq32}
p_r = \frac{1}{8 \pi  r^3}\left(8 \pi  \rho_s {r_s}^3 \left(\tan ^{-1}\left(\frac{r_0}{{r_s}}\right)-\tan ^{-1}\left(\frac{r}{{r_s}}\right)\right)+r_0 \left(\alpha -8 \pi  \left(\eta ^2+\rho_s {r_s}^2\right)\right)+8 \pi  r \left(2 \eta ^2+\rho_s {r_s}^2\right)\right)
\end{multline}
and
\begin{multline}\label{ch2eq33}
p_t = \frac{1}{16 r^3}\left(-\frac{r_0 \alpha }{\pi }+8 \left(\tan ^{-1}\left(\frac{r}{{r_s}}\right)-\tan ^{-1}\left(\frac{r_0}{{r_s}}\right)\right)\rho_s {r_s}^3 +8 r_0 \left(\eta ^2+\rho_s {r_s}^2\right)-\frac{8 \rho_s r {r_s}^4}{r^2+{r_s}^2}\right)\,.
\end{multline}
Furthermore, the NEC at the WH's throat, where $r = r_0$, is represented as
\begin{equation}
\left(\rho + p_r\right)_{\text{at}\,\, r=r_0}=\frac{\frac{\alpha }{8 \pi }+\eta ^2}{r_0^2}+\frac{\rho_s r_s^2}{r_0^2+r_s^2}\
\end{equation}
and
\begin{equation}
\left(\rho + p_t\right)_{\text{at}\,\, r=r_0}=\frac{3 \rho_s r_s^2}{2 \left(r_0^2+r_s^2\right)}-\frac{\alpha -8 \pi  \eta ^2}{16 \pi  r_0^2}\,.
\end{equation}
Additionally, we present the graphical illustration of this condition in Fig. \ref{ch2fig.2}, offers a visual representation of the radial and tangential pressures at the throat of the WH.
\subsubsection{Fractional redshift function $\phi(r)=\frac{\psi}{r}$}\label{subsubsec2}
The function $\phi(r) = \frac{\psi}{r}$ \cite{S. Kar, F. Parsaei}, we choose constant $\psi=1$ for the simplicity of the calculation, is another appropriate option as a redshift function due to its regular behavior for $r > 0$, effectively avoiding the event horizon beyond the WH's throat. Utilizing this redshift function, the expressions for the radial and tangential pressures can be formulated as follows
\begin{multline}\label{ch2eq34}
p_r = \frac{1}{8 \pi  r^4}\left(r_0 (r-2) \left(\alpha -8 \pi  \left(\eta ^2+\rho_s {r_s}^2\right)\right)+8 \pi \rho_s {r_s}^3(r-2)  \left(\tan ^{-1}\left(\frac{r_0}{{r_s}}\right)-\tan ^{-1}\left(\frac{r}{{r_s}}\right)\right)+2 r   
\right.\\ \left.
\times \left(\alpha +8 \pi \eta ^2 (r-1)+4 \pi  \rho_s (r-2) {r_s}^2\right)\right)
\end{multline}
and
\begin{multline}\label{ch2eq35}
p_t = \frac{1}{16 r^5}\left(\frac{r_0}{\pi } \left((r-3) r-2\right) \left(8 \pi  \left(\eta ^2+\rho_s {r_s}^2\right)-\alpha \right)+8 ((r-3) r-2) \left(\tan ^{-1}\left(\frac{r}{{r_s}}\right)-\tan ^{-1}\left(\frac{r_0}{{r_s}}\right)\right)
\right. \\ \left.
\hspace{0.8cm}\times \rho_s {r_s}^3+\frac{8 \rho_s r {r_s}^2 }{r^2+{r_s}^2}\left(2 (r+1) r^2+(2-(r-3) r) {r_s}^2\right)-\frac{2 r }{\pi }(r+1) \left(\alpha -8 \pi  \eta ^2\right)\right)\,.
\end{multline}
Also, NEC for radial and tangential pressures at the throat of the WH is given by
\begin{equation}
\left(\rho + p_r\right)_{\text{at}\,\, r=r_0}=\frac{\frac{\alpha }{8 \pi }+\eta ^2}{r_0^2}+\frac{\rho_s r_s^2}{r_0^2+r_s^2}
\end{equation}
and
\begin{equation}
\left(\rho + p_t\right)_{\text{at}\,\, r=r_0}=\frac{1}{16r_0^3}\left(\frac{8 r_0^2 (3 r_0-1) \rho_s r_s^2}{r_0^2+r_s^2}+\frac{(r_0-1) \left(8 \pi  \eta ^2-\alpha \right)}{\pi }\right)\,.
\end{equation}
Moreover, the concept can be effectively illustrated through a graphical representation, as shown in Fig. \ref{ch2fig.3}.
\subsubsection{Logarithmic redshift function $\phi(r)=\log\left(1+\frac{r_0}{r}\right)$}\label{subsubsec3}
The function $\phi(r)=\log\left(1+\frac{r_0}{r}\right)$ \cite{N. Godani}, which is also suitable as a redshift function, and it avoids the event horizon after the WH's throat. For this redshift function, the radial and tangential pressures can be given as
\begin{multline}\label{ch2eq36}
p_r = \frac{1}{8 \pi  r^3 (r_0+r)}\left(-r_0^2 \alpha +8 \pi  r_0^2 \eta ^2+3 r_0 \alpha  r-8 \pi  r_0 \eta ^2 r+8 \pi  \rho_s {r_s}^2 (r_0-r)^2+8 \pi  \rho_s {r_s}^3 \left(\tan ^{-1}\left(\frac{r}{{r_s}}\right)
\right.\right. \\ \left.\left.
\hspace{0.7cm}\times (r_0-r) -\tan ^{-1}\left(\frac{r_0}{{r_s}}\right)\right)+16 \pi  \eta ^2 r^2\right)
\end{multline}
and
\begin{multline}\label{ch2eq37}
p_t = \frac{1}{16 \pi  r^3 (r_0+r) \left(r^2+{r_s}^2\right)}\left(-8 \pi  \rho_s {r_s}^2 \left(2 r_0^2 \left(r^2+r_s^2\right)-3 r_0 r \left(r^2+{r_s}^2\right)+r^2 {r_s}^2\right)+r_0 \left(\alpha -8 \pi  \eta ^2\right)
\right. \\ \left.
\hspace{0.7cm}\times  (2 r_0-3 r) \left(r^2+{r_s}^2\right)+ \left(\tan ^{-1}\left(\frac{r_0}{{r_s}}\right)-\tan ^{-1}\left(\frac{r}{{r_s}}\right)\right)8 \pi  \rho_s {r_s}^3 (2 r_0-r) \left(r^2+{r_s}^2\right)\right)\,.
\end{multline}
Moreover, NEC at the WH's throat can be given by
\begin{equation}
\left(\rho + p_r\right)_{\text{at}\,\, r=r_0}=\frac{\frac{\alpha }{8 \pi }+\eta ^2}{r_0^2}+\frac{\rho_s r_s^2}{r_0^2+r_s^2}
\end{equation}
and
\begin{equation}
\left(\rho + p_t\right)_{\text{at}\,\, r=r_0}=\frac{5 \rho_s r_s^2}{4 \left(r_0^2+r_s^2\right)}-\frac{\alpha -8 \pi  \eta ^2}{32 \pi  r_0^2}\,.
\end{equation}
Furthermore, the concept can be aptly depicted through a graphical representation, as demonstrated in Fig. \ref{ch2fig.4}.
\begin{figure}[t]
\centering
\includegraphics[width=14.5cm,height=5cm]{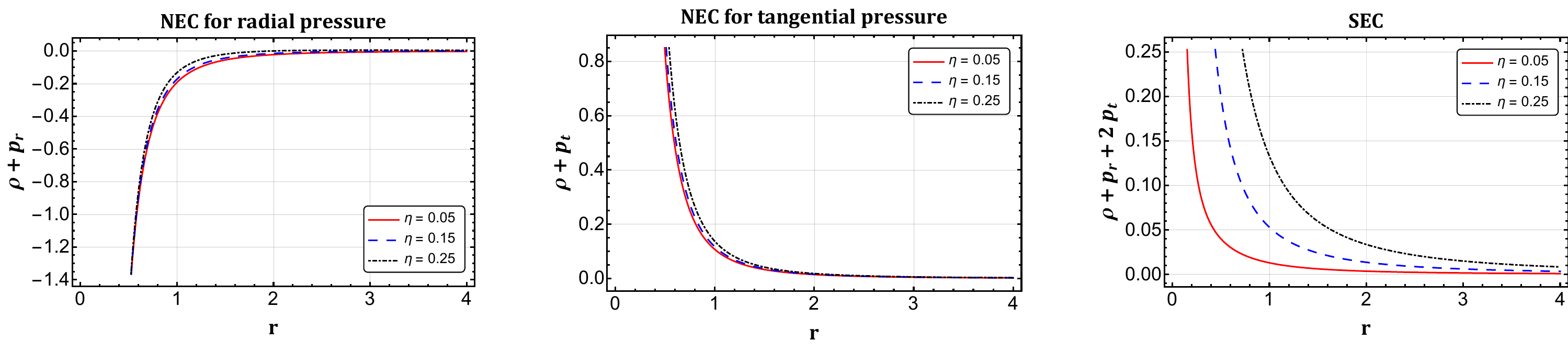}
\caption{The figure displays the PI profile with the variations in the NEC for both pressure and SEC as a function of the radial coordinate `$r$' for various values of `$\eta$ ' under the redshift $\phi(r)=\text{constant}$. Furthermore, we keep other parameters fixed at constant values, including $\alpha=-5,\, r_s=0.5,\, \rho_s=0.02,\, \text{and} \, r_0 = 1$.}
    \label{ch2fig.2}
\end{figure}
\begin{figure}[t]
    \centering
    \includegraphics[width=14.5cm,height=5cm]{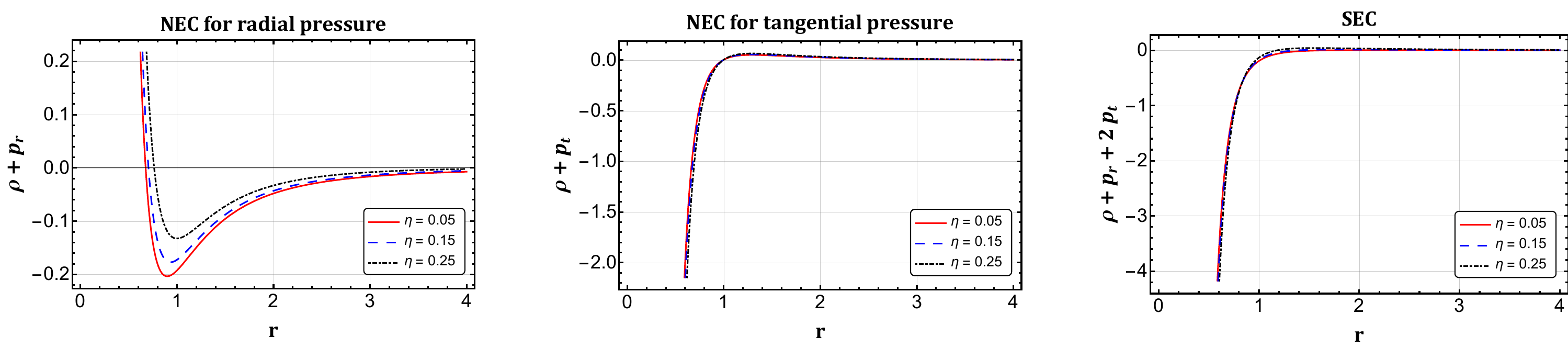}
    \caption{The figure displays the PI profile with the variations in the NEC for both pressure and SEC as a function of the radial coordinate `$r$' for various values of `$\eta$ ' under the redshift $\phi(r)=\frac{1}{r}$. Furthermore, we keep other parameters fixed at constant values, including $\alpha=-5,\, r_s=0.5,\, \rho_s=0.02,\, \text{and} \, r_0 = 1$.}
    \label{ch2fig.3}
\end{figure}
\begin{figure}[t]
    \centering
    \includegraphics[width=14.5cm,height=5cm]{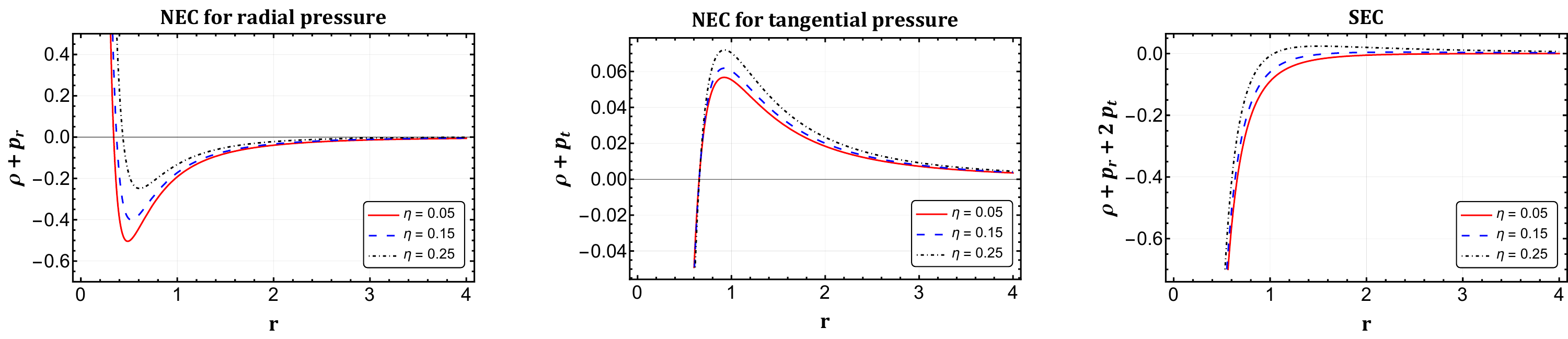}
    \caption{The figure displays the PI profile with the variations in the NEC for both pressure and SEC as a function of the radial coordinate `$r$' for various values of `$\eta$ ' under the redshift $\phi(r)=\log\left(1+\frac{r_0}{r}\right)$. Furthermore, we keep other parameters fixed at constant values, including $\alpha=-5,\, r_s=0.5,\, \rho_s=0.02,\, \text{and} \, r_0 = 1$.}
    \label{ch2fig.4}
\end{figure}
\begin{table}[t]
\begin{tabular}{cllll}
\hline\hline
 & \multicolumn{3}{p{12.5cm}}{Psuedo Isothermal profile ($\alpha=-5,\, r_s=0.5,\, \rho_s=0.02,\, \eta=0.05,\,\text{and} \, r_0 = 1$)} \\ \hline\hline
 &   \multicolumn{1}{l}{\hspace{0.8cm}$\phi(r)=k$}  &  \multicolumn{1}{l}{\hspace{0.5cm}$\phi(r)=\frac{1}{r}$}   &  \multicolumn{1}{l}{\hspace{0.5cm}$\phi(r)=\log\left(1+\frac{r_0}{r}\right)$} \\ \hline
$\rho$  & $>0$ for all $r$   &  $>0$ for all $r$      &  $>0$ for all $r$  \\\\ 
$\rho + p_r$ &  \multicolumn{1}{p{3.3cm}}{$<0$ for $r\in(0,13.9)$ $>0$ for $r\in[13.9,\infty)$}&     \multicolumn{1}{p{4.8cm}}{$>0$ for $r\in(0,0.7)$$\cup [40.7,\infty)$ $<0$ for $r\in[0.7,40.7)$}      &  \multicolumn{1}{p{4.3cm}}{$>0$ for $r\in(0,0.3]$$\cup (40,\infty)$ $<0$ for $r\in(0.3,40]$}     \\\\ 
$\rho + p_t$ &  \multicolumn{1}{p{3cm}}{$>0$ for $r\in(0,\infty)$ }   &    \multicolumn{1}{p{2.8cm}}{$<0$ for $r\in(0,1)$ \hspace{0.5cm} $>0$ for $r\in[1,\infty)$}       &  \multicolumn{1}{p{3cm}}{$<0$ for $r\in(0,0.6]$ \hspace{0.8cm} $>0$ for $r\in(0.6,\infty)$}   \\ \\
$\rho - |p_r|$ &   \multicolumn{1}{p{3.3cm}}{$>0$ for $r\in(0,41.6)$  $<0$ for $r\in[41.6,\infty)$}  &   \multicolumn{1}{p{4.7cm}}{$<0$ for $r\in(0,0.7)$$\cup [124,\infty)$ $>0$ for $r\in[0.7,124)$}        &  \multicolumn{1}{p{4.7cm}}{$<0$ for $r\in(0,0.34)$$\cup (122,\infty)$ $>0$ for $r\in[0.34,122]$}  \\ \\
$\rho - |p_t|$ &  \multicolumn{1}{p{3.3cm}}{$<0$ for $r\in(0,20.8)$  $>0$ for $r\in[20.8,\infty)$}   &    \multicolumn{1}{p{4.4cm}}{$>0$ for $r\in(0,1]$$\cup (61.6,\infty)$ $<0$ for $r\in(1,61.6]$}       &   \multicolumn{1}{p{4.8cm}}{$>0$ for $r\in(0,0.67]$$\cup (60.3,\infty)$ $<0$ for $r\in(0.67,60.3]$}    \\ \\
$\rho + p_r + 2p_t$ &  \multicolumn{1}{p{3cm}}{$>0$ for $r\in(0,\infty)$ }   &   \multicolumn{1}{p{3cm}}{$<0$ for $r\in(0,1.6]$ \hspace{0.5cm} $>0$ for $r\in(1.6,\infty)$}         &    \multicolumn{1}{p{3cm}}{$<0$ for $r\in(0,3.2]$ \hspace{0.5cm} $>0$ for $r\in(3.2,\infty)$}   \\ \hline\hline
\end{tabular}
\caption{Summary for results of PI profile for three different redshift functions}
\label{ch2table:1}
\end{table}
\subsection{Navarro-Frenk-White profile}\label{subsec2}
\begin{figure}[t]
    \centering
    \includegraphics[width=14.5cm,height=6cm]{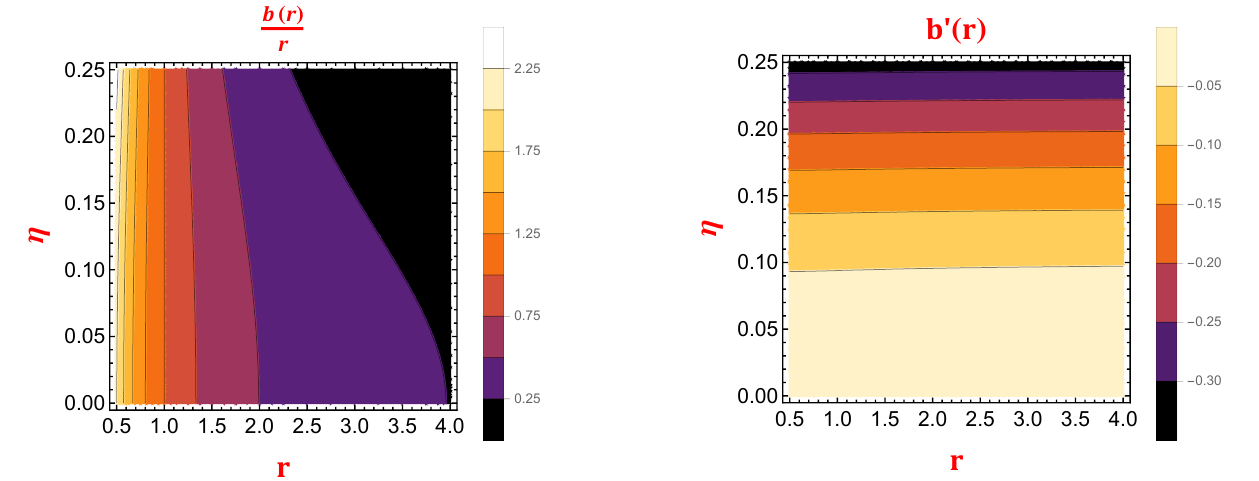}
    \caption{The figure displays NFW profile with the variations in the asymptotic flatness condition \textit{(on the left)} and the flare-out condition \textit{(on the right)} as a function of the radial coordinate `$r$'. Furthermore, we keep other parameters fixed at constant values, including $\alpha=-5,\, r_s=0.5,\, \rho_s=0.02,\, \text{and} \, r_0 = 1$.}
    \label{ch2fig.5}
\end{figure}

\begin{figure}[t]
\centering
\includegraphics[width=13.5cm,height=6cm]{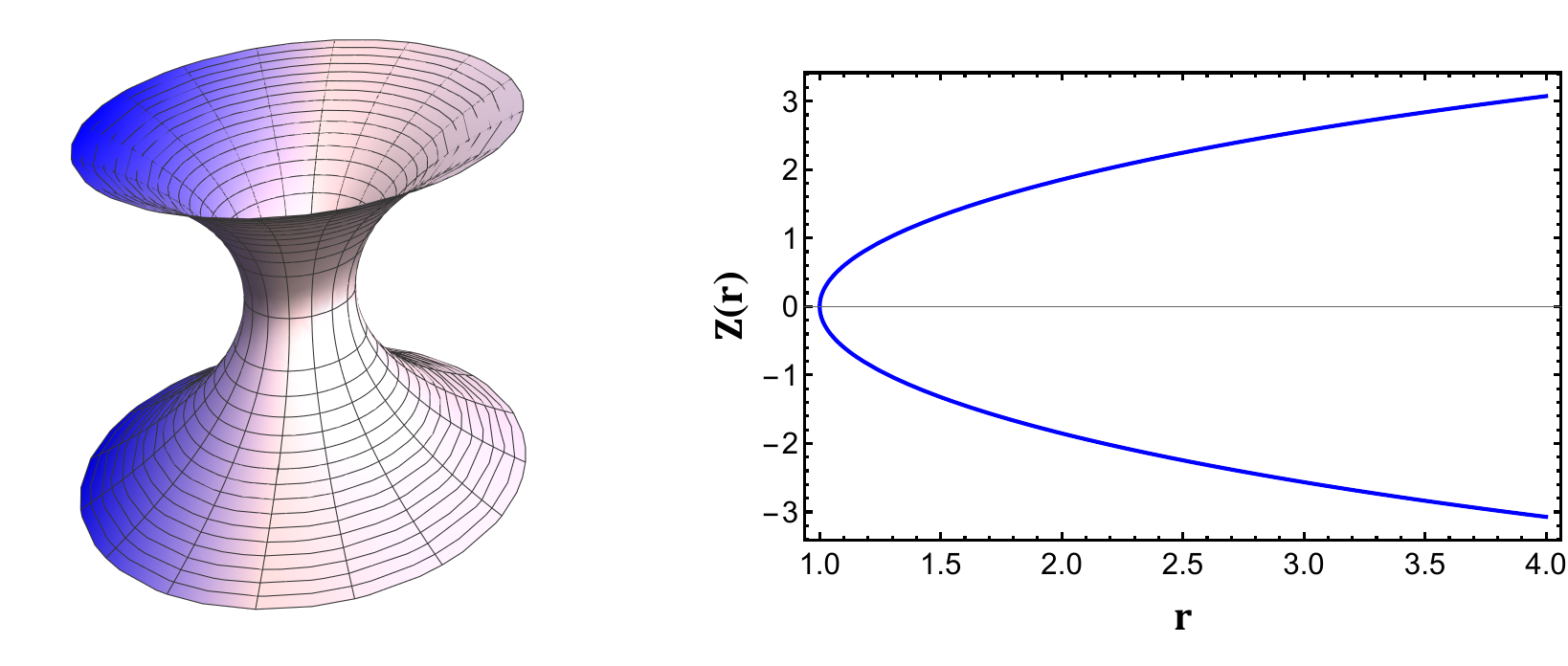}
\caption{The figure displays the embedding diagram for the NFW profile. Furthermore, we keep other parameters fixed at constant values, including $\alpha=-5,\, r_s=0.5,\, \rho_s=0.02,\, \eta=0.15,\,\text{and} \, r_0 = 1$.}
\label{ch2fig.10}
\end{figure}

The potential density model introduced by Hernquist \cite{dm51} aimed to investigate both theoretical and observational aspects of elliptical galaxies. Subsequently, Navarro and his colleagues \cite{dm52} analyzed equilibrium density profiles of DM halos in Universes that exhibit hierarchical clustering through high resolution N-body simulations. This model delineates how the density of DM varies in relation to the distance from the center of a halo. Their studies demonstrated that the structure of these profiles is consistent regardless of the halo mass, the spectral shape of the initial density fluctuations, or the cosmological parameter values. The CDM halo model for X-ray clusters and elliptical galaxies, as defined by Navarro et al \cite{dm52}, is given by Eq. \eqref{4aaa1}.
% \begin{equation}\label{4aaa1}
% \rho = \frac{\rho_s r_s}{r \left(\frac{r}{{r_s}}+1\right)^2}\,,
% \end{equation}
% where $r_s$ and $\rho_s$ denote the characteristic radius and central density of the Universe, respectively.
Now, we compare the energy densities given by Eqs. \eqref{ch2eq25} and \eqref{4aaa1}, the exact form of the shape function using the throat condition $b(r_0)=r_0$ can be given by
\begin{multline}\label{ch2eq39}
b(r)=-\frac{8 \pi}{\alpha }  \left(\frac{\rho_s {r_s}^4}{{r_0}+{r_s}}+\rho_s {r_s}^3 \log ({r_0}+{r_s})+\eta ^2 ({r_0}+{r_s})\right)
\\
\hspace{0.5cm}+{r_0}+\frac{8 \pi  }{\alpha }\left(\frac{\rho_s {r_s}^4}{r+{r_s}}+\rho_s {r_s}^3 \log (r+{r_s})+\eta ^2 (r+{r_s})\right)\,.
\end{multline}
We will now elucidate the graphical representation of the derived shape function and the requirements required for a WH to exist. To achieve this, we carefully select the appropriate parameters. Initially, we will investigate the shape function conditions for the NFW profile. Fig. \ref{ch2fig.5} visualizes the asymptotic flatness and the flare-out conditions related to the parameter $\eta$. The plot on the left shows the asymptotic behavior of the shape function, showing that the ratio $\frac{b(r)}{r}$ approaches zero as the radial distance increases. The plot on the right verifies the flare-out condition, which states that $b'(r_0) < 1$ at the throat of the WH. Here, we consider the throat of the WH at $r_0=1$.\\
Moving on, for the NFW profile WH shape function \eqref{ch2eq39}, we present the embedding diagram $Z(r)$ and its comprehensive visualization in Fig. \ref{ch2fig.10}. Again, we explore three distinct forms of the redshift function for this profile to study the energy conditions.
\subsubsection{Constant redshift function $\phi(r)=k$}\label{subsubsec4}
Using this particular form of the redshift function $\phi(r)=k$ and the shape function provided by \eqref{ch2eq39}, the radial \eqref{ch2eq26} and tangential \eqref{ch2eq27} pressures are given by
\begin{multline}\label{ch2eq40}
p_r =\frac{1}{8 r^3}\left(\frac{r_0 \alpha }{\pi }-8 \eta ^2 (r_0-2 r)+\frac{8 \rho_s {r_s}^4 (r_0-r)}{(r_0+{r_s}) (r+{r_s})}+8 \rho_s {r_s}^3 (\log (r+{r_s})-\log (r_0+{r_s}))\right)
\end{multline}
and
\begin{multline}\label{ch2eq41}
p_t = \frac{1}{16 r^3}\left(\frac{8 \rho_s {r_s}^3 \left(r_0 \left(r^2-r {r_s}-{r_s}^2\right)+r {r_s} (2 r+{r_s})\right)}{(r_0+{r_s}) (r+{r_s})^2}+r_0 \left(8 \eta ^2-\frac{\alpha }{\pi }\right)+8 \left(\log (r_0+{r_s})
\right.\right. \\ \left.\left.
-\log (r+{r_s})\right)\rho_s {r_s}^3 \right)\,.
\end{multline}
Furthermore, at the WH's throat, radial and tangential pressure components for the NEC are given by
\begin{equation}
\left(\rho + p_r\right)_{\text{at}\,\, r=r_0}=\frac{1}{8r_0^2}\left(8 \left(\frac{r_0 \rho_s r_s^3}{(r_0+r_s)^2}+\eta ^2\right)+\frac{\alpha }{\pi }\right)
\end{equation}
and
\begin{equation}
\left(\rho + p_t\right)_{\text{at}\,\, r=r_0}=\frac{1}{16r_0^2}\left(8 \left(\frac{3r_0 \rho_s r_s^3}{(r_0+r_s)^2}+\eta ^2\right)-\frac{\alpha }{\pi }\right)\,.
\end{equation}
And its graphical representation is shown in Fig. \ref{ch2fig.6}.
\subsubsection{Fractional redshift function $\phi(r)=\frac{1}{r}$}\label{subsubsec5}
Utilizing this specific form of the redshift function $\phi(r)=\frac{1}{r}$ and the shape function provided by Eq. \eqref{ch2eq39}, the expressions for the radial \eqref{ch2eq26} and tangential \eqref{ch2eq27} pressures are given by
\begin{multline}\label{ch2eq42}
p_r = \frac{1}{8 r^4}\left(\frac{r_0}{\pi } (r-2) \left(\alpha -8 \pi  \eta ^2\right)+\frac{2 r}{\pi } \left(\alpha +8 \pi  \eta ^2 (r-1)\right)+\frac{8 \rho_s (r-2) {r_s}^4 (r_0-r)}{(r_0+{r_s}) (r+{r_s})}+8 \rho_s (r-2) {r_s}^3 
\right. \\ \left.
\times (\log (r+{r_s})-\log (r_0+{r_s}))\right)
\end{multline}
and
\begin{multline}\label{ch2eq43}
p_t = \frac{1}{16 \pi  r^5 (r_0+{r_s}) (r+{r_s})^2}\left(-r_0^2 ((r-3) r-2) \left(\alpha -8 \pi  \eta ^2\right)(r+{r_s})^2+r_0 \left(8 \pi  \rho_s {r_s}^3 \left((r-1) r^3+(2
\right.\right.\right. \\ \left.\left.\left.
\hspace{1.7cm}-(r-3) r) {r_s}^2+(2-(r-3) r) r {r_s}\right)-\left(\alpha -8 \pi  \eta ^2\right)  (r+{r_s})^2 (((r-3) r-2){r_s}+2 r (r+1))\right)
\right. \\ \left.
\hspace{1.5cm}+8 \pi  \rho_s  ((r-3) r-2) {r_s}^3 (r_0+{r_s}) (r+{r_s})^2 (\log (r_0+{r_s})-\log (r+{r_s}))+8 \pi  \rho_s r {r_s}^4 (r (r (2 r
\right. \\ \left.
+{r_s}-4)-3 {r_s}-2)-2 {r_s})-2 r (r+1) {r_s} \left(\alpha -8 \pi  \eta ^2\right) (r+{r_s})^2\right)\,.
\end{multline}
Additionally, it is crucial to evaluate the radial and tangential pressure components for the NEC at the throat of the WH and they are given as
\begin{equation}
\left(\rho + p_r\right)_{\text{at}\,\, r=r_0}=\frac{1}{8r_0^2}\left(8 \left(\frac{r_0 \rho_s r_s^3}{(r_0+r_s)^2}+\eta ^2\right)+\frac{\alpha }{\pi }\right)
\end{equation}
and
\begin{equation}
\left(\rho + p_t\right)_{\text{at}\,\, r=r_0}=\frac{1}{16r_0^3}\left(\frac{(r_0-1) \left(8 \pi  \eta ^2-\alpha \right)}{\pi }+\frac{8 r_0 (3 r_0-1) \rho_s r_s^3}{(r_0+r_s)^2}\right)\,.
\end{equation}
Additionally, Fig. \ref{ch2fig.7} shows the graphical illustration of the radial and tangential pressure components of the NEC at the throat of the WH.
\subsubsection{Logarithmic redshift function $\phi(r)=\log\left(1+\frac{r_0}{r}\right)$}\label{subsubsec6}
With the help of the specific redshift function $\phi(r)=\log\left(1+\frac{r_0}{r}\right)$ alongside the shape function provided by Eq. \eqref{ch2eq39}, we can derive the expressions for the radial \eqref{ch2eq26} and tangential \eqref{ch2eq27} pressures as follows
\begin{multline}\label{ch2eq44}
p_r = -\frac{1}{8 \pi  r^3 (r_0+r) (r_0+{r_s}) (r+{r_s})}\left((r_0+{r_s}) (r+{r_s}) \left(r_0 \alpha (r_0-3 r)-8 \pi  \eta ^2 \left(r_0^2-r_0 r+2 r^2\right)\right)+8 \pi  \rho_s 
\right. \\ \left.
\times {r_s}^4 (r_0-r)^2-8 \pi  \rho_s {r_s}^3(\log (r_0+{r_s})-\log (r+{r_s}))(r_0-r) (r_0+{r_s}) (r+{r_s}) \right)
\end{multline}
and
\begin{multline}\label{ch2eq45}
p_t = \frac{1}{16 \pi  r^3 (r_0+r) (r_0+{r_s}) (r+{r_s})^2}\left(8 \pi  \rho_s {r_s}^3 \left(2 r_0^2 {r_s} (r+{r_s})+r_0 r \left(r^2-3 r {r_s}-3 {r_s}^2\right)+r^2 {r_s} (2 r+{r_s})\right) 
\right. \\ \left.
\hspace{1.5cm}-8 \pi \rho_s {r_s}^3 (2 r_0-r) (r_0+{r_s}) (\log (r_0+{r_s})-\log (r+{r_s}))(r+{r_s})^2+r_0 \left(\alpha -8 \pi  \eta ^2\right) (2 r_0-3 r) 
\right. \\ \left.
 \times(r_0+{r_s}) (r+{r_s})^2\right)\,.
\end{multline}
Also, it is essential to examine the pressure components for the NEC at the throat of the WH and given by
\begin{equation}
\left(\rho + p_r\right)_{\text{at}\,\, r=r_0}=\frac{1}{8r_0^2}\left(8 \left(\frac{r_0 \rho_s r_s^3}{(r_0+r_s)^2}+\eta ^2\right)+\frac{\alpha }{\pi }\right)
\end{equation}
and
\begin{equation}
\left(\rho + p_t\right)_{\text{at}\,\, r=r_0}=\frac{1}{32r_0^2}\left(8 \left(\frac{5 r_0 \rho_s r_s^3}{(r_0+r_s)^2}+\eta ^2\right)-\frac{\alpha }{\pi }\right)\,.
\end{equation}
Further, it can be demonstrated through a graphical representation, as depicted in Fig. \ref{ch2fig.8}.
\begin{figure}[t]
    \centering
    \includegraphics[width=14.5cm,height=5cm]{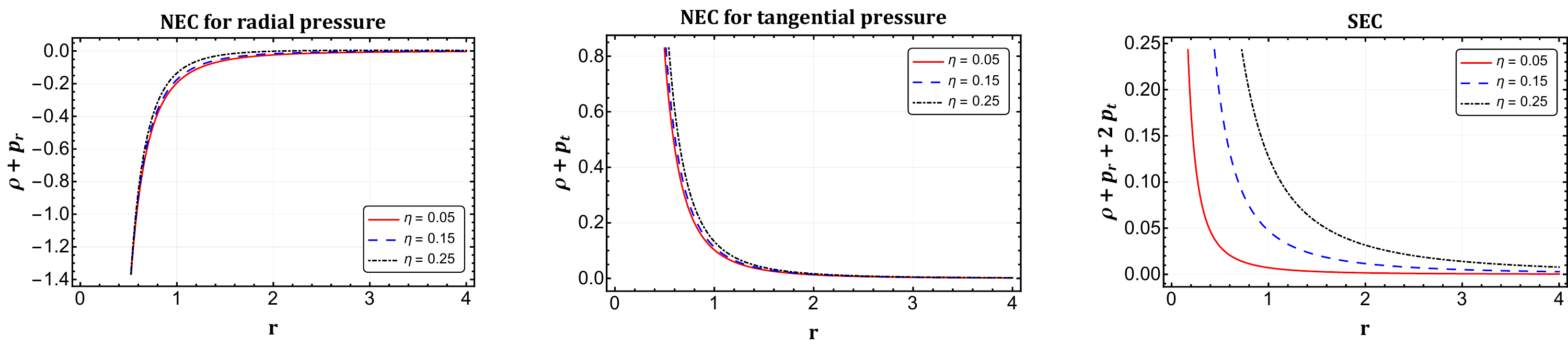}
    \caption{The figure displays the NFW profile with the variations in the NEC for both pressure and SEC as a function of the radial coordinate `$r$' for various values of `$\eta$ ' under the redshift $\phi(r)=\text{constant}$. Furthermore, we keep other parameters fixed at constant values, including $\alpha=-5,\, r_s=0.5,\, \rho_s=0.02,\, \text{and} \, r_0 = 1$.}
    \label{ch2fig.6}
\end{figure}
\begin{figure}[t]
    \centering
    \includegraphics[width=14.5cm,height=5cm]{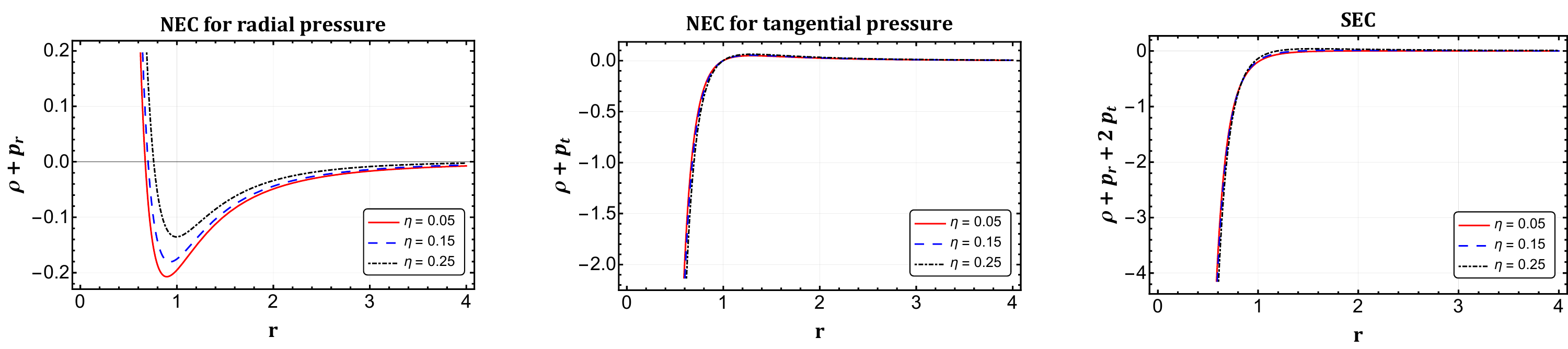}
    \caption{The figure displays the NFW profile with the variations in the NEC for both pressure and SEC as a function of the radial coordinate `$r$' for various values of `$\eta$ ' under the redshift $\phi(r)=\frac{1}{r}$. Furthermore, we keep other parameters fixed at constant values, including $\alpha=-5,\, r_s=0.5,\, \rho_s=0.02,\, \text{and} \, r_0 = 1$.}
    \label{ch2fig.7}
\end{figure}
\begin{figure}[t]
    \centering
    \includegraphics[width=14.5cm,height=5cm]{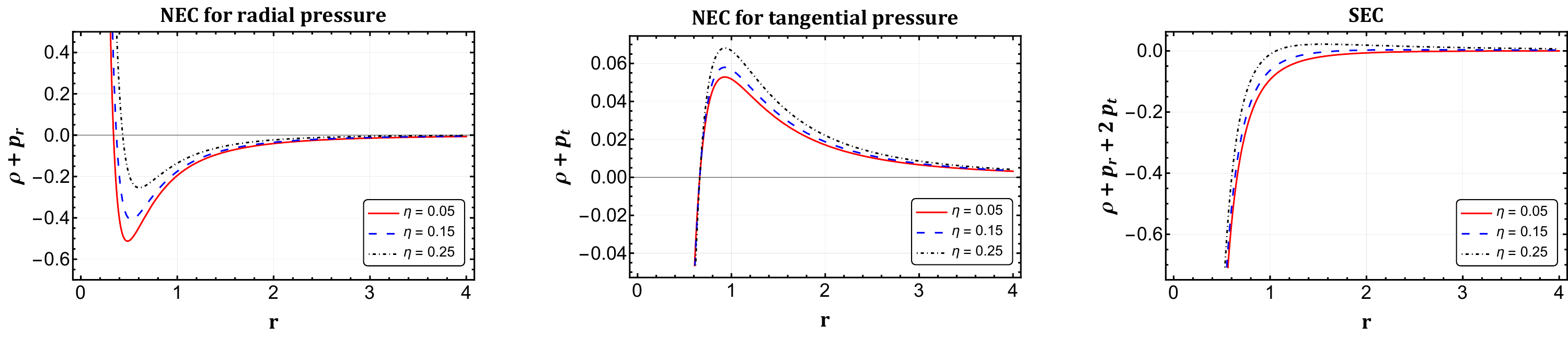}
    \caption{The figure displays the NFW profile with the variations in the NEC for both pressure and SEC as a function of the radial coordinate `$r$' for various values of `$\eta$ ' under the redshift $\phi(r)=\log\left(1+\frac{r_0}{r}\right)$. Furthermore, we keep other parameters fixed at constant values, including $\alpha=-5,\, r_s=0.5,\, \rho_s=0.02,\, \text{and} \, r_0 = 1$.}
    \label{ch2fig.8}
\end{figure}
\begin{table}[t]
\begin{tabular}{cllll}
\hline\hline
 & \multicolumn{3}{p{12.5cm}}{\hspace{1cm}NFW profile ($\alpha=-5,\, r_s=0.5,\, \rho_s=0.02,\, \eta=0.05,\,\text{and} \, r_0 = 1$)} \\ \hline\hline
 &   \multicolumn{1}{l}{\hspace{0.5cm}$\phi(r)=k$}  &  \multicolumn{1}{l}{\hspace{0.5cm}$\phi(r)=\frac{1}{r}$}   &  \multicolumn{1}{l}{\hspace{0.5cm}$\phi(r)=\log\left(1+\frac{r_0}{r}\right)$} \\ \hline
$\rho$  & $>0$ for all $r$   &  $>0$ for all $r$      &  $>0$ for all $r$  \\\\ 
$\rho + p_r$ &  \multicolumn{1}{p{3.3cm}}{$<0$ for $r\in(0,38.3]$ $>0$ for $r\in(38.3,\infty)$}&     \multicolumn{1}{p{4.9cm}}{$>0$ for $r\in(0,0.67]$$\cup (117.7,\infty)$ $<0$ for $r\in(0.67,117.7]$}      &  \multicolumn{1}{p{4.8cm}}{$>0$ for $r\in(0,0.33]$$\cup (117,\infty)$ $<0$ for $r\in(0.33,117]$}     \\\\ 
$\rho + p_t$ &  \multicolumn{1}{p{3.3cm}}{$>0$ for $r\in(0,\infty)$ }   &    \multicolumn{1}{p{3.4cm}}{$<0$ for $r\in(0,0.99]$ \hspace{0.5cm} $>0$ for $r\in(0.99,\infty)$}       &  \multicolumn{1}{p{3.3cm}}{$<0$ for $r\in(0,0.66]$ \hspace{0.5cm} $>0$ for $r\in(0.66,\infty)$}   \\ \\
$\rho - |p_r|$ &   \multicolumn{1}{p{3.3cm}}{$>0$ for $r\in(0,39.2]$  $<0$ for $r\in(39.2,\infty)$}  &   \multicolumn{1}{p{4.9cm}}{$<0$ for $r\in(0,0.67]$$\cup (118.6,\infty)$ $>0$ for $r\in(0.67,118.6]$}        &  \multicolumn{1}{p{4.6cm}}{$<0$ for $r\in(0,0.33]$$\cup (118,\infty)$ $>0$ for $r\in(0.33,118]$}  \\ \\
$\rho - |p_t|$ &  \multicolumn{1}{p{3cm}}{$<0$ for $r\in(0,\infty)$}   &    \multicolumn{1}{p{2.8cm}}{$>0$ for $r\in(0,1]$ $<0$ for $r\in(1,\infty)$}       &   \multicolumn{1}{p{3.3cm}}{$>0$ for $r\in(0,0.67]$ $<0$ for $r\in(0.67,\infty)$}    \\ \\
$\rho + p_r + 2p_t$ &  \multicolumn{1}{p{3cm}}{$>0$ for $r\in(0,\infty)$ }   &   \multicolumn{1}{p{3cm}}{$<0$ for $r\in(0,1.8]$ \hspace{0.5cm} $>0$ for $r\in(1.8,\infty)$}         &    \multicolumn{1}{p{3cm}}{$<0$ for $r\in(0,5.4]$ \hspace{0.5cm} $>0$ for $r\in(5.4,\infty)$}   \\ \hline\hline
\end{tabular}
\caption{Summary for results of NFW profile for three different redshift functions}
\label{ch2table:2}
\end{table}
\subsection{Discussion}\label{subsec3}
In this specific subsection, we will discuss the behavior exhibited by energy conditions within the context of both the PI and the NFW models for three different redshift functions. We will thoroughly examine how these models adhere to various energy conditions and will elucidate any distinctions or similarities. We have standardized specific free parameters, set $\alpha=-5,\, r_s=0.5,\, \rho_s=0.02,\, \text{and} \, r_0 = 1$ according to our analysis of shape functions, and vary the parameter $\eta$ for the required results. These parameter values were selected to align with the specific requirements of our study and to ensure consistency throughout our analysis.
\begin{itemize}
\item \textbf{Constant redshift function} $\phi(r)=k$\,\textbf{:} \\
For the specified redshift function, we present graphical representations of the NEC and SEC for different values of $\eta$, particularly $0.05,\, 0.15,\,\text{and} \,0.25$ plotted against the radial coordinate $r$. These representations are depicted in Figs. \ref{ch2fig.2} and \ref{ch2fig.6}. In these figures, the quantity $\rho+p_r$ shows the violation of the radial NEC in the assumed throat $r_0=1$. Furthermore, we have examined the behavior of $\rho+p_t$ for various values of $\eta$ and confirmed the satisfaction of the tangential NEC. Moreover, our investigation extends to the SEC, where satisfaction is observed near the throat for different $\eta$ values.
\item \textbf{Fractional redshift function} $\phi(r)=\frac{1}{r}$\,\textbf{:}\\
In this case, we illustrate graphical representations of the NEC and SEC at different values of $\eta$: specifically, $0.05$, $0.15$ and $0.25$, plotted against the radial coordinate $r$. These representations are shown in Figs. \ref{ch2fig.3} and \ref{ch2fig.7}. Within these visualizations, the quantity $\rho+p_r$ indicates potential violations of the radial NEC, particularly in the assumed throat $r_0=1$. Furthermore, we have analyzed $\rho+p_t$ in various values $\eta$, confirming the adherence to the tangential NEC. Furthermore, our investigation extends to the SEC, where we observe violations in the vicinity of the throat for different values of $\eta$.
\item \textbf{Logarithmic redshift function} $\phi(r)=\log\left(1+\frac{r_0}{r}\right)$\,\textbf{:}\\
Similarly, in this scenario, we provide graphical representations of the NEC and SEC across various values of $\eta$ as shown in Figs. \ref{ch2fig.4} and \ref{ch2fig.8}. Remarkably, our observations mirror those obtained with the second redshift function. 

The redshift function determines the gravitational redshift experienced by signals traversing the wormhole. Three distinct redshift functions are chosen to ensure that no event horizon forms, as they keep $\phi(r)$ finite everywhere, satisfying a key criterion for traversable wormholes. These depictions offer valuable insights into the behavior of these energy conditions within the framework of the given model. For a complete summary of the energy conditions, refer to Tables \ref{ch2table:1} and \ref{ch2table:2}.
\end{itemize}
\section{Nonlinear $f(Q)$ model}\label{ch2sec5}
\begin{figure}[t]
\centering
\includegraphics[width=14.5cm,height=6cm]{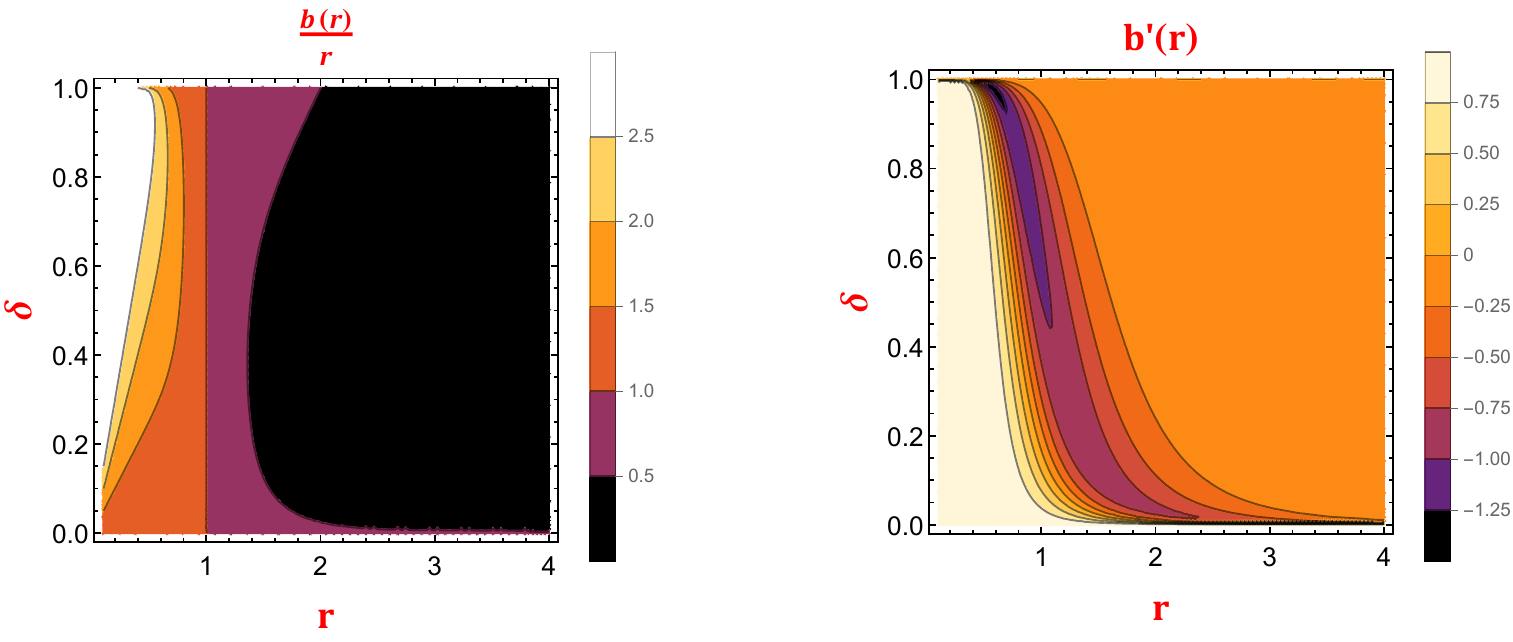}
\caption{The figure displays PI profile with the variations in the asymptotic flatness condition \textit{(on the left)} and the flare-out condition \textit{(on the right)} as a function of the radial coordinate `$r$' under the redshift $\phi(r)=\frac{1}{r}$. Additionally, we consider $r_0 = 1$.}
\label{ch2fig.11}
\end{figure}

In this specific section, we leverage the Karmakar condition \cite{M. F. Shamir, G. Mustafa 1} as our guiding principle. Employing a similar methodology, we proceed to derive the shape function applicable to non-constant redshift functions. Specifically, for the redshift function $\phi(r) = \frac{1}{r}$, the corresponding shape function is expressed as follows
\begin{equation}\label{ch2eq46}
b(r)=r-\frac{r^5}{\frac{r_0^4 (r_0-\delta ) e^{\frac{2}{r}-\frac{2}{r_0}}}{\delta }+r^4}+\delta ,\;\; 0<\delta<r_{0}\,.
\end{equation}
We consider a nonlinear form for $f(Q)$, as articulated in the work \cite{T. Harko1,  R. Solanki}. This form is represented as follows
\begin{equation}\label{ch2eq48}
f(Q)=Q + m Q^n\,.
\end{equation}
In this formulation, $m$ and $n\neq1$ denote free model parameters, offering flexibility within the model. Consequently, the field equations can be expressed as follows
\begin{multline}\label{ch2eq49}
\rho=\frac{-1}{16 \pi  r^6 (r-b)^2}\left(\left(\frac{b \left(-r b'+2 r (b-r) \phi '+b\right)}{r^3 (r-b)}\right)^{n-2} 2 m (n-1) n b  \left(r^2 b \left(r \left(b''-4 \phi '+2 r \phi ''\right)-b' \left(4
\right.\right.\right.\right. \\ \left.\left.\left.\left.
\hspace{0.6cm}\times  r \phi '+5\right)\right)+r^3 b' \left(b'+2 r \phi '\right)+r b^2 \left(-r \left(b''-8 \phi '+4 r \phi ''\right)+b' \left(2 r \phi '+3\right)+4\right)+b^3 \left(2 r^2 \phi ''-4 r \phi '
\right.\right.\right. \\ \left.\left.\left.
\hspace{0.6cm} -3\right)\right)+r^3 (r-b) \left(b \left(r b'+2 r (b-r) \phi '+b\right)-2 r^2 b'\right)\left(1+m n \left(\frac{b \left(-r b'+2 r (b-r) \phi '+b\right)}{r^3 (r-b)}\right)^{n-1}\right)
\right.\\ \left.
\hspace{0.6cm} -r^3 (r-b) \left(b \left(-r b'+2 r (b-r) \phi '+b\right)+\left(\frac{b \left(\frac{b-r b'}{r^2-r b}-2 \phi '\right)}{r^2}\right)^n m r^3 (r-b) \right)+16 \pi  \eta ^2 r^4 (r-b)^2\right)\,,
\end{multline}
\begin{multline}\label{ch2eq50}
p_r = \frac{1}{16 \pi }\left(\frac{-m n}{b \left(-r b'+2 r (b-r) \phi '+b\right)^2} \left( \left(r^2 b \left(2 r \left((n-1) \left(b''+2 r \phi ''\right)-4 (n-2) \phi '+14 r \phi '^2\right)+2 b' 
\right.\right.\right.\right.\\ \left.\left.\left.\left.
\hspace{1.2cm}\times \left(2 (5 -2 n) r \phi '-5 n+6\right)+b'^2\right)+2 r^3 \left(b'+2 r \phi '\right) \left((n-1) b'-2 r \phi '\right)+2 r b^2 \left(r \left(-(n-1) \left(b''
\right.\right.\right.\right.\right.\right. \\ \left.\left.\left.\left.\left.\left.
\hspace{1.1cm}  +4 r\phi ''\right)+2 (4 n-9) \phi '-16 r \phi '^2\right)+b' \left(2 (n-3) r \phi '+3 n-5\right)+4 n-5\right)+b^3 \left(4 r \left(\phi ' \left(-2 n+3
\right.\right.\right.\right.\right.\right. \\ \left.\left.\left.\left. \left.\left. 
\hspace{1cm} \times  r\phi '+5\right)+(n-1) r \phi ''\right)-6 n+9\right)\right)\left(\frac{b}{r^2} \left(\frac{b-r b'}{r^2-r b}-2 \phi '\right)\right)^n\right)
+\left(\frac{b }{r^2}\left(\frac{b-r b'}{r^2-r b}-2 \phi '\right)\right)^n m
\right.\\ \left.
+\frac{2 }{r^3} \left(2 r(b-r) \phi '+b+8 \pi  \eta ^2 r\right)\right)\,,
\end{multline}
and
\begin{multline}\label{ch2eq51}
p_t = \frac{1}{16 \pi  r^4 (r-b)^2}\left(\left(1-\frac{b}{r}\right) \left(2 m (n-1) n \phi ' \left(\frac{b \left(-r b'+2 r (b-r) \phi '+b\right)}{r^3 (r-b)}\right)^{n-2} \left(r^2 b \left(r \left(b'' -4\phi '
\right.\right.\right.\right.\right. \\ \left.\left.\left.\left.\left.
\hspace{0.6cm} +2 r \phi ''\right)-b' \left(4 r \phi '+5\right)\right)+r^3 b' \left(b'+2 r \phi '\right) +r b^2 \left(-r \left(b''-8 \phi '+4 r \phi ''\right)+b' \left(2 r \phi '+3\right)+4\right)
\right.\right.\right.\\ \left.\left.\left. 
\hspace{0.7cm} +b^3 \left(2  r^2  \phi ''-4 r \phi '-3\right)\right)-r^2 \left(\left(b-r b'\right) \left(r (r-b) \phi '+r\right)+2 r^2 (r-b)^2 \phi '^2+2(r-b) (r-2 b)
\right.\right.\right. \\ \left.\left.\left.
\hspace{0.8cm} \times \phi ' r  +2 r^2 (r-b)^2 \phi ''\right) \left(m n \left(\frac{b \left(-r b'+2 r (b-r) \phi '+b\right)}{r^3 (r-b)}\right)^{n-1}+1\right)+r^2 b \left(-r b'+2 r (b-r) 
\right.\right.\right. \\ \left.\left.\left. 
\hspace{0.4cm} \times \phi '+b\right)+ m r^5 (r-b) \left(\frac{b}{r^2} \left(\frac{b-r b'}{r^2-r b}-2 \phi '\right)\right)^n\right)\right)\,.
\end{multline}
We will now discuss the graphical representation of the derived shape function, as well as the requirements required for the occurrence of a WH. For this, we carefully choose suitable parameters. Initially, our focus lies on examining the nature of the shape function conditions for the PI profile. The plot depicted in Fig. \ref{ch2fig.11} showcases both the flare-out and the asymptotic flatness condition with respect to the parameter $\delta$. The left plot in Fig. \ref{ch2fig.11} offers insights into the satisfaction of the shape function's asymptotic behavior. The corresponding graph on the right illustrates the fulfillment of the flare-out condition at the WH's throat. Here, we designate the WH's throat at $r_0=1$.\\
\begin{figure}[t]
\centering
\includegraphics[width=14.5cm,height=5cm]{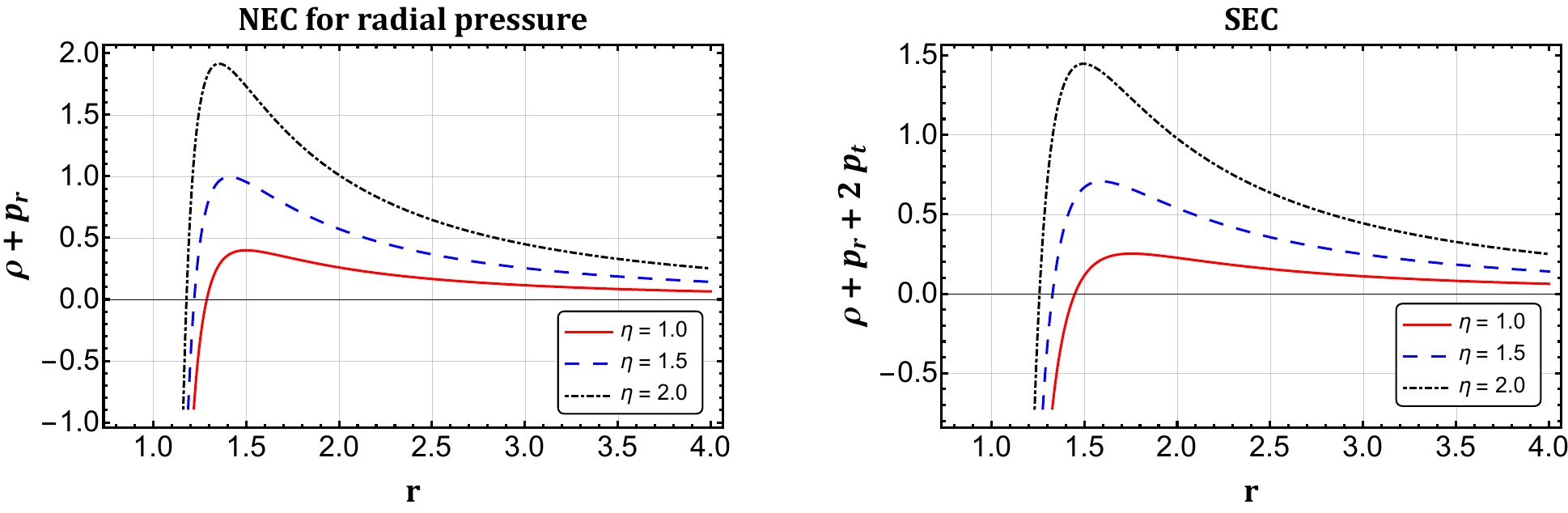}
\caption{The figure displays PI profile with the variations in the NEC for radial pressure and SEC as a function of the radial coordinate `$r$' under the redshift $\phi(r)=\frac{1}{r}$. Furthermore, we keep other parameters fixed at constant values, including $m=2,\,n=2.01,\,\delta=0.5,\, r_s=0.5,\, \rho_s=0.02,\, \text{and} \, r_0 = 1$.}
\label{ch2fig.12}
\end{figure}
Moreover, within this specific context, it becomes evident that the NEC along both radial and tangential directions becomes indeterminate precisely at the WH's throat, where $r=r_0$ and can be visualized using Fig. \ref{ch2fig.12}. This observation strongly indicates the insufficiency to achieve viable WH solutions utilizing the prescribed shape function \eqref{ch2eq46}. Consequently, we arrive at the conclusion that proposing the nonlinear form \eqref{ch2eq48} proves unsuitable for generating WH solutions when paired with the shape function \eqref{ch2eq46}, applicable to both the PI and NFW profiles.\\
Subsequently, we extend our investigation to alternative nonlinear form $Q+\frac{\beta}{Q}$ yielding identical outcomes; the NEC remains undefined precisely at the WH's throat for these models as well. Despite these limitations, it is imperative to note the existence of further options for shape functions, which we may investigate further.

\section{Amount of exotic matter}\label{ch2sec6}
\begin{figure}[t]
\centering
\includegraphics[width=14.5cm,height=5.5cm]{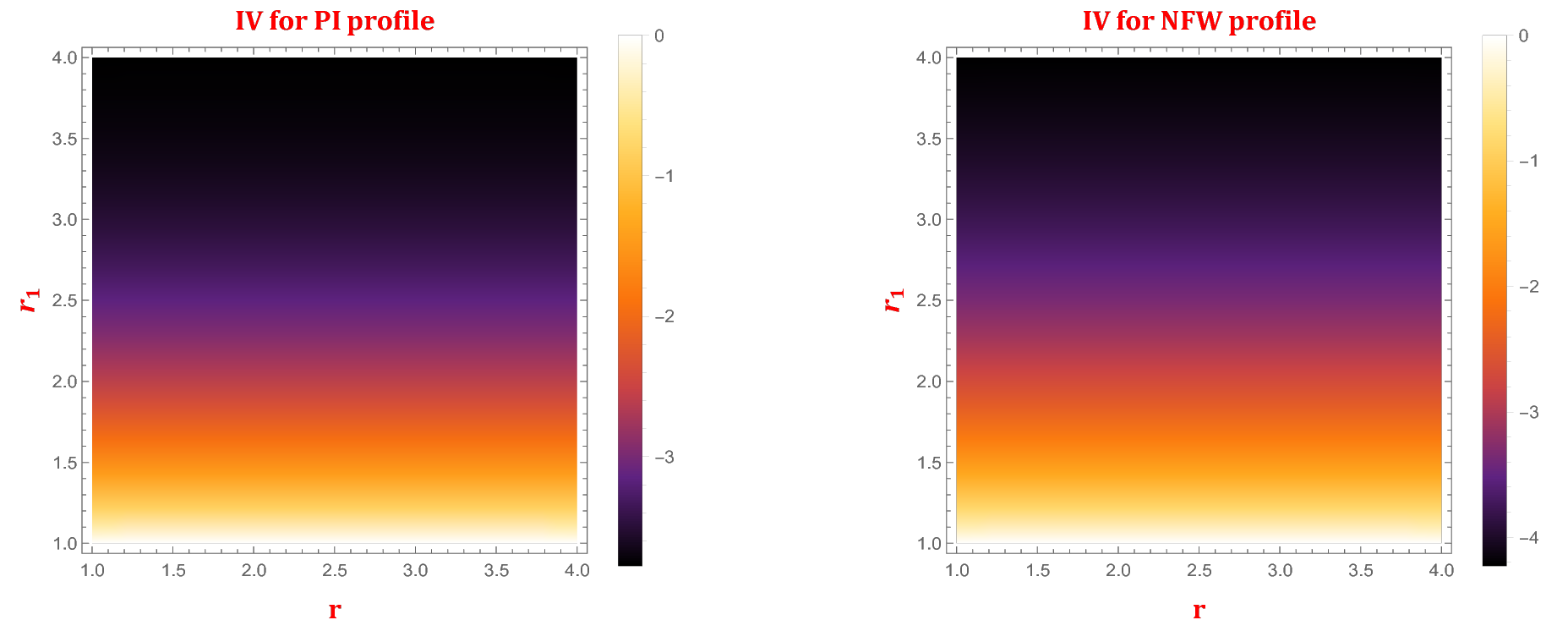}
\caption{The figure displays the variations in the VIQ under the redshift $\phi(r)=\text{constant}$. Furthermore, we keep other parameters fixed at constant values, including $\alpha=-5,\, r_s=0.5,\, \rho_s=0.02,\, \eta=0.15,\,\text{and} \, r_0 = 1$.}
\label{ch2fig.13}
\end{figure}

\begin{figure}[t]
\centering
\includegraphics[width=14.5cm,height=5.5cm]{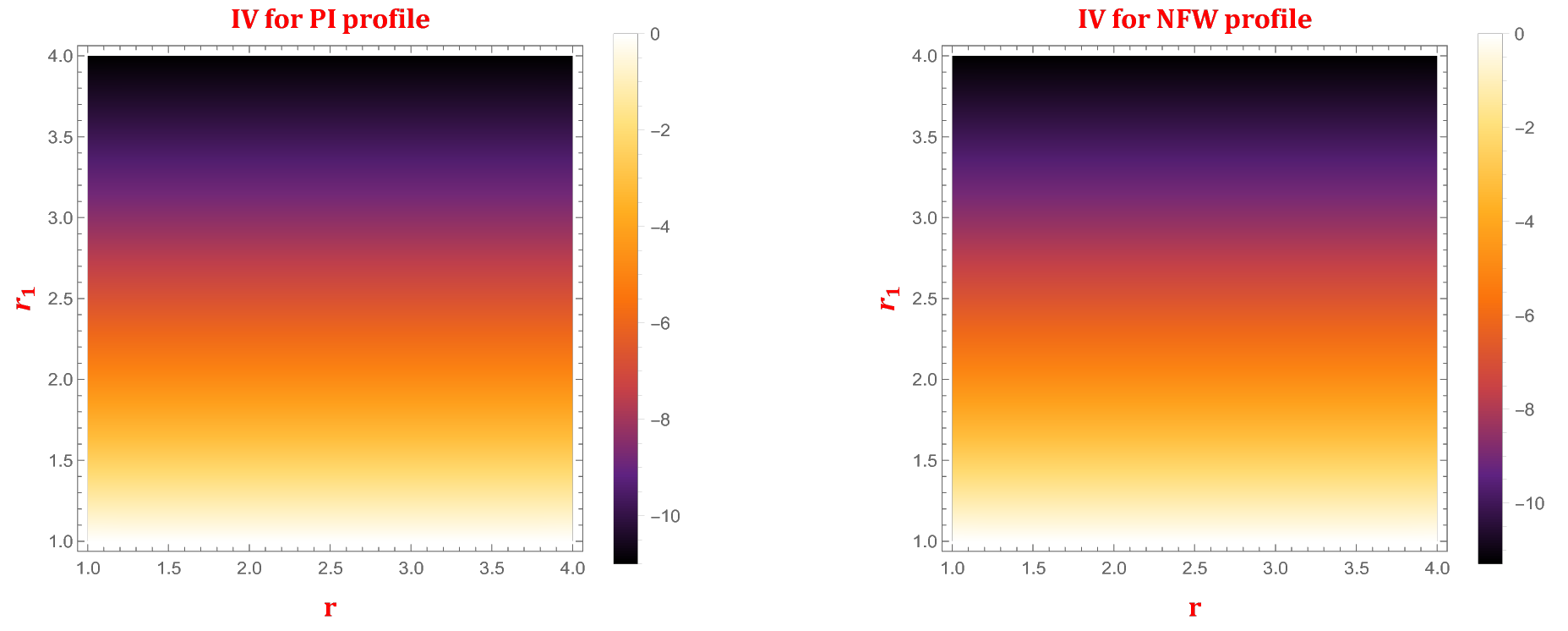}
\caption{The figure displays the variations in the VIQ under the redshift $\phi(r)=\frac{1}{r}$. Furthermore, we keep other parameters fixed at constant values, including $\alpha=-5,\, r_s=0.5,\, \rho_s=0.02,\, \eta=0.15,\,\text{and} \, r_0 = 1$.}
\label{ch2fig.14}
\end{figure}

\begin{figure}[t]
\centering
\includegraphics[width=14.5cm,height=5.5cm]{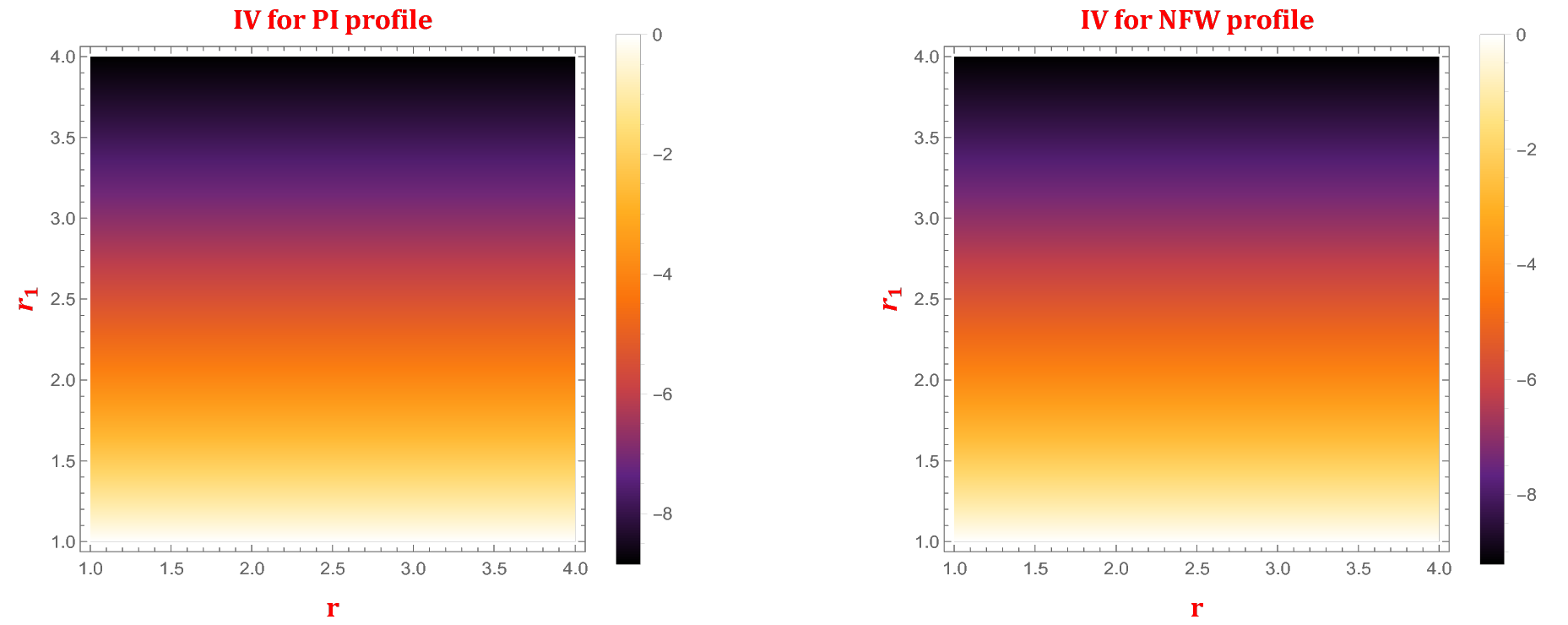}
\caption{The figure displays the variations in the VIQ under the redshift $\phi(r)=\log\left(1+\frac{r_0}{r}\right)$. Furthermore, we keep other parameters fixed at constant values, including $\alpha=-5,\, r_s=0.5,\, \rho_s=0.02,\, \eta=0.15,\,\text{and} \, r_0 = 1$.}
\label{ch2fig.15}
\end{figure}

Here, we shall determine the exact amount of exotic matter required for the stability of WH. This task relies on the Volume Integral Quantifier (VIQ) method, first introduced by Visser et al. \cite{M. Visser 2}. This method offers a structured framework for measuring the average amount of matter within the space-time continuum that violates the NEC. The VIQ method can be formally expressed as follows
\begin{equation}\label{ch2eq52}
IV=\oint \left[\rho+p_r\right]dV\,.
%=2\int_{r_0}^{\infty}(\rho+P_r)dV,
\end{equation}
The volume is defined as $dV=r^2\,dr\,d\Omega$ having the solid angle $d\Omega$. Since $\oint dV=2\int_{r_0}^{\infty}dV=8\pi \int_{r_0}^{\infty}r^2dr,$ we have
% which can also be written as
\begin{equation}\label{ch2eq53}
IV=8\pi \int_{r_0}^{\infty}(\rho+p_r)r^2dr.
\end{equation}
In the equation presented above, the integration bounds extend indefinitely to infinity. This characteristic has significant implications, as studied in the work by Lobo et al. \cite{F.S.N. Lobo}. Their research indicates that for a WH to possess asymptotically flat properties, the VIQ, with its bounds extending to infinity over the radial coordinate, should exhibit divergence. To address this, it becomes advantageous to introduce a cut-off scale for the energy-momentum tensor at a designated position denoted as $r_1$. Expanding upon this notion, we propose incorporating an energy-momentum tensor cut-off scale at a specific radial location, $r_1$, within the context of the volume integral formulation for a WH. Here, the WH is characterized by a field that varies from its throat at $r_0$ to a specified radius $r_1$, ensuring that $r_1$ is greater than or equal to $r_0$. With this condition in mind, we take the volume integral for such a WH as follows
\begin{equation}\label{1122}
IV=8\pi \int_{r_0}^{r_1}(\rho+p_r)r^2dr.
\end{equation}
By using Eq. \eqref{1122}, we have conducted an in-depth exploration of the volume integral, analyzing its behavior through visual representations depicted in Figs. \ref{ch2fig.13}-\ref{ch2fig.15} across various redshift functions. The graphical representations unmistakably indicate that as the parameter $r_1$ converges towards $r_0$, the value of $IV$ progressively tends towards zero. This observation prompts the inference that a relatively minimal exotic matter can serve to stabilize a traversable WH. Furthermore, our investigations have revealed that the judicious selection of WH geometry can potentially minimize the total quantity of matter violating the Average NEC. These findings underscore the significance of carefully tailored geometries in the context of WH stability. For those intrigued by further insights and potential applications of the VIQ, we recommend delving into the works of Jusufi \cite{K. Jusufi 2} and Sokoliuk \cite{O. Sokoliuk}, which offers additional perspectives and intriguing avenues of exploration.

\section{Conclusions}\label{ch2sec7}
In our study, we have explored how the monopole charge influences the WH solutions in $f(Q)$ gravity. This comprehensive study involved the use of two distinct models for DM halos by computing the energy density within the framework of gravity $f(Q)$. In addition, our analysis has yielded consistent solutions to the field equations to describe the Morris-Thorne WH (Eq. \eqref{1ch1}) surrounded by a DM environment. These models encompass the PI model (Eq. \eqref{4aa1}) and CDM halos utilizing the NFW profile (Eq. \eqref{4aaa1}), each chosen for its efficacy in modeling the galactic halo. In addition, shape functions were derived using inverse tangent and logarithmic functions with notable accuracy. The key findings are summarized below.
\begin{enumerate}
    \item The shape functions derived under the PI model and the CDM halo with the profiles of the NFW model for the linear form of $f(Q)$ (Eq. \eqref{ch2eq24}), as illustrated in Figs. \ref{ch2fig.1} and \ref{ch2fig.5}, satisfy both the flare-out condition $b'(r_0)<1$ and the asymptotic flatness condition $\frac{b(r)}{r}\to  0$ as $r \to \infty$. These findings highlight the consistency and relevance of the derived shape functions within these frameworks, and the observed behavior of the shape functions indicates that the solutions obtained confirm the Morris-Thorne WH criterion. This confirms that our solutions in WH physics are viable.
    \item From Figs. \ref{ch2fig.2} and \ref{ch2fig.6}, it is evident that the radial NEC is violated while the tangential NEC is satisfied, leading to an overall violation of both NEC and WEC, which shows that exotic matter may be present at the throat of the WH. Furthermore, it should be noted that the SEC is satisfied for both DM profiles when considering $\phi(r)=\text{constant}$ for various values of $\eta$.
    \item Also, for the cases $\phi(r)=\frac{1}{r}$ and $\phi(r)=\log\left(1+\frac{r_0}{r}\right)$, it is clear from Figs. \ref{ch2fig.3}-\ref{ch2fig.4} and \ref{ch2fig.7}-\ref{ch2fig.8} that the radial NEC is violated, while the tangential NEC is satisfied, resulting in the violation of NEC and WEC. However, in this case, the SEC is violated for both DM profiles across various values of $\eta$.
    \item Information on the energy conditions for both DM profiles in varying redshift functions is available in Tables \ref{ch2table:1} and \ref{ch2table:2}. In this context, we assume the location of the WH's throat to be at $r_0=1$.
    \item The nonlinear form of $f(Q)$ (as described in Eq. \eqref{ch2eq48}) leads to a highly nonlinear differential equation. Consequently, it becomes impractical to find an exact analytical solution. In such cases, the Karmakar condition is employed to obtain the shape function (Eq. \eqref{ch2eq46}), enabling us to navigate the complexities of the differential equation effectively.
    \item For the nonlinear Sec. \ref{ch2sec5}, it becomes evident that the NEC along both the radial and tangential directions becomes indeterminate precisely at the throat of the WH, where $r=r_0$ and can be visualized using Fig. \ref{ch2fig.12}.
    \item In Sec. \ref{ch2sec6}, we used a VIQ method to evaluate the quantity of exotic matter essential to maintain the WH's throat open. It is found that relatively minimal exotic matter can serve to stabilize a traversable WH and can be observed from Figs. \ref{ch2fig.13}-\ref{ch2fig.15}.
\end{enumerate}
In summary, the solutions we derived for the PI model and the CDM halo with the profiles of the NFW model are considered physically viable in the framework of $f(Q)$ gravity with monopole charge and the context of DM halos. In the GR context, WH solutions within the galactic halo involving a monopole charge have been explored in Ref. \cite{P. Das}. They considered various DM profiles, including NFW, PI and URC models, confirming that the NFW profile, combined with monopole charge, provides sufficient support for WH structures in the Milky Way galaxy's halo. Furthermore, WH solutions using NFW, PI and Thomas-Fermi DM models have been examined in the framework of $f(R)$ gravity \cite{Abdelghani}, demonstrating that the energy conditions, particularly the DEC and SEC, depend on the DM models, while the NEC and WEC remain model independent near the throat of the WH. Furthermore, WH solutions with the NFW profile in $f(T)$ gravity have also been studied in Ref. \cite{M. Sharif}, while discussions on WHs surrounded by NFW DM in the context of $f(Q)$ gravity have been discussed in Ref. \cite{G. Mustafa 2}. However, WH solutions incorporating the monopole charge within the context of DM have not yet been explored in any modified gravity theory. This gap motivates the study of WHs with a monopole charge supported by DM in the context of $f(Q)$ gravity. Our analysis indicates that monopole charges exert a minimal influence on the violation of energy conditions.\\
In the next chapter, we will study static spherically symmetric WHs in the extension of $f(Q)$, that is, $f(Q, T)$ gravity with barotropic and anisotropic EoS cases.
%%%%%%%%%%%%%%%%%%%%%%%%%%%%%%%%%%%%%%%%%%%%%%%%%%%%%%%%%%%%%%%%%%%%%%%%%%%%%%%%%%%%%%%%%%%%%%%%%%%%%%%%%%%%%%%%%%%%%%%%%%%%%%%%%%%%%%%%%%%%%%%%%%%%%%%%%%%%%%%%%%%%%%%%%%%%

% Chapter 3

\chapter{Static spherically symmetric wormholes in $f(Q, T)$ gravity} % Main chapter title
\label{Chapter3} 
% For referencing the chapter elsewhere, use \ref{Chapter1} 

\lhead{Chapter 3. \emph{Static spherically symmetric wormholes in $f(Q, T)$ gravity}} % This is for the header on each page - perhaps a shortened title
% \vspace{8 cm}
\blfootnote{*The work in this chapter is covered by the following publication:\\
\textit{Static spherically symmetric wormholes in $f(Q, T)$ gravity}, Chinese Physics C \textbf{46}, 115101 (2022).}

%\blindtext
%----------------------------------------------------------------------------------------
This chapter examines WH solutions in $f(Q, T)$ gravity, an extension of $f(Q)$ gravity, where the gravitational Lagrangian $\mathcal{L}_m$ is defined as an arbitrary function of $Q$ and $T$. The detailed study of the work is described as follows:
\begin{itemize}
    \item Field equations are derived for an Morris-Thorne WH metric within the framework of $f(Q, T)$ gravity, with solutions analyzed under two EoS cases: (i) a barotropy and (ii) an anisotropy.
    \item Two functional forms of $f(Q, T)$ are considered: a linear form $f(Q, T)=\alpha Q+\beta T$ and a nonlinear form $f(Q, T)=Q+\lambda_1 Q^2+\eta_1 T$, to investigate WH solutions.
    \item The energy conditions are studied for the derived solutions, revealing that the NEC is violated for both EoS scenarios within the context of the linear form $f(Q, T)$.
    \item In addition, the TOV tool is utilized to study stability analysis.
\end{itemize}

\section{Introduction}\label{ch3sec1}
The $f(Q, T)$ gravity, which is an extension of $f(Q)$, was presented by Yixin et al. \cite{fqt}. Within this theory, the effects from the quantum field arise from $T$, while the gravitational interaction is mediated through the non-metricity function. Although recently introduced, significant research has already been conducted on this gravity, covering both theoretical \cite{Najera, Bhattacharjee} and observational \cite{Arora1} aspects. The new interactions between non-metricity and matter were explored in Ref. \cite{T.Harko2} and further examination was carried out with matter coupling within modified $Q$ gravity by Harko et al. \cite{T. Harko1}. Delhom \cite{Delhom} analyzed minimal coupling in the context of torsion and non-metricity. Additionally, Sotiriou \cite{Sotiriou} explored $f(R)$ gravity along with torsion and non-metricity. Based on these studies, our aim was to investigate Morris-Thorne WH in a $f(Q, T)$ gravity framework. Further, our analysis is extended by examining (i) a barotropic EoS and (ii) a relationship between radial and tangential pressures in the anisotropic scenario. We then seek exact WH solutions for both linear and nonlinear $f(Q, T)$ models.\\
The overview of this chapter is organized as follows. Sec. \ref{ch3sec2} presents the field equations for WHs in $f(Q, T)$ gravity. In Sec. \ref{ch3sec3}, WH solutions are explored considering different forms of $f(Q, T)$ models. The stability of the obtained solutions is analyzed using the TOV equation in Sec. \ref{ch3sec4}, followed by the concluding remarks in Sec. \ref{ch3sec5}.\\

\section{Wormhole field equation in $f(Q, T)$ gravity}
\label{ch3sec2}
In this particular section, we assume a Morris-Thorne WH metric given by Eq. \eqref{1ch1}.  
% In the present study, we assume the matter content described by an anisotropic energy-momentum tensor to analyze the WH solutions which is given by \cite{a17,a2}
% \begin{equation}\label{10}
% \end{equation}
% where, $\rho$ denotes the energy density. $u_{\mu}$ and $v_{\mu}$ are the four velocity vector and unitary space-like vectors, respectively. Also both are satisfy the conditions $u_{\mu}u^{\nu}=-v_{\mu}v^{\nu}=1$. $p_r$ and $p_t$ denotes the radial and tangential pressures and both are function of radial coordinate $r$. 
Also, for this WH metric, the trace of the energy-momentum tensor is found to be $T=\rho-p_r-2p_t$.\\
Also, we consider the matter Lagrangian $\mathcal{L}_m=-P$ \cite{Correa}.
% and thus Eq. \eqref{fqt4} can be read as [Y. Xu et al., \textit{Eur. Phys. J. C} \textbf{79}, 708 (2019).]
% \begin{equation}\label{ch3eq1}
% \Theta_{\mu}^{\,\,\,\nu}=-\delta_{\mu}^{\,\,\,\nu}\,P-2\,T_{\mu}^{\,\,\,\nu},
% \end{equation}
Here, $P$ represents the total pressure and is written as $P=\frac{p_r+2\,p_t}{3}$.\\
% The non-metricity scalar $Q$ for the metric \eqref{8} is given by
% \begin{equation}\label{9}
% Q=-\frac{b}{r^2}\left[\frac{rb^{'}-b}{r(r-b)}+2\phi^{'}\right].
% \end{equation}
% \\
Now, inserting the Eqs. \eqref{1ch1}, \eqref{ch2eq18a} and \eqref{ch2eq20} into the motion Eq. \eqref{fqt2}, the field equations for $f(Q, T)$ gravity framework are given by
\begin{equation}\label{ch3eq2}
\frac{2 (r-b)}{(2 r-b) f_Q}\left[\rho-\frac{(r-b)}{8 \pi  r^3} \left(\frac{b r f_{\text{QQ}} Q'}{r-b}+b f_Q \left(\frac{ r \phi '+1}{r-b}-\frac{2 r-b}{2 (r-b)^2}\right)+\frac{f r^3}{2 (r-b)}\right)+\frac{f_T (P+\rho )}{8 \pi }\right]=\frac{b'}{8 \pi  r^2},
\end{equation}
\begin{multline}\label{ch3eq3}
\hspace{-0.6cm}\frac{2 b}{f r^3}\left[p_r +\frac{(r-b)}{16 \pi r^3} \left(f_Q \left(\frac{b \left(\frac{r b'-b}{r-b}+2 r \phi '+2\right)}{r-b}-4 r \phi '\right)+\frac{2 b r f_{\text{QQ}} Q'}{r-b}\right)+\frac{fr^3 (r-b)\phi '}{8\pi b  r^2}-\frac{f_T \left(P-p_r\right)}{8 \pi }\right]\\
=\frac{1}{8 \pi}\left[2\left(1-\frac{b}{r}\right)\frac{\phi '}{r}-\frac{b}{r^3}\right],
\end{multline}
and
\begin{multline}\label{ch3eq4}
\frac{1}{f_Q \left(\frac{r}{r-b}+r \phi '\right)}\left[p_t +\frac{(r-b)}{32 \pi r^2}\left(f_Q \left(\frac{4 (2 b-r) \phi '}{r-b}-4 r \left(\phi '\right)^2-4 r \phi ''\right)+\frac{2 f r^2}{r-b}-4 r f_{\text{QQ}} Q' \phi '\right) 
\right.\\ \left.
+\frac{(r-b)}{8\pi r}\left(\phi '' +{\phi '}^2-\frac{(rb'-b)\phi '}{2r(r-b)}+\frac{\phi '}{r}\right)f_Q \left(\frac{r}{r-b}+r \phi '\right)-\frac{f_T \left(P-p_t\right)}{8 \pi }\right]\\
=\frac{1}{8\pi}\left(1-\frac{b}{r}\right)\left[\phi '' +{\phi '}^2-\frac{(rb'-b)\phi '}{2r(r-b)}-\frac{rb'-b}{2r^2 (r-b)}+\frac{\phi '}{r}\right].
\end{multline}\\
As established, the field equations for an Morris-Thorne traversable WH in GR can be expressed as
\begin{equation}\label{13a}
\frac{b'}{8 \pi  r^2}= \tilde{\rho},
\end{equation}
\begin{equation}\label{13b}
\frac{1}{8 \pi}\left[2\left(1-\frac{b}{r}\right)\frac{\phi '}{r}-\frac{b}{r^3}\right] = \tilde{p_r},
\end{equation}
and
\begin{equation}\label{13c}
\frac{1}{8\pi}\left(1-\frac{b}{r}\right)\left[\phi '' +{\phi '}^2-\frac{(rb'-b)\phi '}{2r(r-b)}-\frac{rb'-b}{2r^2 (r-b)}+\frac{\phi '}{r}\right] = \tilde{p_t}\,.
\end{equation}
Here, $\tilde{\rho}$ is the effective energy density, $\tilde{p_r}$ denotes the effective radial pressure, and $\tilde{p_t}$ is the effective tangential pressure. Now, comparing Eqs. \eqref{ch3eq2}-\eqref{ch3eq4} with Eqs. \eqref{13a}-\eqref{13c}, we get
\begin{equation}\label{ch3eq5}
\hspace{-0.5cm}\tilde{\rho}=\frac{2 (r-b)}{(2 r-b) f_Q}\left[\rho-\frac{1}{8 \pi  r^2}\left(1-\frac{b}{r}\right) \left(\frac{b r f_{\text{QQ}} Q'}{r-b}+b f_Q \left(\frac{ r \phi '+1}{r-b}-\frac{2 r-b}{2 (r-b)^2}\right)+\frac{f r^3}{2 (r-b)}\right)+\frac{f_T (P+\rho )}{8 \pi }\right],
\end{equation}
\begin{multline}\label{ch3eq6}
\tilde{p_r}=\frac{2 b}{f r^3}\left[p_r +\frac{1}{16 \pi r^2}\left(1-\frac{b}{r}\right) \left(f_Q \left(\frac{b \left(\frac{r b'-b}{r-b}+2 r \phi '+2\right)}{r-b}-4 r \phi '\right)+\frac{2 b r f_{\text{QQ}} Q'}{r-b}\right)+\frac{fr^3 (r-b)\phi '}{8\pi b  r^2}
\right. \\ \left.
-\frac{f_T \left(P-p_r\right)}{8 \pi }\right],
\end{multline}
and
\begin{multline}\label{ch3eq7}
\hspace{-0.7cm}\tilde{p_t}=\frac{1}{f_Q \left(\frac{r}{r-b}+r \phi '\right)}\left[p_t +\frac{1}{32 \pi r}\left(1-\frac{b}{r}\right) \left(f_Q \left(\frac{4 (2 b-r) \phi '}{r-b}-4 r \left(\phi '\right)^2-4 r \phi ''\right)+\frac{2 f r^2}{r-b}-4 r f_{\text{QQ}} Q' \phi '\right) \right.\\ \left.
+\frac{1}{8\pi}\left(1-\frac{b}{r}\right)\left(\phi '' +{\phi '}^2-\frac{(rb'-b)\phi '}{2r(r-b)}+\frac{\phi '}{r}\right)f_Q \left(\frac{r}{r-b}+r \phi '\right)-\frac{f_T \left(P-p_t\right)}{8 \pi }\right].
\end{multline}\\
The above equations reveal additional terms beyond those in GR. Moreover, by using Eqs. \eqref{ch3eq2}-\eqref{ch3eq4}, the nonzero components of the field equations for $f(Q, T)$ gravity can be expressed as follows.
\begin{equation}\label{ch3eq18}
8 \pi  \rho =\frac{(r-b)}{2 r^3} \left[f_Q \left(\frac{(2 r-b) \left(r b'-b\right)}{(r-b)^2}+\frac{b \left(2 r \phi '+2\right)}{r-b}\right)+\frac{2 b r f_{\text{QQ}} Q'}{r-b}+\frac{f r^3}{r-b}-\frac{2r^3 f_T (P+\rho )}{(r-b)}\right],
\end{equation}
\begin{equation}\label{ch3eq19}
8 \pi  p_r=-\frac{(r-b)}{2 r^3} \left[f_Q \left(\frac{b }{r-b}\left(\frac{r b'-b}{r-b}+2 r \phi '+2\right)-4 r \phi '\right)+\frac{2 b r f_{\text{QQ}} Q'}{r-b}+\frac{f r^3}{r-b}-\frac{2r^3 f_T \left(P-p_r\right)}{(r-b)}\right],
\end{equation}
and
\begin{multline}\label{ch3eq20}
\hspace{-0.6cm}8 \pi  p_t=-\frac{(r-b)}{4 r^2} \left[f_Q \left(\frac{\left(r b'-b\right) \left(\frac{2 r}{r-b}+2 r \phi '\right)}{r (r-b)}+\frac{4 (2 b-r) \phi '}{r-b}-4 r \left(\phi '\right)^2-4 r \phi ''\right)-4 r f_{\text{QQ}} Q' \phi '+\frac{2 f r^2}{r-b}
\right. \\ \left.
-\frac{4r^2 f_T \left(P-p_t\right)}{(r-b)}\right].
\end{multline}\\
These field equations allow for the investigation of WH solutions by exploring various models in $f(Q, T)$ gravity.\\

% Let us dedicate a few lines to classical energy conditions developed from the Raychaudhuri equations. These conditions are used to discuss the physically realistic matter configuration. The four energy conditions the NEC (NEC), weak energy condition (WEC), dominant energy condition (DEC),
% and strong energy condition (SEC), are expressed as:\\
% $\bullet$ Weak energy conditions (WEC) if $\rho\geq0$, $\rho+p_j\geq0$, $\forall j$.\\
% $\bullet$ NEC (NEC) if $\rho+p_j\geq0$, $\forall j$.\\
% $\bullet$ Dominant energy conditions (DEC) if $\rho\geq0$, $\rho \pm p_j\geq0$, $\forall j$.\\
% $\bullet$ Strong energy conditions (SEC) if $\rho+p_j\geq0$, $\rho+\sum_jp_j\geq0$, $\forall j$,\\
% where $j=r,\,t$.\\
% It is known that, in GR, The NEC is significant because the violation of the NEC may confirm that there must be present exotic matter in WH's throat. Hence NEC is usually studied for WH solutions in GR. Moreover, energy density also needs to be positive for a realistic matter source that maintains the WH solutions.\\
Solving Eqs. \eqref{ch3eq18} to \eqref{ch3eq20}, one can find
\begin{multline}\label{ch3eq21}
\rho =-\frac{f_Q f_T \left(-r b' \left(2 r (r-b) \phi '+b+2 r\right)+3 b^2+4 r (b-r) \left(\phi ' \left(r (b-r) \phi '+3 b-2 r\right)+r (b-r) \phi ''\right)\right)}{48 \pi  r^3 (r-b) \left(f_T+8 \pi \right)}\\
-\frac{24 \pi  f_Q \left(r (b-2 r) b'+b \left(2 r (b-r) \phi '+b\right)\right)}{48 \pi  r^3 (r-b) \left(f_T+8 \pi \right)}\\
-\frac{r (b-r) \left(f_T \left(2 f_{\text{QQ}} Q' \left(2 r (b-r) \phi '+b\right)+3 f r^2\right)+24 \pi  \left(2 b f_{\text{QQ}} Q'+f r^2\right)\right)}{48 \pi  r^3 (r-b) \left(f_T+8 \pi \right)},
\end{multline}
\begin{multline}\label{ch3eq22}
p_r=\frac{f_Q f_T \left(r b' \left(2 r (r-b) \phi '+b+2 r\right)-3 b^2-4 r (b-r) \left(\phi ' \left(r (b-r) \phi '+3 b-2 r\right)+r (b-r) \phi ''\right)\right)}{48 \pi  r^3 (b-r) \left(f_T+8 \pi \right)}\\
+\frac{24 \pi  f_Q \left(b r b'-(3 b-2 r) \left(2 r (b-r) \phi '+b\right)\right)}{48 \pi  r^3 (b-r) \left(f_T+8 \pi \right)}\\
-\frac{r (b-r) \left(f_T \left(2 f_{\text{QQ}} Q' \left(2 r (b-r) \phi '+b\right)+3 f r^2\right)+24 \pi  \left(2 b f_{\text{QQ}} Q'+f r^2\right)\right)}{48 \pi  r^3 (b-r) \left(f_T+8 \pi \right)},
\end{multline}
and
\begin{multline}\label{ch3eq23}
p_t=-\frac{f_Q f_T \left(-r b' \left(2 r (r-b) \phi '+b+2 r\right)+3 b^2+4 r (b-r) \left(\phi ' \left(r (b-r) \phi '+3 b-2 r\right)+r (b-r) \phi ''\right)\right)}{48 \pi  r^3 (b-r) \left(f_T+8 \pi \right)}\\
-\frac{24 \pi  r f_Q \left(r \left(b' \left((b-r) \phi '-1\right)+2 (b-r)^2 \left(\left(\phi '\right)^2+\phi ''\right)+(2 r-5 b) \phi '\right)+b \left(3 b \phi '+1\right)\right)}{48 \pi  r^3 (b-r) \left(f_T+8 \pi \right)}\\
-\frac{r (b-r) \left(f_T \left(2 b f_{\text{QQ}} Q'+3 f r^2\right)+4 r (b-r) f_{\text{QQ}} \left(f_T+12 \pi \right) Q' \phi '+24 \pi  f r^2\right)}{48 \pi  r^3 (b-r) \left(f_T+8 \pi \right)}.
\end{multline}

\section{Wormhole solutions with distinct $f(Q, T)$ models}\label{ch3sec3}
In this particular section, we are going to study the WH solutions using distinct forms of the $f(Q, T)$ model. Using energy conditions, we will study how the WH solution behaves. In addition, we use $\phi(r)=constant$ to attain the asymptotic behavior of de Sitter and anti-de Sitter. In our whole study of this chapter, we consider the constant redshift function.

\subsubsection{Barotropic wormhole solutions}
\begin{figure}[h]
\centering
\subfigure[]{\includegraphics[width=6.5cm,height=4cm]{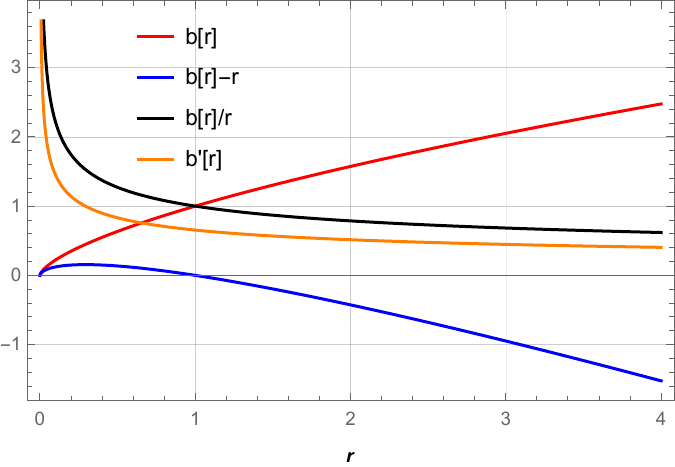}}\,\,\,\,\,\,\,\,\,\,\,\,\,\,\,\,\,\,\,\,\,\,\,\,\,
\subfigure[]{\includegraphics[width=6.5cm,height=4cm]{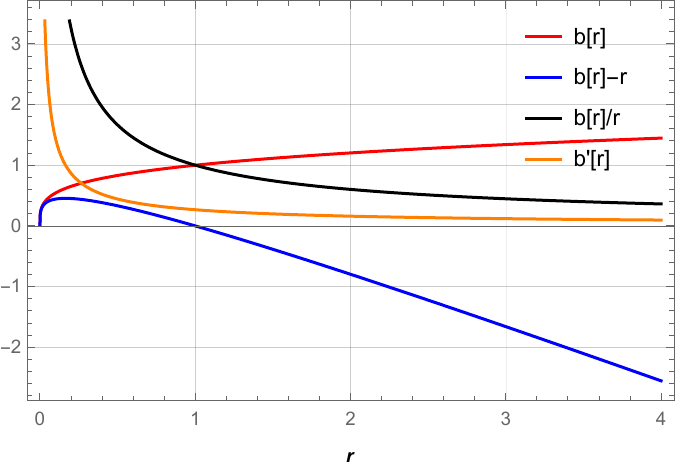}}
\caption{The figure displays how the shape function $b(r)$, alongside the conditions for flare-out ($b'(r)<1$), the throat ($b(r)-r<0$), and asymptotic flatness ($\frac{b(r)}{r}\rightarrow 0$ as $r\rightarrow \infty$) vary with $r$. Specifically, it examines these properties for two cases: (a) $\omega= -1.5$ with $\beta=1$ and (b) $\omega= -0.5$ with $\beta=14$. The parameter $\alpha=1$ and the radial distance is measured in kilometers (km).}
\label{ch3fig1}
\end{figure}

This subsection focuses on the barotropic EoS, which defines a connection between the elements of the energy-momentum tensor. We utilize this framework to obtain solutions to the field equations and develop feasible WH models. In the context of GR, this method is commonly used to derive precise WH configurations. However, in modified gravity theories, the complexity of the field equations often prevents researchers from deriving precise analytical solutions. As a result, they either resort to numerical techniques or proceed without assuming a specific EoS. Among the various choices in the literature, the barotropic EoS is commonly adopted to investigate WH structures. In this study, we focus on WH solutions governed by the following form of EoS \cite{Lobo5, Jusufi1}
\begin{equation}
\label{ch3eq24}
p_r=\omega\rho\,.
\end{equation}
Here, $\omega$ represents the EoS parameter. The authors \cite{F.S.N. Lobo} noted that WH solutions, which are asymptotically flat and have $\omega \leq -1$, are called phantom region EoS. In symmetric teleparallel gravity, deriving WH solutions with a linear EoS is challenging, which is stated in Ref. \cite{Zinnat 1}. However, in this study, our goal is to derive an analytical solution for WH in the context of $f(Q, T)$ gravity.\\
Using Eqs. \eqref{ch3eq21}, \eqref{ch3eq22} in Eq. \eqref{ch3eq24}, we get
%\begin{widetext}
\begin{multline}
\label{ch3eq25}
f_Q \left[f_T \left(r (b+2 r) b'-3 b^2\right)+24 \pi  \left(b r b'-b (3 b-2 r)\right)\right]-r (b-r) \left[f_T \left(2 b f_{\text{QQ}} Q'+3 f r^2\right)
\right. \\ \left.
+24 \pi  \left(2 b f_{\text{QQ}} Q'+f r^2\right)\right]=\omega  \left[f_Q \left(f_T \left(3 b^2-r (b+2 r) b'\right)+24 \pi  \left(r (b-2 r) b'+b^2\right)\right)
\right. \\ \left.
+r (b-r) \left(f_T \left(2 b f_{\text{QQ}} Q'+3 f r^2\right)+24 \pi  \left(2 b f_{\text{QQ}} Q'+f r^2\right)\right)\right].
\end{multline}
%\end{widetext}
Solving the aforementioned equation for an arbitrary $f(Q, T)$ form is complex; therefore, we will focus on some particular forms of $f(Q, T)$ to obtain the shape function $b(r)$. For this study, we assume two distinct forms of $f(Q, T)$ such as linear ($f(Q, T)=\alpha Q+\beta T$ \cite{fqt}) and nonlinear ($f(Q, T)=Q+\lambda_1 Q^2+\eta_1 T$ \cite{Dixit}) and try to obtain an analytical solution for WH. Through our calculations, we determined that obtaining an analytical solution for the WH is not feasible in the nonlinear case. However, for the linear case, such solutions are achievable. Therefore, we assume
\begin{equation}
\label{ch3eq26}
f(Q, T)=\alpha\,Q+\beta\,T
\end{equation}
to study the WH solutions with EoS cases. Here, $\alpha$ and $\beta$ are model parameters.\\
Apply the above form \eqref{ch3eq26} to the field Eqs. \eqref{ch3eq18}-\eqref{ch3eq20}, we derive
 \begin{equation}\label{ch3eq27}
 \rho =\frac{\alpha  (12 \pi -\beta ) b'}{3 (4 \pi -\beta ) (\beta +8 \pi ) r^2},
 \end{equation}
 \begin{equation}\label{ch3eq28}
 p_r=-\frac{\alpha  \left(2 \beta  r b'-3 \beta  b+12 \pi  b\right)}{3 (4 \pi -\beta ) (\beta +8 \pi ) r^3},
 \end{equation}
 and
 \begin{equation}\label{ch3eq29}
 p_t=-\frac{\alpha  \left((\beta +12 \pi ) r b'+3 b (\beta -4 \pi )\right)}{6 (4 \pi -\beta ) (\beta +8 \pi ) r^3}.
 \end{equation}
Using the EoS from Eq. \eqref{ch3eq24} in conjunction with Eqs. \eqref{ch3eq27} and \eqref{ch3eq28}, we derive the shape function as follows
\begin{equation}\label{ch3eq30}
b\left(r\right)= c_1 r^{\lambda_2}\,,
\end{equation}
where
\begin{equation*}
\label{ch3eq31}
\lambda_2={-\frac{3 (\beta -4 \pi )}{\beta  (\omega -2)-12 \pi  \omega }}
\end{equation*}
and $c_1$ represents the constant of integration. Also, without loss of generality, we can set $c_1=1$.\\
To meet the conditions for asymptotic flatness, $\lambda_2$ must be less than $1$, that is, $\lambda_2 < 1$. Therefore, the ranges allowed for $\omega$ and $\beta$ are as follows:\\
\begin{table}[H]
\begin{center}
\begin{tabular}{ |c|c|c| }
 \hline
 \multicolumn{2}{|c|}{Parameters} \\
 \hline
 $\omega$ & $\beta$ \\
 \hline
 $(-\infty, -1)$   & $(-\infty, \frac{12 \pi  \omega }{\omega -2})\cup (12 \pi, \infty)$\\
 \hline
$(-1, 2)$ &  $(\frac{12 \pi  \omega }{\omega -2}, 12 \pi)$ \\ 
\hline
 $2$ & $(-\infty, 12 \pi)$\\
 \hline
$(2, \infty)$ & $(-\infty, 12 \pi)\cup (\frac{12 \pi  \omega }{\omega -2}, \infty)$ \\
\hline
\end{tabular}
\caption{Permissible ranges for $\omega$ and $\beta$.}
\label{ch3table:1}
\end{center}
\end{table}
Based on the Table \ref{ch3table:1}, we can observe that $\omega<-1$ shows the phantom region, and $-1<\omega<0$ represents the quintessence region. Further, we are neglecting the region $\omega\geq0$ as the Universe is accelerating. Taking into account the ranges of $\omega$ and $\beta$, we select specific values and illustrate how the shape function behaves. One can observe that Fig. \ref{ch3fig1} indicates a positively increasing behavior of the shape function and satisfaction with the
flare-out condition at the throat. In this scenario, the radius of the throat is $r_0=1$. Therefore, we can conclude that the shape function meets all the essential conditions for traversability.\\
Now, considering Eq. \eqref{ch3eq30}, we can express the field Eqs. \eqref{ch3eq27}-\eqref{ch3eq29} as follows.\\
% $\bullet$ \textbf{NEC :} $\rho+p_r=-\frac{\alpha  (12 \pi -\beta ) (\omega +1) r^{\eta-3}}{\Lambda}$ and $\rho+p_t=\frac{\alpha \left( 12 \pi  (\omega -1)-\beta  (\omega -5)\right) r^{\eta-3}}{2\,\Lambda}$.\\
% $\bullet$ \textbf{DEC :} $\rho-p_r=\frac{\alpha  (12 \pi -\beta )(\omega -1) r^{\eta-3}}{\Lambda}$ and $\rho-p_t=-\frac{\alpha\,\left( \beta\,(1-\omega ) +12 \pi  (\omega +3)\right) r^{\eta-3}}{2\,\Lambda}$.\\
% $\bullet$ \textbf{SEC :} $\rho+p_r+2p_t=\frac{4\,\alpha\,\beta\,r^{\eta-3}}{\Lambda}$,\\
\begin{equation}
\label{ch3eq32a}
\rho=\frac{\alpha  (\beta -12 \pi ) r^{\lambda_2-3}}{\Lambda},
\end{equation}
\begin{equation}
\label{ch3eq32b}
p_r=\frac{\alpha  (\beta -12 \pi ) \omega  r^{\lambda_2-3}}{\Lambda},
\end{equation}
and
\begin{equation}
\label{ch3eq32c}
p_t=\frac{\alpha (12 \pi  (\omega +1)-\beta  (\omega -3)) r^{\lambda_2-3}}{2\Lambda},
\end{equation}
where $\Lambda=(\beta +8 \pi ) \left(12 \pi  \omega -\beta  (\omega -2)\right)$.\\
\begin{center}
\begin{figure}[h]
\centering
\includegraphics[width=7cm,height=4.5cm]{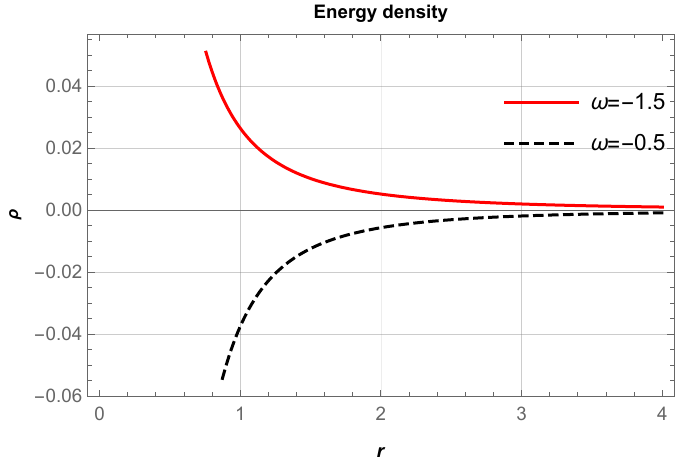}
\caption{The figure displays how the energy density $\rho$  varies with $r$ for particular values of $\omega=-1.5,-0.5$ corresponding to $\beta=1,14$, respectively. The parameter $\alpha=1$ and the radial distance is measured in km.}
\label{ch3fig2}
\end{figure}
\end{center}
Considering some specific values of $\omega$ and $\beta$ from the derived range, we plotted the graph for the energy density in Fig. \ref{ch3fig2}. This shows that the energy density exhibits a monotonically decreasing behavior throughout the space-time for the region $\omega<-1$. However, it shows a violation for the region $-1<\omega<0$. Therefore, we have considered some specific value of $\omega\in(-\infty, -1)$ and plotted the graph for the energy conditions in Fig. \ref{ch3fig3}. This figure indicates that $\rho + p_t$ is positive but with decreasing behavior, while the radial NEC, that is, $\rho + p_r$, shows an increasing negative behavior in the vicinity of the throat. Furthermore, the DEC holds for both radial and tangential pressure components, while violation of the SEC is observed. The NEC violation implies the existence of exotic matter at the WH's throat.
\begin{table}[h]
\begin{center}
\begin{tabular}{ |c|c|c|c|c| }
 \hline
  \multicolumn{1}{|c|}{Terms} &
  \multicolumn{2}{|c|}{Interpretations}\\
 \hline
 $\omega$ & $(-\infty, -1)$ & $(-1, 2)$ \\
 \hline
 $\beta$   & $(-\infty, \frac{12 \pi  \omega }{\omega -2})\cup (12 \pi, \infty)$ & $(\frac{12 \pi  \omega }{\omega -2}, 12 \pi)$\\
 \hline
$\rho$ &  $satisfied$ & $violated$\\ 
 \hline
$\rho + p_r$ & $violated$ & $violated$\\
 \hline
$\rho + p_t$ & $satisfied$ & $satisfied$\\
 \hline
$\rho - |p_r|$ & $satisfied$ & $violated$\\
 \hline
$\rho - |p_t|$ & $satisfied$ & $violated$\\
 \hline
$\rho + p_r + 2p_t$ & $violated$ & $satisfied$\\
 \hline
\end{tabular}
\caption{An overview of the energy conditions for $p_r=\omega\,\rho$.}
\label{ch3table:2}
\end{center}
\end{table}
\begin{figure}[H]
\centering
\includegraphics[width=12cm,height=8cm]{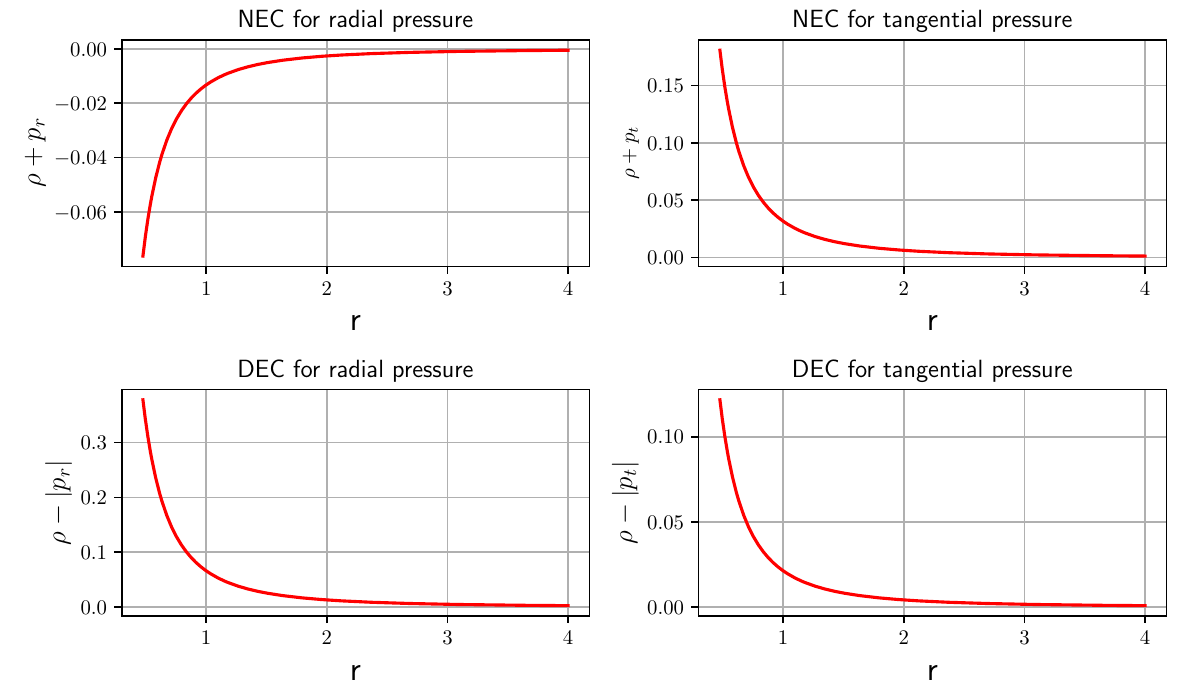}\,\,
\includegraphics[width=7cm,height=4cm]{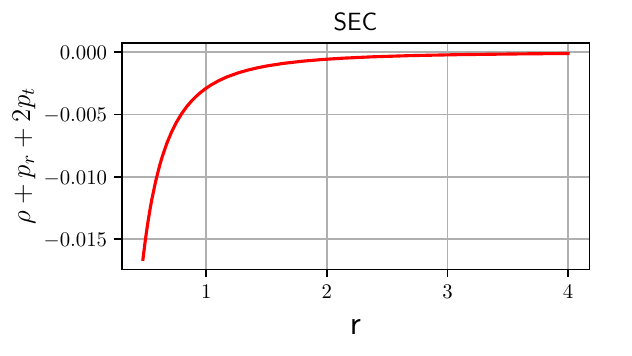}
\caption{The figure displays how the NEC, DEC and SEC vary with $r$ for $\omega=-1.5$ and $\beta=1$. The parameter $\alpha=1$ and the radial distance is measured in km.}
\label{ch3fig3}
\end{figure}
\subsubsection{Anisotropic wormhole solutions}
\begin{figure}[H]
\centering
\subfigure[]{\includegraphics[width=6.5cm,height=4cm]{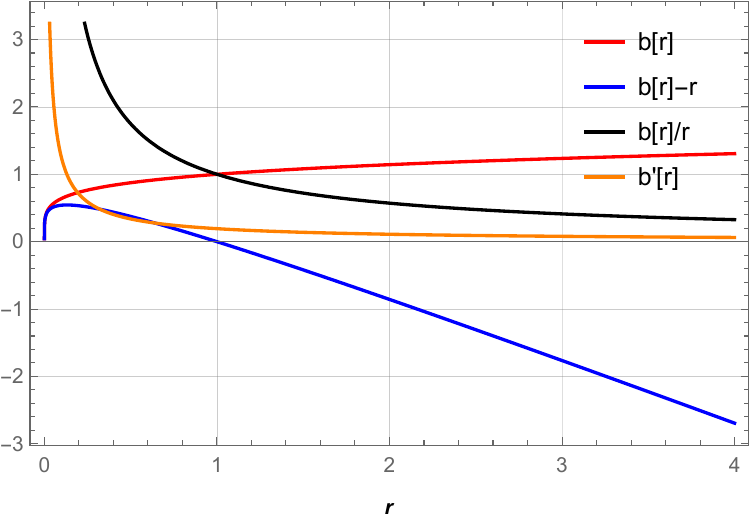}}\,\,\,\,\,\,\,\,\,\,\,\,\,\,\,\,\,\,\,\,\,\,\,\,\,
\subfigure[]{\includegraphics[width=6.5cm,height=4cm]{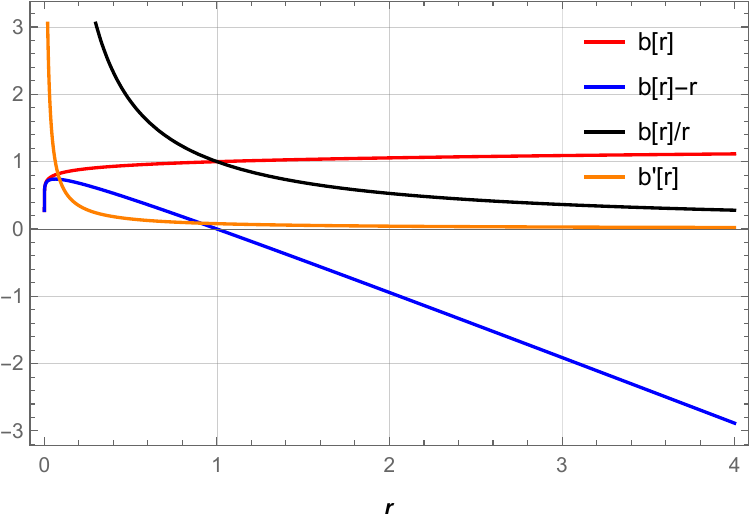}}
\caption{The figure displays how the shape function $b(r)$ alongside the conditions for flare-out ($b'(r)<1$), the throat ($b(r)-r<0$) and asymptotic flatness ($\frac{b(r)}{r}\rightarrow 0$ as $r\rightarrow \infty$) vary with $r$. Specifically, it examines these properties for (a) $n=-2.5$ with $\beta=16$ and (b) $n= -2$ with $\beta=14$. The parameter $\alpha=1$ and the radial distance is measured in km.}
\label{ch3fig6}
\end{figure}

\begin{figure}[H]
\centering
\subfigure[]{\includegraphics[width=6.5cm,height=4cm]{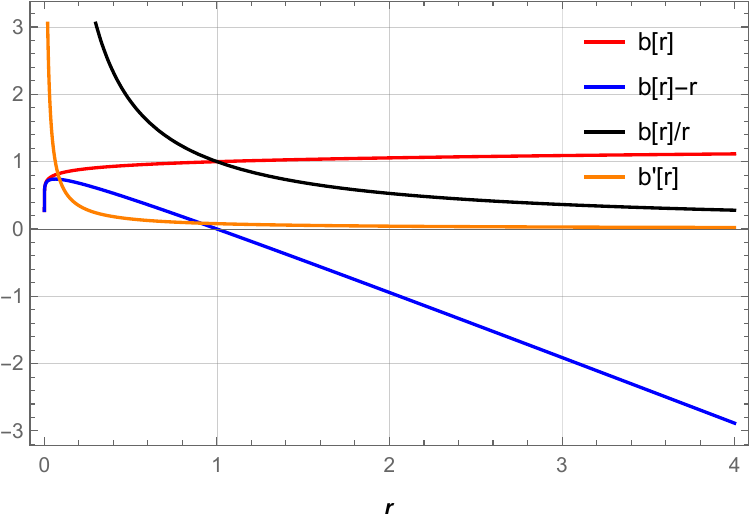}}\,\,\,\,\,\,\,\,\,\,\,\,\,\,\,\,\,\,\,\,\,\,\,\,\,
\subfigure[]{\includegraphics[width=6.5cm,height=4cm]{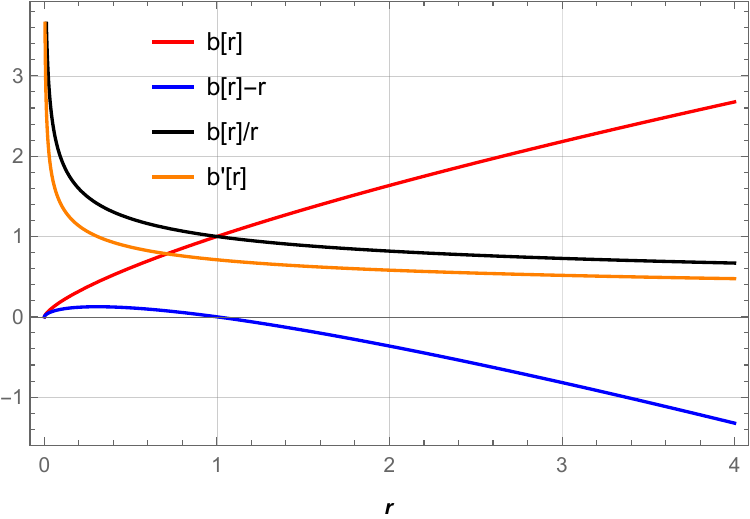}}
\caption{The figure displays how the shape function $b(r)$ alongside the conditions for flare-out ($b'(r)<1$), the throat ($b(r)-r<0$) and asymptotic flatness ($\frac{b(r)}{r}\rightarrow 0$ as $r\rightarrow \infty$) vary with $r$. Specifically, it examines these properties for (a) $n= -1.5$ with $\beta=14$ and (b) $n= -0.1$ with $\beta=1$. The parameter $\alpha=1$ and the radial distance is measured in km.}
\label{ch3fig7}
\end{figure}

\begin{figure}[h]
\centering
\subfigure[]{\includegraphics[width=6.5cm,height=4cm]{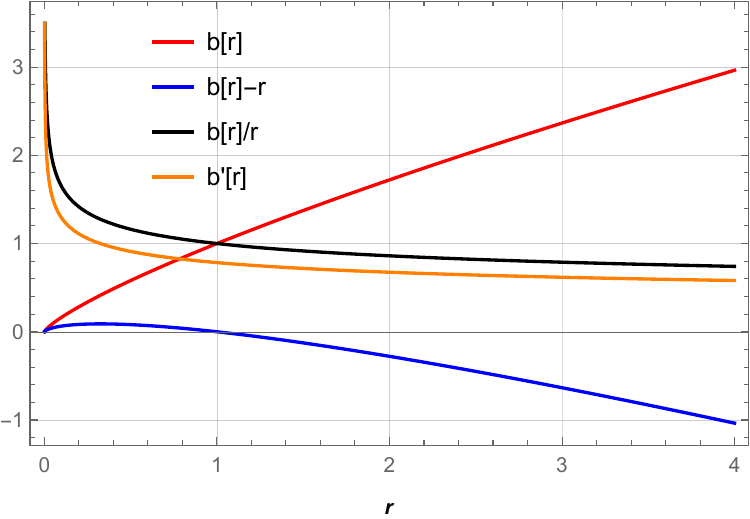}}\,\,\,\,\,\,\,\,\,\,\,\,\,\,\,\,\,\,\,\,\,\,\,\,\,
\subfigure[]{\includegraphics[width=6.5cm,height=4cm]{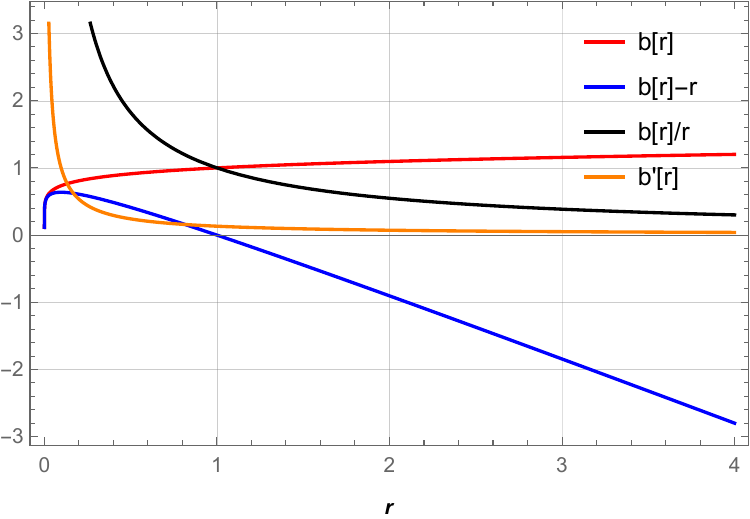}}
\caption{The figure displays how the shape function $b(r)$ alongside the conditions for flare-out ($b'(r)<1$), the throat ($b(r)-r<0$) and asymptotic flatness ($\frac{b(r)}{r}\rightarrow 0$ as $r\rightarrow \infty$) vary with $r$. Specifically, it examines these properties for (a) $n= 0.25$ with $\beta=6$ and (b) $n= 0.5$ with $\beta=12$. The parameter $\alpha=1$ and the radial distance is measured in km.}
\label{ch3fig8}
\end{figure}

\begin{figure}[h]
\centering
\includegraphics[width=6.5cm,height=4cm]{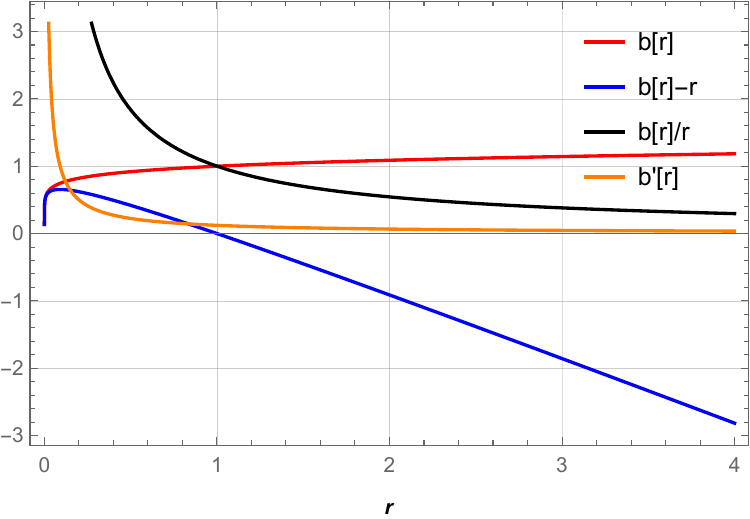}
\caption{The figure displays how the shape function $b(r)$ alongside the conditions for flare-out ($b'(r)<1$), the throat ($b(r)-r<0$) and asymptotic flatness ($\frac{b(r)}{r}\rightarrow 0$ as $r\rightarrow \infty$) vary with $r$. Specifically, it examines these properties for $n= 2$ with $\beta=13$. The parameter $\alpha=1$ and the radial distance is measured in km.}
\label{ch3fig9}
\end{figure}

\begin{figure}[h]
\centering
\includegraphics[width=7.5cm,height=5cm]{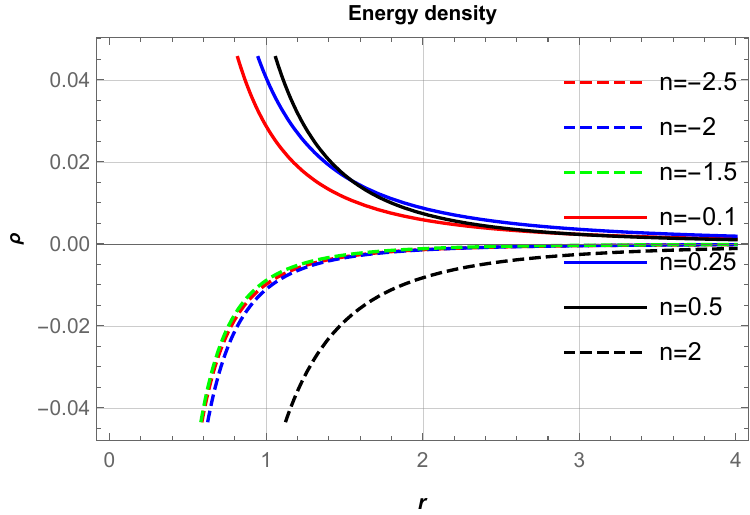}
\centering
\caption{The figure displays how the $\rho$ varies with $r$. Specifically, it examines these properties for $n=-2.5,\,-2,\,-1.5,\,-0.1,\,0.25,\,0.5,$ and $2$ corresponding to $\beta=16,\,14,\,14,\,1,\,6,\,12,$ and $13$, respectively. The parameter $\alpha=1$ and the radial distance is measured in km.}
\label{ch3fig10}
\end{figure}

\begin{figure}[h]
\centering
\includegraphics[width=6.5cm,height=4cm]{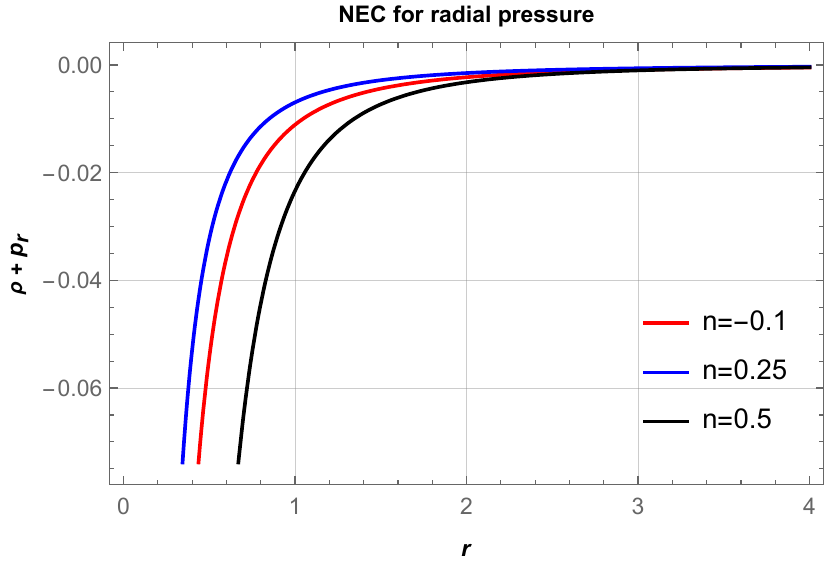}\,\,\,\,\,\,\,\,\,\,\,\,\,\,\,\,\,\,\,\,\,\,\,\,\,
\includegraphics[width=6.5cm,height=4cm]{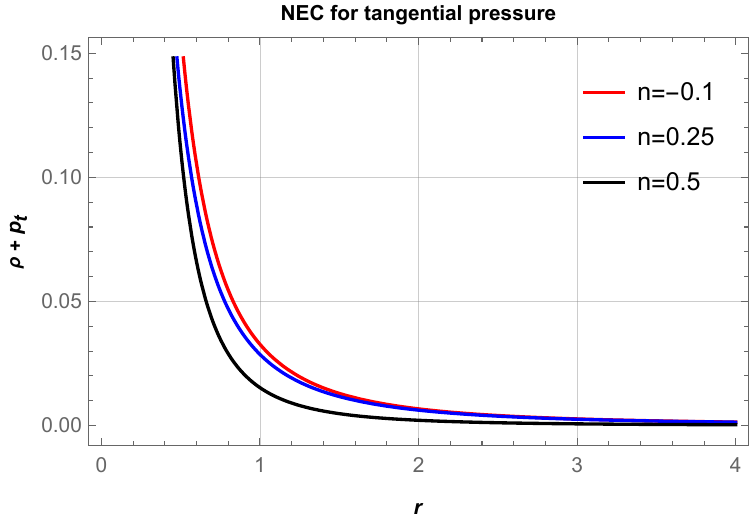}
\caption{The figure displays how the NEC varies with $r$. Specifically, it examines these properties for $n=-0.1,\,0.25,$ and $0.5$ corresponding to $\beta=1,\,6,$ and $12$, respectively. The parameter $\alpha=1$ and the radial distance is measured in km.}
\label{ch3fig11}
\end{figure}

\begin{figure}[h]
\centering
\includegraphics[width=6.5cm,height=4cm]{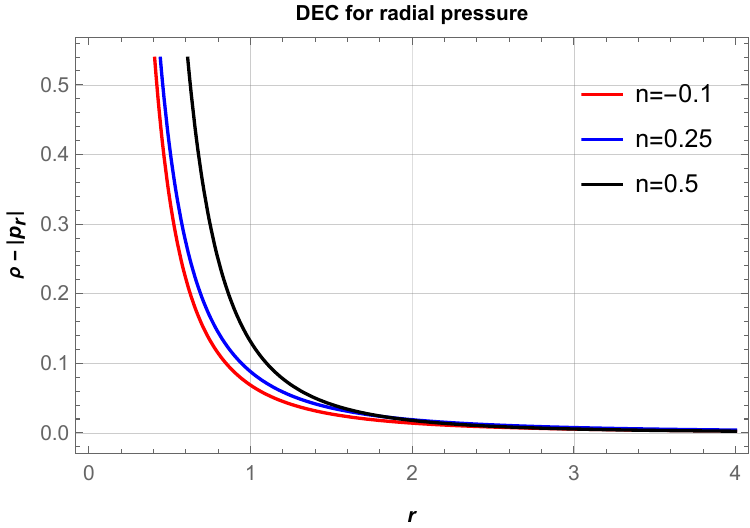}\,\,\,\,\,\,\,\,\,\,\,\,\,\,\,\,\,\,\,\,\,\,\,\,\,
\includegraphics[width=6.5cm,height=4cm]{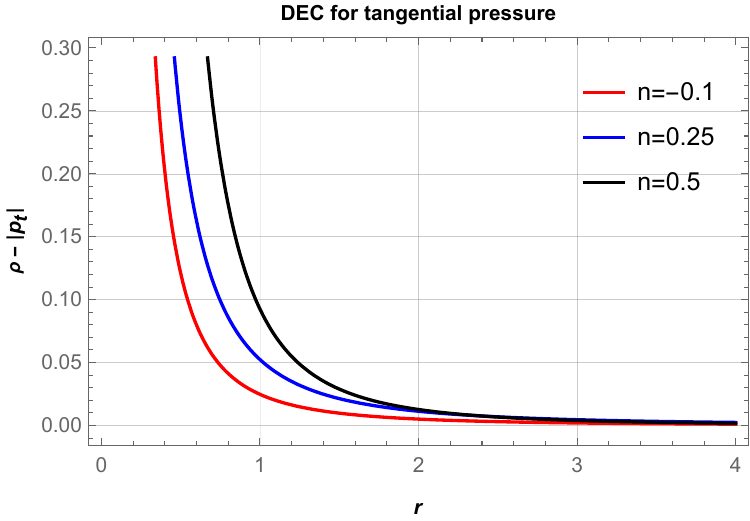}
\caption{The figure displays how the DEC varies with $r$. Specifically, it examines these properties for $n=-0.1,\,0.25,$ and $0.5$ corresponding to $\beta=1,\,6,$ and $12$, respectively. The parameter $\alpha=1$ and the radial distance is measured in km.}
\label{ch3fig12}
\end{figure}

\begin{figure}[h]
\centering
\includegraphics[width=6.5cm,height=4cm]{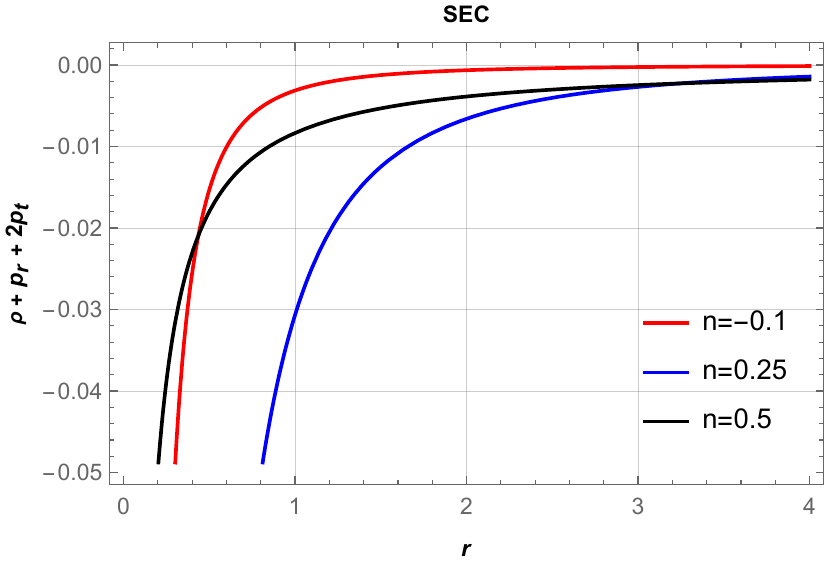}
\centering
\caption{The figure displays how the SEC varies with $r$. Specifically, it examines these properties for $n=-0.1,\,0.25,$ and $0.5$ corresponding to $\beta=1,\,6,$ and $12$, respectively. The parameter $\alpha=1$ and the radial distance is measured in km.}
\label{ch3fig13}
\end{figure}

This section is dedicated to the anisotropic relation, i.e., $p_r\neq p_t$, to study the geometry of the WH. For this scenario, we establish the relationship between $p_r$ and $p_t$ as follows \cite{Sarker,Moraes1}
\begin{equation}
\label{ch3eq32}
p_t = n\,p_r\,.
\end{equation}
Here, $n$ is any parameter. Since our analysis involves an anisotropic fluid, therefore $n$ should not be $1$ (that is, $n \neq 1$), as this would reduce the system to an isotropic fluid.\\
Using Eqs. \eqref{ch3eq22}, \eqref{ch3eq23} in Eq. \eqref{ch3eq32}, we will get
\begin{multline}
\label{ch3eq33}
f_Q \left[(n-1) f_T \left(3 b^2-r (b+2 r) b'\right)+24 \pi  \left(r b' (r-b n)+b (3 b n-2 n r-r)\right)\right]+\\
r (b-r) \left[2 b f_{\text{QQ}} Q' \left((n-1) f_T+24 \pi  n\right)+3 f (n-1) r^2 \left(f_T+8 \pi \right)\right]=0.
\end{multline}
In addition, in this case, we could not find the analytical solution for WH with the nonlinear model $f(Q, T)=Q+\lambda_1 Q^2+\eta_1 T$. Thus, using the same $f(Q, T)$ model (that is, Eq. \eqref{ch3eq26}) for this anisotropic case, we try to find the analytical solution for $b(r)$.\\
Substituting Eqs. \eqref{ch3eq28} and \eqref{ch3eq29} in Eq. \eqref{ch3eq32}, we obtain the shape function $b(r)$ as follows
\begin{equation}
\label{ch3eq33}
b(r)=c_2 r^{\eta_2},
\end{equation}
where
\begin{equation*}
\label{ch3eq34}
\eta_2=\frac{3 (4 \pi -\beta ) (2 n+1)}{\beta -4 \beta  n+12 \pi },
\end{equation*}
and $c_2$ is the constant of integration. Without loss of generality, one can set $c_2=1$.\\
To meet the condition for asymptotic flatness, $\eta_2$ must be less than $1$, that is, $\eta_2 < 1$. Therefore, the ranges allowed for $n$ and $\beta$ are as follows:\\
\begin{table}[H]
\begin{center}
\begin{tabular}{ |c|c|c| }
 \hline
 \multicolumn{2}{|c|}{Parameters} \\
 \hline 
 $n$ & $\beta$ \\
 \hline
 $(-\infty, -2)$   & $(\frac{12 \pi}{4n-1}, \frac{12 \pi n}{n+2})$\\
 \hline
 $-2$ & $(\frac{-4 \pi}{3}, \infty)$ \\
\hline
$(-2, \frac{-1}{2}]$ & $(-\infty, \frac{12 \pi n}{n+2}))\cup (\frac{12 \pi}{4n-1}, \infty)$ \\
\hline
$(\frac{-1}{2}, \frac{1}{4})$ & $(-\infty, \frac{12 \pi}{4n-1}) \cup (\frac{12 \pi n}{n+2}, \infty)$  \\
\hline
$\frac{1}{4}$ & $(\frac{4 \pi}{3}, \infty)$ \\
\hline
$(\frac{1}{4}, 1)$ & $(\frac{12 \pi n}{n+2}, \frac{12 \pi}{4n-1})$ \\
\hline
$(1, \infty)$ & $(\frac{12 \pi}{4n-1}, \frac{12 \pi n}{n+2})$  \\
\hline
\end{tabular}
\caption{Permissible ranges for $\beta$ and $n$.}
\label{ch3table:3}
\end{center}
\end{table}
Based on the table above, we picked up some specific values of $\beta$ and $n$ from each allowed range and plotted the conditions of the shape function in Figs. \ref{ch3fig6}-\ref{ch3fig9}. One can observe from the figures that the shape function meets all
the essential conditions for traversability. In this scenario, the radius of the throat is $r_0=1$.\\
Again, by considering the Eq. \eqref{ch3eq33}, we can express the field Eqs. \eqref{ch3eq27}-\eqref{ch3eq29} as follows.\\
% $\bullet$ \textbf{NEC :} $\rho+p_r=\frac{r^{k} (24 \pi  \alpha   n-2 \alpha  \beta   (n+2))}{\Lambda_{1}}$ and $\rho+p_t=\frac{\alpha   (12 \pi  (n+1)-\beta  (5 n+1)) r^{k}}{\Lambda_{1}}$.\\
% $\bullet$ \textbf{DEC :} $\rho-p_r=\frac{2 \alpha   (\beta -\beta  n+12 \pi  (n+1)) r^{k}}{\Lambda_{1}}$ and $\rho-p_t=\frac{\alpha   (\beta  (n-1)+12 \pi  (3 n+1)) r^{k}}{\Lambda_{1}}$.\\
% $\bullet$ \textbf{SEC :} $\rho+p_r+2p_t=-\frac{4 \alpha  \beta   (2 n+1) r^{k}}{\Lambda_{1}}$,\\
% where $\Lambda_{1}=(\beta +8 \pi ) (\beta -4 \beta  n+12 \pi )$ and $k=-\frac{6 (\beta +4 \pi ) (n-1)}{\beta  (4 n-1)-12 \pi }$.\\
\begin{equation}
\label{ch3eq35a}
\rho=\frac{\alpha  (12 \pi -\beta )  (2 n+1) r^k}{\Lambda_1},
\end{equation}
\begin{equation}
\label{ch3eq35b}
p_r=-\frac{3 \alpha  (\beta +4 \pi ) c r^{k}}{\Lambda_1},
\end{equation}
and
\begin{equation}
\label{ch3eq35c}
p_t=-\frac{3 \alpha  (\beta +4 \pi ) c n r^{k}}{\Lambda_1},
\end{equation}
where $\Lambda_{1}=(\beta +8 \pi ) (\beta -4 \beta  n+12 \pi )$ and $k=-\frac{6 (\beta +4 \pi ) (n-1)}{\beta  (4 n-1)-12 \pi }$.\\
Now, considering some specific values from each allowed range in the Table \ref{ch3table:3} and plotting the graph for the energy density, which is depicted in Fig. \ref{ch3fig10}, one can check that the energy density is positive for $\frac{-1}{2}<n<\frac{1}{4}$, $n=\frac{1}{4}$, $\frac{1}{4}<n<1$. However, it shows a violation for other regions of $n$. Thus, we have considered some specific values of $n$ from $\frac{-1}{2}<n<\frac{1}{4}$, $n=\frac{1}{4}$, $\frac{1}{4}<n<1$ and plotted the graph for energy conditions in Figs. \ref{ch3fig11}-\ref{ch3fig13}. Fig. \ref{ch3fig11} indicates that $\rho + p_t$ is positive with decreasing behavior, whereas the radial NEC, that is, $\rho + p_r$, shows increasing negative behavior in the vicinity of the throat. Moreover, from Figs. \ref{ch3fig12} and \ref{ch3fig13}, the DEC is maintained for both pressure components, while the violation of the SEC is observed.
\begin{table}[h]
\begin{center}
\resizebox{16cm}{3cm}{
\begin{tabular}{ |c|c|c|p{2cm}|p{2cm}|c|c|c| }
 \hline
  \multicolumn{1}{|c|}{Terms} &
  \multicolumn{7}{|c|}{Interpretations}\\
 \hline
 $n$ & $(-\infty, -2)$ & $-2$ & $(-2, \frac{-1}{2}]$ & $(\frac{-1}{2}, \frac{1}{4})$ & $\frac{1}{4}$ & $(\frac{1}{4}, 1)$ & $(1, \infty)$ \\
 \hline
 $\beta$   & $(\frac{12 \pi}{4n-1}, \frac{12 \pi n}{n+2})$ & $(\frac{-4 \pi}{3}, \infty)$ & $(-\infty, \frac{12 \pi n}{n+2})\cup (\frac{12 \pi}{4n-1}, \infty)$ & $(-\infty, \frac{12 \pi}{4n-1}) \cup (\frac{12 \pi n}{n+2}, \infty)$ & $(\frac{4 \pi}{3}, \infty)$ & $(\frac{12 \pi n}{n+2}, \frac{12 \pi}{4n-1})$ & $(\frac{12 \pi}{4n-1}, \frac{12 \pi n}{n+2})$\\
 \hline
$\rho$ &  $violated$ & $violated$ & $violated$ & $satisfied$ & $satisfied$ & $satisfied$ & $violated$\\ 
 \hline
$\rho + p_r$ &  $violated$ & $violated$ & $violated$ & $violated$ & $violated$ & $violated$ & $violated$\\
 \hline
$\rho + p_t$ &  $satisfied$ & $satisfied$ & $satisfied$ & $satisfied$ & $satisfied$ & $satisfied$ & $satisfied$\\
 \hline
$\rho - |p_r|$ &  $violated$ & $satisfied$ & $satisfied$ & $satisfied$ & $satisfied$ & $satisfied$ & $violated$\\
 \hline
$\rho - |p_t|$ &  $violated$ & $violated$ & $violated$ & $satisfied$ & $satisfied$ & $satisfied$ & $violated$\\
 \hline
$\rho + p_r + 2p_t$ &  $satisfied$ & $satisfied$ & $satisfied$ & $violated$ & $violated$ & $violated$ & $satisfied$\\
 \hline
\end{tabular}}
\caption{An overview of the energy conditions for $p_t=n\,p_r$.}
\label{ch3table:4}
\end{center}
\end{table}

\section{Equilibrium conditions}
\label{ch3sec4}
This section is dedicated to check the WH stability using the generalized Tolman-Oppenheimer-Volkoff (TOV) equation \cite{Oppenheimer, Gorini, Kuhfittig}
\begin{eqnarray}\label{ch3eq36}
\frac{\varpi^{'}}{2}(\rho+p_r)+\frac{dp_r}{dr}+\frac{2}{r}(p_r-p_t)=0,
\end{eqnarray}
where $\varpi=2\phi(r)$.\\
Due to anisotropic matter distribution, the hydrostatic, gravitational and anisotropic forces are defined as follows
\begin{eqnarray}\label{ch3eq37}
F_h=-\frac{dp_r}{dr}, ~~~~~F_g=-\frac{\varpi^{'}}{2}(\rho+p_r), ~~~~F_a=\frac{2}{r}(p_t-p_r).
\end{eqnarray}
To stabilize the WH solutions, it is necessary that $F_h+F_g+F_a=0$ hold. The gravitational contribution $F_g$ in the equilibrium equation will vanish as we have considered $\phi(r)=constant$. Hence, the equilibrium equation becomes
\begin{equation}
\label{ch3eq38}
F_h+F_a=0.
\end{equation}
Using Eqs. \eqref{ch3eq32a}-\eqref{ch3eq32c}, we get the following expressions for $F_h$ and $F_a$ under the EoS $p_r=\omega \rho$ as
\begin{equation}
F_h=-\frac{(\beta -12 \pi )  \omega  \left(\frac{3 (\beta -4 \pi )}{12 \pi  \omega -\beta  (\omega -2)}-3\right) r^{\frac{3 (\beta -4 \pi )}{12 \pi  \omega -\beta  (\omega -2)}-4}}{(\beta +8 \pi ) (12 \pi  \omega -\beta  (\omega -2))}
\end{equation}
and
\begin{equation}
F_a=\frac{3 \alpha   (\beta  (-\omega )+\beta +4 \pi  (3 \omega +1)) r^{\frac{3 (\beta -4 \pi )}{12 \pi  \omega -\beta  (\omega -2)}-4}}{(\beta +8 \pi ) (12 \pi  \omega -\beta  (\omega -2))}.
\end{equation}
Also, for the relation $p_t=n p_r$, considering Eqs. \eqref{ch3eq35a}-\eqref{ch3eq35c}, we get
\begin{equation}
F_h=-\frac{18 \alpha  (\beta +4 \pi )^2  (n-1) r^{-\frac{6 (\beta +4 \pi ) (n-1)}{\beta  (4 n-1)-12 \pi }-1}}{(\beta +8 \pi ) (\beta -4 \beta  n+12 \pi ) (\beta  (4 n-1)-12 \pi )}
\end{equation}
and
\begin{equation}
F_a=-\frac{6 \alpha  (\beta +4 \pi )  (n-1) r^{-\frac{6 (\beta +4 \pi ) (n-1)}{\beta  (4 n-1)-12 \pi }-1}}{(\beta +8 \pi ) (\beta -4 \beta  n+12 \pi )}.
\end{equation}
\begin{figure}[h]
\centering
\includegraphics[width=6.5cm,height=4cm]{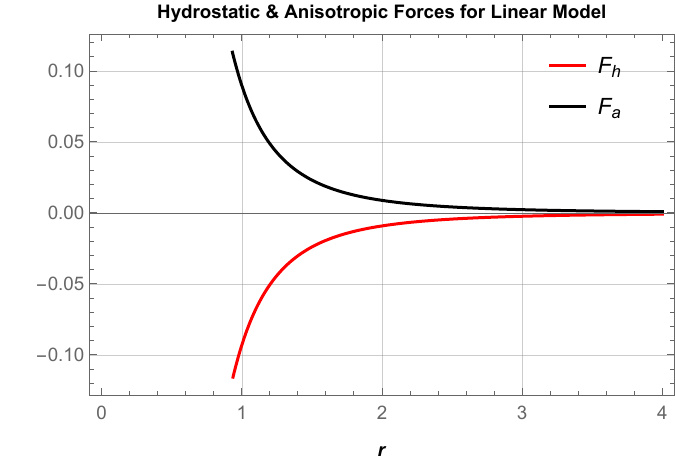}
\caption{The figure displays how the hydrostatic and anisotropic forces vary with $r$. Specifically, it examines these properties for $p_r= \omega \rho$ with $\omega=-1.5$. The parameter $\alpha=1$ and the radial distance is measured in km.}
\label{ch3fig14}
\end{figure}
\begin{figure}[h]
\includegraphics[width=6.5cm,height=4cm]{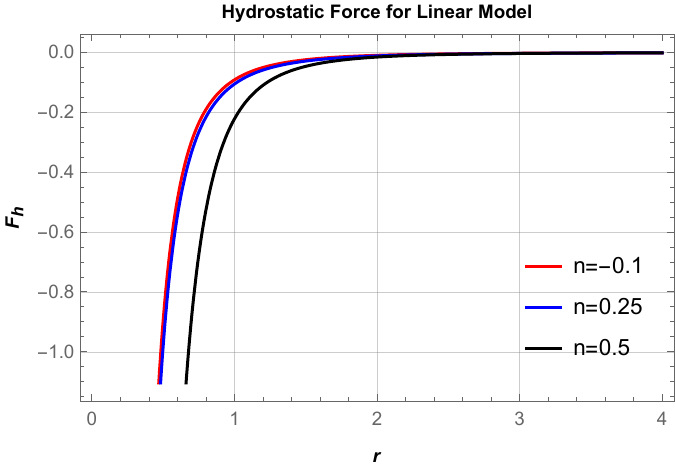}\,\,\,\,\,\,\,\,\,\,\,\,\,\,\,\,\,\,\,\,\,\,\,\,\,
\includegraphics[width=6.5cm,height=4cm]{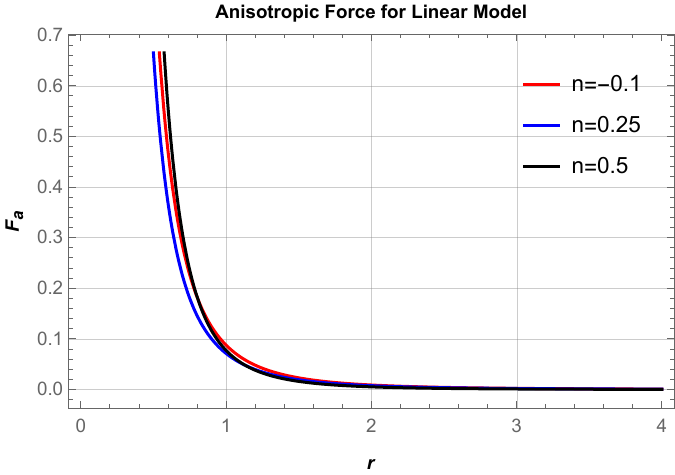}\\
\centering
\includegraphics[width=6.5cm,height=4cm]{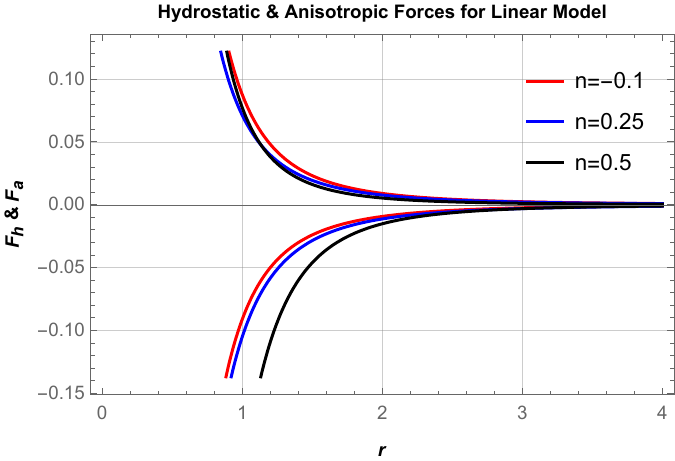}
\caption{The figure displays how the hydrostatic and anisotropic forces vary with $r$. Specifically, it examines these properties for $p_t=n p_r$. The parameter $\alpha=1$ and the radial distance is measured in km.}
\label{ch3fig15}
\end{figure}
The graphs illustrating the forces $F_h$ and $F_a$ for both EoS cases are shown in Figs. \ref{ch3fig14} and \ref{ch3fig15}. It is evident that, while these forces exhibit the same pattern, they act in opposite directions. This equilibrium suggests that the derived solutions for the WHs are stable.

\section{Conclusions}
\label{ch3sec5} 
%WHs can act as tunnels that connects two different spatially regions separated by a space-like interval of the same space-time manifold. It is as of now a theoretical possibility that has not been observed yet. Based on the advances in gravity wave astronomy, one may be able to conjecture a scenario where the signatures of astrophysical WH may be detected. In GR, these WH solutions that are of interest (stable, traversable) suffer from a known pathology where energy conditions are violated. The requirement for exotic matter where one explores standard model extensions in the context of classical GR is a big hurdle. One can now turn to modified versions of gravity to solve the issue regarding the WHs. The study of WHs in modified theory of gravity makes it possible to derive stable solutions without compromising the energy conditions.\\
In this chapter, we have examined WH solutions in $f(Q, T)$ gravity, an extension of symmetric teleparallel gravity, where the gravitational Lagrangian $\mathcal{L}_m$ is defined as an arbitrary function of $Q$ and $T$. The late-time acceleration of the Universe \cite{Arora1} and the early cosmological perturbation \cite{Najera} have been effectively utilized in $f(Q, T)$ gravity. It is a compelling question to explore whether modified gravity theories play a significant role in the study of astrophysical structures such as WHs. If so, it becomes essential to determine which specific modifications allow for the existence of stable and traversable WHs. This serves as motivation to investigate the solution of a WH within the framework of the recently introduced $f(Q, T)$ gravity theory. As a recently developed framework, $f(Q, T)$ gravity has the potential to offer fresh insights into astrophysical phenomena, including black holes and WHs. In this chapter, we formulate the field equations within the framework of $f(Q, T)$ gravity. Our analysis was carried out in two EoS cases: (i) barotropy and (ii) anisotropy for exploring the solution of a WH. Furthermore, considering two distinct forms of the $f(Q, T)$ gravity models (linear $f(Q, T)=\alpha\,Q+\beta\,T$ and nonlinear $f(Q, T)=Q+\lambda_1\,Q^2+\eta_1\,T$), we examined these solutions of a WH. Our objective is to derive an analytical solution for a WH in both $f(Q, T)$ gravity models. Due to the increased complexity of the field equations in $f(Q, T)$ gravity compared to classical GR, obtaining analytical solutions under barotropic and anisotropic conditions presents a significant challenge for both models. However, we have successfully derived exact analytical solutions for both cases by assuming a linear form of $f(Q, T)$. In contrast, for the nonlinear model, obtaining exact analytical solutions was not feasible for either of the EoS. \\
For the linear form of $f(Q, T)$, we have explicitly obtained solutions in a barotropic case. Our analysis revealed that the resulting shape function follows a power-law form. Furthermore, to ensure compliance with the flatness condition, specific ranges have been identified for $\omega$ and $\beta$, as presented in Table \ref{ch3table:1}. By selecting specific values of $\omega$ and $\beta$ from this table, we analyze the behavior of the shape function. Our findings indicate that all the essential criteria for the shape functions are met, confirming the viability of a traversable WH. Additionally, we examined the energy density behavior in both the quintessence and phantom regions. Our analysis revealed that in the phantom region, the energy density remains positive and exhibits a decreasing trend, whereas in the quintessence region it violates expected conditions. Taking this into account, we picked up some values of $\omega$ and plotted the corresponding energy condition graphs to further analyze their behavior. Our analysis showed that the radial component of NEC is violated but the tangential component is satisfied. The DEC remains satisfied, whereas the violation of the SEC is detected. A summary of the behavior of the energy condition is provided in Table \ref{ch3table:2}.\\
Furthermore, we have studied the solution of a WH for the anisotropic EoS under both $f(Q, T)$ gravity models. Our analysis revealed that obtaining the exact analytical solution for the nonlinear $f(Q, T)$ model is a great challenge. In contrast, for the linear model, we successfully obtained the solution for a WH, in power-law form. Table \ref{ch3table:3} presented all the allowed ranges of $n$ and $\beta$ that satisfy flatness conditions. Under each range from this table, the satisfaction of the essential conditions of the shape function can be verified from Figs. \ref{ch3fig6}-\ref{ch3fig9}. Furthermore, from Fig. \ref{ch3fig10}, one can observe that the energy density is showing a positive nature for some specific values of $n$. On the basis of these values, we have plotted the energy conditions. From our analysis, the violation of NEC and SEC was revealed, while the satisfaction of DEC was revealed. The Table \ref{ch3table:4} shows an overview of these energy conditions.\\
In addition, we analyze the stability of the derived WH solutions in both EoS cases. Based on Figs. \ref{ch3fig14}-\ref{ch3fig15}, we concluded that the derived WH solutions are stable. Throughout this chapter, we assumed $\phi(r)=constant$, to prevent the formation of an event horizon.\\
In the next chapter, we will examine a static spherically symmetric WH with the MIT Bag model by assuming two specific shape functions and check the stability of the WH using the TOV equation.
% Chapter 4

\chapter{Wormhole solutions in $f(Q, T)$ gravity with a radial dependent $B$ parameter} % Main chapter title
\label{Chapter4} 
% For referencing the chapter elsewhere, use \ref{Chapter1} 

\lhead{Chapter 4. \emph Wormhole solutions in $f(Q, T)$ gravity with a radial dependent $B$ parameter}% This is for the header on each page - perhaps a shortened title
% \vspace{10 cm}
\blfootnote{*The work in this chapter is covered by the following publication:\\
\textit{Wormhole solutions in $f(Q, T)$ gravity with a radial dependent $B$ parameter}, European Physical Journal Plus, \textbf{138}, 539 (2023).}
 
%\blindtext
%----------------------------------------------------------------------------------------
This chapter explores spherically symmetric WH solutions in the framework of $f(Q, T)$ gravity with the MIT bag model. The detailed study of the work is described as follows:
\begin{itemize}
    \item We apply the embedding method to determine the energy and equilibrium conditions for the existence of WH.
    \item Two specific shape functions are assumed to find the bag parameter $B$ and to study the energy conditions.
    \item With these specific shape functions, the NEC is found to be violated at the WH's throat.
    \item In addition, the TOV equation is used for stability analysis, offering insight into the physical stability of WH configurations.
\end{itemize}
%\clearpage
%-------------------------------------------------------------
\section{Introduction}\label{ch4sec1}
Cutting-edge experiments in astrophysics and cosmology are driving a new era of discovery. Collaborations like Laser Interferometer Gravitational-Wave Observatory (LIGO) \cite{Gourgoulhon/2019}, Virgo \cite{Abuter/2020}, Event Horizon Telescope (EHT) \cite{Chael/2016, Akiyama/2019}, International Gamma-Ray Astrophysics Laboratory (INTEGRAL) \cite{Winkler/2003}, Advanced Telescope for High-Energy Astrophysics (ATHENA) \cite{Barcons/2017}, Imaging x-ray Polarimetry mission (IXPE) \cite{Soffitta/2013}, Swift \cite{Burrows/2005} and Canadian Hydrogen Intensity Mapping Experiment (CHIME) \cite{Chime/2018} are testing gravity and astrophysical objects with unprecedented precision. Meanwhile, upcoming surveys such as the Laser Interferometer Space Antenna (LISA) \cite{Lisa/2013}, Baryon Acoustic Oscillations from Integrated Neutral Gas Observations (BINGO) \cite{Abdalla/2021} and Square Kilometer Array (SKA) \cite{Hall/2004} promise to place stringent constraints on gravity theories, refining the vast array of theoretical models proposed in recent years. \\
In this study, we employ the embedding procedure to obtain traversable WH solutions in $f(Q, T)$ gravity with a radially dependent Bag parameter. This approach incorporates the nontrivial effects of $f(Q, T)$ gravity into the effective density and pressure equations. Using the Raychaudhuri equations, we derive constraints on the $f(Q, T)$ families of traversable WHs. Furthermore, we examine the WEC, NEC, DEC and SEC in the presence of an anisotropic fluid.\\
In our analysis, we explore a specific family of $f(Q, T)$ gravity, deriving WH solutions using two distinct shape functions: $r_0 \gamma  \left(1-\frac{r_0}{r}\right)+r_0$ and $r_0\,\left(\frac{r}{r_0}\right)^{\gamma_1}$. We present scenarios where traversable WHs can exist while satisfying at least two sets of energy conditions in the presence of strange matter. We also analyzed how the Bag parameter varies with the radial coordinate for these solutions. Moreover, we confirmed the stability of these WHs using equilibrium conditions derived from the generalized TOV equation. \\
The key findings of this chapter are summarized as follows. In Sec. \ref{ch4sec4}, we examined the energy conditions for WH solutions in $f(Q, T)$ gravity, demonstrating the existence of traversable WH that satisfy at least two sets of energy conditions under the equation of motion of the MIT bag model. In Sec. \ref{ch4sec5}, we discussed the equilibrium conditions for these WH families. Finally, our concluding remarks and prospects for future research are outlined in Sec. \ref{ch4sec6}.
 
\section{Wormholes in $f(Q, T)$ gravity}
\label{ch4sec4}
In order to approach the generalities on the WH solutions in $f(Q, T)$ gravity, we examine the WH metric in the Schwarzschild coordinates $(t,\,r,\,\theta,\,\varphi)$ following the seminal work of Morris and Thorne \cite{a2,a17}, whose specific form is given by Eq. \eqref{1ch1}. Here, we consider that the WH solutions for $f(Q, T)$ are compatible with Birkhoff's theorem, since we are writing the line element in Schwarzschild's form. Our hypothesis on the viability of Birkhoff's theorem is based on the work of Meng and Wang \cite{Meng/2011} and also on the recent review by Bahamond et al. \cite{Bahamonde/2023}. In their investigation \cite{Meng/2011}, they proved the viability of Birkhoff's theorem for teleparallel gravity. Recently, Birkhoff's theorem in generalized teleparallel gravity, including boundary term and scalar-field couplings, was reexamined by Bahamonde et al. \cite{Bahamonde/2023}. They found that the theorem holds unless the scalars depend on $t$ or $r$. Bahamonde et al. also unveiled in Eq. (5.49) of \cite{Bahamonde/2023}, a transformation that maps the torsion to the non-metricity scalar. This mapping, along with discussions of Birkhoff's theorem in teleparallel gravity, reinforces our conjecture regarding the theorem's validity for $f(Q, T)$ gravity solutions.\\
Here, we follow the embedding procedure introduced by Mandal et al. \cite{S. Mandal} in determining the energy conditions for $f(Q)$ gravity. For this purpose, we take into account Eqs. \eqref{ch3eq5}-\eqref{ch3eq7}. Furthermore, to constrain energy conditions in $f(Q)$ gravity, Hassan et al. \cite{Hassan/2022} applied this method to traversable WHs.\\ 
% Then, from Eqs. \eqref{24a1}-\eqref{24a3}, one can find the following field equations for $f(Q, T)$ gravity \cite{Tayde 4}:
% \begin{equation}\label{24a10}
% 8 \pi  \rho =\frac{(r-b)}{2 r^3} \left[f_Q \left(\frac{(2 r-b) \left(r b'-b\right)}{(r-b)^2}+\frac{b \left(2 r \phi '+2\right)}{r-b}\right)+\frac{2 b r f_{\text{QQ}} Q'}{r-b}+\frac{f r^3}{r-b}-\frac{2r^3 f_T (P+\rho )}{(r-b)}\right],
% \end{equation}
% \begin{equation}\label{24a11}
% 8 \pi  p_r=-\frac{(r-b)}{2 r^3} \left[f_Q \left(\frac{b }{r-b}\left(\frac{r b'-b}{r-b}+2 r \phi '+2\right)-4 r \phi '\right)+\frac{2 b r f_{\text{QQ}} Q'}{r-b}+\frac{f r^3}{r-b}-\frac{2r^3 f_T \left(P-p_r\right)}{(r-b)}\right],
% \end{equation}
% \begin{equation}\label{24a12}
% 8 \pi  p_t=-\frac{(r-b)}{4 r^2} \left[f_Q \left(\frac{\left(r b'-b\right) \left(\frac{2 r}{r-b}+2 r \phi '\right)}{r (r-b)}+\frac{4 (2 b-r) \phi '}{r-b}-4 r \left(\phi '\right)^2-4 r \phi ''\right)-4 r f_{\text{QQ}} Q' \phi '+\frac{2 f r^2}{r-b}-\frac{4r^2 f_T \left(P-p_t\right)}{(r-b)}\right].
% \end{equation}\\
As Morris and Thorne noted, $\phi(r)$ must be finite to prevent horizons in traversable WHs \cite{a17}. A simple choice satisfying this and allowing constraints of the analytic energy condition is $\phi(r) = constant$, leading to effective density and pressures from Eqs. \eqref{ch3eq5}-\eqref{ch3eq7} as
\begin{equation}\label{ch4eq11}
\tilde{\rho}=\frac{2 (r-b) }{(2 r-b) f_Q}\left(\rho-\frac{\left(r-b\right)}{8 \pi  r^2} \left(\frac{b r f_{\text{QQ}} Q'}{r-b}+\frac{b f_Q}{r-b}-\frac{b (2 r-b) f_Q}{2 (r-b)^2}+\frac{f r^3}{2 (r-b)}\right)+\frac{f_T (P+\rho )}{8 \pi }\right)\,,
\end{equation}
\begin{equation}\label{ch4eq12}
\tilde{p_r}=\frac{2 b }{f r^3}\left(p_r- \frac{f_T \left(P-p_r\right)}{8 \pi }+\frac{\left(r-b\right) }{16 \pi  r^3}\left(\frac{b f_Q \left(\frac{r b'-b}{r-b}+2\right)}{r-b}+\frac{2 b r f_{\text{QQ}} Q'}{r-b}\right)\right)\,,
\end{equation}
and
\begin{equation}\label{ch4eq13}
\tilde{p_t}=\frac{(r-b)}{r f_Q}\left(p_t-\frac{f_T \left(P-p_t\right)}{8 \pi }+ \frac{f r \left(1-\frac{b}{r}\right)}{16 \pi  (r-b)}\right)\,.
\end{equation}

As is well known, all the energy conditions are not obeyed in standard GR for traversable WH solutions. Especially the NEC violation indicates the presence of exotic matter at the throat of the WH, moreover, for a realistic matter source in WH solutions, the energy density must also be positive. \\
In order to test the viability of traversable WH solutions that also obey the EoS of the MIT bag, we shall work with the particular function $f(Q, T)$ given by Eq. \eqref{ch3eq26} \cite{fqt}.
Using the linear functional form of $f(Q, T)$ (Eq. \eqref{ch3eq26}) and assuming the redshift function $\phi(r)=constant$ in the field Eqs. \eqref{ch3eq21}-\eqref{ch3eq23},  we obtain the reduced-field Eqs. \eqref{ch3eq27}-\eqref{ch3eq29}. We will use these reduced-field equations for further study.
 % \begin{equation}\label{ch4eq14}
 % \rho =\frac{\alpha  (12 \pi -\beta ) b'}{3 (4 \pi -\beta ) (\beta +8 \pi ) r^2}\,,
 % \end{equation}
 % \begin{equation}\label{ch4eq15}
 % p_r=-\frac{\alpha  \left(2 \beta  r b'-3 \beta  b+12 \pi  b\right)}{3 (4 \pi -\beta ) (\beta +8 \pi ) r^3}\,,
 % \end{equation}
 % \begin{equation}\label{ch4eq16}
 % p_t=-\frac{\alpha  \left((\beta +12 \pi ) r b'+3 b (\beta -4 \pi )\right)}{6 (4 \pi -\beta ) (\beta +8 \pi ) r^3}\,.
 % \end{equation}
 
\section{Wormhole solutions with different shape functions}
In this subsection, we consider a constant redshift function and that the radial pressure obeys a generalized version of the EoS for the MIT bag model, whose form is
\begin{equation}\label{ch4eq17}
p_r = \omega (\rho - 4B)\,,
\end{equation}
 where $\omega$ is EoS parameter and $B$ is the bag parameter. We can observe that this equation recovers the standard MIT bag model if $\omega=1/3$. Consequently, by taking Eqs. \eqref{ch3eq27} and \eqref{ch3eq28} in Eq. \eqref{ch4eq17}, we obtain
 \begin{equation}\label{ch4eq18}
B=\frac{\alpha  \left(-\beta  r \omega  b'+2 \beta  r b'+12 \pi  r \omega  b'-3 \beta  b+12 \pi  b\right)}{12 (4 \pi -\beta ) (\beta +8 \pi ) r^3 \omega }\,,
 \end{equation}
as a bag parameter whose form depends on the shape function $b(r)$. From the previous Eq. \eqref{ch4eq18}, we realize that the bag parameter indicates different energy domains for stabilizing the strange-quark matter in different regions of the WH.  \\

\begin{figure}[H]
\centering
\includegraphics[width=7.5cm,height=5cm]{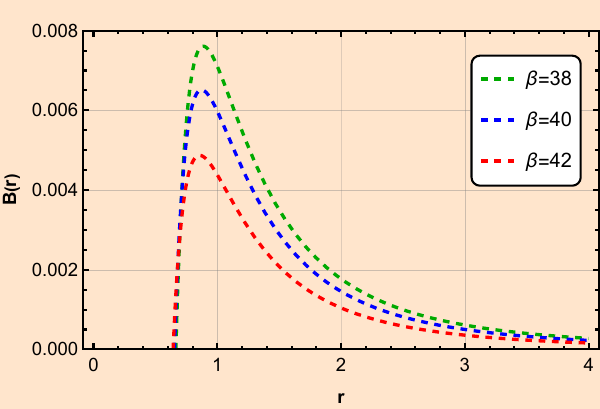}
\caption{The figure displays how the Bag parameter $B$  varies with $r$ for (a) $\beta=38$ with $\omega =0.28$,\,\,\,\,(b) $\beta=40$ with $\omega =0.333$\,\,\, and (c) $\beta=42$ with $\omega =0.455$. Also, we consider $\alpha=1,\, \gamma=0.5\,\,\text{and}\,\,r_0=1$.}
\label{ch4fig1}
\end{figure}

 \textbf{Shape Function- I :} As a first example, we consider a shape function introduced by Harko et al. \cite{T. Harko3} on the study of electromagnetic signatures of the accretion disks of WHs, whose form is 
 \begin{equation}\label{ch4eq19}
 b(r)= r_0 \gamma  \left(1-\frac{r_0}{r}\right)+r_0\,,
 \end{equation}
 where $\gamma<1$ in order to satisfy conditions of shape function. From Eqs. \eqref{ch4eq18} and \eqref{ch4eq19} we determine the following bag parameter
 \begin{equation}\label{ch4eq20}
 B=\frac{r_0 \alpha  (12 \pi  (r_0 \gamma  (\omega -1)+(\gamma +1) r)-\beta  (r_0 \gamma  (\omega -5)+3 (\gamma +1) r))}{12 (4 \pi -\beta ) (\beta +8 \pi ) r^4 \omega }\,,
 \end{equation}
 which is depicted in detail in Fig \ref{ch4fig1}. We can observe that $B$ approaches zero inside the throat of WH $\left(r=r_0\right)$; it also goes to constant values for the big values of $r$. Moreover, as $\beta$ decreases, we have smaller maximum values for the bag parameter, resulting in smaller binding energy for the matter. This behavior of $B(r)$ was observed by Deb et al. \cite{Deb/2022} for $f(R, T)$ gravity with a hybrid exponential shape function. It is relevant to say that $\omega=0.455$ and $\omega=0.333$ were used by Harko and Cheng \cite{Harko/2002} in their investigation on equations of state for strange stars and $\omega=0.28$ is connected with a constraint over the mass of strange quarks, as pointed out by Deb et al. \cite{Deb/2022}.

\begin{figure}[H]
\centering
\includegraphics[width=7.5cm,height=5cm]{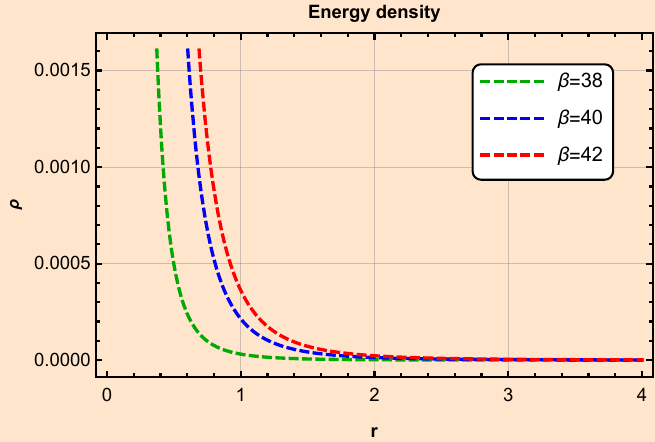}
\centering
\caption{The figure displays how the $\rho$  varies with $r$ for different valus of $\beta$. We consider $\alpha=1,\, \gamma=0.5\,\,\text{and}\,\,r_0=1$.}
\label{ch4fig2}
\end{figure}

\begin{figure}[H]
\centering
\includegraphics[width=14.5cm,height=5cm]{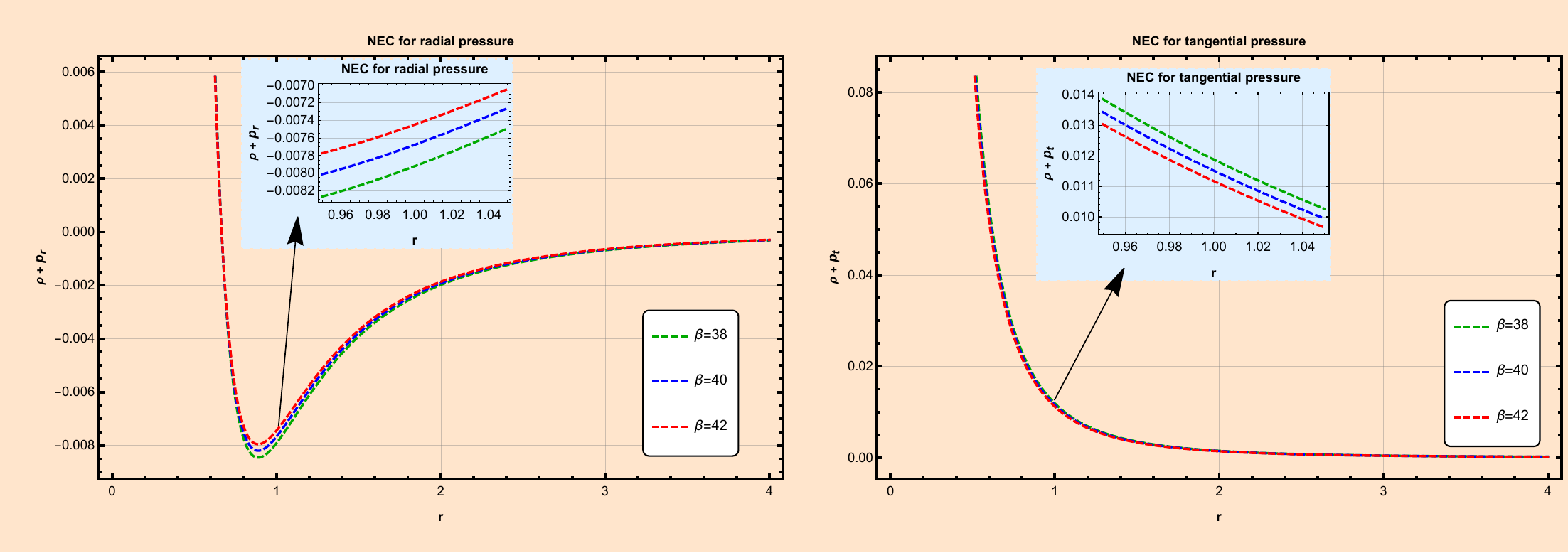}
\caption{The figure displays how the NEC  varies with $r$ for different values of $\beta$. We consider $\alpha=1,\, \gamma=0.5\,\,\text{and}\,\,r_0=1$.}
\label{ch4fig3}
\end{figure}

\begin{figure}[H]
\centering
\includegraphics[width=14.5cm,height=5cm]{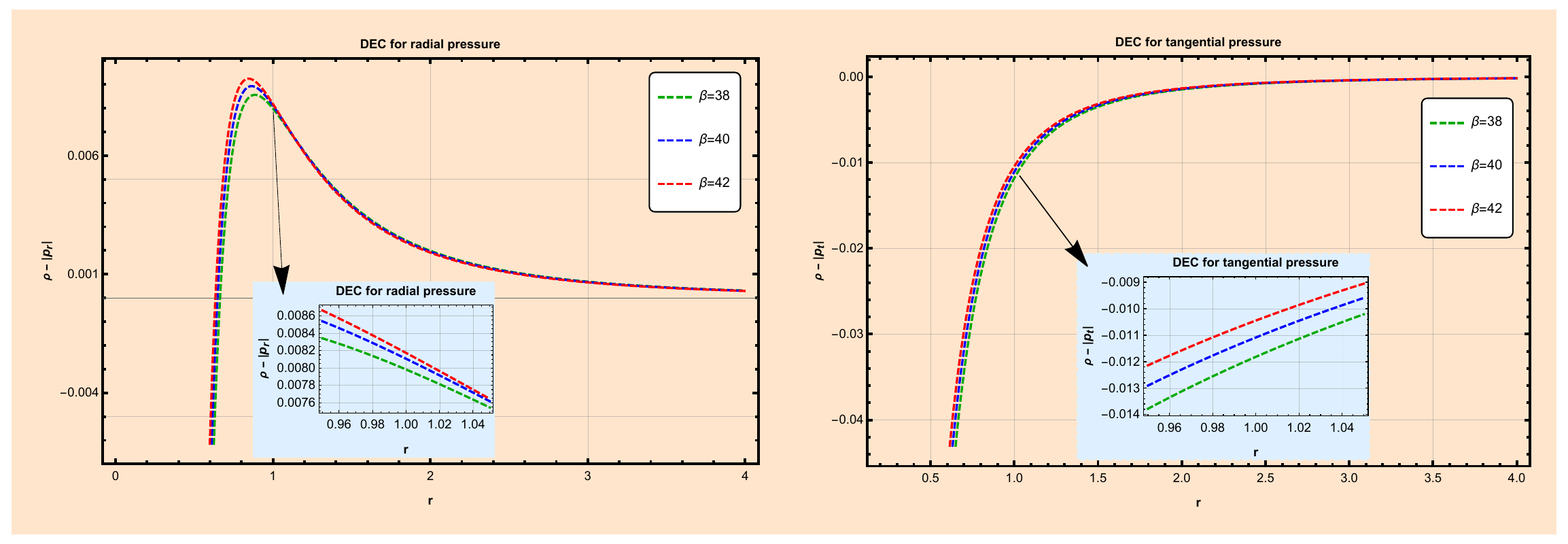}
\caption{The figure displays how the DEC varies with $r$ for different values of $\beta$. We consider $\alpha=1,\, \gamma=0.5\,\,\text{and}\,\,r_0=1$.}
\label{ch4fig4}
\end{figure}

\begin{figure}[H]
\centering
\includegraphics[width=6.5cm,height=4cm]{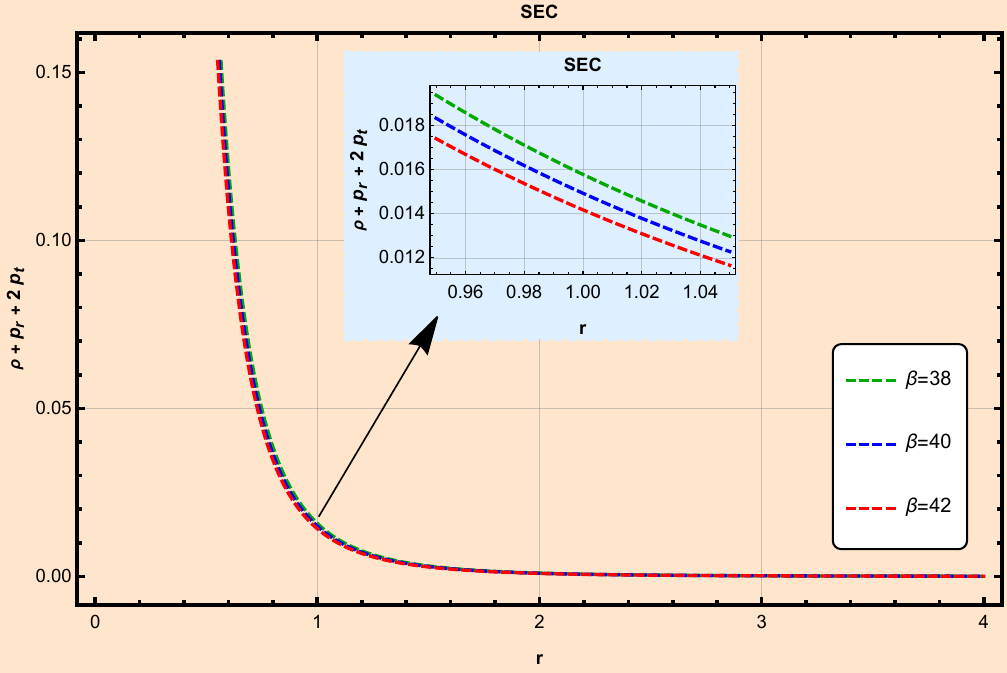}
\centering
\caption{The figure displays how the SEC  varies with $r$ for different values of $\beta$. We consider $\alpha=1,\, \gamma=0.5\,\,\text{and}\,\,r_0=1$.}
\label{ch4fig5}
\end{figure}

Now, considering Eq. \eqref{ch4eq19}, we obtain the following set of field equations from Eqs. \eqref{ch3eq27}-\eqref{ch3eq29} as
\begin{equation}\label{ch4eq20a}
    \rho=\frac{r_0^2 \alpha  (12 \pi -\beta ) \gamma }{3 (4 \pi -\beta ) (\beta +8 \pi ) r^4},
\end{equation}
\begin{equation}\label{ch4eq20b}
    p_r=\frac{r_0 \alpha  (-5 r_0 \beta  \gamma -12 \pi  (-r_0 \gamma +\gamma  r+r)+3 \beta  (\gamma +1) r)}{3 (4 \pi -\beta ) (\beta +8 \pi ) r^4},
\end{equation}
and
\begin{equation}\label{ch4eq20c}
    p_t=\frac{r_0 \alpha  (2 r_0 \beta  \gamma +12 \pi  (-2 r_0 \gamma +\gamma  r+r)-3 \beta  (\gamma +1) r)}{6 (4 \pi -\beta ) (\beta +8 \pi ) r^4}.
\end{equation}
% $\bullet$ \textbf{NEC :} $\rho+p_r=-\frac{r_0 \alpha  (-2 r_0 \gamma +\gamma  r+r)}{(\beta +8 \pi ) r^4}$ and $\rho+p_t=\frac{r_0 \alpha  (\gamma +1)}{2 (\beta +8 \pi ) r^3}$\,.\\
% $\bullet$ \textbf{DEC :} $\rho-p_r=\frac{r_0 \alpha  (4 r_0 \beta  \gamma -3 \beta  (\gamma +1) r+12 \pi  (\gamma +1) r)}{3 (4 \pi -\beta ) (\beta +8 \pi ) r^4}$ and $\rho-p_t=\frac{r_0 \alpha  (-4 r_0 \beta  \gamma -12 \pi  (-4 r_0 \gamma +\gamma  r+r)+3 \beta  (\gamma +1) r)}{6 (4 \pi -\beta ) (\beta +8 \pi ) r^4}$\,.\\
% $\bullet$ \textbf{SEC :} $\rho+p_r+2p_t=-\frac{4 r_0^2 \alpha  \beta  \gamma }{3 (4 \pi -\beta ) (\beta +8 \pi ) r^4}$\,.\\
In addition, features of energy conditions are presented in Figs. \ref{ch4fig2} - \ref{ch4fig5}. There, we can see that the energy density is positive throughout the variation of $r$, satisfying the WEC. We find that the radial pressure component of the NEC is violated, while the tangential pressure component and the SEC are satisfied. Moreover, DEC is not validated inside the WH's throat for both radial and tangential pressures. The violation of DEC in this region is consistent with the presence of exotic matter within the throat of the WH, and they are compatible with the features presented on WHs in $f(R, T)$ gravity \cite{Elizalde/2018}. In addition, this partial violation of NEC can be justified by the behavior of Eq. \eqref{ch4eq17} for positive values of $B$ and $\rho$. Once $B>\rho$ is inside the WH's throat, then $p_r$ is going to be negative in this region, resulting in the partial violation of NEC. Therefore, we can connect such a violation with the existence of strange matter in the inner region of the WH.\\
It is relevant to point out that these constraints over the energy conditions are also observed for $\tilde{\rho}$, $\tilde{p}_r$ and $\tilde{p}_t$. The violations of NEC and DEC are compatible with the existence of strange matter inside the WH's throat, which may be formed by strange quarks.


\begin{figure}[H]
\centering
\includegraphics[width=7.5cm,height=5cm]{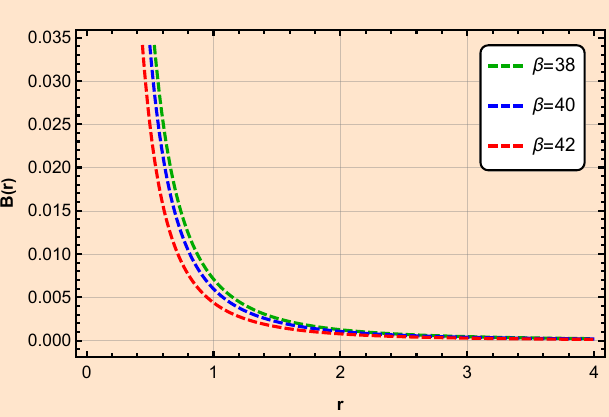}
\caption{The figure displays how the Bag parameter $B$  varies with $r$ for (a) $\beta=38$ with $\omega =0.28$,\,\,\,\,(b) $\beta=40$ with $\omega =0.333$\,\,\, and (c) $\beta=42$ with $\omega =0.455$. Also, we consider $\alpha=1,\, r_0=1$.}
\label{ch4fig6}
\end{figure}

\textbf{Shape Function- II :}  We consider a shape function discussed in the seminal paper of Lobo et al. \cite{Lobo/2009}, where the authors investigated the viability of WH solutions for different $f(R)$ theories of gravity. This shape function is given by 
\begin{equation}\label{ch4eq21}
b(r) = r_0\,\left(\frac{r}{r_0}\right)^{\gamma_1}\,,
\end{equation}
 where $\gamma_1 <1$ in order to satisfy the conditions of the shape function. In order to find the energy constraints, we work with  $\gamma_1=\frac{1}{2}$, yielding the shape function
 \begin{equation}\label{ch4eq22}
   b(r) = r_0 \sqrt{\frac{r}{r_0}}\,.
 \end{equation}
 With these ingredients in hand, we can determine the following form for the bag parameter
 \begin{equation}\label{ch4eq23}
 B=\frac{\alpha  (12 \pi  (\omega +2)-\beta  (\omega +4))}{24 (4 \pi -\beta ) (\beta +8 \pi ) r^2 \omega  \sqrt{\frac{r}{r_0}}}\,,
 \end{equation}
 whose features are presented in Fig. \ref{ch4fig6}. There we observe that $B(r)$ is always positive for the different values of $\beta$ and $\omega$ chosen here. We also realize that $B(r)$ increases near the WH's throat, raising the binding energy to keep the strange matter stable. Moreover, the bag parameter increases to a constant value for large values of $r$. These features also corroborate the investigations of Deb et al. in Ref. \cite{Deb/2022}.
 
\begin{figure}[H]
\centering
\includegraphics[width=7.5cm,height=5cm]{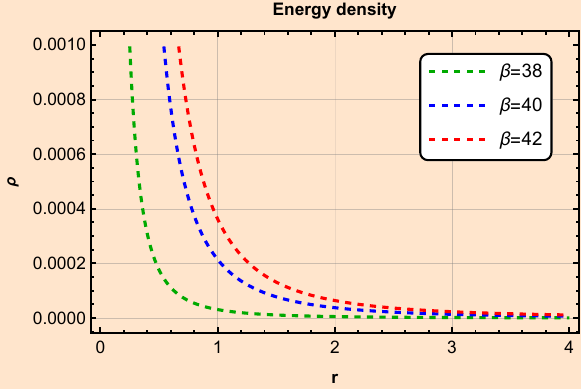}
\centering
\caption{The figure displays how the $\rho$ varies with $r$ for different values of $\beta$. We consider $\alpha=1,\, r_0=1$.}
\label{ch4fig7}
\end{figure}

\begin{figure}[H]
\centering
\includegraphics[width=14.5cm,height=5cm]{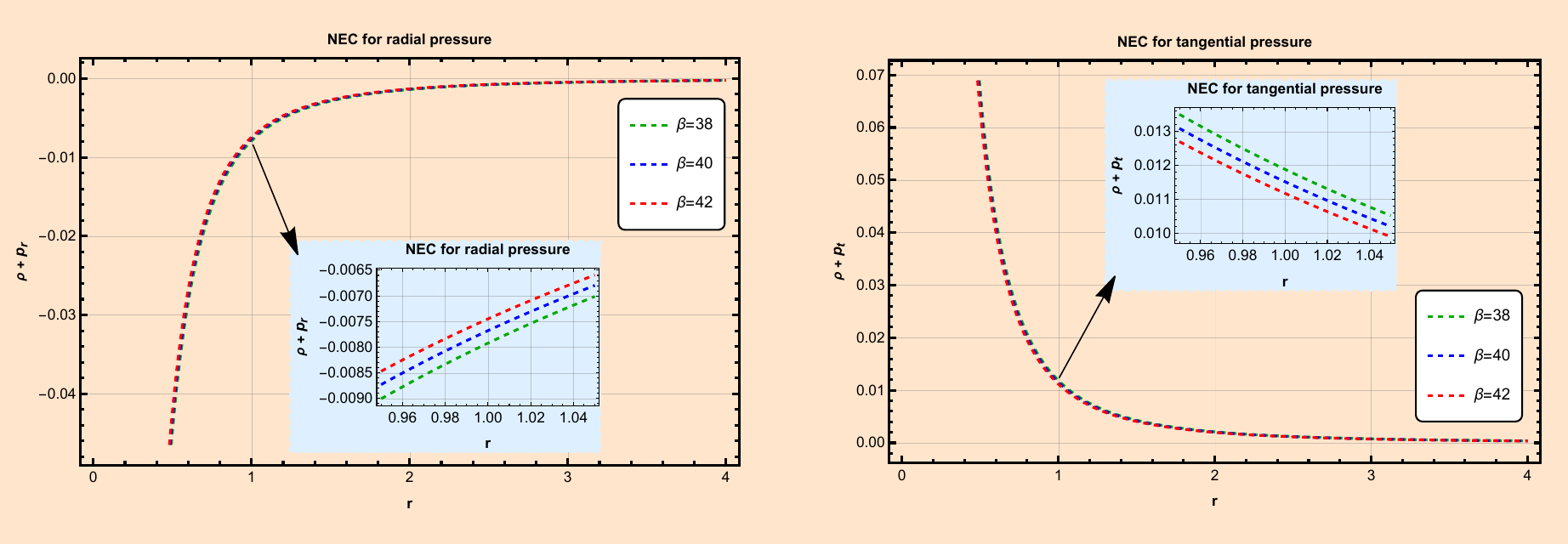}
\caption{The figure displays how the NEC  varies with $r$ for different values of $\beta$. We consider $\alpha=1,\,r_0=1$.}
\label{ch4fig8}
\end{figure}

\begin{figure}[h]
\centering
\includegraphics[width=14.5cm,height=5cm]{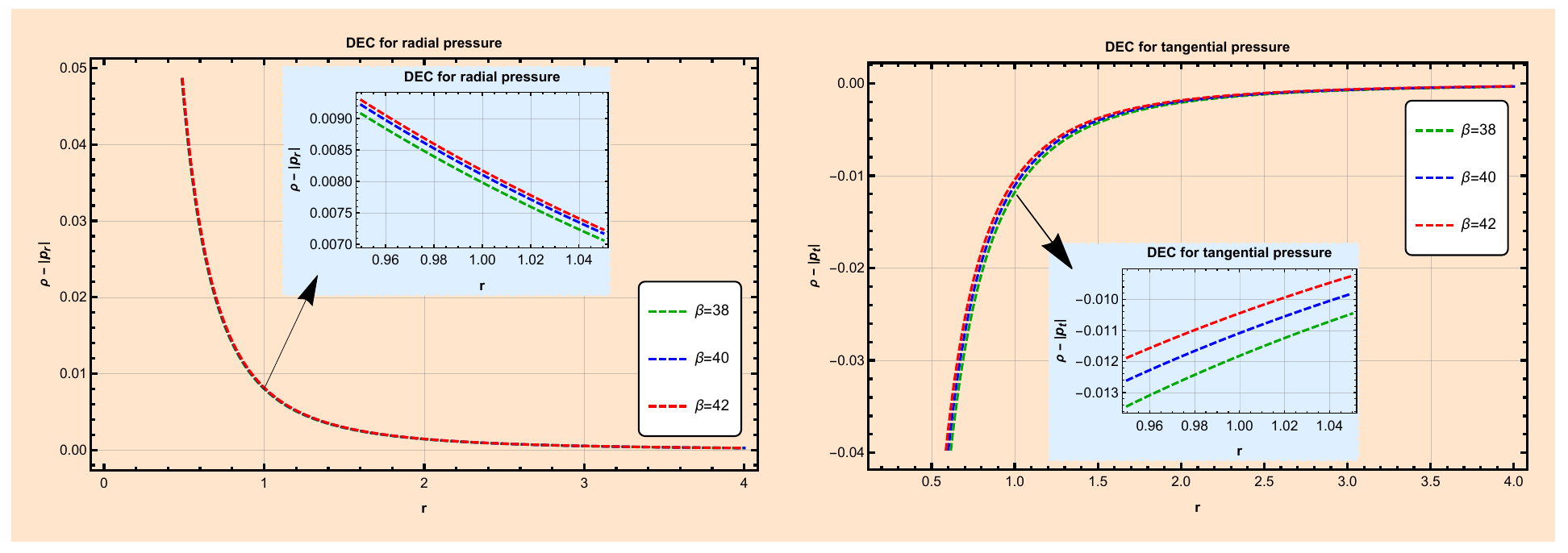}
\caption{The figure displays how the DEC  varies with $r$ for different values of $\beta$. We consider $\alpha=1,\,r_0=1$.}
\label{ch4fig9}
\end{figure}

\begin{figure}[h]
\centering
\includegraphics[width=6.5cm,height=4.7cm]{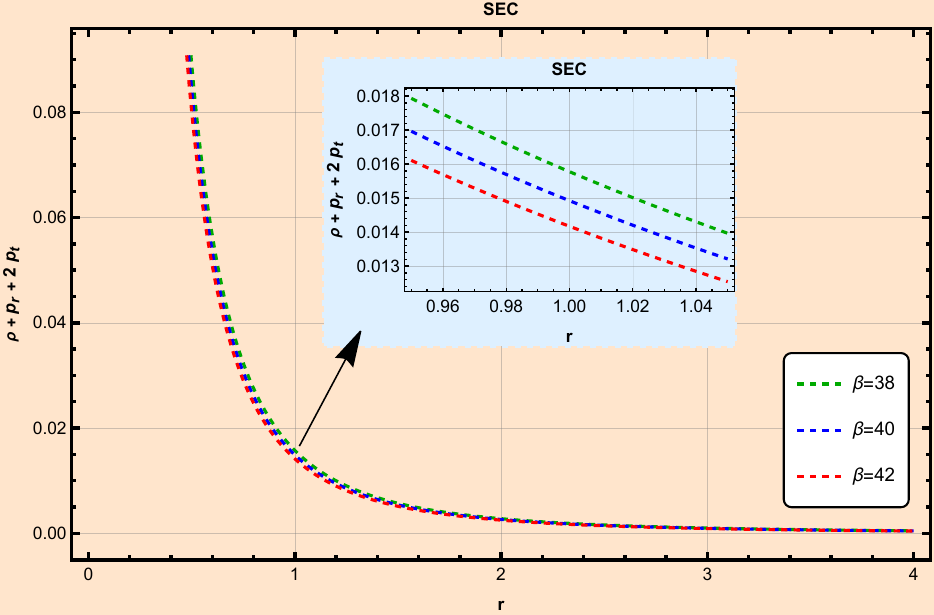}
\centering
\caption{The figure displays how the SEC  varies with $r$ for different values of $\beta$. We consider $\alpha=1,\,r_0=1$.}
\label{ch4fig10}
\end{figure}

Taking Eq. \eqref{ch4eq22}, we derive the following set of field equations from Eq. \eqref{ch3eq27}-\eqref{ch3eq29} as
\begin{equation}\label{ch4eq23a}
    \rho=\frac{12 \pi -\beta }{6 (4 \pi -\beta ) (\beta +8 \pi ) r^2 \sqrt{\frac{r}{r_0}}},
\end{equation}
\begin{equation}\label{ch4eq23b}
    p_r=-\frac{2 (6 \pi -\beta )}{3 (4 \pi -\beta ) (\beta +8 \pi ) r^2 \sqrt{\frac{r}{r_0}}},
\end{equation}
and
\begin{equation}\label{ch4eq23c}
    p_t=\frac{12 \pi -7 \beta }{12 (4 \pi -\beta ) (\beta +8 \pi ) r^2 \sqrt{\frac{r}{r_0}}}.
\end{equation}
% $\bullet$ \textbf{NEC :} $\rho+p_r=-\frac{\alpha}{2 (\beta +8 \pi ) r^2 \sqrt{\frac{r}{r_0}}}$ and $\rho+p_t=\frac{3\alpha}{4 (\beta +8 \pi ) r^2 \sqrt{\frac{r}{r_0}}}$\,.\\
% $\bullet$ \textbf{DEC :} $\rho-p_r=\frac{\alpha(36 \pi -5 \beta) }{6 (4 \pi -\beta ) (\beta +8 \pi ) r^2 \sqrt{\frac{r}{r_0}}}$ and $\rho-p_t=\frac{\alpha(5 \beta +12 \pi )}{12 (4 \pi -\beta ) (\beta +8 \pi ) r^2 \sqrt{\frac{r}{r_0}}}$\,.\\
% $\bullet$ \textbf{SEC :} $\rho+p_r+2p_t=-\frac{2 \alpha \beta }{3 (4 \pi -\beta ) (\beta +8 \pi ) r^2 \sqrt{\frac{r}{r_0}}}$\,.\\
In addition, the energy conditions are described in detail in Figs. \ref{ch4fig7} - \ref{ch4fig10}. As in our first example, we observe that the energy density is always positive, which means that WEC is satisfied. Again, we also observe that the radial NEC is violated while the tangential NEC is satisfied, unveiling the presence of strange matter in the WH's throat. Moreover, DEC is satisfied by the radial pressure and violated by the tangential pressure for $r<1$. Furthermore, SEC is satisfied, so WEC and SEC allow the existence of stable traversable WHs with strange matter on their throats. We also highlight that these constraints on the energy conditions are observed for $\tilde{\rho}$, $\tilde{p}_r$ and $\tilde{p}_t$ as well. Moreover, imposing that the parameter $B$ and the energy density $\rho$ are positively defined, in addition to the required energy conditions, we find that $\beta>12\pi$ for shape functions \eqref{ch4eq19} and \eqref{ch4eq22}, if we consider $\alpha=1,\, \gamma=0.5,\, \text{and}\,\,r_0=1$.


\section{Equilibrium conditions}
\label{ch4sec5}
This section is dedicated to checking the stability of WH using the generalized TOV equation \cite{Oppenheimer, Gorini, Kuhfittig} in $f(Q, T)$ gravity. The generalized TOV equation can be given by Eq. \eqref{ch3eq36}, and here we replace the energy density with the effective energy density and the pressures by the effective pressures for our study.
% Due to anisotropic matter distribution, the hydrostatic, gravitational, and anisotropic forces are defined as follows
% \begin{eqnarray}\label{39}
% F_h=-\frac{d \tilde{p_r}}{dr}, ~~~~~F_g=-\frac{\varpi^{'}}{2}(\tilde{\rho}+\tilde{p_r}), ~~~~F_a=\frac{2}{r}(\tilde{p_t}-\tilde{p_r}).
% \end{eqnarray}
% The stability of the WH solutions, imposes the constraint $F_h+F_g+F_a=0$. Since in this study, we have assumed the redshift function $\phi(r)=constant$, the gravitational contribution $F_g$ is null, so the stability constraint is rewritten as
% \begin{equation}
% \label{40}
% F_h+F_a=0.
% \end{equation}
Then, by taking Eqs. \eqref{ch4eq20a}-\eqref{ch4eq20c}, we get the following equations for the hydrostatic and anisotropic forces when assuming the constant redshift function for the shape function- I
\begin{equation}\label{ch4eq24}
F_h=\frac{r_0 (4 r_0 \gamma -3 (\gamma +1) r)}{8 \pi  r^5}
\end{equation}
and
\begin{equation}\label{ch4eq25}
F_a=\frac{r_0 (3 (\gamma +1) r-4 r_0 \gamma )}{8 \pi  r^5}\,.
\end{equation}
Moreover, by using Eqs. \eqref{ch4eq23a}-\eqref{ch4eq23c} for the shape function- II, we yield to equations
\begin{equation}\label{ch4eq26}
F_h=-\frac{5}{16 \pi  r^3 \sqrt{\frac{r}{r_0}}}
\end{equation}
and
\begin{equation}\label{ch4eq27}
F_a=\frac{5}{16 \pi  r^3 \sqrt{\frac{r}{r_0}}}\,,
\end{equation}
for the hydrostatic and anisotropic forces, respectively. We can verify that the hydrostatic and anisotropic forces are independent of $\omega$ and $\beta$ for both shape functions. 
\begin{figure}[H]
\centering
\includegraphics[width=16cm,height=5cm]{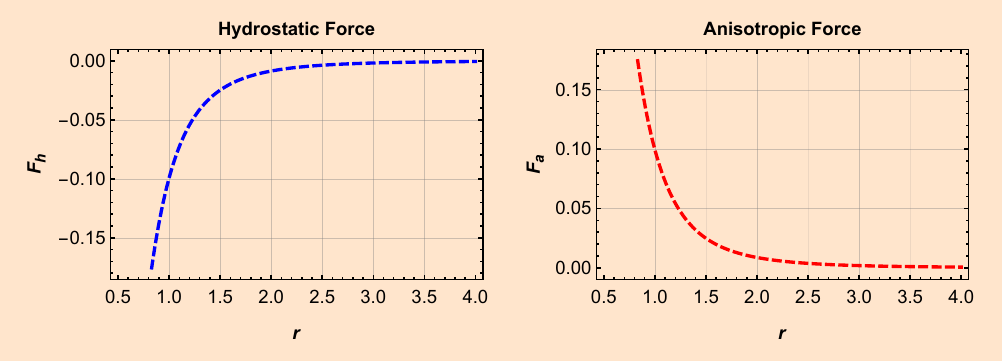}
\includegraphics[width=7.3cm,height=4.7cm]{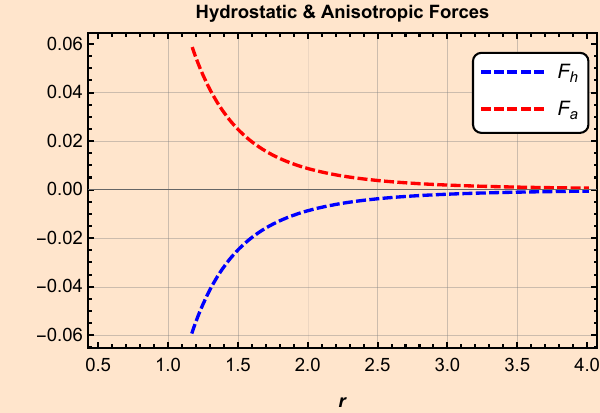}
\caption{The figure displays how the hydrostatic forces and anisotropic forces vary with $r$ for shape function- I. We consider $\gamma=0.5,\,r_0=1$.}
\label{ch4fig11}
\end{figure}
\begin{figure}[h]
\centering
\includegraphics[width=14.5cm,height=5cm]{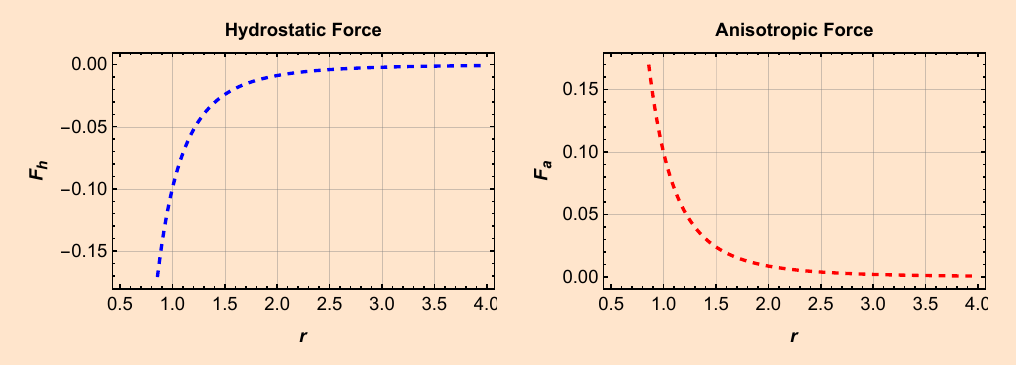}
\includegraphics[width=7.2cm,height=4.6cm]{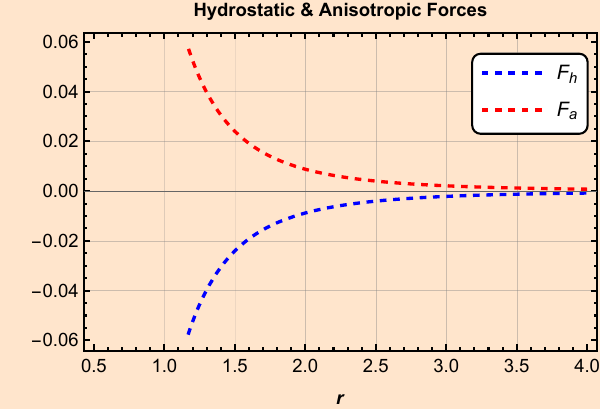}
\caption{The figure displays how the hydrostatic forces and anisotropic forces vary with $r$ for shape function- II. We consider $r_0=1$.}
\label{ch4fig12}
\end{figure}
The hydrostatic and anisotropic forces from shape functions 1 and 2 are illustrated in Figs. \ref{ch4fig11} and \ref{ch4fig12}. These forces exhibit equal magnitude, but act in opposite directions, suggesting the stability of the WHs.

\section{Conclusions}
\label{ch4sec6}
This study proposes a new approach to evaluate the viability of WH in $f(Q, T)$ gravity with a radially dependent Bag parameter. We derived various WH families using two shape functions widely studied in recent literature on alternative gravity theories. Furthermore, we obtained the energy conditions using Raychaudhuri constraints and the embedding procedure introduced by Mandal et al. \cite{S. Mandal}. By combining energy conditions with the EoS model of the MIT bag and the limits on $\omega$, we derived WH families for different values of $\beta$, ensuring solutions that satisfy at least two energy conditions with positive energy densities.\\
WEC and SEC were satisfied by both shape functions (Eqs. \eqref{ch4eq19} and \eqref{ch4eq22}) with positive energy density. Moreover, both shape functions partially violated the NEC and DEC for $r<1$, corroborating the existence of strange matter at the throat of the WH. The connection between the partial violation of NEC and strange matter is evident through the analysis of Eq. \eqref{ch4eq17}. By imposing the conditions that $\rho$, $\omega$ and $B$ are positive and that $B>\rho$ within the throat of the WH, we confirm the violation of the NEC. Thus, our method suggests that stable traversable WH solutions could be constructed using strange matter that adheres to both the SEC and WEC. We also performed a thorough stability analysis of our solutions by applying the TOV equation.\\
The following chapter examines WH geometry within the $f(Q, T)$ gravity framework, considering two distributions, namely Gaussian and Lorentzian, under noncommutative geometry from string theory.
%%%%%%%%%%%%%%%%%%%%%%%%%%%%%%%%%%%%%%%%%%%%%%%%%%%%%%%%%%%%%%%%%%%%%%%%%%%%%%%%%%%%%%%%%%%%%%%%%%%%%%%%%%%%%%%%%%%%%%%%%%%%%%%%%%%%%%%%%%%%%%%%%%%%%%%%%%%%%%%%%%%%%%
% Chapter 5

\chapter{Existence of wormhole solutions in $f(Q, T)$ gravity under noncommutative geometries} % Main chapter title
\label{Chapter5} 
% For referencing the chapter elsewhere, use \ref{Chapter1} 

\lhead{Chapter 5. \emph{Existence of wormhole solutions in $f(Q, T)$ gravity under noncommutative geometries}} % This is for the header on each page - perhaps a shortened title
% \vspace{10 cm}
\blfootnote{*The work in this chapter is covered by the following publication:\\
\textit{Existence of wormhole solutions in $f(Q, T)$ gravity under noncommutative geometries}, Physics of the Dark Universe \textbf{42}, 101288 (2023).}

%\blindtext
%----------------------------------------------------------------------------------------
This chapter examines spherically symmetric and static WH solutions within the $f(Q, T)$ gravity framework, considering two distributions, namely Gaussian and Lorentzian, under noncommutative geometry from string theory. The detailed study of the work is described as follows:
\begin{itemize}
    \item Two $f(Q, T)$ models are considered: a linear model $f(Q, T)=\alpha Q+\beta T$ and a nonlinear model $f(Q, T)=Q+\lambda_1 Q^2+\eta_1 T$, with both analytic and numerical solutions obtained under noncommutative distributions.
    \item WH solutions for the linear model are discussed analytically, while numerical and graphical methods are employed for the nonlinear model, demonstrating the behavior of the solutions with selected free parameter values.
    \item The derived shape function satisfies the flare-out conditions under asymptotic backgrounds, and the NEC is found to be violated at the WH's throat for both models in the presence of noncommutative geometry.
    \item Gravitational lensing is analyzed for a specific WH model, showing a divergent deflection angle at the throat, providing insights into its observational properties.
\end{itemize}

%\clearpage
%----------------------------------------------------------------------------------------
\section{Introduction}\label{sec1}
Noncommutative geometry aims to develop a unified framework that merges space-time gravitational forces with strong and weak interactions. Non-commutativity replaces point-like structures with smeared objects, offering a discretization of space-time and the commutator described by $\left[x^\alpha,\,x^\beta\right]= i\,\Theta^{\alpha\beta}$ \cite{Doplicher, Spallucci, Nicolini}, where $\Theta^{\alpha\beta}$ represented as a second-order anti-symmetric matrix of dimension $(\text{length})^2$. It is analogous to the discretization of the phase space by the Planck constant $\hbar$ \cite{Smailagic1/2003}. The standard concept of mass density in the form of the Dirac delta function does not hold true in noncommutative space. Thus, Gaussian and Lorentzian distributions with negligible length $\sqrt{\Theta}$ are used instead of the Dirac delta function to demonstrate the impact of this distribution. The static, spherically symmetric, smeared, and particle-like gravitational source handles the geometry of the Gaussian distribution with a noncommutative nature with the maximum mass $M$ retained. For the Lorentzian distribution, the density capacity of the particle-like mass $M$ can be assumed. In this case, the entire mass $M$ can be regarded as a form of the diffused unified object. This kind of matter distribution was used by Nicolini and Spalluci \cite{Nicolini2} to specify the physical substances of short-separated divergences of noncommutating coordinates in probing black holes. In Ref. \cite{Islam}, the authors investigated WH solutions for a higher dimension under Gaussian distributions and observed that WH solutions could exist only in the fourth and fifth dimensions. In addition, Lorentzian distributed noncommutative WHs in $f(R)$ gravity has been studied in Ref. \cite{Banerjee}. Recently, WH solutions have been investigated for both distributions in $f(Q)$ gravity \cite{Sokoliuk, Hassan1} as well as other modified theories of gravity \cite{Shi-Qin Song, Nilofar Rahman, Yihu Feng, Anshuman Baruah, M. Sharif 1, Banerjee, Mubasher Jamil, Shamaila Rani}.\\
Gravitational lensing is one of the initial applications of GR that have been investigated \cite{dm38}. It occurs when an astronomical object is enormous enough to bend the falling light into a lens, allowing the observer to gather more data regarding the source than feasible. It has grown in interest among researchers after the observational verification of the theoretical prediction of light bending \cite{Dyson, Eddington}. Not only did it make us study stars and galaxies, it also enabled us to investigate extrasolar planets, dark energy, and DM. Recently, a group of researchers successfully performed the first detection of astrometric microlensing to calculate the mass of the Stein 2051 B white dwarf \cite{Sahu}. A charming characteristic of gravitational lensing is that in unstable light rings, light can undergo an unboundedly large (theoretically infinite) amount of bending \cite{Bozza1, Virbhadra, Bozza2}. As a result of such strong gravitational lensing, a large number of relativistic images are formed. Also, it is well-known that strong and weak gravitational lensing is a powerful tool for analyzing gravitational fields around different astrophysical objects such as black holes and WHs. Some works in the literature where WHs were inspected by gravitational lensing on a large scale can be viewed in theoretical physics as well as in astrophysics \cite{A1, A2, A3, A4, A5, A6, A7, A8, A9, A10}. \\
This chapter is organized as follows. In Sec. \ref{sec2}, we consider the WH field equations for $f(Q, T)$ gravity. A brief review of the noncommutative geometry with linear form of $f(Q, T)$ is presented in Sec. \ref{sec3} and a nonlinear model of $f(Q, T)$ is presented in Sec. \ref{sec4}. To examine the deflection angle at the WH's throat, Sec. \ref{sec5} will be dedicated to analyzing the gravitational lensing for the specific model of Sec. \ref{sec3}. Finally, in the concluding Sec. \ref{sec6}, our findings are analyzed, accompanied by an in-depth discussion of the implications of the results of this study.

\section{Wormhole field equations in $f(Q, T)$ gravity}\label{sec2}
In this section, we analyze the WH field equations within the framework of $f(Q, T)$ gravity, as defined by Eqs. \eqref{ch3eq18}-\eqref{ch3eq20}, corresponding to the metric \eqref{1ch1}. Using these particular field equations, various WH solutions can be examined within the $f(Q, T)$ gravity framework. For this study, we consider linear and nonlinear $f(Q, T)$ models.

\section{Wormhole solutions with $f(Q, T)=\alpha\,Q+\beta\,T$}
\label{sec3}
In this section, we shall consider a specific and interesting $f(Q, T)$ model given by Eq. \eqref{ch3eq26}. Xu et al. \cite{fqt} introduced this model, and it naturally describes an exponentially expanding Universe, with $\rho \propto e^{-H_0 t}$ \cite{fqt}. In addition, this model has been used to be constrained by observational data of the Hubble parameter in Ref. \cite{Arora1}. With the same model, Loo et al. \cite{Loo1} investigated Bianchi type-I cosmology with observational data sets (Hubble and Type Ia supernovae). Moreover, WH solutions for this model have been studied with the different EoS relations in Ref. \cite{Tayde 4}. In this study, we will check the strength of this model under noncommutative geometries. Using the linear functional form \eqref{ch3eq26} with constant redshift function, the field Eqs. \eqref{ch3eq21}-\eqref{ch3eq23} can be read as Eqs. \eqref{ch3eq27}-\eqref{ch3eq29}.
%  \begin{equation}\label{18}
%  \rho =\frac{\alpha  (12 \pi -\beta ) b'}{3 (4 \pi -\beta ) (\beta +8 \pi ) r^2}\,,
%  \end{equation}
%  \begin{equation}\label{19}
%  p_r=-\frac{\alpha  \left(2 \beta  r b'-3 \beta  b+12 \pi  b\right)}{3 (4 \pi -\beta ) (\beta +8 \pi ) r^3}\,,
%  \end{equation}
%  \begin{equation}\label{20}
%  p_t=-\frac{\alpha  \left((\beta +12 \pi ) r b'+3 b (\beta -4 \pi )\right)}{6 (4 \pi -\beta ) (\beta +8 \pi ) r^3}\,.
%  \end{equation}
 In the following subsections, we will study WH solutions under Gaussian and Lorentzian distributions. We will investigate the effect of these noncommutative geometries on WH solutions through the behaviors of shape functions and energy conditions.
 
 \subsection{Gaussian distribution}
 \label{subsubsec1}
\begin{figure}[h]
\centering
\includegraphics[width=14.5cm,height=6cm]{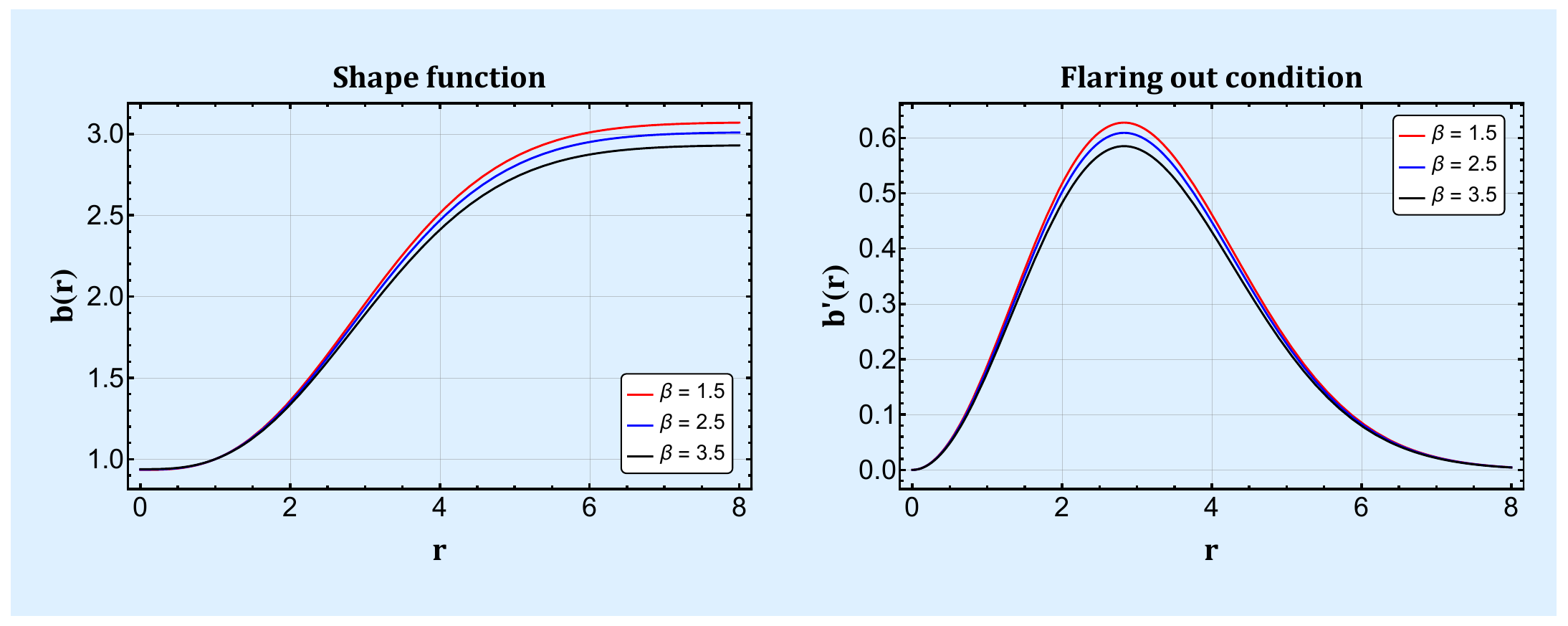}
\caption{The figure displays how the shape function (\textit{left}) and flare-out condition (\textit{right}) vary with radial coordinate $r$ for different values of $\beta$. Also, we consider $\alpha=1,\, \Theta=2,\, M=1.1,\, \text{and} \, r_0 = 1$. It is clear that $b(r)$ shows a positively increasing behavior and $b'(r)$ satisfies the flare-out condition at the throat.}
\label{fig1}
\end{figure}

\begin{figure}[h]
\centering
\includegraphics[width=14.5cm,height=6cm]{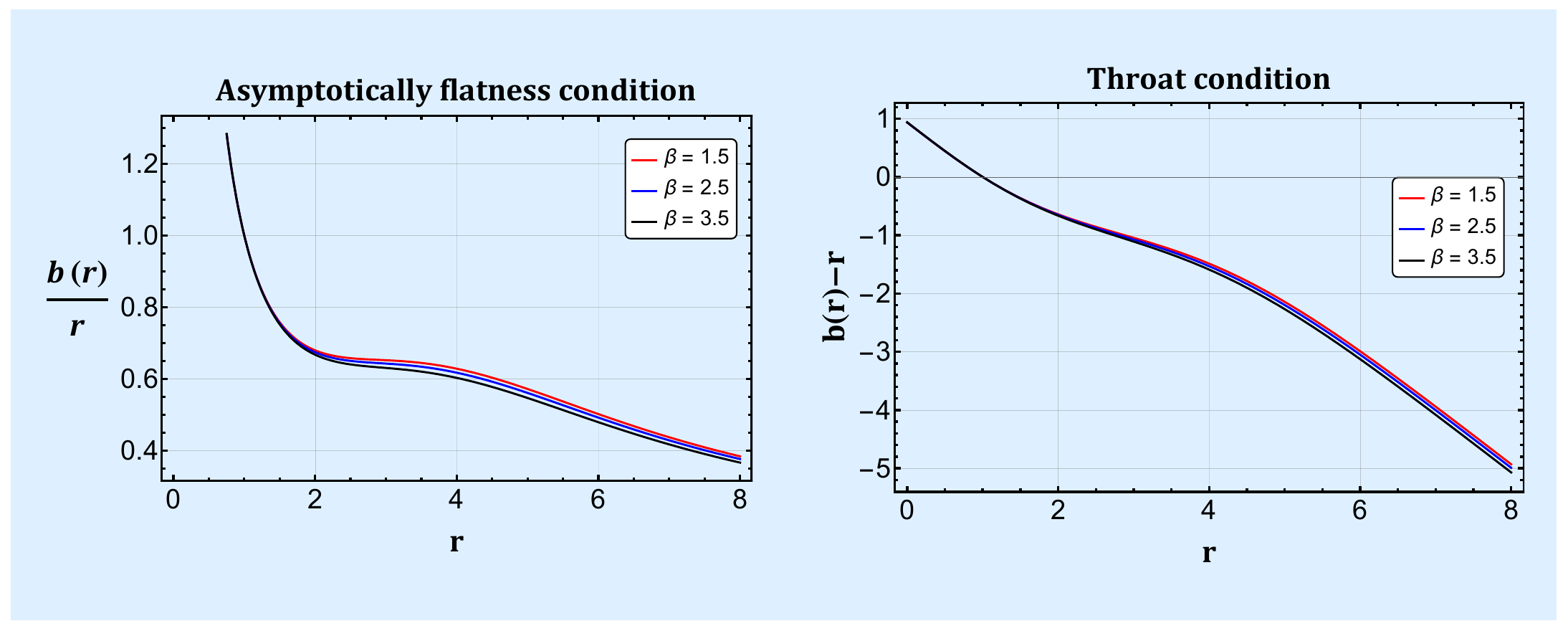}
\caption{The figure displays how the asymptotically flatness (\textit{left}) and throat conditions (\textit{right}) vary with radial coordinate $r$ for different values of $\beta$. Also, we consider $\alpha=1,\, \Theta=2,\, M=1.1,\, \text{and} \, r_0 = 1$. It is clear that $\frac{b(r)}{r}\rightarrow 0$ as $r\rightarrow \infty$ and $b(r)-r$  satisfies the throat condition.}
\label{fig2}
\end{figure}

The energy density for a spherically symmetric, static, smeared, particle-like gravitational source is provided in the work \cite{Nicolini/2006, Smailagic1/2003} by Eq. \eqref{21}.
 % \begin{equation}\label{21}
 % \rho =\frac{M e^{-\frac{r^2}{4 \Theta }}}{8 \pi ^{3/2} \Theta ^{3/2}}\,.
 % \end{equation}
 % The particle mass $M$, instead of being concentrated at a single point, is spread over a region with a characteristic scale of $\sqrt{\Theta}$.
This arises from the uncertainty encoded in the coordinate commutator.\\
 By comparing the Eqs. \eqref{ch3eq27} and \eqref{21} , the differential equation under Gaussian distribution is given by 
 \begin{equation}\label{23}
 \frac{\alpha  (12 \pi -\beta ) b'(r)}{3 (4 \pi -\beta ) (\beta +8 \pi ) r^2}=\frac{M e^{-\frac{r^2}{4 \Theta }}}{8 \pi ^{3/2} \Theta ^{3/2}}\,.
\end{equation}
 By solving the above expressions, we obtain
 \begin{equation}\label{a1}
    b(r)=c_1+\mathcal{K}_1 \left(2 \sqrt{\pi } \Theta ^{3/2} \text{erf}\left(\frac{r}{2 \sqrt{\Theta }}\right)-2 \Theta  r e^{-\frac{r^2}{4 \Theta }}\right)\,,
 \end{equation}
where $\mathcal{K}_1=\frac{3 \left(-\beta ^2-4 \pi  \beta +32 \pi ^2\right) M}{8 \pi ^{3/2} \alpha  (12 \pi -\beta ) \Theta ^{3/2}}$ and $c_1$ is the integrating constant. Also, erf is an error function and can be defined by $\text{erf}(\Theta)=\frac{2}{\sqrt{\pi}}\bigintsss\limits_{0}^{\Theta}e^{-t^2}dt$. Now, to calculate $c_1$, we use the throat condition $b(r_0)=r_0$ in the above expression \eqref{a1}, and obtained as
 \begin{equation}\label{a2}
    c_1=r_0-\mathcal{K}_1 \left(2 \sqrt{\pi } \Theta ^{3/2} \text{erf}\left(\frac{r_0}{2 \sqrt{\Theta }}\right)-2 \Theta  r_0 e^{-\frac{r_0^2}{4 \Theta }}\right)\,.
 \end{equation}
Now, the shape function \eqref{a1} can be read as
 \begin{equation}\label{a3}
    b(r)=r_0+2\,\Theta \mathcal{K}_1\left[e^{-\frac{r^2+r_0^2}{4 \Theta }} \left(e^{\frac{r^2}{4 \Theta }} \left(\sqrt{\pi } \sqrt{\Theta } e^{\frac{r_0^2}{4 \Theta }} \left(\text{erf}\left(\frac{r}{2 \sqrt{\Theta }}\right)-\text{erf}\left(\frac{r_0}{2 \sqrt{\Theta }}\right)\right)+r_0\right)-r e^{\frac{r_0^2}{4 \Theta }}\right)\right].
 \end{equation}
Next, we present the graphical representation of the acquired shape function with the essential conditions required for a WH to exist. Also, we set appropriate choices for the concerned free parameters. First, we examine how the shape functions for the Gaussian distribution behave when the redshift function is assumed to be constant. The behavior of the shape function and flare-out condition for different values of $\beta$ are shown in Fig. \ref{fig1}. One can notice that the shape function exhibits a favorably increasing behavior. However, increasing the value of the model parameter $\beta$, the shape function shows a decreasing behavior. Also, the right graph of Fig. \ref{fig1} depicts that the flare-out condition $b'(r_0)<1$ is satisfied at the WH's throat. Furthermore, the graph on the left of Fig. \ref{fig2} indicates the asymptotic behavior of the shape function for different values of $\beta$. It shows that for increasing larger values of the radial distance, the ratio $\frac{b(r)}{r}$ closes to zero, which guarantees the asymptotic behavior of the shape function. Also, in this case, we consider the WH's throat $r_0=1$, and its graphical representation is shown in the right panel of Fig. \ref{fig2}.

Now to investigate the energy conditions, especially NEC and SEC, we substitute the obtained shape function \eqref{a3} in Eqs. \eqref{ch3eq28} and \eqref{ch3eq29}, and obtained the pressure components

\begin{multline}\label{26}
p_r = \frac{e^{-\frac{r^2+r_0^2}{4 \Theta }} }{\mathcal{K}_2\,(\beta +8 \pi )}\left[(\beta +8 \pi ) M e^{\frac{r_0^2}{4 \Theta }} \left(3 \sqrt{\pi } (4 \pi -\beta ) \Theta ^{3/2} e^{\frac{r^2}{4 \Theta }} \mathcal{K}_3-\beta  r^3-3 (\beta -4 \pi ) \Theta  r\right)+\Theta  r_0 e^{\frac{r^2}{4 \Theta }} 
\right.\\\left.
\times \left(-3 (4 \pi -\beta ) (\beta +8 \pi ) M-4 \pi ^{3/2} \alpha  (12 \pi -\beta ) \sqrt{\Theta } 
e^{\frac{r_0^2}{4 \Theta }}\right)\right]
\end{multline} 
and
\begin{multline}\label{27}
p_t = \frac{1}{4\mathcal{K}_2}\left[-6 \sqrt{\pi } (4 \pi -\beta ) \Theta ^{3/2} M \mathcal{K}_3+M e^{-\frac{r^2}{4 \Theta }}\left(6 (\beta -4 \pi ) \Theta  r-(\beta +12 \pi ) r^3\right)+6 (4 \pi -\beta ) 
\right.\\\left.
\times \Theta  M r_0 e^{-\frac{r_0^2}{4 \Theta }}+\frac{8 \pi ^{3/2} \alpha  (12 \pi -\beta ) \Theta ^{3/2} r_0}{\beta +8 \pi }\right],
\end{multline}
where $\mathcal{K}_2=4 \pi ^{3/2} (12 \pi -\beta ) \Theta ^{3/2} r^3$ and $\mathcal{K}_3=\left(\text{erf}\left(\frac{r_0}{2 \sqrt{\Theta }}\right)-\text{erf}\left(\frac{r}{2 \sqrt{\Theta }}\right)\right)$.\\
In this case, the NEC for radial and tangential pressures can be obtained as
\begin{multline}\label{28}
\rho + p_r = \frac{e^{-\frac{r^2+r_0^2}{4 \Theta }}}{2 \mathcal{K}_2 (\beta +8 \pi )}\left[3 (4 \pi -\beta ) (\beta +8 \pi ) M e^{\frac{r_0^2}{4 \Theta }} \left(2 \sqrt{\pi } \Theta ^{3/2} e^{\frac{r^2}{4 \Theta }} \mathcal{K}_3+r^3+2 \Theta  r\right)-2 \Theta  r_0 e^{\frac{r^2}{4 \Theta }} 
\right.\\\left.
\times \left(3 (4 \pi -\beta ) (\beta +8 \pi ) M+4 \pi ^{3/2} \alpha  (12 \pi -\beta ) \sqrt{\Theta } e^{\frac{r_0^2}{4 \Theta }}\right)\right]
\end{multline}
and
\begin{multline}\label{29}
\rho + p_t = \frac{1}{16 \pi ^{3/2} \Theta ^{3/2} r^3}\left[\frac{3 (4 \pi -\beta ) M e^{-\frac{r^2}{4 \Theta }}}{12 \pi -\beta } \left(-2 \sqrt{\pi } \Theta ^{3/2} e^{\frac{r^2}{4 \Theta }} \mathcal{K}_3+r^3-2 \Theta  r\right)+2 \Theta  r_0 
\right.\\\left.
\times \left(\frac{4 \pi ^{3/2} \alpha  \sqrt{\Theta }}{\beta +8 \pi }+\frac{3 (4 \pi -\beta ) M e^{-\frac{r_0^2}{4 \Theta }}}{12 \pi -\beta }\right)\right],
\end{multline}
and at the WH's throat i.e. at $r=r_0$, the aforementioned expressions for NEC reduce to
\begin{equation}
\rho+p_r\bigg\vert_{r=r_0}= \frac{3 (4 \pi -\beta ) M e^{-\frac{r_0^2}{4 \Theta }}}{8 \pi ^{3/2} (12 \pi -\beta ) \Theta ^{3/2}}-\frac{\alpha }{(\beta +8 \pi ) r_0^2}
\end{equation}
and
\begin{equation}
\rho+p_t\bigg\vert_{r=r_0}=\frac{3 (4 \pi -\beta ) M e^{-\frac{r_0^2}{4 \Theta }}}{16 \pi ^{3/2} (12 \pi -\beta ) \Theta ^{3/2}}+\frac{\alpha }{2 (\beta +8 \pi ) r_0^2}\,.
\end{equation}
Also, the SEC for this case is given by
\begin{equation}\label{32}
\rho + p_r + 2 p_t = -\frac{\beta  M e^{-\frac{r^2}{4 \Theta }}}{\pi ^{3/2} (24 \pi -2 \beta ) \Theta ^{3/2}}.
\end{equation}

\begin{figure}
\centering
\includegraphics[width=14.5cm,height=5cm]{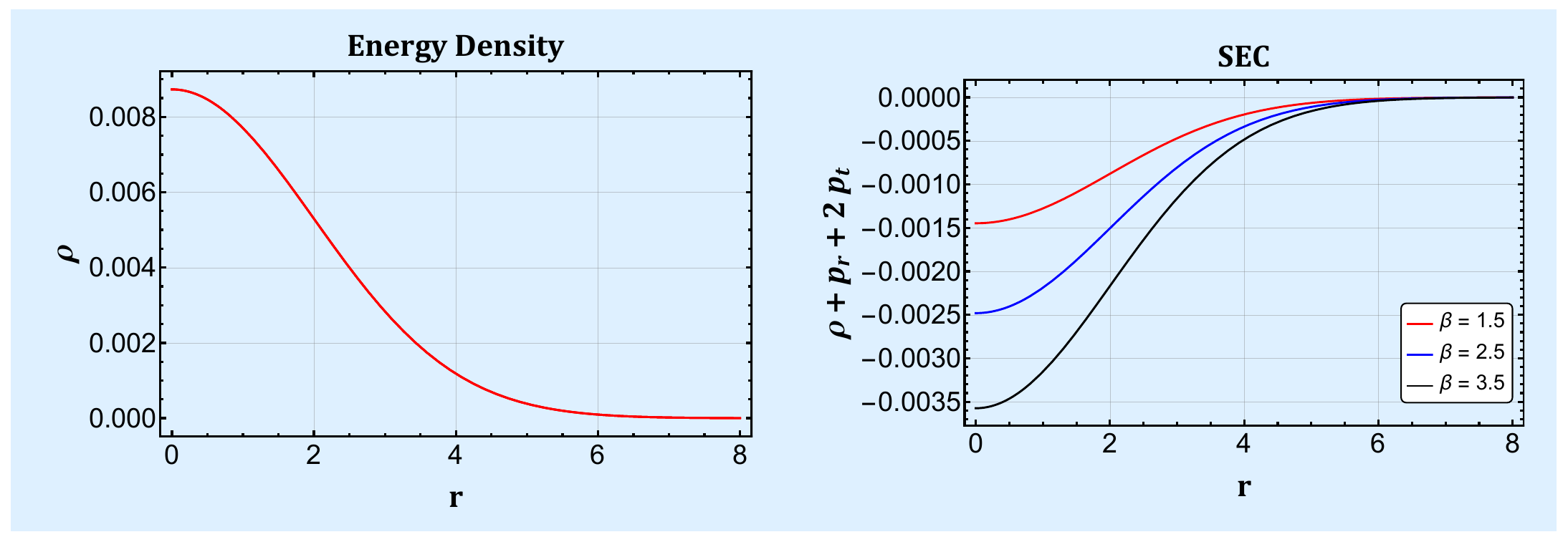}
\caption{The figure displays how the energy density (\textit{left}) and SEC (\textit{right}) vary with radial coordinate $r$ for different values of $\beta$. Also, we consider $\Theta=2\, \text{and} \, M=1.1$. It is clear that $\rho$ shows positively decreasing behavior and $\rho+p_r+2p_t$  shows negative behavior.}
\label{fig3}
\end{figure}

\begin{figure}
\centering
\includegraphics[width=14.5cm,height=5cm]{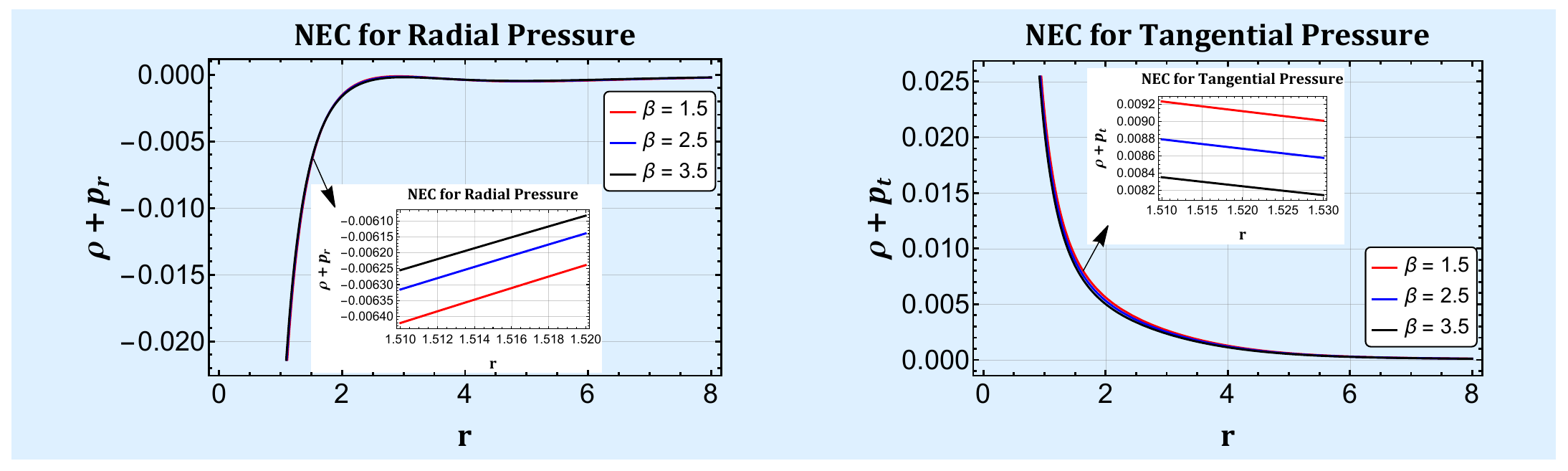}
\caption{The figure displays how the NEC for radial (\textit{left}) and tangential (\textit{right}) pressure vary with radial coordinate $r$ for different values of $\beta$. Also, we consider $\alpha=1,\, \Theta=2,\, M=1.1,\, \text{and} \, r_0 = 1$. It is clear that $\rho+p_r$ shows negative behavior and $\rho+p_t$  shows positive behavior.}
\label{fig4}
\end{figure}

We noticed from the above expressions that $\beta\neq -8\pi\, \text{and}\, 12\pi$. Also, for $\beta=4\pi$, the effect of noncommutative geometry will no longer be available at the throat. Keeping these in mind, we have plotted the graphs for the energy conditions in Figs. \ref{fig3}-\ref{fig4}. The graph for energy density versus $r$ is depicted in the left graph of Fig. \ref{fig3}, indicating the positively decreasing behavior throughout the space-time. In contrast, the right graph shows the behavior of SEC for different values of $\beta$, representing the negative behavior. One can notice that the expression for SEC is independent of the model parameter $\alpha$, and hence for vanishing $\beta$, i.e., $\beta\rightarrow 0$, the Eq. \eqref{32} aligns with the result of \cite{Hassan1}. The graph on the left of Fig. \ref{fig4} shows the negative behavior of radial NEC, leading to a violation of NEC. In contrast, the right graph confirms the positive behavior of tangential NEC in a decreasing way. Thus exotic matter may preserve WH solutions in the situation of non-metricity-based gravity with noncommutative geometry, which is the case for GR.


\subsection{Lorentzian distribution}
In this subsection,  we will discuss the WH solution under noncommutative geometry using a Lorentzian distribution. The energy density \cite{Nicolini/2006, Mehdipour11} for this case is given by Eq. \eqref{22}.
 % \begin{equation}\label{22}
 % \rho =\frac{\sqrt{\Theta } M}{\pi ^2 \left(\Theta +r^2\right)^2}\,,
 % \end{equation}
 % where $M$ is the total mass and $\Theta$is the non-commutativity parameter.\\

Now, by comparing the Eqs. \eqref{ch3eq27} and \eqref{22}, the differential equation under Lorentzian distribution is given by 
 \begin{equation}\label{33}
 \frac{\alpha  (12 \pi -\beta ) b'(r)}{3 (4 \pi -\beta ) (\beta +8 \pi ) r^2}=\frac{\sqrt{\Theta } M}{\pi ^2 \left(\Theta +r^2\right)^2}\,.
 \end{equation}
 Incorporating the above equation \eqref{33} for the shape function $b(r)$, we get
 \begin{equation}\label{4b1}
b(r)=\frac{\mathcal{M}_1}{\left(\Theta +r^2\right)}  \left(\left(\Theta +r^2\right) \tan ^{-1}\left(\frac{r}{\sqrt{\Theta }}\right)-\sqrt{\Theta } r\right)+c_2\,,
 \end{equation}
where $c_2$ represents the constant of integration and $\mathcal{M}_1=\frac{3 (4 \pi -\beta ) (\beta +8 \pi ) M}{2 \pi ^2 \alpha  (12 \pi -\beta )}$. Now we impose the throat condition to the above equation and obtain $c_2$ as
\begin{equation}
c_2=r_0-\frac{\mathcal{M}_1}{\left(\Theta +r_0^2\right)}\left(\left(\Theta +r_0^2\right) \tan ^{-1}\left(\frac{r_0}{\sqrt{\Theta }}\right)-\sqrt{\Theta } r_0\right).
\end{equation}
Substituting the value of $c_2$ in Eq. \eqref{4b1}, we get
 \begin{equation}\label{34}
 b(r) = r_0+ \frac{\mathcal{M}_1}{\left(\Theta +r^2\right) \left(\Theta +r_0^2\right)}\left(\left(\Theta +r^2\right) \left(\Theta +r_0^2\right) \mathcal{M}_2 +\sqrt{\Theta } (r-r_0) (r r_0-\Theta )\right)\,,
 \end{equation}
where $\mathcal{M}_2=\left(\tan ^{-1}\left(\frac{r}{\sqrt{\Theta }}\right)-\tan ^{-1}\left(\frac{r_0}{\sqrt{\Theta }}\right)\right)$.\\

\begin{figure}
\centering
\includegraphics[width=14.5cm,height=6cm]{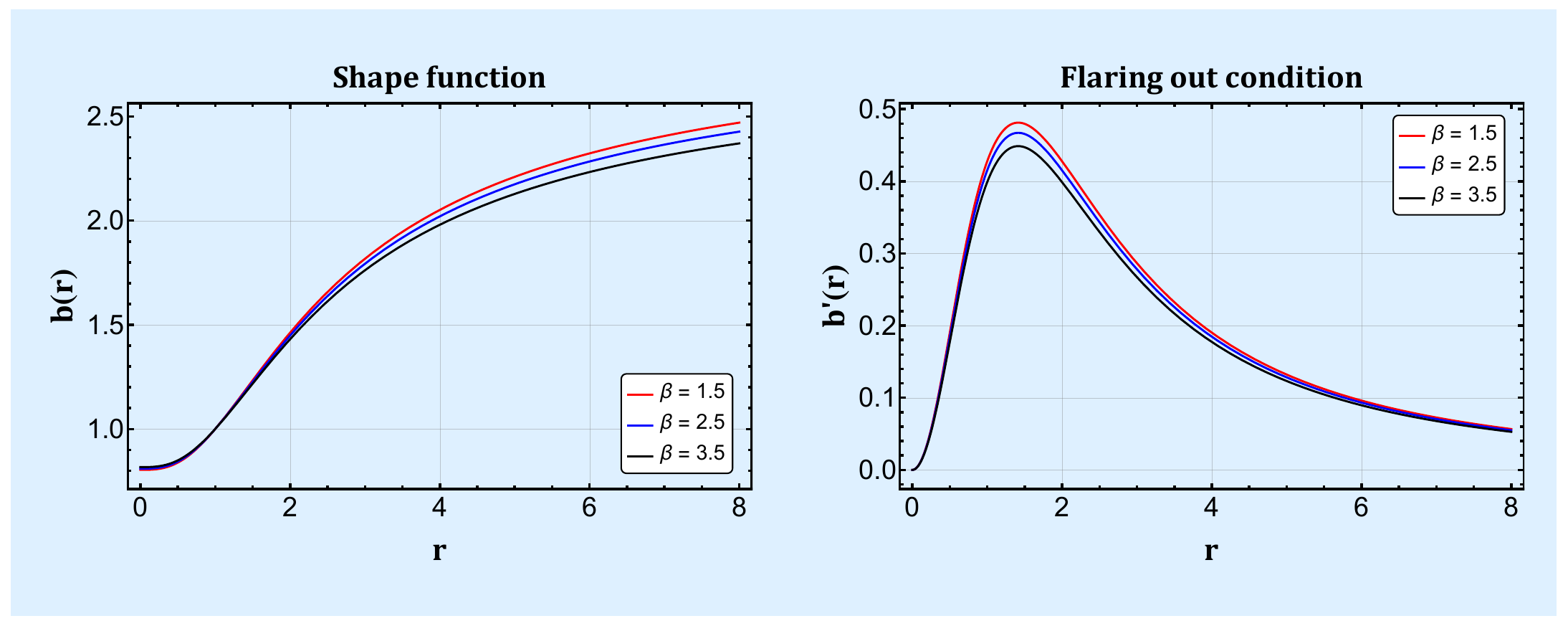}
\caption{The figure displays how the shape function (\textit{left}) and flare-out condition (\textit{right}) vary wih radial coordinate $r$ for different values of $\beta$. Also, we consider $\alpha=1,\, \Theta=2,\, M=1.1,\, \text{and} \, r_0 = 1$. It is clear that $b(r)$ shows a positively increasing behavior, and $b'(r)$ satisfies the flare-out condition at the throat.}
\label{fig7}
\end{figure}

\begin{figure}
\centering
\includegraphics[width=14.5cm,height=6cm]{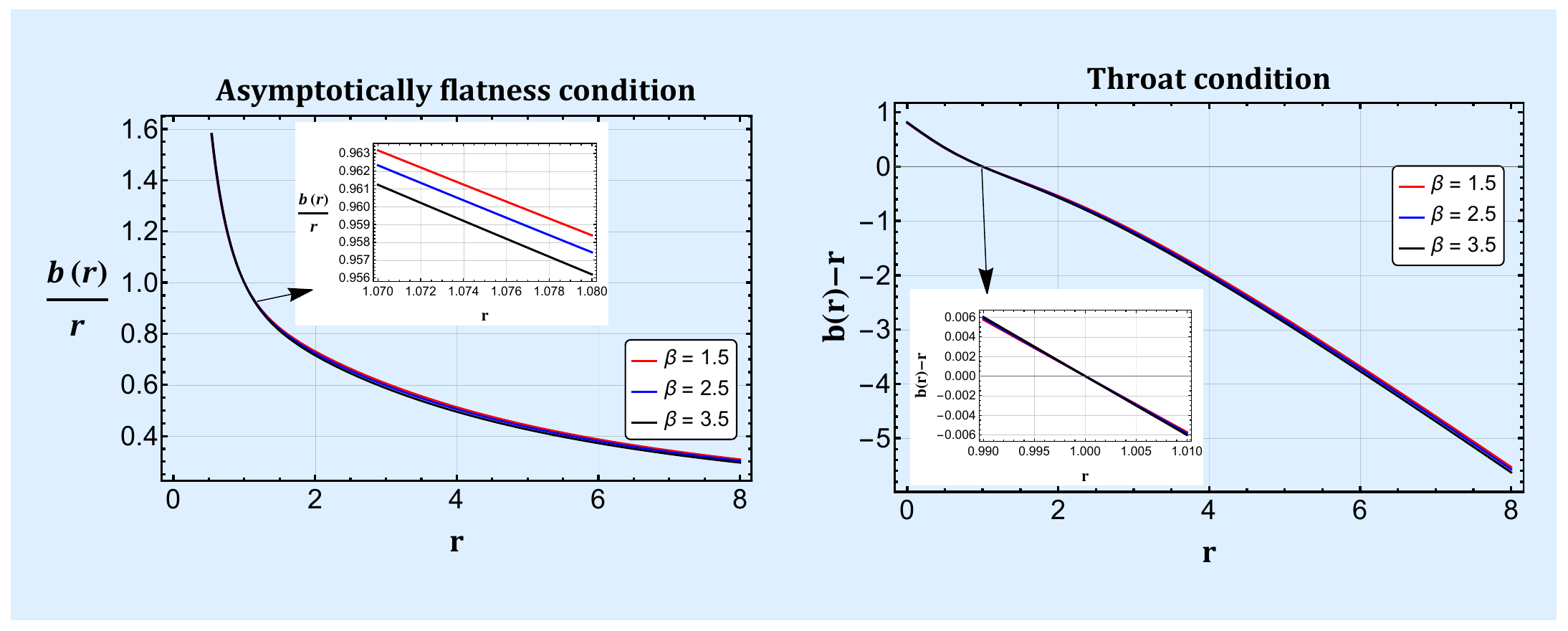}
\caption{The figure displays how the asymptotically flatness (\textit{left}) and throat conditions (\textit{right}) vary with radial coordinate $r$ for different values of $\beta$. Also, we consider $\alpha=1,\, \Theta=2,\, M=1.1,\, \text{and} \, r_0 = 1$. It is clear that $\frac{b(r)}{r}\rightarrow 0$ as $r\rightarrow \infty$ and $b(r)-r$  satisfies the throat condition.}
\label{fig8}
\end{figure}

The graphical representation of the derived shape function \eqref{34} and its related conditions for the Lorentzian distribution are illustrated in Figs. \ref{fig7} and \ref{fig8}. The left graph of Fig. \ref{fig7} indicates the positively increasing behavior of the shape function, while the right graph confirms that the flare-out condition is satisfied at the throat. In this case, we also noticed that, for increasing the model parameter $\beta$, the shape function behaves in a decreasing way. In the left curve of Fig. \ref{fig8}, we have plotted $\frac{b(r)}{r}$ versus radial coordinate $r$ for different values of $\beta$. It is seen that as we increase the values of the radial distance $r$, the ratio $\frac{b(r)}{r}$ converges to zero, which guarantees the asymptotic behavior of the shape function. Here, we also consider the radius of the throat $r_0=1$. In conclusion, we can deduce that our obtained shape function obeyed all the conditions for its traversability.\\
For investigating the behavior of energy conditions under the Lorentzian distribution, we will use the shape function \eqref{34} in the Eqs. \eqref{ch3eq28} and \eqref{ch3eq29}. The pressure components obtained are
% \begin{equation}\label{31}
% \rho = \frac{\sqrt{\Theta } M}{\pi ^2 \left(\Theta +r^2\right)^2}
% \end{equation}
\begin{multline}\label{35}
\hspace{-0.3cm}p_r =\frac{1}{2 r^3}\left[\frac{M}{\pi ^2 (12 \pi -\beta ) \left(\Theta +r^2\right)^2} \left(\sqrt{\Theta } r \left(12 \pi  \left(\Theta +r^2\right)-\beta  \left(3 \Theta +7 r^2\right)\right)-3 (4 \pi -\beta ) \left(\Theta +r^2\right)^2 \mathcal{M}_2\right)
\right.\\ \left.
+r_0 \left(\frac{3 (\beta -4 \pi ) \sqrt{\Theta } M}{\pi ^2 (12 \pi -\beta ) \left(\Theta +r_0^2\right)}-\frac{2 \alpha }{\beta +8 \pi }\right)\right]
\end{multline} 
and
\begin{multline}\label{36}
p_t =-\frac{1}{\mathcal{M}_3}\left[\left(\Theta +r^2\right) \left(3 (4 \pi -\beta ) (\beta +8 \pi ) M \left(\Theta +r_0^2\right)\left(\sqrt{\Theta } r-\left(\Theta +r^2\right) \tan ^{-1}\left(\frac{r}{\sqrt{\Theta }}\right)\right)-3
\right.\right.\\\left.\left.
 \hspace{0.5cm}\times (4 \pi -\beta ) (\beta +8 \pi ) M\left(\Theta +r^2\right) \left(\sqrt{\Theta } r_0-\left(\Theta +r_0^2\right) \tan ^{-1}\left(\frac{r_0}{\sqrt{\Theta }}\right)\right)-2 \pi ^2 \alpha (12 \pi -\beta )
 \right.\right.\\\left.\left.
\times  r_0 \left(\Theta +r^2\right) \left(\Theta +r_0^2\right)\right)
+2 (\beta +8 \pi ) (\beta +12 \pi ) \sqrt{\Theta } M r^3 \left(\Theta +r_0^2\right)\right],
\end{multline}
where $\mathcal{M}_3=4\pi^2(12\pi-\beta)(\beta +8\pi ) r^3\left(\Theta +r^2\right)^2 \left(\Theta+r_0^2\right)$.\\

\begin{figure}
\centering
\includegraphics[width=14.5cm,height=5cm]{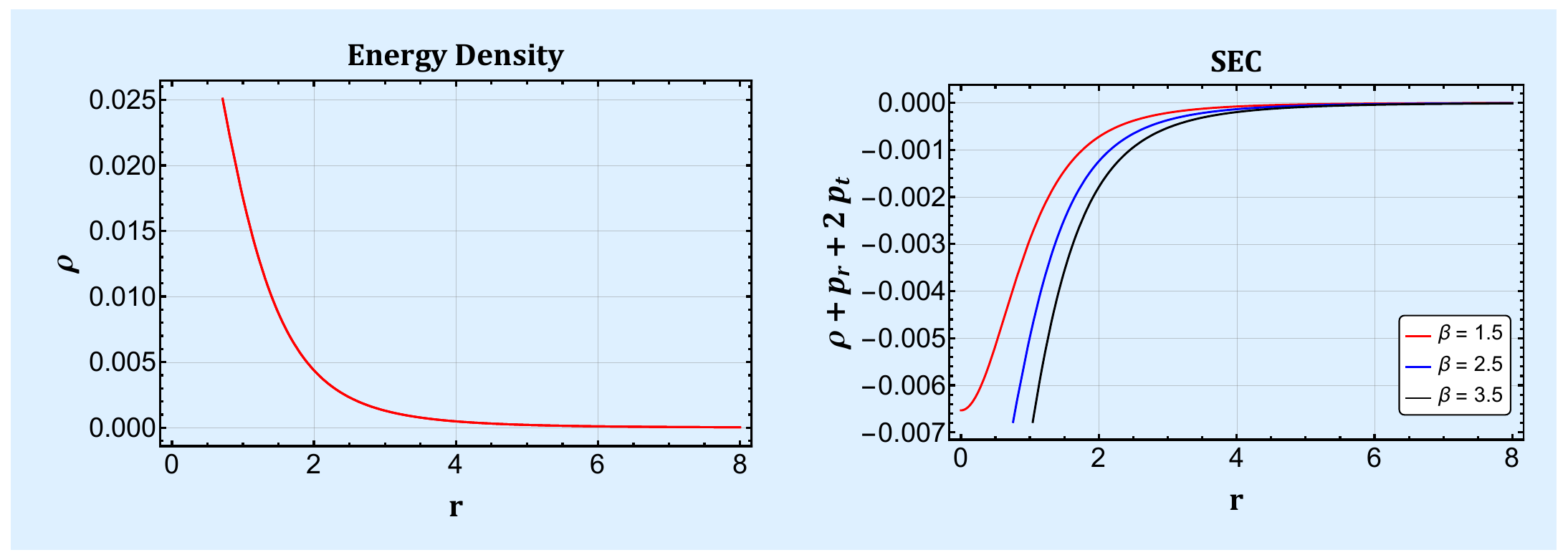}
\caption{The figure displays how the energy density (\textit{left}) and SEC (\textit{right}) vary with radial coordinate $r$ for different values of $\beta$. Also, we consider $\Theta=2\, \text{and} \, M=1.1$. It is clear that $\rho$ shows positively decreasing behavior and $\rho+p_r+2p_t$  shows negative behavior.}
\label{fig9}
\end{figure}

\begin{figure}
\centering
\includegraphics[width=14.5cm,height=5cm]{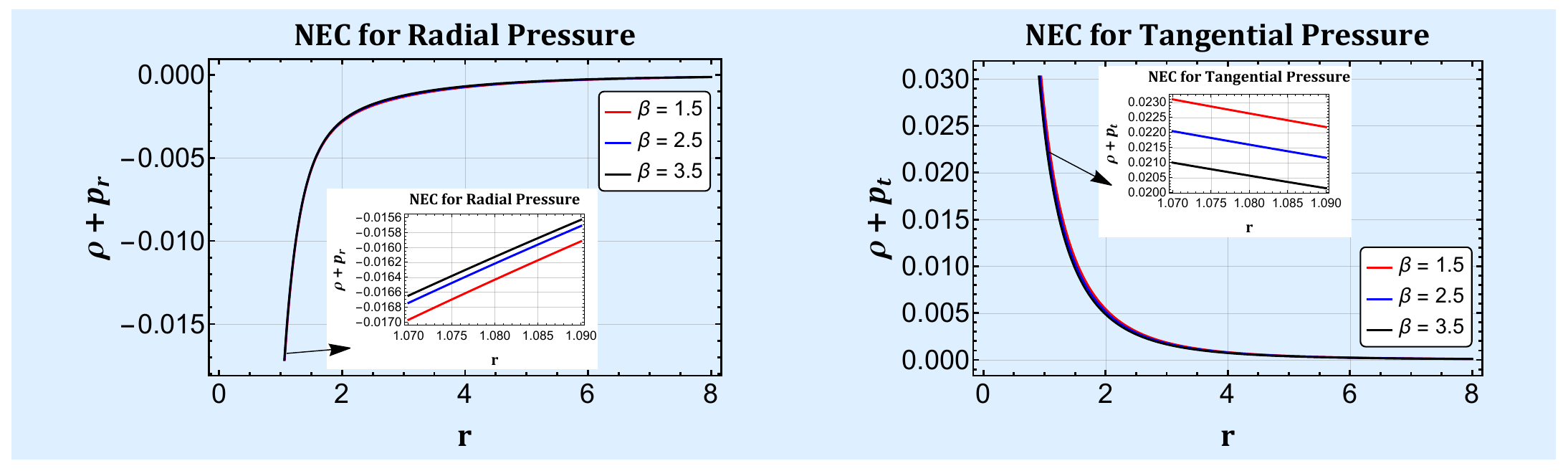}
\caption{The figure displays how the NEC for radial (\textit{left}) and tangential (\textit{right}) pressure vary with radial coordinate $r$ for different values of $\beta$. Also, we consider $\alpha=1,\, \Theta=2,\, M=1.1,\, \text{and} \, r_0 = 1$. It is clear that $\rho+p_r$ shows negative behavior and $\rho+p_t$  shows positive behavior.}
\label{fig10}
\end{figure}

At WH's throat, the expressions for NEC can be read as
\begin{equation}
\rho+p_r\bigg\vert_{r=r_0}=\frac{3 (4 \pi -\beta ) \sqrt{\Theta } M}{\pi ^2 (12 \pi -\beta ) \left(\Theta +r_0^2\right)^2}-\frac{\alpha }{(\beta +8 \pi ) r_0^2}
\end{equation}
and
\begin{equation}
\rho+p_t\bigg\vert_{r=r_0}=\frac{1}{2} \left(\frac{3 (4 \pi -\beta ) \sqrt{\Theta } M}{\pi ^2 (12 \pi -\beta ) \left(\Theta +r_0^2\right)^2}+\frac{\alpha }{(\beta +8 \pi ) r_0^2}\right).
\end{equation}
From the above expressions, it is clear that $\beta\neq -8\pi \,\text{and}\,12\pi$. We graphically explained the behavior of the energy conditions in Figs. \ref{fig9} and \ref{fig10} for different values of $\beta$. As usual, the energy density is positive throughout space-time.  Fig. \ref{fig10} represents the NEC versus the radial distance $r$, graph indicating that $\rho+p_t$ is a positive but decreasing behavior, whereas the radial NEC, i.e. $\rho+p_r$ shows a negative increase in the vicinity of the throat. Also, SEC is violated as shown in Fig. \ref{fig9}. This shows that the NEC was violated in this case, allowing the WH to exist. Consequently, the provided solutions are physically viable in both Gaussian and Lorentzian cases because they satisfy all the necessary characteristics of the shape function for the existence of the WH.\\


\section{Wormhole solutions with $f(Q, T)=Q+\,\lambda_1\,Q^2+\eta_1\,T$}
\label{sec4}
In this section, we assume a nonlinear form of $f(Q, T)$ defined in Ref. \cite{Tayde 4}
\begin{equation}
\label{41}
f(Q, T)=Q+\,\lambda_1\,Q^2+\eta_1\,T\,,
\end{equation}
where $\lambda_1$ and $\eta_1$ are model parameters. It is important to note that the above nonlinear model will reduce to the linear model \eqref{ch3eq26} for the model parameter $\lambda_1=0$; also, $\lambda_1=\eta_1=0$ gives the equivalent to the GR solution. This model successfully describes the de Sitter-type expansion of the Universe \cite{Loo}. In addition, an investigation of WH was attempted; unfortunately, the authors did not find the exact solution due to the high nonlinearity of the field equations \cite{Tayde 4}. Here, we are going to check the strength of the model in WH solutions under noncommutative geometries. Using the above model \eqref{41}, the energy-momentum tensor components can be written from Eqs. \eqref{ch3eq21}-\eqref{ch3eq23} as
 \begin{multline}\label{42}
 \rho =\frac{1}{\mathcal{N}}\left(r b' \left(b \lambda_1  r b' (-11 b \eta_1 +72 \pi  b+8 \eta_1  r-96 \pi  r) +2 b^2 
 \right.\right. \\ \left.\left.
 \lambda_1(15 b (\eta_1 -8 \pi )+16 (12 \pi -\eta_1 ) r)+4 (12 \pi -\eta_1 ) r^3 (b-r)^2\right)
 \right.\\\left.
 +b^3 \lambda_1 \left(3 b (88 \pi -9 \eta_1 )+32 (\eta_1 -12 \pi ) r\right)\right),
 \end{multline}
 \begin{multline}\label{43}
 p_r=\frac{1}{\mathcal{N}}\left(r b' \left(b \lambda_1  r b' (-13 b \eta_1 +24 \pi  b+16 \eta_1  r)+2 b^2 \lambda_1  (3 b 
 \right.\right. \\ \left.\left.
 (3 \eta_1 +8 \pi )-8 (\eta_1 +12 \pi ) r)-8 \eta_1  r^3 (b-r)^2\right)
 +b^3 \lambda_1  (3 b 
 \right. \\ \left.
 (\eta_1 -56 \pi )+8 (36 \pi -\eta_1 ) r)-12 b (4 \pi -\eta_1 ) r^3 (b-r)^2\right),
 \end{multline}
 and
 \begin{multline}\label{44}
p_t= \frac{1}{\mathcal{N}}\left(r b' \left(b \lambda_1  r b' (4 (\eta_1 +12 \pi ) r-b (\eta_1 +24 \pi ))+2 b^2
\right.\right. \\ \left.\left.
\lambda_1  (9 b \eta_1 +24 \pi  b-20 \eta_1  r-48 \pi  r)-2 (\eta_1 +12 \pi ) r^3 
\right.\right. \\ \left.\left.
(b-r)^2\right)+b^3 \lambda_1  (-3 b (11 \eta_1 +8 \pi )+52 \eta_1  r+48 \pi  r)
 \right. \\ \left.
  +6 b (4 \pi -\eta_1 ) r^3(b-r)^2\right)\,,
 \end{multline}
 where $\mathcal{N}=12 (4 \pi -\eta_1 ) (\eta_1 +8 \pi ) r^6 (b-r)^2$\,.\\
We will study the above modified field equations with numerical approaches under both noncommutative distributions in the following consecutive subsections.
 
 \subsection{Gaussian distribution}
 \label{section4}
\begin{figure}[h]
\centering
\includegraphics[width=14.5cm,height=6cm]{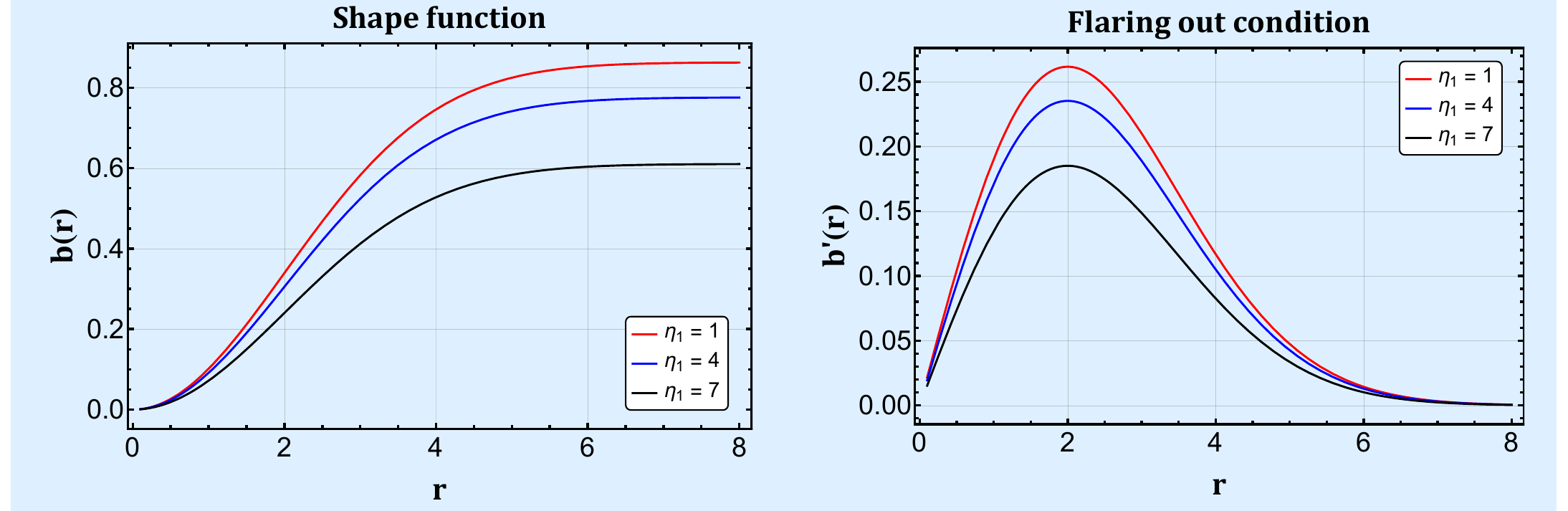}
\caption{The figure displays how the shape function (\textit{left}) and flare-out condition (\textit{right}) vary with radial coordinate $r$ for different values of $\eta_1$. Also, we consider $\lambda_1=0.01,\, \Theta=2,\, \text{and} \, M=1.1$. It is clear that $b(r)$ shows a positively increasing behavior and $b'(r)$ satisfies the flare-out condition at the throat.}
\label{fig13}
\end{figure}

\begin{figure}[h]
\centering
\includegraphics[width=14.5cm,height=6cm]{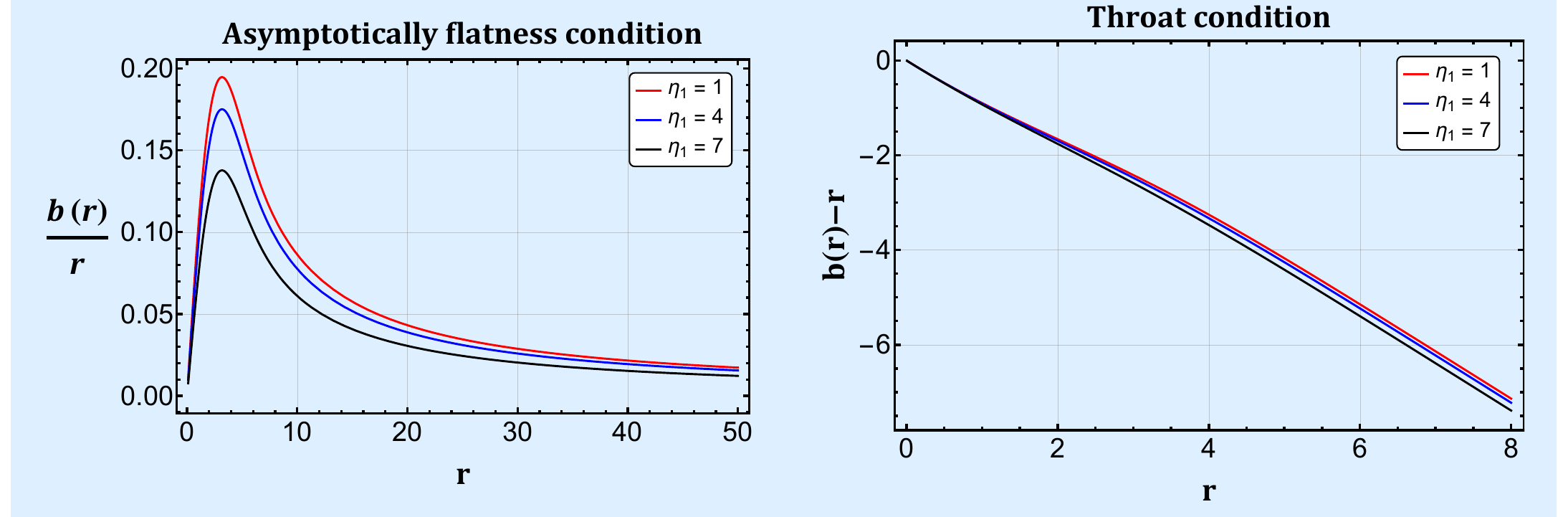}
\caption{The figure displays how the asymptotically flatness condition (\textit{left}) and throat condition (\textit{right}) vary with radial coordinate $r$ for different values of $\eta_1$. Also, we consider $\lambda_1=0.01,\, \Theta=2,\, \text{and} \, M=1.1$. It is clear that $\frac{b(r)}{r}\rightarrow 0$ as $r\rightarrow \infty$ and $b(r)-r$ shows the location of the throat at 0.001 (approximately).}
\label{fig14}
\end{figure}

\begin{figure}
\centering
\includegraphics[width=14.5cm,height=5cm]{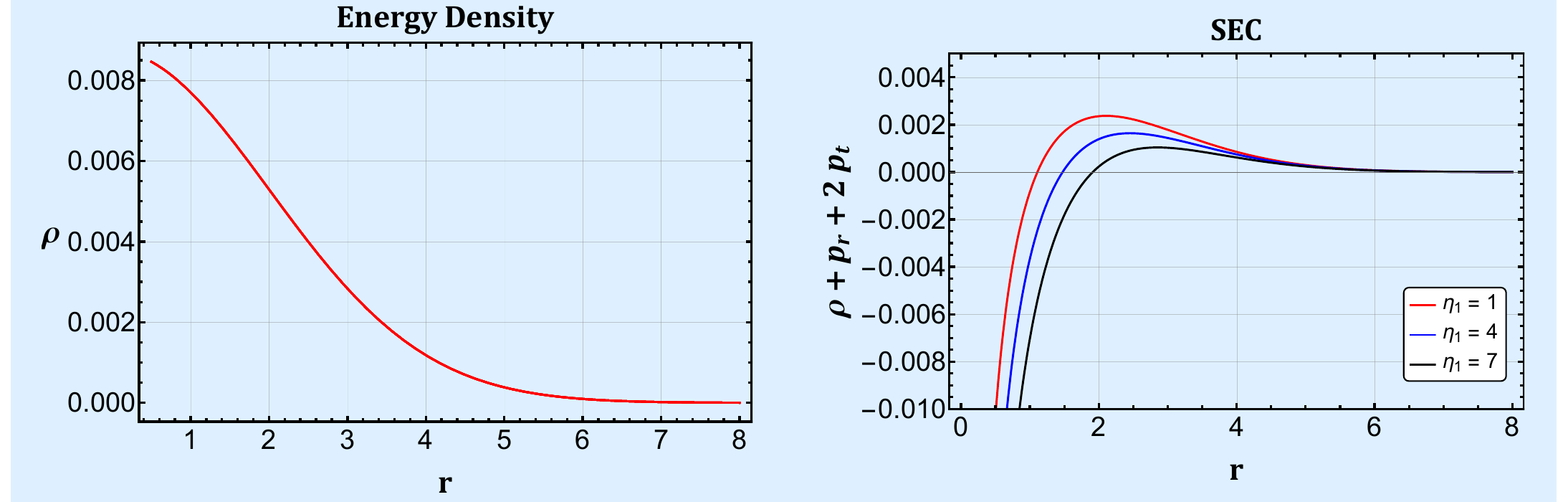}
\caption{The figure displays how the energy density (\textit{left}) and SEC (\textit{right}) vary with radial coordinate $r$ for different values of $\eta_1$. Also, we consider $\lambda_1=0.01,\, \Theta=2,\, \text{and} \, M=1.1$. It is clear that $\rho$ shows positively decreasing behavior and $\rho+p_r+2p_t$  shows negative behavior at the throat.}
\label{fig15}
\end{figure}

\begin{figure}
\centering
\includegraphics[width=14.5cm,height=5cm]{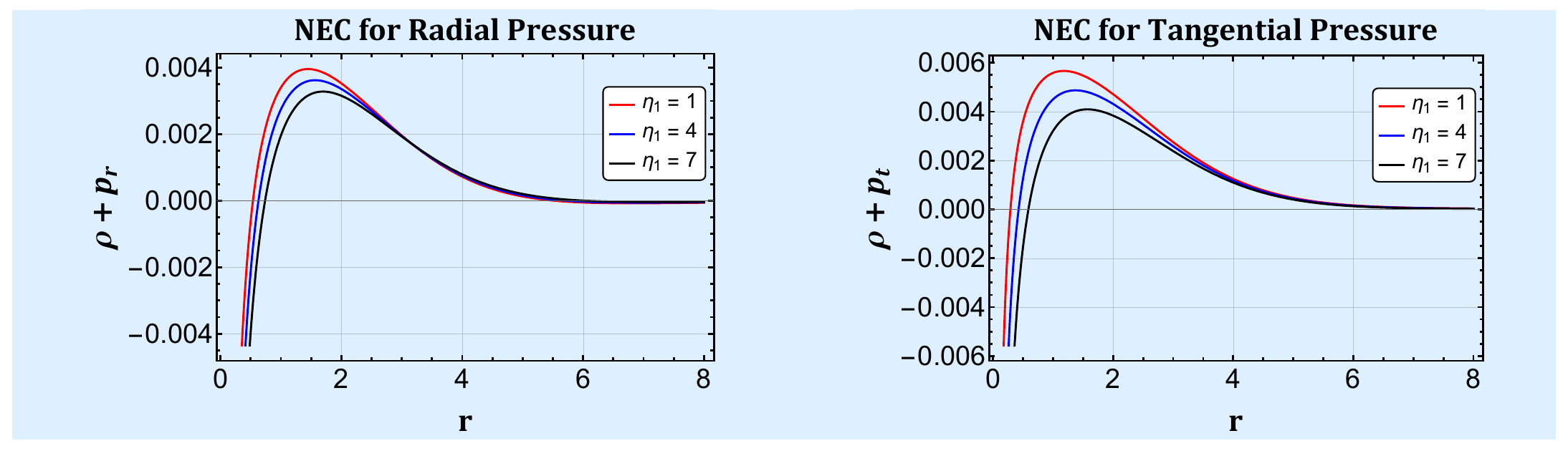}
\caption{The figure displays how the NEC for radial (\textit{left}) and tangential (\textit{right}) pressure vary with radial coordinate $r$ for different values of $\eta_1$. Also, we consider $\lambda_1=0.01,\, \Theta=2,\, \text{and} \, M=1.1$. It is clear that NEC for both pressures shows negative behavior at the throat.}
\label{fig16}
\end{figure}

We compare the Eq. \eqref{42} with the Gaussian distributed energy density \eqref{21} and obtain the following differential equation
 \begin{multline}\label{45}
     \frac{1}{\mathcal{N}}\left(r b' \left(b \lambda_1  r b' (-11 b \eta_1 +72 \pi  b+8 \eta_1  r-96 \pi  r) +2 b^2 
 \right.\right. \\ \left.\left.
 \lambda_1(15 b (\eta_1 -8 \pi )+16 (12 \pi -\eta_1 ) r)+4 (12 \pi -\eta_1 ) r^3 (b-r)^2\right)
 \right.\\\left.
 +b^3 \lambda_1 \left(3 b (88 \pi -9 \eta_1 )+32 (\eta_1 -12 \pi ) r\right)\right)=\frac{M e^{-\frac{r^2}{4 \Theta }}}{8 \pi ^{3/2} \Theta ^{3/2}}\,,
 \end{multline}
 which is challenging to solve analytically; therefore, we approach it numerically to determine the shape function $b(r)$. Here, we consider the initial condition
 \begin{equation}\label{4a1}
   b(1\times 10^{-2})=1\times 10^{-5} . 
 \end{equation}
We find this initial condition in such a way that this condition satisfies all the important requirements of shape functions. We used the \textit{mathematica} tool with code \textit{NDSolve} and solved the Eq. \eqref{45} with the help of the initial condition \eqref{4a1} and discussed the essential conditions for the WH to exist, and it is shown graphically in Figs. \ref{fig13} and \ref{fig14}. In addition, we have considered different possible values of the model parameters for the graphical views. The plot on the left of Fig. \ref{fig13} depicts the evolutions of the shape function, indicating the positively increasing behavior, while for increasing the value of the model parameter $\eta_1$, the behavior of the shape function is decreasing. The derivative of the shape function versus $r$ is represented in the right plot of Fig. \ref{fig13}. We noticed that $b^{'}(r)<1$ at $r=r_0$, confirming that the flare-out condition is met at the throat. Furthermore, we plotted the graph for $\frac{b(r)}{r}$ with respect to $r$ for different positive values of the model parameter $\eta_1$ in the left panel of Fig. \ref{fig14}, which shows that $\frac{b(r)}{r}\rightarrow 0$ as the radial distance $r\rightarrow \infty$, for any positive values of $\eta_1$. However, the right plot corresponds to the function $b(r)-r$ versus $r$, which provides the location of the WH's throat at $r_0 = 0.001$ (approximately). This kind of result can be found in Refs. \cite{Zubair2, Zubair1}. Furthermore, we analyzed the energy conditions for various values of $\eta_1$ shown in Figs. \ref{fig15} and \ref{fig16}. As usual, the energy density shows positively decreasing behavior while the SEC is violated in the vicinity of the throat. In addition, SEC decreases when the model parameter $\eta_1$ increases, and there is no effect of $\eta_1$ on the energy density as it is not present in the equation of energy density for the Gaussian distribution. Fig. \ref{fig16} indicates the violation of NEC near the throat for positive values of $\eta_1$, and far from the throat, NEC is satisfied. However, very far from the throat, the NEC could be violated.


\subsection{Lorentzian distribution}
Similar to the Gaussian distribution, we compare Eq. \eqref{42} with the Lorentzian distributed energy density \eqref{22}, and obtained a nonlinear differential equation
\begin{multline}\label{46}
    \frac{1}{\mathcal{N}}\left(r b' \left(b \lambda_1  r b' (-11 b \eta_1 +72 \pi  b+8 \eta_1  r-96 \pi  r) +2 b^2 
 \right.\right. \\ \left.\left.
 \lambda_1(15 b (\eta_1 -8 \pi )+16 (12 \pi -\eta_1 ) r)+4 (12 \pi -\eta_1 ) r^3 (b-r)^2\right)
 \right.\\\left.
 +b^3 \lambda_1 \left(3 b (88 \pi -9 \eta_1 )+32 (\eta_1 -12 \pi ) r\right)\right)=\frac{\sqrt{\Theta } M}{\pi ^2 \left(\Theta +r^2\right)^2},
\end{multline}
whose exact solution is also not possible. Thus, we use the numerical technique to evaluate the possible form of the shape function. Here, we consider the same initial condition \eqref{4a1} to solve the above Eq. \eqref{46}.\\

 \begin{figure}[h]
\centering
\includegraphics[width=14.5cm,height=6cm]{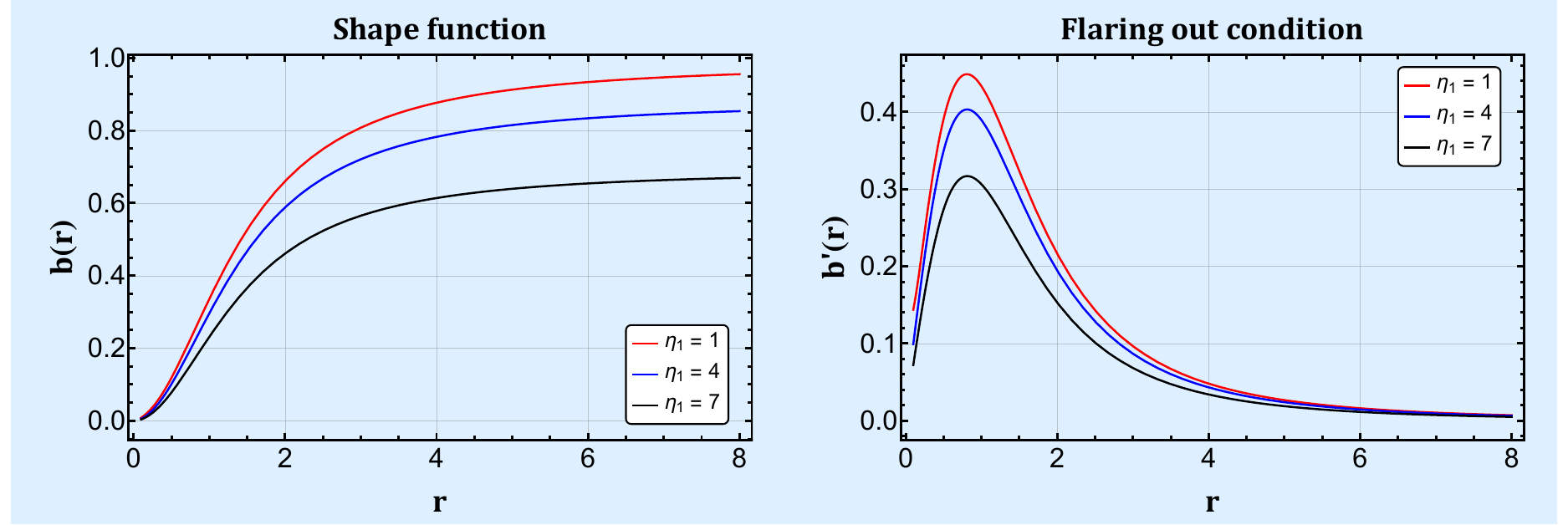}
\caption{The figure displays how the shape function (\textit{left}) and flare-out condition (\textit{right}) vary with radial coordinate $r$ for different values of $\eta_1$. Also, we consider $\lambda_1=0.01,\, \Theta=2,\, \text{and} \, M=1.1$. It is clear that $b(r)$ shows a positively increasing behavior, and $b'(r)$ satisfies the flare-out condition at the throat.}
\label{fig19}
\end{figure}
\begin{figure}[h]
\centering
\includegraphics[width=14.5cm,height=6cm]{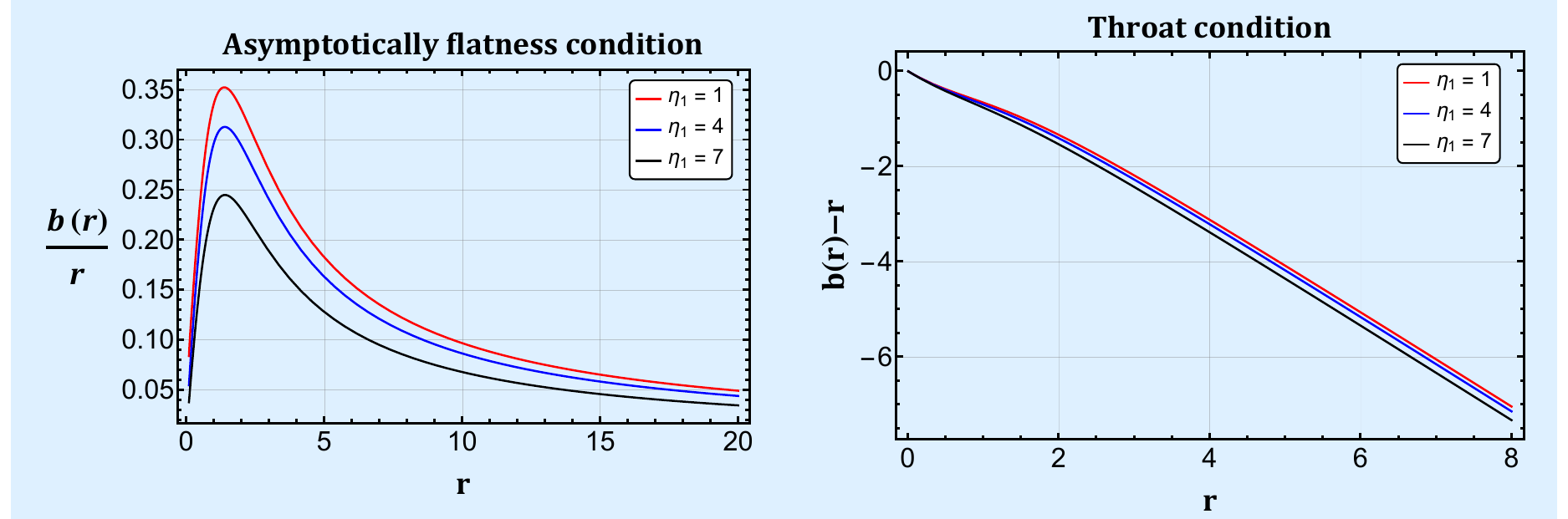}
\caption{The figure displays how the asymptotically flatness condition (\textit{left}) and throat condition (\textit{right}) vary with radial coordinate $r$ for different values of $\eta_1$. Also, we consider $\lambda_1=0.01,\, \Theta=2,\, \text{and} \, M=1.1$. It is clear that $\frac{b(r)}{r}\rightarrow 0$ as $r\rightarrow \infty$, and $b(r)-r$ shows the location of the throat at 0.001 (approximately).}
\label{fig20}
\end{figure}
\begin{figure}
\centering
\includegraphics[width=14.5cm,height=5cm]{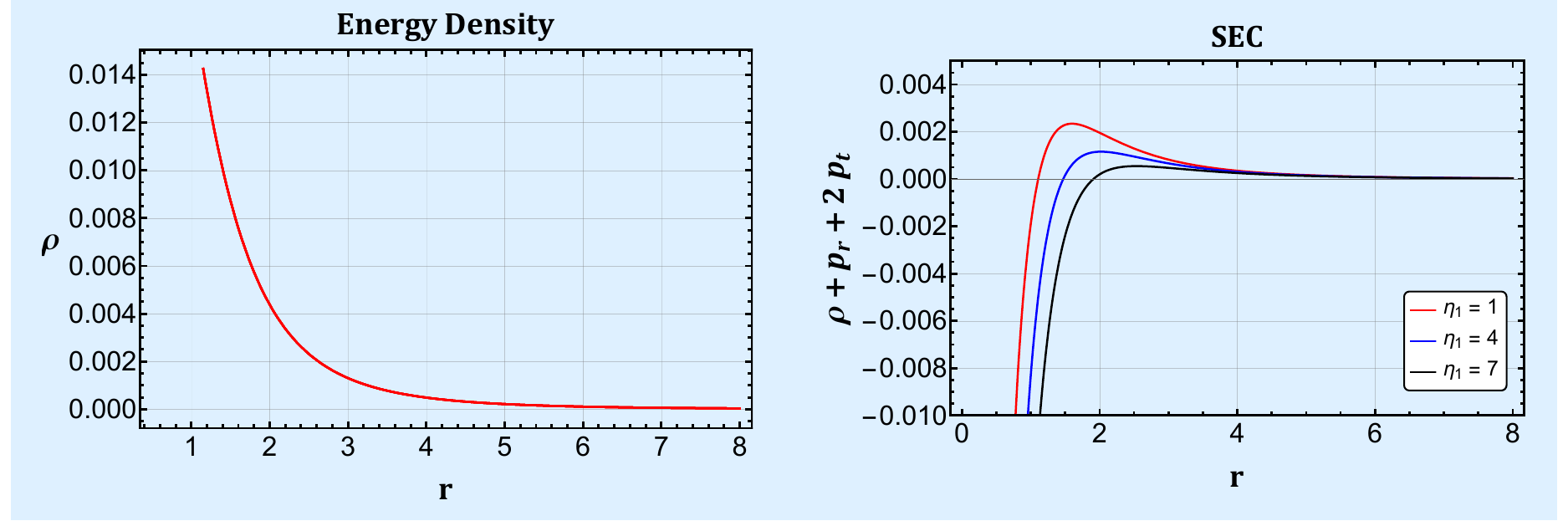}
\caption{The figure displays how the energy density (\textit{left}) and SEC (\textit{right}) vary with radial coordinate $r$ for different values of $\eta_1$. Also, we consider $\lambda_1=0.01,\, \Theta=2,\, \text{and} \, M=1.1$. It is clear that $\rho$ shows positively decreasing behavior, and $\rho+p_r+2p_t$  shows negative behavior at the throat.}
\label{fig21}
\end{figure}
\begin{figure}
\centering
\includegraphics[width=14.5cm,height=5cm]{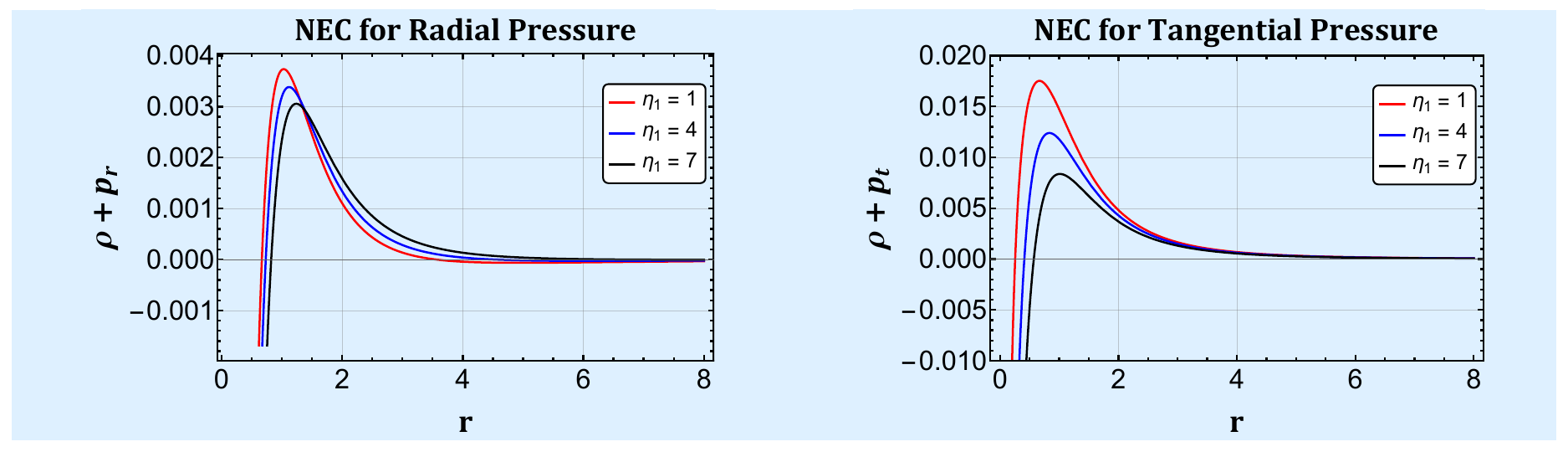}
\caption{The figure displays how the NEC for radial (\textit{left}) and tangential (\textit{right}) pressure vary with radial coordinate $r$ for different values of $\eta_1$. Also, we consider $\lambda_1=0.01,\, \Theta=2,\, \text{and} \, M=1.1$. It is clear that NEC for both pressures shows negative behavior at the throat.}
\label{fig22}
\end{figure}
Following the same procedure as for the Gaussian distribution, we conducted a numerical solution of Eq. \eqref{46} and analyzed the corresponding properties of the WH structure. The initial condition \eqref{4a1} was also utilized for this purpose. The graphical representations in Figs. \ref{fig19} and \ref{fig20} were used to illustrate our findings. In addition, various values of the model parameters were considered in order to present a comprehensive view. The plot on the left of Fig. \ref{fig19} demonstrates the evolution of the shape function, indicating an increasing positive behavior. One can notice that as the model parameter $\eta_1$ increases, the shape function shows a decreasing trend. The plot on the right of Fig. \ref{fig19} represents the derivative of the shape function with respect to $r$, confirming the satisfaction of the flare-out condition ($b'(r) < 1$) at $r=r_0$. In Fig. \ref{fig20}, the left panel shows the graph of $\frac{b(r)}{r}$ as a function of $r$ for various positive values of the model parameter $\eta_1$. It can be seen that as the radial distance $r$ tends toward infinity, $\frac{b(r)}{r}$ approaches zero. On the other hand, the right plot in Fig. \ref{fig20} shows the function $b(r)-r$ plotted against $r$, providing the location of the WH's throat at approximately $r_0 \approx 0.001$. Furthermore, we investigated the energy conditions under Lorentzian distribution, depicted in Figs. \ref{fig21} and \ref{fig22}. As expected, the energy density exhibited a positively decreasing trend, while the SEC was violated in the vicinity of the throat. Moreover, with increasing values of the parameter $\eta_1$, the violation of SEC became more prominent. Fig. \ref{fig22} confirms the violation of the NEC for both pressures in the throat for the values chosen of $\eta_1$. However, it is worth mentioning that NEC violations may still occur far from the throat. During this study, we noticed that the effect of the model parameter and the noncommutative parameter is responsible for the violation of the energy conditions.

In conclusion, the numerical WH solutions obtained for the nonlinear $f(Q, T)$ model are interesting from a physical perspective in both noncommutative distributions.

\section{Gravitational lensing}
\label{sec5}
In this section, we will utilize gravitational lensing phenomena to explore the possible detection of a traversable WH. The interest of researchers in gravitational lensing, particularly in strong gravitational lensing, has increased over time after the work of Virbhadra et al. \cite{Virbhadra1, Virbhadra, Virbhadra3}. Also, for any spherically symmetric space-time, Bozza \cite{Bozza2} developed an analytic technique for calculating gravitational lensing in strong-field limits. Later, this technique was used in many studies such as in Refs. \cite{Quinet, Tejeiro}. This motivates us to implement this method in this study. For this purpose, we consider the spherically symmetric line element involving the radius in Schwarzschild units, i.e., $x = \frac{r}{2\mathcal{M}}$ given by
\begin{equation}\label{5a}
ds^2=A(x)dt^2-B(x)dr^2-C(x)\,\left(d\theta^2+\sin^2\theta\,d\varphi^2\right)\,.
\end{equation}

Here, $\bar{x} = \frac{\bar{r}}{2\mathcal{M}}$ denotes the closest path of the light ray. We will also take into account the exact shape function for this simple linear $f(Q, T)=\alpha Q+\beta T$ model given by Eq. \eqref{4b1}. Here, one can clearly observe that the model parameters $\alpha\neq0$ and $\beta\neq 4\pi, -8\pi \,\, \text{and}\,\, 12\pi$ exist for the shape function \eqref{4b1}. In essence, we take into account the WH metric \eqref{1ch1}, where $e^{2\phi(r)}=\left(\frac{r}{b_0}\right)^m$, $b_0$ is a constant of integration and $m=2(v^\varphi)^2$ is the rotational velocity. It is noted in Refs. \cite{K. K. Nandi, Tejeiro, Kuhfittig2, Kuhfittig3} that $m \approx 0.000001$, a very tiny number (nearly zero), renders the redshift function constant (as we assumed in previous sections). The following connections are produced by comparing these metrics
\begin{equation}
    A(x)=\left(\frac{r}{b_0}\right)^m;\quad B(x)= \left(1-\frac{b(r)}{r}\right)^{-1}; \quad C(x)=r^2\,.
\end{equation}
It is discussed in Ref. \cite{Tejeiro}, the deflection angle $\alpha(\bar{x})$ consists of the sum of two terms
\begin{equation}
\alpha(\bar{x})=\alpha_e\,+I(\bar{x}),
\end{equation}
where 
\begin{equation}
\alpha_e=-2 \ln \left(\frac{2a}{3}-1\right)-0.8056\,, 
\end{equation}

is due to the external Schwarzschild metric outside the WH's mouth $r=a$ and $I(\bar{x})$ is the contribution from the internal metric, which is given by
\begin{equation}
    I(\bar{x}) = 2\bigintss_{\bar{x}}^{\infty} \frac{\sqrt{B(x)}}{\sqrt{C(x)}\sqrt{\frac{C(x)A(\bar{x})}{C(\bar{x})A(x)}-1}} \,\,dx.
\end{equation}
%Now to investigate the convergence or divergence
We can simplify the above relation for the line element \eqref{5a} with the shape function \eqref{34}
\begin{equation}
    I(\bar{x}) = \int_{\bar{x}}^{a} R(x) dx\,,
\end{equation}
where 
\begin{multline}
\hspace{-0.5cm}R(x)=\frac{2}{\sqrt{x^2 \left[1-\frac{1}{x}\Biggl\{\frac{\mathcal{N}_1 \left(\left(\Theta +(2\mathcal{M})^2 x^2\right) \tan ^{-1}\left(\frac{2\mathcal{M}}{\sqrt{\Theta }}x\right)-2\mathcal{M}\sqrt{\Theta } x\right)}{ \left(\Theta +(2\mathcal{M})^2 x^2\right)}-\frac{\mathcal{N}_1 \left(\left(\Theta +(2\mathcal{M})^2 x_0^2\right) \tan ^{-1}\left(\frac{2\mathcal{M}}{\sqrt{\Theta }} x_0\right)-2\mathcal{M} \sqrt{\Theta } x_0\right)}{ \left(\Theta +(2\mathcal{M})^2 x_0^2\right)}+x_0\Biggl\}\right]}}\\
\frac{1}{\sqrt{\left(\frac{x^{2-m}}{{\bar{x}}^{2-m}}-1\right)}}\,,
\end{multline}
and $\mathcal{N}_1=\frac{3 (4 \pi -\beta ) (\beta +8 \pi ) M}{4\mathcal{M} \pi ^2 \alpha  (12 \pi -\beta )
}$. To see where this integral diverges, we make the change of variable $y = \frac{x}{\bar{x}}$, so that
$x = \bar{x}y$ and $ x_0= \bar{x}y_0$
\begin{equation}
     I(\bar{x}) = \int_{1}^{\frac{a}{\bar{x}}} \frac{2}{\sqrt{H(y)}} dy\,,
\end{equation}
where
%\Biggl\{  \Biggl\} This bracket is not counted in the bracket while breaking the brackets in multline code.
\begin{multline}
H(y)=\left[1-\frac{1}{\bar{x}y}\Biggl\{\frac{\mathcal{N}_1 \left(\left(\Theta +(2\mathcal{M})^2 (\bar{x}y)^2\right) \tan ^{-1}\left(\frac{2\mathcal{M}}{\sqrt{\Theta }}\bar{x}y\right)-2\mathcal{M}\sqrt{\Theta } \bar{x}y\right)}{ \left(\Theta +(2\mathcal{M})^2 (\bar{x}y)^2\right)}-
\right. \\ \left.
\frac{\mathcal{N}_1 \left(\left(\Theta +(2\mathcal{M})^2 (\bar{x}y_0)^2\right) \tan ^{-1}\left(\frac{2\mathcal{M}}{\sqrt{\Theta }} \bar{x}y_0 \right)-2\mathcal{M} \sqrt{\Theta } \bar{x}y_0 \right)}{ \left(\Theta +(2\mathcal{M})^2 (\bar{x}y_0)^2\right)}+\bar{x}y_0\Biggl\}\right]\left(y^{4-m}-y^2\right)\,.
\end{multline}

Letting $H(y)=g(y)\,f(y)$, where
\begin{multline}
g(y)= 1-\frac{1}{\bar{x}y}\Biggl\{\frac{\mathcal{N}_1 \left(\left(\Theta +(2\mathcal{M})^2 (\bar{x}y)^2\right) \tan ^{-1}\left(\frac{2\mathcal{M}}{\sqrt{\Theta }}\bar{x}y\right)-2\mathcal{M}\sqrt{\Theta } \bar{x}y\right)}{ \left(\Theta +(2\mathcal{M})^2 (\bar{x}y)^2\right)}-\\
\frac{\mathcal{N}_1 \left(\left(\Theta +(2\mathcal{M})^2 (\bar{x}y_0)^2\right) \tan ^{-1}\left(\frac{2\mathcal{M}}{\sqrt{\Theta }} \bar{x}y_0 \right)-2\mathcal{M} \sqrt{\Theta } \bar{x}y_0 \right)}{ \left(\Theta +(2\mathcal{M})^2 (\bar{x}y_0)^2\right)}+\bar{x}y_0\Biggl\}
\end{multline}
and
\begin{equation}
    f(y)= \left(y^{4-m}-y^2\right)\,.
\end{equation}
The function $H(y)$ can be expanded around $y = 1$ using Taylor's series as
\begin{multline}
    H(y) = (2-m)g(1)(y-1)+\left[\frac{1}{2}(5-m)(2-m)g(1)+(2-m)g'(1)\right](y-1)^2\\
    + \text{higher powers}.
\end{multline}
Here, the Taylor expansion is truncated to the second order. Note that the leading term in the aforementioned equation indicates whether the integral $I(\bar{x})$ converges or diverges. According to Ref. \cite{Kuhfittig2}, the integral converges due to the leading term $(y-1)^{\frac{1}{2}}$ where $g(1)\neq 0$ whereas if $g(1)= 0$, the integral diverges because the second term leads to $\ln (y-1)$. If we choose the WH's throat to be the light ray's closest approach, which is $\bar{r} = r_0$, then naturally $y_0 = \frac{x_0}{\bar{x}} $ and consequently $y_0 = 1$. Thus, $g(y)$ is reduces to
\begin{multline}
g(y)= 1-\frac{1}{\bar{x}y}\Biggl\{\frac{\mathcal{N}_1 \left(\left(\Theta +(2\mathcal{M})^2 (\bar{x}y)^2\right) \tan ^{-1}\left(\frac{2\mathcal{M}}{\sqrt{\Theta }}\bar{x}y\right)-2\mathcal{M}\sqrt{\Theta } \bar{x}y\right)}{ \left(\Theta +(2\mathcal{M})^2 (\bar{x}y)^2\right)}-\\
\frac{\mathcal{N}_1 \left(\left(\Theta +(2\mathcal{M})^2 (\bar{x})^2\right) \tan ^{-1}\left(\frac{2\mathcal{M}}{\sqrt{\Theta }} \bar{x} \right)-2\mathcal{M} \sqrt{\Theta } \bar{x} \right)}{ \left(\Theta +(2\mathcal{M})^2 (\bar{x})^2\right)}+\bar{x}\Biggl\}.
\end{multline}

It is simple to confirm from the above equation that $g(1) = 0$, and hence the integral diverges. Thus, a photon sphere can be found at the throat, and such a photon sphere can be detected.

\section{Conclusions}
\label{sec6}
 In GR, the existence of WH solutions with some exotic matter has always fascinated researchers. One of the necessary conditions for WH formation is the availability of exotic matter because it results in NEC violation and hence allows the WH to exist. This violation is necessary for the theoretical concept of WHs, as the negative energy associated with exotic matter stabilizes and maintains the WH structure. The Casimir effect is one such example in which vacuum fluctuations in quantum fields in ground stages create the attracting force between two parallel plates \cite{K. Jusufi 2}. The formation of WHs in modified theories has gained interest because of their incorporation of the effective energy-momentum tensor that disrespects the NEC without separately involving any exotic matter. Essentially, they provide alternative frameworks in which conventional energy conditions are relaxed or modified, enabling the exploration of phenomena such as WHs without explicitly relying on exotic matter. In this study, we have constructed WH solutions in the non-metricity-based modified gravity under the background of Gaussian and Lorentzian distributed noncommutative geometries. In addition, in our study, we assume two functional forms of $f(Q, T)$ gravity such as linear $f(Q, T)=\alpha\,Q+\beta\,T$ and nonlinear $f(Q, T)=Q+\lambda_1\,Q^2+\eta_1\,T$ models. The key theoretical observations are described below.\\
 First, we discuss the WH solutions for the linear model with Gaussian and Lorentzian distributions. We compared the modified gravity energy density with the noncommutative one and integrated it to derive the WH shape function. We plotted $b(r)$ versus the radial distance to analyze the physical behavior of these acquired solutions. We noticed that the shape function respects the flare-out condition under an asymptotic background under both distributions. Also, we have graphically shown the effect of the model parameter on shape functions. It was depicted that an increase in the model parameter $\beta$ decreases the shape function for both noncommutative distributions. However, in the throat, this impact is insignificant. Further, we checked both distributions' energy density, NEC and SEC. It is noticed that the energy density shows a positively decreasing trend throughout the space-time for both distributions. SEC is violated in the vicinity of the throat. Radial NEC indicates negative behavior, whereas tangential NEC exhibits positive behavior near the throat. This shows that the NEC was violated in this case, allowing the WH to exist. Therefore, the acquired solutions are valid, supporting the existence of a WH in $f(Q, T)$ gravity under these noncommutative geometries.\\
 Later, we tried to investigate the existence of WH solutions for the model $f(Q, T)=Q+\lambda_1 Q^2+\eta_1 T$ in the framework of Gaussian and Lorentzian distributions. Due to the difficulty of analytic solutions in this case, we were forced to solve the highly complex nonlinear differential equations for $b(r)$ numerically. In this case, we consider the initial condition $b(1\times 10^{-2})=1\times 10^{-5}$ for this study. We have graphically represented the numerical solutions of the shape functions and the energy conditions for both distributions in Figs. (\ref{fig13}-\ref{fig22}). One can observe that the shape function shows a positively increasing behavior; however, as we increase the values of the model parameter $\eta_1$, the shape function decreases. Also, one of the necessary conditions is the flare-out condition, which is compatible with an asymptotic background under both distributions. Further, we located the WH's throat for these distributions at $r_0=0.001$ (approximately). This kind of result can be found in Refs. \cite{Zubair2, Zubair1}. Also, we noticed that the energy density shows an entirely positive behavior. SEC is disrespected near the throat; while it will hold away from the throat. NEC exhibits negative behavior, making the solution incompatible. Hence, all the requirements for the existence of a WH are fulfilled for specific values of the free parameters, ensuring the viability of the obtained WH solutions.\\
Finally, we have examined the effects of gravitational lensing for the WH solutions under a noncommutative background. In order to do this, we adopted the methodology outlined in the Refs. \cite{Tejeiro,Kuhfittig2,Kuhfittig3} and investigated the convergence of the deflection angle. This technique was first introduced by Bozza et al. \cite{Bozza1} to investigate black hole physics in the strong-field limit. In addition, in Ref. \cite{Iftikhar}, the authors studied the behavior of the deflection angle and the location of the WH through its shape function. Following the same procedure, we have studied the deflection angle convergence with the exact shape function solutions. The obtained results show the divergence of the deflection angle of the outward light ray, which occurs exactly at the throat of the WH representing the correspondence of the surface to the photon sphere. Thus, the integral divergence of the deflection angle at the throat indicates the potential presence of a photon sphere. The detection of a photon sphere in close proximity to a WH's throat would have considerable significance due to its implications. This would ensure a strong gravitational field and validate WH predictions. Observationally, it offers the opportunity to directly study these mysterious structures, enhancing our understanding of gravity and WHs.\\
In summary, $f(Q, T)$ gravity allows us to study WH solutions under the effect of noncommutative geometries that violate energy conditions in some regions of space-time. We first consider a linear model that provides analytically exact WH solutions for both distributions. The obtained shape functions obey asymptotic flatness as well as flare-out conditions, which are the essential requirements for a traversable WH. In addition, these kinds of solution violate energy conditions, especially NEC and SEC in the vicinity of the throat. In addition, we numerically studied the WH solutions for the nonlinear model under both distributions. In this case, we found that all the energy conditions are disrespected only in some regions of the space-time. We also noticed the effect of the model parameter and the noncommutative parameter in the shape functions and energy conditions under both Gaussian and Lorentzian distributions. In a recent study, it was noticed that due to noncommutative geometry, the asymptotically flat behavior of the shape function could not be achieved in $f(Q)$ gravity with conformal symmetry \cite{G. Mustafa}. Interestingly, in this study, we found that asymptotic flatness behavior is achieved for both distributions. Since the main issue of WH geometry is the exotic matter, the modified gravity allows us to avoid or minimize the usage of the exotic matter. Thus, in that case, the model parameter and noncommutative geometry may become the source of violation of NEC in the symmetric teleparallel with matter coupling case. Similar behavior can also be found in the teleparallel case \cite{R10}. Note that the current investigation has been done using the constant redshift function.
%%%%%%%%%%%%%%%%%%%%%%%%%%%%%%%%%%%%%%%%%%%%%%%%%%%%%%%%%%%%%%%%%%%%%%%%%%%%%%%%%%%%%%%%%%%%%%%%%%%%%%%%%%%%%%%%%%%%%%%%%%%%%%%%%%%%%%%%%%%%%%%%%%%%%%%%%%%%%%%%%%%%%%%%%%%
%\input{Chapters/Chapter2}
% Chapter 10

\chapter{Concluding remarks and future perspectives} % Main chapter title

\label{Chapter6} % For referencing the chapter elsewhere, use \ref{Chapter1} 

\lhead{Chapter 6. \emph{Concluding Remarks and Future Perspectives}} % This is for the header on each page - perhaps a shortened title

%----------------------------------------------------------------------------------------
%\section{Summary of Results}
Let us summarize the key findings of this thesis. The primary aim of this thesis is to explore the geometry of WH within the framework of symmetric teleparallel theories of gravity. This thesis provides an in-depth review of critical concepts in WH physics, as well as the latest methodologies used to construct these fascinating structures. The results presented in the previous chapters are discussed in four specific works from Chapters \ref{Chapter2} to \ref{Chapter5}.\\
In Chapter \ref{Chapter1}, we begin with an overview of the motivations behind this study, the historical evolution of WH geometry, and energy conditions. This chapter also delved into the profiles of DM. In addition, it explores fundamental concepts, mathematical frameworks, and the core theories of gravity. Although GR is a cornerstone in understanding gravity, it falls short of addressing certain challenges, such as fine-tuning and the flatness problem. Consequently, modifications and generalizations of gravity have been proposed to address these issues more effectively. Specifically, the chapter concluded with an overview of the symmetric teleparallel theory of gravity.\\
In Chapter \ref{Chapter2}, we explore WH solutions in $f(Q)$ gravity under the influence of the monopole charge, employing the PI model and the NFW profile for DM halos. The derived shape functions, expressed via inverse tangent and logarithmic functions, satisfy the flare-out condition and asymptotic flatness, confirming their compatibility with the Morris-Thorne WH framework. The energy conditions indicate that, while the NEC and WEC were violated overall, exotic matter is required at the throat for stability. For the constant redshift function, the SEC was satisfied, but it was violated when using variable forms such as $\phi(r) = \frac{1}{r}$ and $\phi(r)=\log\left(1+\frac{r_0}{r}\right)$. In addition, we face challenges for the nonlinear forms of the $f(Q)$ model. To resolve this issue, we used the Karmakar condition. Analysis based on the VIQ parameter suggested that minimal exotic matter suffices to stabilize the WH. In conclusion, our findings established that the PI model and the NFW profile produce viable WH solutions in $f(Q)$ gravity with monopole charge, with monopole effects minimally impacting violations of the energy condition. This study bridged the gaps in understanding WHs in modified gravity with DM environments.\\
Chapter \ref{Chapter3} examined WH solutions in $f(Q, T)$ gravity under two EoS: barotropy and anisotropy. We analyze solutions for both linear ($f(Q, T) = \alpha Q + \beta T$) and nonlinear ($f(Q, T) = Q + \lambda_1 Q^2 + \eta_1 T$) functional forms. However, we could not find an exact solution for the nonlinear model. Furthermore, for the linear $f(Q, T)$ model under barotropic EoS, exact solutions were obtained, with the shape function following a power law form. The domains of specific parameters ($\omega$, $\beta$) that satisfy asymptotic flatness were identified (Table \ref{ch3table:1}). The energy density was positive and decreased in the phantom region, but was violated in the quintessence region. NEC was violated for the radial pressure, but was satisfied for the tangential pressure, whereas DEC was obeyed and SEC was violated throughout. In addition, for the linear model $f(Q,T)$ in the anisotropic case, exact solutions were derived for the linear model, with power law shape functions satisfying all necessary conditions. The domains of $n$ and $\beta$ that meet asymptotic flatness are shown in table \ref{ch3table:3}. The energy density was positive for certain values $n$, with the NEC and SEC violated. Stability analysis using the TOV equation confirmed that the obtained solutions are stable. Constant redshift functions $\phi(r)$ were used to avoid event horizons, ensuring traversability.\\
In Chapter \ref{Chapter4}, we studied the viability of WH solutions in $f(Q, T)$ gravity with a radial-dependent bag parameter. To achieve this, we derived families of WH solutions based on two widely adopted shape functions. The energy conditions were analyzed by applying constraints derived from the Raychaudhuri equations and the embedding procedure in Ref. \cite{S. Mandal}. By combining these conditions with the EoS for the MIT bag model and the limits on $\omega$, we identified WH families for various values of the free parameter $\beta$. Using this approach, we were able to identify WH solutions that adhere to at least two energy conditions under positive energy density assumptions. Specifically, both shape functions produced solutions characterized by positive energy densities that satisfied the WEC and the SEC. However, it was observed that the NEC and DEC were only partially fulfilled for $r<1$, which points to the existence of exotic matter concentrated at the throat of the WH. This partial violation highlighted the intricate balance required to construct physically plausible WH models within the framework of these energy conditions. This connection was evident from Eq. \eqref{ch4eq17}, where $\rho$, $\omega$ and $B$ are positive and $B > \rho$ within the throat, which leads to a violation of the NEC. The findings suggested that stable and traversable WHs can form with the help of exotic matter. A detailed stability analysis using the TOV equation further supported the robustness of the proposed solutions.\\
In Chapter \ref{Chapter5}, the $f(Q, T)$ gravity framework enabled the study of WH solutions influenced by noncommutative geometries that violate energy conditions in specific space-time regions. Initially, we examined a linear model that provides exact and analytical WH solutions for both Gaussian and Lorentzian distributions. The derived shape functions satisfy the flare-out and asymptotic flatness conditions, which are essential for traversable WHs. However, these solutions exhibited violations of the NEC and SEC near the WH's throat. For the nonlinear model, WH solutions were studied numerically under both distributions. In this case, the energy conditions were violated only in localized space-time regions. Additionally, we analyzed the influence of the model and noncommutative parameters on the shape functions and energy conditions for Gaussian and Lorentzian profiles. Finally, we investigated the convergence of the deflection angle using the exact shape function solutions. The results revealed a divergence of the deflection angle for outward light rays precisely at the WH's throat, corresponding to the presence of a photon sphere. This integral divergence at the throat suggested the existence of a potential photon sphere, offering further insights into the geometry of the WH.\\
%In regard to future perspectives of these works, we find that the exploration of more general forms of action, different variational formalisms, and more general modified theories are some of the routes to be pursued.  Also, there are a lot of scopes in extending the work done in this thesis. For example, in the third chapter, the GUP correction up to the first order on the minimal length scale has been applied, and this approach can also be used to compute the corrections of Casimir energy up to the next leading order and investigate its significance on the WH geometry. Further, in the last chapters, we have assumed the parameter values of DM models. Thus, one can extend these works using astrophysical observational data from the $M87$ galaxy and Milky Way galaxy datasets. Construction of traversable WHs is a field of physics in which, despite the recent progress and development occurring constantly, there are still many questions left unanswered. It is an exciting time for physicists working in this domain, and one can only hope that the coming decades will bring even more wonderful revelations.
Looking ahead, future research into WHs and their connection to symmetric teleparallel gravity theory could prioritize the development of generalized forms of action and the exploration of alternative variational principles. By broadening the theoretical frameworks used to describe space-time, researchers could uncover novel insights into the mechanisms enabling stable and traversable WHs while addressing the limitations of current models. Moreover, expanding the scope of symmetric teleparallel gravity theory could help to resolve outstanding challenges in understanding exotic solutions, such as WHs within GR and beyond. In the first chapter, we have established the parameter values for various DM models based on theoretical assumptions. However, there is significant potential to enhance this work by incorporating astrophysical observational data. Specifically, data sets from the $M87$ galaxy and the Milky Way galaxy (one can refer to \cite{MustafaM87}) can provide crucial insight for refining and validating these models. In the third chapter, the MIT bag model with constant redshift function and linear model has been used. This work can be extended with a variable redshift function and the nonlinear model to check the viability of the WH solution.

% For example, in the third chapter, the GUP correction up to the first order on the minimal length scale has been applied. This approach can also be used to calculate the corrections of Casimir energy up to the next leading order and investigate its significance on the WH geometry. Furthermore, in the last two chapters, we have assumed the parameter values of DM models. Therefore, these works can be extended using astrophysical observational data from the $M87$ galaxy and Milky Way galaxy datasets. The construction of traversable WHs is a field of physics that, despite the recent progress and ongoing development, still has many unanswered questions. It is an exciting time for physicists working in this area, and one can only hope that the coming decades will bring even more wonderful revelations.

%\end{part}
%%%
%\part{Non-minimal Coupling}
%\input{Chapters/Chapter6}
%\input{Chapters/Chapter2}
%\end{part}
%\input{Chapters/Chapter8}
%\input{Chapters/Chapter9}
%-------------------------------------------------------------------------------
%	THESIS CONTENT - APPENDICES
%-------------------------------------------------------------------------------

%\addtocontents{toc}{\vspace{2em}} % Add a gap in the Contents, for aesthetics

%\appendix % Cue to tell LaTeX that the following 'chapters' are Appendices

% Include the appendices of the thesis as separate files from the Appendices
% folder
% Uncomment the lines as you write the Appendices

%\input{Appendices/AppendixA}
%\input{Appendices/AppendixB}
%\input{Appendices/AppendixC}

\addtocontents{toc}{\vspace{2em}} % Add a gap in the Contents, for aesthetics

\backmatter

%-------------------------------------------------------------------------------
%	BIBLIOGRAPHY
%-------------------------------------------------------------------------------

%\label{Bibliography}
\label{References}
%%%\lhead{\emph{Bibliography}} % Change the page header to say "Bibliography"
\lhead{\emph{References}}
%\input{Chapters/Bibliography}
%%%\input{Bibliography.bib}

%\bibliographystyle{amsplain}
%\printbibliography
%%%\bibliography{Bibliography.bib}
%\bibliography{References.bib}
\cleardoublepage
%%\backmatter
\pagestyle{fancy}

\label{Publications}

\lhead{\emph{List of Publications}}
\chapter{List of Publications}
\section*{Thesis publications}
\begin{enumerate}

\item \textbf{Moreshwar Tayde} and P.K. Sahoo, ``Wormhole formations in the galactic halos supported by dark matter models and monopole charge within $f(Q)$ gravity”, \textcolor{blue}{European Physical Journal Plus} \textbf{84},  643 (2024).

\item \textbf{Moreshwar Tayde}, Z. Hassan, P.K. Sahoo and S. Gutti, ``Static spherically symmetric wormholes in $f(Q, T)$ gravity”, \textcolor{blue}{Chinese Physics C} \textbf{46}, 115101 (2022).

\item \textbf{Moreshwar Tayde}, J.R.L. Santos, J.N. Araujo and P.K. Sahoo, ``Wormhole solutions in $f(Q, T)$ gravity with a radial dependent B parameter”, \textcolor{blue}{European Physical Journal Plus} \textbf{138}, 539 (2023).

\item \textbf{Moreshwar Tayde}, Z. Hassan and P.K. Sahoo, ``Existence of wormhole solutions in $f(Q, T)$ gravity under noncommutative geometries”, \textcolor{blue}{Physics of Dark Universe} \textbf{42}, 101288 (2023).

\end{enumerate}

\section*{Other publications}
\begin{enumerate}

\item \textbf{Moreshwar Tayde}, S. Ghosh and P.K. Sahoo, ``Non-exotic static spherically symmetric thin- shell wormhole solution in $f(Q, T)$ gravity”, \textcolor{blue}{Chinese Physics C} \textbf{47}, 075102 (2023).

\item \textbf{Moreshwar Tayde}, Z. Hassan and P.K. Sahoo, ``Conformally symmetric wormhole solutions supported by noncommutative geometries in the context of $f(Q, T)$ gravity”, \textcolor{blue}{Chinese Journal of Physics} \textbf{89}, 195-209  (2024).

\item \textbf{Moreshwar Tayde}, Z. Hassan and P.K. Sahoo, ``Impact of dark matter galactic halo models on wormhole geometry within $f(Q, T)$ gravity”, \textcolor{blue}{Nuclear Physics B} \textbf{1000}, 116478  (2024).

\item L. V. Jaybhaye, \textbf{Moreshwar Tayde} and P.K. Sahoo, ``Wormhole solutions under the effect of dark matter in $f(R, \mathcal{L}_m)$ gravity”, \textcolor{blue}{Communication in Theoretical Physics} \textbf{76}, 055402 (2024).

\item \textbf{Moreshwar Tayde} and P.K. Sahoo, ``Exploring wormhole solutions with monopole charge in the context of $f(Q)$ gravity”, \textcolor{blue}{European Physical Journal C} \textbf{84}, 643 (2024).

\item N. Loewer, \textbf{Moreshwar Tayde} and P.K. Sahoo, ``Formation of stable wormhole solution with noncommutative geometry in the framework of $f(R, \mathcal{L}_m, T)$ gravity”, 	\textcolor{blue}{European Physical Journal C} \textbf{84}, 1196 (2024).

\item D. Mohanty, \textbf{Moreshwar Tayde} and P.K. Sahoo, ``Noncommutative gravastar configuration in  $f(R, \mathcal{L}_m, T)$ gravity”, \textcolor{blue}{Nuclear Physics B} \textbf{1016}, 116914 (2025).

\item \textbf{Moreshwar Tayde}, D. Mohanty and P.K. Sahoo, “A study of wormhole solution using MIT bag parameter $B$ in the framework of $f(R, \mathcal{L}_m, T)$  gravity”, \textcolor{blue}{Physics of Dark Universe} \textbf{48}, 101937 (2025).

\end{enumerate}

\cleardoublepage
%\addtocontents{toc}{\vspace{2em}} % Add a gap in the Contents, for aesthetics
%\backmatter
\pagestyle{fancy}
\lhead{\emph{Conferences/Workshops/Schools}}

\chapter{Conferences/Workshops/Schools}
\section*{Papers presented}
 \begin{enumerate}
\item Presented research paper entitled “\textit{Static spherically symmetric wormholes in $f(Q, T)$ gravity}” at the “\textbf{International Conference on Mathematical Sciences and Its Applications}” organized by the \textbf{School of Mathematical Sciences, Swami Ramanand Teerth Marathwada University, Nanded, Maharashtra, India} during \textcolor{blue}{28th -30th July, 2022}.

\item Presented poster with a flash talk entitled “\textit{Wormhole solutions in $f(Q, T)$ gravity with a radial dependent $B$ parameter}” in the “\textbf{32nd meeting of Indian Association for General Relativity and Gravitation (IAGRG32)}” organized by the \textbf{Indian Institute of Science Education and Research, Kolkata, India} during \textcolor{blue}{19th -21st December, 2022}.

\item Presented research paper entitled “\textit{Exploring wormhole solutions with monopole charge in the context of $f(Q)$ gravity}” at the “\textbf{BCVSPIN Conference 2024: Particle Physics and Cosmology in the Himalayas}” organized by the \textbf{Tribhuvan and Kathmandu University, Kathmandu, Nepal} during \textcolor{blue}{9th -13th December, 2024}.
 \end{enumerate}
 \section*{Attended}
 \begin{enumerate}
 \item Attended National Workshop on  “\textit{Python}” organized by \textbf{Department of Mathematics, Indira Gandhi University Meerpur, Rewari, Haryana, India} during \textcolor{blue}{18th -22nd  October, 2021}.

\item Attended International Workshop on “\textit{Mathematical Foundations and Applications of Gravity (MFAG-2022)}” organized by \textbf{Department of Basic Sciences and Social Sciences, North-Eastern Hill University, Shillong, Meghalaya, India} during \textcolor{blue}{25th - 26th August, 2022}.

\item Attended Workshop on “\textit{General Relativity and Cosmology}” organized by \textbf{Centre for Cosmology, Astrophysics and Space Science GLA University, Mathura (U.P.), India} during \textcolor{blue}{24th -26th November, 2022}.

\item Attended International Workshop on “\textit{Astronomical Data Analysis with Python}” organized by \textbf{Maulana Azad National Urdu University, Hyderabad, India} during \textcolor{blue}{5th - 8th  September, 2023}.

\item Participated International Conference on “\textit{89th Annual Conference of the Indian Mathematical Society}” organized by \textbf{Department of Mathematics, Birla Institute of Technology and Science, Pilani, Hyderabad Campus, Telangana, India} during \textcolor{blue}{22nd - 25th December, 2023}.

\item Attended National Workshop on “\textit{Contemporary Issues in Astronomy and Astrophysics-2024 (CIAA-2024)}” organized by \textbf{Department of Physics, Shivaji University, Kolhapur, Maharashtra, India} during \textcolor{blue}{13th - 15th September, 2024}.

\item Attended Second School on “\textit{Black Holes and Gravitational Waves}” organized by \textbf{Center for Strings, Gravitation and Cosmology, IIT Madras, Tamilnadu, India} during \textcolor{blue}{10th - 14th February, 2025}.

\item Attended School on “\textit{Beyond the Horizon: Testing the Black Hole Paradigm}” organized by \textbf{International Centre for Theoretical Sciences (ICTS), Karnataka, India} during \textcolor{blue}{24th March - 4th April, 2025}.
 \end{enumerate}

\cleardoublepage
%\addtocontents{toc}{\vspace{2em}} % Add a gap in the Contents, for aesthetics
%\backmatter
\pagestyle{fancy}
\lhead{\emph{Biography}}

\chapter{Biography}

\section*{Brief biography of the candidate:}
\textbf{Mr. Moreshwar Jagadeorao Tayde} earned his Bachelor's degree in Mathematics from the Government Vidarbha Institute of Science and Humanities, Amravati, under Sant Gadge Baba Amravati University in 2017, followed by a Master's degree in Mathematics from the same University in 2019. He has demonstrated an exemplary academic record, further enhanced by notable research achievements and exceptional performance as a research scholar. In 2021, he qualified for the National Eligibility Test (NET), the Maharashtra State Eligibility Test (MH-SET) and the Graduate Aptitude Test in Engineering (GATE). He was also awarded the prestigious National Fellowship for Scheduled Caste Students (NFSC) by the University Grants Commission (UGC), Ministry of Education, Government of India, to support his doctoral studies from 2021 to 2025. During his four-year research career, he has published 12 research articles in highly reputed international journals like European Physical Journal C, European Physical Journal Plus, Nuclear Physics B, Physics of the Dark Universe, etc. Also, he presented his work at several national and international conferences, showcasing his dedication and contributions to the field of mathematics. Moreover, he has been awarded an International Travel Grant by BITS Pilani, Hyderabad Campus, for presenting his work at the BCVSPIN Conference 2024 in Kathmandu, Nepal.

%Institute of Mathematics and Applications (IMA) scholarship in 2016, University Rank Holder (URH-UGC) fellowship from 2016 to 2018 for his Master's degree, DST Inspire Fellowship, Govt. of India from 2019-2024 for Ph.D., Best Paper Award (in International conference) in 2022. In his three and half years of research career, he has published 21 research articles in various renowned international journals. He has presented research papers at several national and international conferences (such as Brazil, South Africa, and Taiwan).

\section*{Brief biography of the supervisor:}
\textbf{ Prof. Pradyumn Kumar Sahoo} is a distinguished academic with over 24 years of experience in Applied Mathematics, Cosmology, Astrophysical Objects, the General Theory of Relativity and Modified Theories of Gravity. He earned his Ph.D. from Sambalpur University, Odisha, in 2004, and joined the Department of Mathematics at BITS Pilani, Hyderabad Campus, in 2009 as an Assistant Professor. He currently serves as a Professor in the Department and is an Associate Member of the Inter-University Center for Astronomy and Astrophysics (IUCAA), Pune. In recognition of his academic contributions, he received the “Prof. S. Venkateswaran Faculty Excellence Award” from BITS Pilani in 2022. In 2023, he was recognized as a Distinguished Referee for Europhysics Letters (EPL) by the European Physical Society (EPS) and named an Outstanding Reviewer for the Canadian Journal of Physics, highlighting his critical role in maintaining the accuracy and quality of scientific research. He was also awarded a visiting professor fellowship at Transilvania University of Brașov, Romania. A survey by researchers from Stanford University has ranked him among the top 2\% of scientists globally in the field of Nuclear and Particle Physics over the past five years. Over his prolific career, Prof. Sahoo has published more than 250 research articles in reputed national and international journals and has participated in numerous conferences worldwide as an invited speaker. He has also visited the European Organization for Nuclear Research (CERN) in Geneva, Switzerland, as a visiting scientist. Beyond his research, he has led and contributed to multiple sponsored projects, including initiatives funded by the University Grants Commission (UGC), DAAD Research Internships in Science and Engineering (RISE) Worldwide, the Council of Scientific and Industrial Research (CSIR), the National Board for Higher Mathematics (NBHM) and the Science and Engineering Research Board (SERB), under India’s Department of Science and Technology (DST). His active involvement extends to serving as an expert reviewer for government research projects and as an editorial board member for several prominent journals. Through his research collaborations at national and international levels, Prof. Sahoo has made significant contributions to advancing knowledge in his field.

\end{document}